%% file: pub.tex
\documentclass[titlepage, a4paper, 12pt, reqno]{amsbook}
\usepackage{graphicx}
\usepackage{graphics}
\usepackage{latexsym}
\usepackage{amssymb}
\usepackage{array}
\usepackage{portland}
\usepackage{lgrind}
\usepackage{oldgerm}
\numberwithin{equation}{chapter}

\title{Fast Computational Algorithms for the Discrete Wavelet
  Transform and Applications of Localized Orthonormal Bases in Signal
  Classification}
\author{Eirik Fossgaard}
\address{Department of mathematics, Faculty of Science,\newline 
  \indent University of Troms{\o}, 9037 Troms{\o}.}
\email{eirikf@math.uit.no}

\newcommand{\bR}{\ensuremath{\mathbf R}}
\newcommand{\bZ}{\ensuremath{\mathbf Z}}
\newcommand{\bN}{\ensuremath{\mathbf N}}
\newcommand{\IP}[2]{\ensuremath{\left\langle #1, #2 \right\rangle}}

\newtheorem{theorem}{Theorem}[chapter]
\newtheorem{corollary}{Corollary}[chapter]
\newtheorem{definition}{Definition}[chapter]
\newtheorem{proposition}{Proposition}[chapter]
\newtheorem{algorithm}{Algorithm}[chapter]

\begin{document}
\maketitle

\begin{abstract}

In the first part of this paper we construct an algorithm for
implementing the discrete wavelet transform by means 
of matrices in $SO_{2}({\mathbf R})$ for orthonormal compactly supported 
wavelets and matrices in $SL_{m}({\mathbf R}), m \geq 2,$ for compactly 
supported biorthogonal wavelets. We show that in 1 dimension the 
total operation count using this algorithm can be reduced to about 50\% of the conventional
convolution and downsampling by 2-operation for both orthonormal and 
biorthogonal filters. In the special case of biorthogonal symmetric odd-odd filters, 
we show an implementation yielding a total operation count 
of about 38\% of the conventional method. In 2 dimensions
we show an implementation of this algorithm yielding a reduction in
the total operation count of about 70\% when the filters are 
orthonormal, a reduction of about 62\% for general biorthogonal
filters, and a reduction of about 70\% if the filters are symmetric
odd-odd length filters. We further extend these results to 3 dimensions.

In the second part of the paper we show how the $SO_{2}({\mathbf R})$-method
for implementing the discrete wavelet transform may be exploited to
compute short FIR filters, and we construct edge mappings
where we try to improve upon the preservation of regularity due to
conventional methods.

In the third part of the paper we consider the problem of 
discriminating two classes of radar signals generated from some number
of point sources distributed randomly in a bounded plane domain. 
A statistical space-frequency analysis is performed on a 
set of training signals using the LDB-algorithm of N.Saito and R.Coifman. In 
this analysis we consider several dictionaries of orthonormal bases. The 
resulting most discriminating basis functions are used to construct 
classifiers. The success of different dictionaries is measured by computing 
the misclassification rates of the classifiers on a set of test signals.

\end{abstract}
\newpage 

\include{side2mal}

\tableofcontents

\thanks{
  \begin{center}
    {\Large{\bf Acknowledgements}}
  \end{center} \vspace{5mm}
  
  I want to express my special thanks to my advisor, Professor 
  Jan-Olov Str\"{o}mberg at Troms{\o} University Math Department, for
  inviting me to come with him during his research year at 
  Yale University Math Department 1995/1996, and for his helpful and
  patient guidance through my work with this paper. 
  
  I thank Yale University Math Department and especially 
  Professor R.R. Coifman for providing space for me during 
  my stay there. 

  And finally, but not least, I thank my fellow students at
  the Faculty of Mathematical Sciences at Troms{\o} University for 
  many helpful discussions and for being exceptionally 
  social humans.}


\include{mathfacts}
\include{algorithms}

\include{edgemap}

\include{features}
\appendix
\include{alphas}

\include{edgematrices}
\include{plots}
\include{dummyappendix}

\end{document}

%% file: side2mal.tex



\begin{center}
{\Large {\bf Fast Computational Algorithms for the\vspace{1mm}
  
  \noindent Discrete Wavelet Transform and\vspace{1mm}
  
  \noindent Applications of Localized Orthonormal\vspace{3mm}
  
  \noindent Bases in Signal Classification}}\vspace{30mm}

\noindent {\em {\large Eirik Fossgaard}}\vspace{10mm}

\noindent january 1999\vspace{10cm}

\noindent \underline{University {\vrule height20pt width0pt depth10pt}
  of Troms{\o}}\vspace{3mm}

\noindent 1999
\end{center}

\thispagestyle{empty}
\newpage
\begin{tabbing}
Author \hspace{20mm}\=: \hspace{20mm} \= Eirik Fossgaard \\
Publisher \> : \> University of Troms{\o}\\
ISBN \> : \> 82-90487-93-2\\
\end{tabbing}
\thispagestyle{empty}
\newpage


%% file: mathfacts.tex
\chapter{Preliminaries and Concepts}

\section{A library of bases for $L^{2}(\bR)$} 

Let $L^{2}(\bR)$ denote the space of all square integrable
functions of a single real variable. We present some well-known bases for 
this space.

\subsection{Wavelet bases.}

Expressed shortly, these bases consist
exclusively of all dyadic dilations and integer translations of a
single {\em mother wavelet} $\psi$ with zero integral, that is 

\[ f \in L^{2}({\bR}) \Longrightarrow f = \sum_{j,k \in \bZ}
d_{j,k}\psi_{j,k} \mbox{ where } 
\psi_{j,k}(x) = 2^{j/2}\psi(2^{j}x-k), 
\]

\noindent with the series converging in $L^{2}$ sense. 
In this paper we will only be concerned with real and 
compactly supported wavelets. 
In practice, $\psi$ will possess some localization in both time and 
frequency, and the basis will have to be stable, thus asserting the
existence of positive numbers $A$ and $B$ such that   

\[A \|f\|^{2} \leq \sum_{j,k \in \bZ} \left|d_{j,k}\right|^{2} 
\leq B\|f\|^{2},
\]

\noindent and $\psi$ will possess some {\em vanishing moments}, that
is 

\[ \int \psi(x)x^{l} dx = 0, \mbox{ } \mbox{ } 0\leq l \leq M,
\mbox{ } M \geq 1.
\]
\noindent We can further separate this family of bases into 
two subfamilies:

\begin{itemize}
  
\item {\em Orthonormal Wavelet Bases.} The mother wavelet $\psi$ is  
  orthonormal to all its dyadic dilations and integer translates, 
  thus asserting

  \[ \IP{\psi_{j,k}}{\psi_{j^{\prime},k^{\prime}}} = 
  \delta_{j, j^{\prime}}\delta_{k,k^{\prime}} \mbox{ , } 
  d_{j,k} = \IP{f}{\psi_{j,k}}. 
  \]
  
\item {\em Biorthogonal Wavelet Bases.} The mother wavelet is
  biorthogonal to all dyadic dilations and integer translates of its
  dual ${\widetilde \psi}$, yielding
  
  \[ \IP{\psi_{j,k}}{\widetilde{\psi}_{j^{\prime},k^{\prime}}} = 
  \delta_{j, j^{\prime}}\delta_{k,k^{\prime}} \mbox{ , } 
  d_{j,k} = \IP{f}{\widetilde{\psi}_{j,k}} . 
  \]
  
  \noindent The $\psi_{j,k}$ are called the analysis wavelets, while the 
  ${\widetilde \psi_{j,k}}$ are called the synthesis wavelets.
  
\end{itemize}

\noindent Wavelet bases are intimately connected to the concept of a 
``Multiresolution Analysis'' (MRA). 

\begin{definition} 
  A Multiresolution Analysis is a nested sequence $\{V_{j}\}_{j \in
    \bZ}$ of closed subspaces of $L^{2}(\bR)$ satisfying
  
  \begin{enumerate}
    
  \item $ V_{j} \subset V_{j+1} \mbox{ , } j \in \bZ. $
    
  \item $ \overline{\bigcup_{j \in \bZ}V_{j}} = L^{2}(\bR). $
    
  \item $ \bigcap_{j \in {\bf Z}}V_{j} = \{0\}. $
    
  \item $ f(x) \in V_{j} \Longrightarrow f(x-k) \in V_{j} 
    \mbox{ , } k \in \bZ. $
    
  \item $ f(x) \in V_{j} \Longrightarrow f(2x) \in V_{j+1}. $
    
  \item There exists a ``scaling function'' $\phi \in V_{0}$ 
    such that $\{\phi_{j,k}\}_{k \in \bZ}$ constitute a Riesz 
    basis (= stable basis) for $V_{j} \mbox{ , } j \in \bZ. $
    
  \end{enumerate}
\end{definition}

\noindent The great triumph of $MRA$ is the following result which proof 
can be found in \cite{ten}:

\begin{theorem}
  Given a MRA, there exists a wavelet $\psi \in V_{1}\cap V^{c}_{0}$ 
  such that $\overline{\operatornamewithlimits{Span}_{j,k}
    \psi_{j,k}} = L^{2}(\bR)$. 
  The wavelet can be chosen to have compact support.
  \label{existence of a wavelet basis}
\end{theorem}

\noindent {\bf Remark}. The wavelets possessing the maximum number of 
vanishing moments compatible with their support width, are called 
Daubechies' wavelets. It is shown in \cite{ten} that a wavelet 
with $N$ vanishing moments has support width $2N-1$. It is also
possible to assign vanishing moments to the scaling function $\phi$,
(except a zeroth vanishing moment). Wavelets where both the 
wavelet and the corresponding scaling function have vanishing moments
are called coiflets.

From the first MRA property we get the ``scaling iteration
equation'' 

\begin{equation}        
  \phi_{j,k} = 2^{1/2}\sum_{l \in {\bf Z}} H(l-2k) \phi_{j+1,l} 
  \mbox{ , } H(l) = \IP{\phi}{\phi_{1,l}}.
  \label{scaling equation}
\end{equation} 

\noindent If the $\{\phi_{0,k}\}_{k \in \bZ}$ form an orthonormal basis 
for $V_{0}$ (if not we can carry out an ``orthonormalization trick'',
see \cite{ten} for details) we get an orthonormal wavelet basis 
from the $MRA$ by defining the {\em wavelet} space $W_{j}$ at scale $j$ 
as the orthogonal complement of $V_{j}$ in $V_{j+1}$, that is 

\begin{equation}
  V_{j+1} = W_{j}\oplus V_{j}.
  \label{the wavelet space} 
\end{equation}

\noindent Now, the properties of the scaling spaces $V_{j}$ together with
the definition (\ref{the wavelet space}) yields the {\em orthogonal}
decomposition 

\begin{equation}
  L^{2}(\bR) = \bigoplus_{j \in \bZ}W_{j}, 
  \label{L2 orthogonal decomposition} 
\end{equation}

\noindent and the ``wavelet scaling equation'' 

\begin{equation}        
  \psi_{j,k} = 2^{1/2}\sum_{l \in \bZ} G(l-2k) \phi_{j+1,l} \mbox{ , }
  G(l) = \IP{\psi}{\phi_{1,l}}.
  \label{wavelet scaling equation}
\end{equation} 

\noindent The vanishing integral of $\psi$ implies a non-vanishing
integral for $\phi$. To see this, assume $\int \phi = 0$ and consider 
the characteristic function $\chi_{[-K,K]} \in L^{2}$. 
By choosing K large enough, we get a contradiction to the fact that 
$\{\phi_{j,k}\}_{j,k \in \bZ}$ constitute a Riesz basis for $L^{2}$.   

Since $\sum_{n}H(n) = \int \phi \neq 0$ and $\sum_{n}G(n) = \int
\psi = 0$, the $H(n)$ make up a {\em lowpass filter}, convolving it 
with ${\bf x}\in \bR^{n}$ produces weighted local averages, 
while the $g_{n}$ make up a {\em highpass filter}, convolving it with
${\bf x}\in \bR^{n}$ produces details by catching up local oscillations. 
Filters with finite support width (i.e finite number of nonzero coefficients)
are called ``finite impulse response filters'' or simply FIR filters.

By multiplying equation (\ref{scaling equation}) on both
sides by $\phi_{j,k}$ and integrating the result, using orthonormality
of $\phi$ to its integer translates, we get orthonormality of the 
lowpass filter to its even translates

\begin{equation}
\delta_{0,k} = \sum_{n \in \bZ}H(n)H(n+2k).
\label{double shift orthogonality}
\end{equation}  

\noindent In \cite{ten} it is shown that the highpass coefficients 
$G(n)$ are uniquely determined by the lowpass coefficients $H(n)$ 
up to

\begin{equation}
G(n) = (-1)^{n}H(2N+1-n) \mbox{ , modulo phase and } N \in \bZ .
\label{orthonormal highpass from lowpass}
\end{equation}

\noindent The finite support of the wavelet $\psi$ and the filters
$\{H(k)\}_{k \in \bZ}$, $\{G(k)\}_{k \in \bZ}$
are thus seen to be simple consequences of the finite support of the
scaling function. We can also see that by relation 
(\ref{orthonormal highpass from lowpass}) the lowpass and highpass
filters have the same length, moreover, it has to be {\em even}
by relation (\ref{double shift orthogonality}).

We remark that by Fourier transforming equation 
(\ref{scaling equation}) and iterating the result, 
we get a (infinite) product formula for 
$\widehat{\phi}$, given by

\begin{eqnarray*}
  \hat{\phi}(\xi)  &=& \frac{1}{\sqrt{2\pi}}\prod_{j=1}^{\infty}m_{0}(2^{-j}\xi)\\
  m_{0}(\xi) &=& \frac{1}{\sqrt{2}}\sum_{n=0}^{L-1}H(n)e^{-i\cdot n\xi},\\
\end{eqnarray*}

\noindent thus $\phi$ (and consequently $\psi$) is uniquely determined
by the filter coefficients $H(n)$. 

Now, given the projection $P_{J}f$ of a function $f$ on some scaling space 
$V_{J}$, there is an easy way of  computing the projection
of $f$ on all ``coarser'' spaces $\{V_{j}, W_{j}\}_{j<J}$ without
integrating. From (\ref{scaling equation}) and  
(\ref{wavelet scaling equation})
we obtain a set of recursion relations:

\begin{eqnarray}
  &&\IP{f}{\phi_{j-1,k}} = 2^{1/2}\sum_{l}H(l-2k)
  \IP{f}{\phi_{j,l}},
  \label{orthogonal lowpass channel} \\ 
  &&\IP{f}{\psi_{j-1,k}} = 2^{1/2}\sum_{l}G(l-2k)
  \IP{f}{\phi_{j,l}}.
  \label{orthogonal highpass channel} 
\end{eqnarray}

\noindent We observe that at each scale $j$ the projection $P_{j}f$ is
decomposed into two parts: 

\begin{itemize}
  
\item {\em Local averages.} The lowpass coefficients $c^{j}_{k} \equiv 
  \IP{f}{\phi_{j,k}}$ are the result of passing
  $\{c^{j-1}_{l}\}_{l \in \bZ}$ through the ``lowpass channel'' 
  determined by (\ref{orthogonal lowpass channel}). This operation can be
  described as ``convolution and downsampling by 2'': First convolve with
  $\{H(-k)\}_{k \in \bZ}$, then throw away the odd numbered
  indicies. This operation is often called ``lowpass filtering''. 
  With the convention $c = \{c_{n}\}_{n \in \bZ}$, we can write a 
  (infinite) matrix equation
  
\begin{equation}
  c^{j-1} = {\bf H}c^{j}, \mbox{  }
  {\bf H}(i,j) = H(i-2j), \mbox{  } 0 \leq i,j < \infty.
  \label{lowpass matrix form}
\end{equation}

\item {\em Local oscillations.}  Similarly, the highpass coefficients 
  $d^{j}_{k}$ result from passing $\{c^{j-1}_{l}\}_{l \in \bZ}$
  through the highpass channel determined by 
  (\ref{orthogonal highpass channel}) and is called ``highpass
  filtering''. We write
  
  \begin{equation}
    d^{j-1} = {\bf G}c^{j} \mbox{ , }
    {\bf G}(i,j) = G(i-2j) \mbox{, }0 \leq i,j < \infty.
    \label{highpass matrix form}
  \end{equation}
  
\end{itemize}

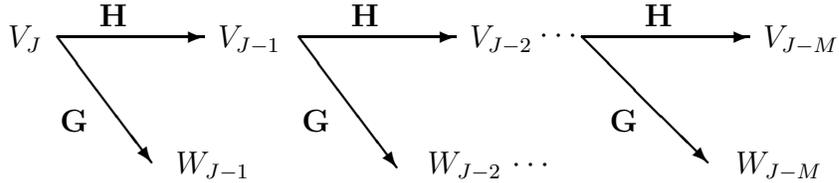
\begin{figure}[h]
  \begin{center}
    \input{thesisfig1.latex}
  \end{center}
  \caption{The ``logarithmic tree'' of a wavelet decomposition in M levels.}
  \label{logarithmic tree of wavelets}
\end{figure}

The decomposition of $P_{J}f$ into averages and details at different
scales is illustrated in Figure \ref{logarithmic tree of wavelets}. 
This decomposition is {\em the discrete wavelet transform} in
dimension 1.

\noindent The inverse operation is in this orthogonal case given by the
transpose matrices:

\begin{eqnarray}
  c^{j}_{k} & = & {\bf H}^{t}c^{j-1}(k) + 
  {\bf G}^{t}d^{j-1}(k) \nonumber \\
  & = & 2^{1/2}\sum_{l \in \bZ}H(k-2l)c^{j-1}_{l} +
  2^{1/2}\sum_{l \in \bZ}G(k-2l)d^{j-1}_{l}.
  \label{orthogonal reconstruction}
\end{eqnarray}

\noindent This operation is illustrated by simply reversing the arrows in 
Figure \ref{logarithmic tree of wavelets} and replacing ${\bf H},{\bf
  G}$ by their transposes.

We emphasize that the expansion of a 
function into a wavelet basis is fast: From the equations 
(\ref{orthogonal lowpass channel}) and  
(\ref{orthogonal highpass channel}) we get that a
complete wavelet analysis of a sequence of length $N$ has a total 
operation cost of no more than $4L\cdot N$ multiply-adds, 
where $L$ is the support width of the filters. Equation
(\ref{orthogonal reconstruction}) yields the same bound on the 
operation cost for reconstruction.

In summary the FIR filters $\{H(n)\}_{n \in \bZ}$,
$\{G(n)\}_{n \in \bZ}$ make up a two-channel orthogonal filter bank
with perfect reconstruction.  

If $\phi$ is not orthogonal to its translates, the sums in (\ref{the
  wavelet space}) and (\ref{L2 orthogonal decomposition}) are {\em
  direct} rather than orthogonal and the wavelet basis is no longer its
own dual. The construction of a {\em biorthogonal} Riesz wavelet basis is 
shown in \cite{ten}. Thus we get two $MRA$'s: The analysis $MRA$ that
is built up from  the analysis scaling functions $\phi_{j,k}$ and the
corresponding wavelets $\psi_{j,k}$, and the synthesis $MRA$ built
from the synthesis scaling functions $\widetilde{\phi}_{j,k}$ and
their wavelets $\widetilde{\psi}_{j,k}$.  

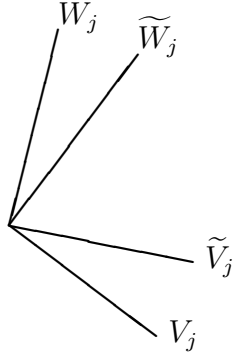
\begin{figure}[h]
  \begin{center}
    \input{thesisfig2.latex}
  \end{center}
  \caption{The scaling- and wavelet spaces $V_{j}$, $W_{j}$
      and their duals $\widetilde{V}_{j}$, $\widetilde{W}_{j}$.}
  \label{dual spaces}
\end{figure}

The relations between the basis functions and filters from the two 
$MRA$'s are given below. A simple picture of the biorthogonality
relations is shown in Figure \ref{dual spaces}.

\begin{eqnarray}
  && \mbox{Filter relations: }  \nonumber \\
  && \nonumber \\
  && \delta_{0,k} = \sum_{n}H(n)\widetilde{H}(n+2k) \mbox{ , } 
  \label{double shift biorthogonality} \\
  && G(n) = (-1)^{n+1}\widetilde{H}(2N+1-n), \nonumber \\
  && \widetilde{G}(n) = (-1)^{n+1}H(2N+1-n), N \in \bZ.  
  \label{biorthogonal highpass from lowpass} \\
  && \nonumber \\
  && \mbox{Biorthogonality relations: }  \nonumber \\
  && \nonumber \\
  && \IP{\psi_{j,k}}{\widetilde{\psi}_{j^{\prime},k^{\prime}}} = 
  \delta_{j,j^{\prime}}\delta_{k,k^{\prime}} \Longleftrightarrow
  \IP{\phi_{0,k}}{\widetilde{\phi}_{0,k^{\prime}}} = 
  \delta_{k,k^{\prime}}. 
  \label{biorthogonal MRA relations}
\end{eqnarray}

\noindent The scaling equations for the biorthogonal scaling- 
and wavelet functions and the expansion and reconstruction equations 
are immediate generalizations of the corresponding equations in the 
orthogonal case. We note that the lengths of the filters 
$H$ and $\tilde{H}$ need not be equal. 

When working in $L^{2}(\bR^{n})$, that is with square integrable 
functions of $n$ variables, it is possible to generate a wavelet basis 
for this space by taking tensor products of  $n$ (possibly different)
MRA's in $L^{2}(\bR)$. This results in $2^{n}-1$ different mother 
functions, the dilations and translates of which make up what we 
call a {\em tensor} wavelet basis in dimension $n$. The iteration is 
done on the {\em pure} lowpass band, that is the
coefficients originating from the {\em pure} tensor scaling function
$\phi(x_{1})\otimes \phi(x_{2})\otimes \cdots \otimes \phi(x_{n})$.

\subsection{Wavelet packet bases} 

These bases are particular linear combinations of elements in a 
wavelet basis. It follows that 
the elements in a wavelet packet basis inherit some of the properties
of the wavelets they are made of, such as compact support and some
localization in both time and frequency. We will see that the wavelet
basis is included as a special case of wavelet packet bases.

Following \cite{wick}, we briefly describe the construction of these 
bases. Define

\begin{eqnarray}
  \psi_{0} = H\psi_{0};  &&
  \int_{\bR}\psi_{0} = 1, \nonumber \\
  \psi_{2n} = H\psi_{n}; && 
  \psi_{2n}(t) = 2^{1/2}\sum_{j \in \bZ}H(j)\psi_{n}(2t-j),
  \nonumber \\
  \psi_{2n+1} = G\psi_{n}; &&
  \psi_{2n+1}(t) = 2^{1/2}\sum_{j \in \bZ}G(j)\psi_{n}(2t-j), 
  \nonumber \\
  \psi_{s,f,p}(t) = 2^{-s/2}\psi_{f}(2^{-s}t-p); &&
  \Lambda_{f} = \overline{\operatornamewithlimits{Span}_{p}
    \psi_{0,f,p}},  \nonumber \\
  \sigma x(t) = 2^{-1/2}x(t/2); &&
  \sigma \Lambda_{f} = \{\sigma x: x \in \Lambda_{f}\},
  \label{wavelet packet definitions}
\end{eqnarray}

\noindent where $H,G$ is a pair of conjugate FIR filters
and $\psi_{s,f,p}$ is called a wavelet packet of scale 
index $s$, frequency index $f$ and position index $p$. 
We immediately have that 
$\sigma^{s}\Lambda_{f}$ is the closure of 
$\operatornamewithlimits{Span}_{p}\psi_{s,f,p}$.
Exploiting the natural one-to-one correspondence 

\begin{eqnarray*}
  I_{s,f} \longleftrightarrow \sigma^{s}\Lambda_{f}; &&
  I_{s,f} = \left[\frac{s}{2^{f}}, \frac{s+1}{2^{f}}\right),
\end{eqnarray*}
we have the following result for which proof 
we refer to \cite{wick}:

\begin{theorem}
  If $\mathcal{I}$ is a dyadic cover of 
  $\bR^{+}$, then\newline 
  $\{\overline{\bigcup_{s,f}\sigma^{s}\Lambda_{f}}:
  I_{s,f} \in \mathcal{I}\} = L^{2}(\bR)$ and if $\mathcal{I}$ is
  disjoint, the wavelet packets\newline
  $\{\psi_{s,f,p}: I_{s,f} \in \mathcal{I},p \in \bZ\}$ form a basis
  for $L^{2}(\bR)$. Furthermore, if the filters $H$, $G$ are 
  orthogonal, this basis is an orthonormal basis.
  \label{completeness of wavelet packets}
\end{theorem}

\noindent To keep the same filtering formulas for sequences as for functions,
make the definitions:

\begin{eqnarray*}
  && \psi_{s,f,p}^{<} = 2^{-s/2}\psi_{f}(p-2^{-s}t), \\
  && \lambda_{s,f,p} = \IP{x}{\psi_{s,f,p}^{<}}, \mbox{ }
  x \in \sigma^{s}\Lambda_{f}.
\end{eqnarray*}
Then it is easy to verify the following recursion relations:

\begin{eqnarray}
&& \lambda_{s+1,2f,p} = H\lambda_{s,f,p}, \nonumber \\
&& \lambda_{s+1,2f+1,p} = G\lambda_{s,f,p}.
\label{recursion relations for wavelet packet coefficients}
\end{eqnarray}

\noindent Thus, identifying $\sigma^{s}\Lambda_{f}$ with $\Omega_{s,f}$, 
the expansion of a function into a collection of wavelet packet
bases can be illustrated as in Figure \ref{binary tree of wavelet
  packets}. We note that every disjoint cover of the top box in this
figure by smaller boxes from levels below, yields a new basis 
for the space $\Omega_{0,0}$.

\begin{figure}[h]
  \begin{center}
    \input{thesisfig3.latex}
  \end{center}
  \caption{The binary tree of a wavelet packet decomposition
      in 4 levels.}
  \label{binary tree of wavelet packets}
\end{figure}
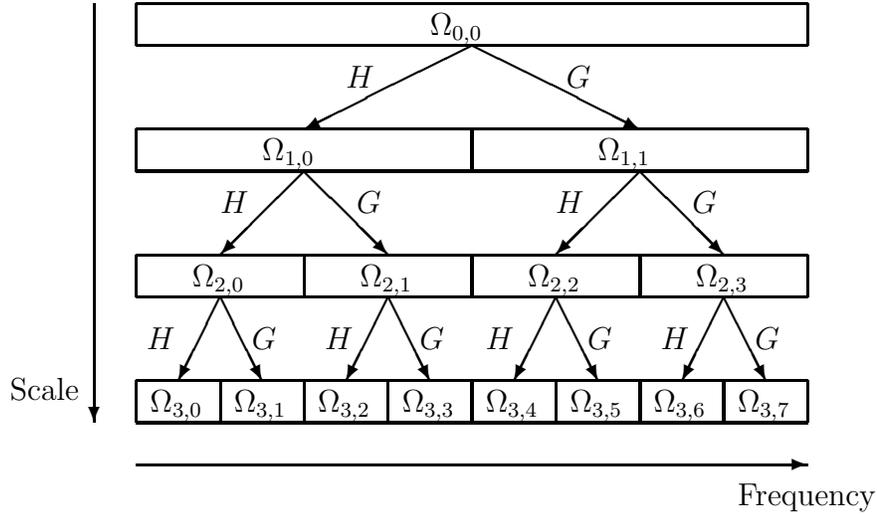

\noindent By comparing Figure \ref{logarithmic tree of wavelets} and 
Figure \ref{binary tree of wavelet packets} it is easy to see that 
the logarithmic tree of expansion into
a wavelet basis is a subtree of the binary tree of expansion into a
collection of wavelet packet bases. We conclude that the wavelet 
basis is included in the collection of wavelet packet bases.

For reconstruction one needs the dual basis of the particular 
wavelet packet basis. It is shown in \cite{wick} that the wavelet
packets defined by the dual filters $H^{\prime}$, $G^{\prime}$
are the duals of the wavelet packets defined by $H$ and $G$. 

Given a sequence of length $N$, a rather crude estimate given
in \cite{wick} shows that a wavelet packet analysis of the sequence
will provide more than $2^{N}$ different bases. The recursion 
relations (\ref{recursion relations for wavelet packet coefficients})
makes the expansion into a wavelet packet analysis 
cheap: The total operation cost will be no more than 
$L\cdot N \log_{2}N$ multiply-adds, where $L$ is the support width 
of the filters.

\subsection{Smooth local trigonometric bases.}

The huge drawback of the Fourier basis for applications in signal 
analysis is its ``non-locality'' in time, thus it is impossible 
to relate specific frequencies of a signal to specific time or 
space windows using this basis. Also, this basis only span the space of
the periodic $L^{2}(\bR)$ functions. What we would like 
is a trigonometric basis for $L^{2}(\bR)$ with the following
desirable properties: 

\begin{itemize}

\item Orthogonality.

\item Smoothness: The basis elements should possess at least a few
  continuous derivatives.

\item Localization in space: Each basis element should have finite support.

\item Efficiency: There should exist a fast algorithm 
  (i.e. $N\log_{2}N$) for expansion into the basis.

\end{itemize}

\noindent Such bases do exist, we refer to \cite{wick} for a 
thorough exposition on 
the subject. We will briefly outline the construction of a Local Sine
Basis and prove that this basis enjoys the properties above.
Define the bell functions $\{b_{k}(x)\}_{k \in \bZ}$ with 
the following properties:

\begin{eqnarray}
  && \sum_{k \in \bZ} b_{k}^{2}(x) = 1, \mbox{ } \forall x \in \bR, 
  \nonumber \\
  && supp(b_{k}) = I_{k} = 
  (\alpha_{k}-\epsilon_{k}, \alpha_{k+1}+\epsilon_{k+1}), \nonumber \\
  && b_{k}b_{k+2} = 0; \mbox{   } b_{k-1}b_{k} \mbox{ is even around } 
  \alpha_{k}.
  \label{properties of bell functions}
\end{eqnarray}   

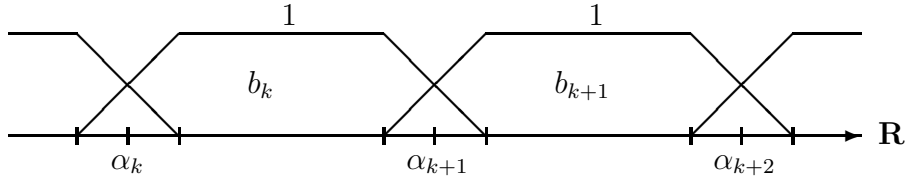
\begin{figure}[h]
  \begin{center}
    \input{thesisfig4.latex}
  \end{center}
  \caption{The bell functions.}
  \label{bells}
\end{figure}

\noindent A schematic picture of these bell functions is shown in Figure
\ref{bells}. 

\noindent Define 

\begin{equation}
  s_{j,k}(x) = \frac{b_{k}(x)}{(\alpha_{k+1}-\alpha_{k})^{1/2}}
  \sin\left(\pi(j+1/2)\frac{x-\alpha_{k}}{\alpha_{k+1}-\alpha_{k}}\right).
  \label{smooth local sines}
\end{equation}

\noindent Then we have the following result:

\begin{theorem}
The $s_{j,k}$ form a smooth orthogonal basis for $L^{2}(\bR)$.
Furthermore, this basis enjoys all the desirable properties
listed above.
\end{theorem}

\noindent {\em Proof:} The smoothness and localization properties of 
this basis are immediately clear. 
To prove orthogonality we only need to consider
\IP{s_{j,k}}{s_{j^{\prime},k^{\prime}}} when 
$\left|k-k^{\prime}\right| \leq 1$ because of the support 
property of the bell functions. Define

\[ \widetilde{s}_{j,k}(x) =  \frac{1}{(\alpha_{k+1}-\alpha_{k})^{1/2}}
\sin\left(\pi(j+1/2)\frac{x-\alpha_{k}}
  {\alpha_{k+1}-\alpha_{k}}\right). 
\]

\noindent For $k-k^{\prime} = 1$ we get  

\[ \IP{s_{j,k}}{s_{j^{\prime},k-1}} = 
  \int_{\alpha_{k}-\epsilon_{k}}^{\alpha_{k}+\epsilon_{k}}
  b_{k-1}(x)b_{k}(x)\widetilde{s}_{j^{\prime},k-1}(x)
  \widetilde{s}_{j,k}(x) dx = 0.
\]

\noindent To see this, observe that since $b_{k-1}b_{k}$ is even around 
$\alpha_{k}$, the integrand is an odd function around the center of 
integration, thus the integral equals zero. For $k= k^{\prime}$ we get 

\begin{eqnarray*}
  \IP{s_{j,k}}{s_{j^{\prime},k}} &=& \int_{\alpha_{k}-\epsilon_{k}}^{
    \alpha_{k+1}+\epsilon_{k+1}}b_{k}^{2}(x)
  \widetilde{s}_{j^{\prime},k}(x)\widetilde{s}_{j,k}(x) dx \\
  &=& \int_{\alpha_{k}}^{\alpha_{k+1}}\widetilde{s}_{j^{\prime},k}(x)
  \widetilde{s}_{j,k}(x) dx = \delta_{j,j^{\prime}}.
\end{eqnarray*}

\noindent To see that the first integral equals the second, observe 
that $\widetilde{s}_{j^{\prime},k}\widetilde{s}_{j,k}$ is even around 
$\alpha_{k}$ and $b_{k}^{2}(x) = 
1-b_{k}^{2}(\alpha_{k}+\epsilon_{k}-x), \mbox{ } x \in  
[\alpha_{k}-\epsilon_{k}, \alpha_{k}+\epsilon_{k}]$.  
The second integral is an elementary calculation. This shows that the 
$s_{j,k}$ are orthonormal over all frequencies $j$ and translates $k$. 

Now, let $f \in L^{2}(\bR)$ and consider $\IP{f}{s_{j,k}}$. We split 
the integral into three parts and change variables on the leftmost 
and rightmost integrals to obtain a formula for the inner product 
involving integration only on the interval $[\alpha_{k},\alpha_{k+1}]$.  
We may express this operation by saying that the part of 
$f$ that lives in 
$[\alpha_{k}-\epsilon_{k}, \alpha_{k}] \cup [\alpha_{k+1},\alpha_{k}
+\epsilon_{k+1}]$ is folded into the interval $[\alpha_{k},
\alpha_{k+1}]$.

\begin{eqnarray*} 
  \IP{f}{s_{j,k}} &=& \int_{\alpha_{k}-\epsilon_{k}}^ 
  {\alpha_{k+1}+\epsilon_{k+1}}f(x)b_{k}(x)\widetilde{s}_{j,k}(x) dx
  \\ &=& \int_{\alpha_{k}-\epsilon_{k}}^{\alpha_{k}}(\cdot) +
  \int_{\alpha_{k}}^{\alpha_{k+1}}(\cdot) +
  \int_{\alpha_{k+1}}^{\alpha_{k+1}+\epsilon_{k+1}}(\cdot) \\ 
  &=& \int_{\alpha_{k}}^{\alpha_{k+1}}b_{k}(x)f(x)
  \widetilde{s}_{j,k}(x) dx \\
  &-& \int_{\alpha_{k}}^{\alpha_{k+1}}
  b_{k}(2\alpha_{k}-x)f(2\alpha_{k}-x)\widetilde{s}_{j,k}(x)dx \\
  &+&\int_{\alpha_{k}}^{\alpha_{k+1}}
  b_{k}(2\alpha_{k+1}-x)f(2\alpha_{k+1}-x)\widetilde{s}_{j,k}(x) dx \\
  &=& \int_{\alpha_{k}}^{\alpha_{k+1}}f^{\sharp}(x)
  \widetilde{s}_{j,k}(x) dx,  \\
  f^{\sharp}(x) &\equiv& 
  b_{k}(x)f(x) - b_{k}(2\alpha_{k}-x)f(2\alpha_{k}-x) \\
  &+&b_{k}(2\alpha_{k+1}-x)f(2\alpha_{k+1}-x).
\end{eqnarray*} 

\noindent To implement the local sine basis, we place the gridpoints
at half-integers, and evaluate the integral by $DST-IV$, meaning a 
discrete sine transform of type $IV$. This transform enjoys the nice 
property of being its own inverse, and has a $N\log_{2}N$ 
implementation.  We refer to \cite{wick} for a proof of these facts.

It only remains to prove completeness. We have $\IP{f}{s_{j,k}} =
\IP{f^{\sharp}}{\widetilde{s}_{j,k}}$. Furthermore, 
$\operatornamewithlimits{Span}_{j}\widetilde{s}_{j,k} 
= L^{2}([\alpha_{k},\alpha_{k+1}))$ 
because of the completeness of the Fourier basis in 
$L^{2}([0,2\pi))$. This yields

\begin{eqnarray*}
  \left. f^{\sharp}\right|_{[\alpha_{k},\alpha_{k+1})}(x) 
  &=& \sum_{j=0}^{\infty}\IP{f^{\sharp}}{\widetilde{s}_{j,k}}
  \widetilde{s}_{j,k}(x) \\
  && \Downarrow \\
  b_{k}(x)f^{\sharp}(x) &=& \sum_{j=0}^{\infty}
  \IP{f^{\sharp}}{\widetilde{s}_{j,k}}b_{k}(x)\widetilde{s}_{j,k}(x) \\
  &=& \sum_{j=0}^{\infty}\IP{f}{s_{j,k}}s_{j,k}(x) = P_{k}(f)(x), \\
  P_{k}(f)(x) &\equiv& b_{k}^{2}(x)f(x) \\
  &-& b_{k}(x)b_{k}(2\alpha_{k}-x)f(2\alpha_{k}-x) \\
  &+& b_{k}(x)b_{k}(2\alpha_{k+1}-x)f(2\alpha_{k+1}-x).
\end{eqnarray*}

Summing $P_{k}(f)$ over all $k \in \bZ$, it is easy to see 
that all the terms cancel except the $b_{k}^{2}(x)f(x)$ terms. Thus
we get 

\[ \sum_{k\in \bZ}\sum_{j=0}^{\infty}\IP{f}{s_{j,k}}s_{j,k}(x) = 
\sum_{k \in \bZ}P_{k}(f)(x) = 
\left(\sum_{k\in \bZ}b_{k}^{2}(x)\right)f(x) = f(x), \]
this confirms the completeness of the $s_{j,k}$. $\Box$

We note that by replacing sines by cosines and $DST-IV$ by
$DCT-IV$ in the construction above, 
we get a local cosine basis for $L^{2}(\bR)$.

\subsection{Local sine/cosine packet analysis.} 

\begin{figure}[h]
  \begin{center}
    \input{thesisfig5.latex}
  \end{center}
  \caption{{\em Dyadic local sine/cosine analysis in three levels.}}
  \label{dyadic local sine/cosine analysis}
\end{figure}
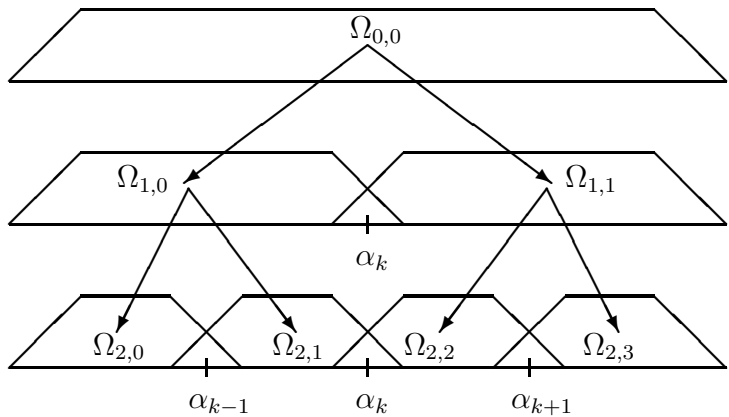

It is immediately clear how the notion of a 
wavelet packet analysis can be transfered to this case of local 
trigonometric bases. We restrict ourselves to bells of a fixed length
at each level of decomposition. Thus, we replace each bell from the level
above by two child bells of half the length of the parent bell. 
A schematic picture of this analysis is shown in Figure 
\ref{dyadic local sine/cosine analysis}. 

\noindent It is easy to see that in analog to the case of wavelet 
packets, every cover of the original interval by our special bell 
functions corresponds to a (unique) basis for this interval.
To get back to the original coordinates, simply perform the 
$DST-IV$ or $DCT-IV$ (they are their own inverses) on each subinterval,
then ``unfold'' the result back into the folding regions
$[\alpha_{k}-\epsilon_{k}, \alpha_{k}+\epsilon_{k}]$.

For a input of length $N$, the number of levels in a Local Sine/Cosine
Analysis cannot exceed $\log_{2}N$, thus the total operation cost 
is bounded by $N(\log_{2}N)^{2}$ multiply-adds.

\section{Signal classification using local feature extraction} 

The notation and ideas presented below are due to \cite{saito}, 
to which we refer for a more thorough review of this subject.
We say that a wavelet packet analysis or a local (co)sine packet analysis 
constitute a {\em dictionary} of bases for $L^{2}$. 
We call a collection of dictionaries 
a {\em library} of bases for $L^{2}$.

\subsection{Local feature extraction.}

We define the problem of signal 
classification as the construction of the map $c$

\[ c: \mathcal{S}\subset \bR^{n} \longmapsto \mathcal{C} = 
\{1,2,...,C\}, \]

\noindent where $\mathcal{S}$ is the set of all signals under 
consideration and $\mathcal{C}$ is the set of all relevant class names.   
We call $\mathcal{S}$ the {\em signal space}, $\mathcal{C}$ the
{\em response space} and $c$ a {\em classifier}. 
Since the dimension $n$ of the signal space is normally very large 
compared to the dimension of the response space, it is important 
to extract only the relevant {\em features} of the signals 
to obtain an efficient and accurate classification. In other words, 
we want to map the information relevant to our problem into a few
coordinates and ignore all the rest. To achieve this, define the map $f$

\[f: \mathcal{S}\longmapsto \mathcal{F}\subset \bR^{k}, 
\mbox{ }k \leq n, \]   

\noindent $\mathcal{F}$ is called the {\em feature space}
and $f$ the {\em feature extractor}. Thus, once we have $f$, we only
have to construct $c: \mathcal{F} \mapsto \mathcal{C}$, 
which in general is much easier because $k << n$.

To construct the feature extractor
we use a training data set $\tau = \{{\bf s}_{i},c_{i}\}_{i=1}^{N} \subset 
\mathcal{S}\times \mathcal{C}$ of $N$ training signals ${\bf s}_{i}$ 
and their class names $c_{i}$. We will denote by $N_{i}$ the number of
training signals belonging to class $i$, so that 
$N = N_{1}+ \cdots + N_{C}$. Once $f$ and $c$ have been constructed, we 
measure the {\em misclassification rate} $r_{error}$ using a test data
set $\tau^{\prime} = \{{\bf s}_{i}^{\prime},
c_{i}^{\prime}\}_{i=1}^{N^{\prime}}$ which has not been used in the 
construction of $f$ and $c$, by 

\[ r_{error} = \frac{1}{N^{\prime}}\sum_{i=1}^{N^{\prime}}
\delta(c_{i}^{\prime}-c({\bf s}_{i}^{\prime})), \mbox{ } \delta(0) = 0, 
\mbox{ } \delta(x) = 1, \mbox{ } x \neq 0,
\]

\noindent We will use feature extractors on the form 
$f = \Theta^{(k)}\circ \Psi$
where $\Psi \in SO_{2}(n)$ with column vectors
${\bf w}_{j,k,l}^{T}$ with the correspondence

\begin{eqnarray}
  &&\operatornamewithlimits{Span}_{l} {\bf w}_{j,k,l}^{T} 
  = \Omega_{j,k}, \nonumber \\
  &&j=0,...,J,\mbox{ }k=0,...,2^{j}-1,\mbox{ }l=0,...,2^{n}-1, 
  \label{orthogonal expansion matrix}
\end{eqnarray}

\noindent where $\Omega_{j,k}$ are the subspaces of a wavelet packet 
or a local sine/cosine analysis.  
$\Theta^{(k)}: \mathcal{S}\longmapsto \mathcal{F}$ is a 
{\em selection rule} picking out the $k$ most relevant coordinates
from $n$ coordinates.

\begin{figure}[h]
  \begin{center}
    \includegraphics[scale = 0.65]{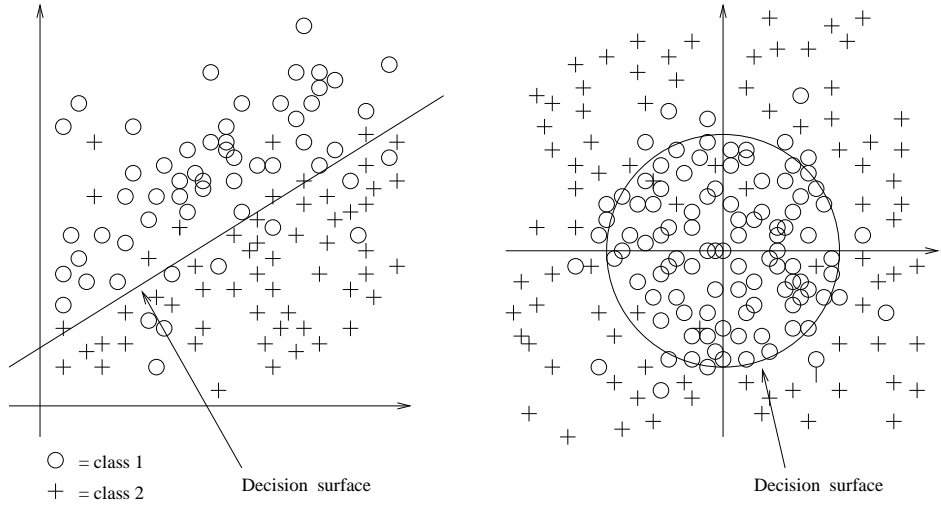}
  \end{center}
  \caption{Example of sample plots of the two most discriminating coordinates in the 
    two-class case.}
  \label{sample plot}
\end{figure}

Figure \ref{sample plot} illustrates two examples of 
$\mathcal{C} = \{1,2\}$ using some $\Psi$ and some $\Theta^{(2)}$. 
The classifiers $c$ are indicated in the figure. 

There are several possible choices for a classifier $c$, such as 
Linear Discriminant Analysis (LDA) and Classification and Regression
Trees (CART), we refer to \cite{saito} for a review of these, as we
will not make use of them in this paper.

\subsection{Entropy and discriminant measures.}

Given a norm $\|\cdot\|_{r}$, we define the entropy $H_{r}$ of a 
sequence ${\bf x}$ as

\[ H_{r}({\bf x}) = -\sum_{i}\frac{\left|x_{i}\right|^{r}}
{\|{\bf x}\|_{r}^{r}}\log_{2}\frac{\left|x_{i}\right|^{r}}
{\|{\bf x}\|_{r}^{r}}, \mbox{ } 1\leq r <\infty.
\]

$H_{r}({\bf x})$ measures the degree of order in the sequence ${\bf
  x}$, that is the information cost in describing ${\bf x}$.

In the two-class case, we need a {\em discriminant} measure 
$d$ that measures how differently two sequences 
are distributed, in other words the relative entropy of
the two sequences. In this paper we will only use 
$d=m_{p}$ defined as  

\begin{equation}
m_{p}({\bf x},{\bf y}) = \|{\bf x}-{\bf y}\|_{p}^{p} = 
\sum_{i}(x_{i}-y_{i})^{p}, 
\label{l^2 measure}
\end{equation}
with $p=2$ in most cases.

In the general case of $C$ classes, we define the discriminant
measure of $C$ sequences as  

\begin{equation}
  d(\{{\bf x}^{(c)}\}_{c=1}^{C}) \equiv \sum_{i=1}^{C-1}
\sum_{j=i+1}^{C}d({\bf x}^{(i)},{\bf x}^{(j)}). 
\end{equation}

\noindent We see that once we have a discriminant measure, we are capable
of evaluating the power of discrimination of the different subspaces
in any dictionary $\mathcal{D}$. This indicates that it is possible 
to choose a basis in $\bR^{n}$ for the set $\mathcal{S}$ of signals 
with the property that no other basis in the dictionary will discriminate more
between classes. Furthermore, the discriminant measure should be
{\em additive} to ensure a fast computational algorithm.

\begin{definition}
A discrimant measure d is said to be additive if
$d({\bf x}, {\bf y}) = \sum_{i}d(x_{i},y_{i}).$
\end{definition}

\subsection{The local discriminant basis algorithm.}

Given a discriminant measure $d$, how do we best evaluate the power of 
discrimination in each subspace of a dictionary $\mathcal{D}$, and 
how do we select the most discriminating basis?
The result should ideally only depend on the characteristic features 
of each class. The ``local discriminant basis algorithm''
(LDB-algorithm) developed in \cite{saito} yields a particular 
basis called a {\em local discriminant basis}, LDB for short, 
as described below.

\begin{definition}
  Let $\{{\bf x}_{i}^{(c)}\}_{i=1}^{N_{c}}$ be a set of training signals
  belonging to class $c$. Then the time-frequency energy map of class $c$,
  $\Gamma_{c}$ , is a table of real values specified by $(j,k,l)$ as 
  \begin{equation}
    \Gamma_{c}(j,k,l) = \sum_{i=1}^{N_{c}}\left({\bf w}_{j,k,l}^{T}\cdot
      {\bf x}_{i}^{(c)}\right)^{2}/\sum_{i=1}^{N_{c}}\|
    {\bf x}_{i}^{(c)}\|^{2},
    \label{time-frequency energy map}
  \end{equation}
  for $j=0,...,J,\mbox{ }k=0,...,2^{j}-1,\mbox{ }l=0,...,2^{n-j}-1.$
\end{definition}

For notational convenience, define

\begin{eqnarray*}
  && d(\{\Gamma_{c}(j,k,\cdot)\}_{c=1}^{C}) =
  \sum_{l=0}^{2^{n-j}-1}d(\Gamma_{1}(j,k,l),...,\Gamma_{C}(j,k,l)), \\
  && B_{j,k} = \operatornamewithlimits{Span}_{0\leq l \leq 2^{n-j}-1}
  {\bf w}_{j,k,l}
  \subset \Omega_{j,k}, \\
  && A_{j,k} = \left. \mbox{LDB}\right|_{B_{j,k}}, \\
  && \Delta_{j,k} = \mbox{ discriminant measure of }\Omega_{j,k}.
\end{eqnarray*}

Then we have the following algorithm:

\begin{algorithm}
  (The Local Discriminant Basis Selection Algorithm). Given a
  training dataset $\tau$ consisting of $C$ classes of signals
  $\{\{{\bf x}_{i}^{(c)}\}_{i=1}^{N_{c}}\}_{c=1}^{C}$, \newline
    
  \noindent {\bf Step 0:}
  Choose a dictionary $\mathcal{D}$ of orthonormal wavelet packets
  or local sine/cosine packets and specify the maximum depth $J$ of
  decomposition and an additive discriminant measure $d$.\newline
  
  \noindent {\bf Step 1:}
  Construct time-frequency energy maps $\Gamma_{c}$ for 
  $c$ = 1,...,C. \newline
    
  \noindent {\bf Step 2:}
  Set $A_{J,k} = B_{J,k}$ and 
  $\Delta_{J,k} = d(\{\Gamma_{c}(J,k,\cdot)\}_{c=1}^{C})$
  for $k=0,...,2^{J}-1$. \newline
   
  \noindent {\bf Step 3:}
  Determine the best subspace $A_{j,k}$ for $j=J-1,...,0,
  \mbox{ }k=0,...,2^{j}-1$ by the following rule: \newline
   
  {\bf Set} $\Delta_{j,k} =
  d(\{\Gamma_{c}(J,k,\cdot)\}_{c=1}^{C})$. \newline
   
  {\bf If} $\Delta_{j,k} \geq \Delta_{j+1,2k}+\Delta_{j+1,2k+1}$,\newline
   
  {\bf then} $A_{j,k} = B_{j,k}$,\newline
   
  {\bf else} $A_{j,k} = A_{j+1,2k}\oplus A_{j+1,2k+1}$ and set
  $\Delta_{j,k} = \Delta_{j+1,2k}+\Delta_{j+1,2k+1}$. \newline
     
  \noindent {\bf Step 4:} 
  Order the basis functions by their power of discrimination
  (explained below). \newline
  \vspace{3mm}
    
  \noindent {\bf Step 5:}
  Use $k \leq n$ most discriminating basis functions for
  constructing classifiers.
  
\end{algorithm}
\vspace{3mm}

We note that Step 3 is fast, $O(n)$, since $d$ is additive. 
In Step 4, we evaluate the power of discrimination of a single basis
function ${\bf w}_{j,k,l}$ by $d(\{\Gamma_{c}(j,k,l)\}_{c=1}^{C})$.
That the basis obtained by the LDB-algorithm indeed has the desired
property, is stated in Proposition \ref{property of LDB basis},
for which (simple) proof we refer to \cite{saito}.

\begin{proposition}
  The basis obtained by the LDB-algorithm maximizes the additive
  discriminant measure $d$ on the time-frequency energy maps 
  $\{\Gamma_{c}\}_{c=1}^{C}$ among all
  the bases in the dictionary $\mathcal{D}$.
  \label{property of LDB basis}
\end{proposition}


%% file: thesisfig1.latex
\setlength{\unitlength}{0.00073300in}%
\begingroup\makeatletter\ifx\SetFigFont\undefined
\def\x#1#2#3#4#5#6#7\relax{\def\x{#1#2#3#4#5#6}}%
\expandafter\x\fmtname xxxxxx\relax \def\y{splain}%
\ifx\x\y   
\gdef\SetFigFont#1#2#3{%
  \ifnum #1<17\tiny\else \ifnum #1<20\small\else
  \ifnum #1<24\normalsize\else \ifnum #1<29\large\else
  \ifnum #1<34\Large\else \ifnum #1<41\LARGE\else
     \huge\fi\fi\fi\fi\fi\fi
  \csname #3\endcsname}%
\else
\gdef\SetFigFont#1#2#3{\begingroup
  \count@#1\relax \ifnum 25<\count@\count@25\fi
  \def\x{\endgroup\@setsize\SetFigFont{#2pt}}%
  \expandafter\x
    \csname \romannumeral\the\count@ pt\expandafter\endcsname
    \csname @\romannumeral\the\count@ pt\endcsname
  \csname #3\endcsname}%
\fi
\fi\endgroup
\begin{picture}(5412,1230)(2401,-1597)
\thicklines
\put(2851,-586){\vector( 3,-4){675}}
\put(4576,-586){\vector( 1, 0){1125}}
\put(4576,-586){\vector( 3,-4){702}}
\put(6601,-586){\vector( 1, 0){1200}}
\put(6601,-586){\vector( 1,-1){900}}
\put(2500,-661){\makebox(0,0)[lb]{\smash{$V_{J}$}}}
\put(4000,-661){\makebox(0,0)[lb]{\smash{$V_{J-1}$}}}
\put(5800,-661){\makebox(0,0)[lb]{\smash{$V_{J-2}$}}}
\put(2851,-586){\vector( 1, 0){1050}}
\put(6300,-630){$\cdots$}
\put(6100,-1550){$\cdots$}
\put(3700,-1561){\makebox(0,0)[lb]{\smash{$W_{J-1}$}}}
\put(6800,-1261){\makebox(0,0)[lb]{\smash{${\bf G}$}}}
\put(5500,-1561){\makebox(0,0)[lb]{\smash{$W_{J-2}$}}}
\put(3151,-511){\makebox(0,0)[lb]{\smash{${\bf H}$}}}
\put(2870,-1261){\makebox(0,0)[lb]{\smash{${\bf G}$}}}
\put(4600,-1261){\makebox(0,0)[lb]{\smash{${\bf G}$}}}
\put(4951,-511){\makebox(0,0)[lb]{\smash{${\bf H}$}}}
\put(7900,-661){\makebox(0,0)[lb]{\smash{$V_{J-M}$}}}
\put(7700,-1561){\makebox(0,0)[lb]{\smash{$W_{J-M}$}}}
\put(7051,-511){\makebox(0,0)[lb]{\smash{${\bf H}$}}}
\end{picture}

%% file: thesisfig2.latex
\setlength{\unitlength}{0.00043300in}%
\begingroup\makeatletter\ifx\SetFigFont\undefined
\def\x#1#2#3#4#5#6#7\relax{\def\x{#1#2#3#4#5#6}}%
\expandafter\x\fmtname xxxxxx\relax \def\y{splain}%
\ifx\x\y   
\gdef\SetFigFont#1#2#3{%
  \ifnum #1<17\tiny\else \ifnum #1<20\small\else
  \ifnum #1<24\normalsize\else \ifnum #1<29\large\else
  \ifnum #1<34\Large\else \ifnum #1<41\LARGE\else
     \huge\fi\fi\fi\fi\fi\fi
  \csname #3\endcsname}%
\else
\gdef\SetFigFont#1#2#3{\begingroup
  \count@#1\relax \ifnum 25<\count@\count@25\fi
  \def\x{\endgroup\@setsize\SetFigFont{#2pt}}%
  \expandafter\x
    \csname \romannumeral\the\count@ pt\expandafter\endcsname
    \csname @\romannumeral\the\count@ pt\endcsname
  \csname #3\endcsname}%
\fi
\fi\endgroup
\begin{picture}(2412,4080)(2689,-4222)
\thicklines
\put(2701,-2761){\line( 3, 4){1575}}
\put(2701,-2761){\line( 4,-3){1800}}
\put(2701,-2761){\line( 5,-1){2250}}
\put(3301,-361){\line(-1,-4){600}}
\put(4276,-586){\makebox(0,0)[lb]{\smash{$\widetilde{W}_{j}$}}}
\put(4651,-4186){\makebox(0,0)[lb]{\smash{$V_{j}$}}}
\put(3301,-286){\makebox(0,0)[lb]{\smash{$W_{j}$}}}
\put(5101,-3286){\makebox(0,0)[lb]{\smash{$\widetilde{V}_{j}$}}}
\end{picture}

%% file: thesisfig3.latex
\setlength{\unitlength}{0.00073300in}%
\begingroup\makeatletter\ifx\SetFigFont\undefined
\def\x#1#2#3#4#5#6#7\relax{\def\x{#1#2#3#4#5#6}}%
\expandafter\x\fmtname xxxxxx\relax \def\y{splain}%
\ifx\x\y   
\gdef\SetFigFont#1#2#3{%
  \ifnum #1<17\tiny\else \ifnum #1<20\small\else
  \ifnum #1<24\normalsize\else \ifnum #1<29\large\else
  \ifnum #1<34\Large\else \ifnum #1<41\LARGE\else
     \huge\fi\fi\fi\fi\fi\fi
  \csname #3\endcsname}%
\else
\gdef\SetFigFont#1#2#3{\begingroup
  \count@#1\relax \ifnum 25<\count@\count@25\fi
  \def\x{\endgroup\@setsize\SetFigFont{#2pt}}%
  \expandafter\x
    \csname \romannumeral\the\count@ pt\expandafter\endcsname
    \csname @\romannumeral\the\count@ pt\endcsname
  \csname #3\endcsname}%
\fi
\fi\endgroup
\begin{picture}(5475,3423)(2176,-3772)
\thicklines
\put(2701,-1261){\line( 0,-1){300}}
\put(2701,-1561){\line( 1, 0){4800}}
\put(7501,-1561){\line( 0, 1){300}}
\put(7501,-1261){\line(-1, 0){4800}}
\put(5101,-1261){\line( 0,-1){300}}
\put(2701,-2161){\line( 0,-1){300}}
\put(2701,-2461){\line( 1, 0){4800}}
\put(2701,-2161){\line( 1, 0){4800}}
\put(7501,-2161){\line( 0,-1){300}}
\put(5101,-2161){\line( 0,-1){300}}
\put(3901,-2161){\line( 0,-1){300}}
\put(6301,-2161){\line( 0,-1){300}}
\put(2701,-3061){\line( 0,-1){300}}
\put(2701,-3361){\line( 1, 0){4800}}
\put(7501,-3361){\line( 0, 1){300}}
\put(7501,-3061){\line(-1, 0){4800}}
\put(3301,-3061){\line( 0,-1){300}}
\put(3901,-3061){\line( 0,-1){300}}
\put(5101,-3061){\line( 0,-1){300}}
\put(4501,-3061){\line( 0,-1){300}}
\put(5701,-3061){\line( 0,-1){300}}
\put(6301,-3061){\line( 0,-1){300}}
\put(6901,-3061){\line( 0,-1){300}}
\put(3301,-2461){\vector(-1,-2){300}}
\put(3301,-2461){\vector( 1,-2){300}}
\put(4501,-2461){\vector(-1,-2){300}}
\put(4501,-2461){\vector( 1,-2){300}}
\put(5701,-2461){\vector(-1,-2){300}}
\put(5701,-2461){\vector( 1,-2){300}}
\put(6901,-2461){\vector(-1,-2){300}}
\put(6901,-2461){\vector( 1,-2){300}}
\put(3901,-1561){\vector(-1,-1){600}}
\put(3901,-1561){\vector( 1,-1){600}}
\put(6301,-1561){\vector(-1,-1){600}}
\put(6301,-1561){\vector( 1,-1){600}}
\put(2701,-361){\line( 0,-1){300}}
\put(2701,-661){\line( 1, 0){4800}}
\put(7501,-661){\line( 0, 1){300}}
\put(7501,-361){\line(-1, 0){4800}}
\put(5101,-661){\vector(-2,-1){1200}}
\put(1800,-3200){\makebox(0,0)[lb]{\smash{Scale}}}
\put(5101,-661){\vector( 2,-1){1200}}
\put(2701,-3661){\vector( 1, 0){4800}}
\put(2401,-361){\vector( 0,-1){3000}}
\put(4801,-586){\makebox(0,0)[lb]{\smash{$\Omega_{0,0}$}}}
\put(3601,-1486){\makebox(0,0)[lb]{\smash{$\Omega_{1,0}$}}}
\put(6001,-1486){\makebox(0,0)[lb]{\smash{$\Omega_{1,1}$}}}
\put(3100,-2386){\makebox(0,0)[lb]{\smash{$\Omega_{2,0}$}}}
\put(4300,-2386){\makebox(0,0)[lb]{\smash{$\Omega_{2,1}$}}}
\put(5500,-2386){\makebox(0,0)[lb]{\smash{$\Omega_{2,2}$}}}
\put(6700,-2386){\makebox(0,0)[lb]{\smash{$\Omega_{2,3}$}}}
\put(2800,-3286){\makebox(0,0)[lb]{\smash{$\Omega_{3,0}$}}}
\put(3400,-3286){\makebox(0,0)[lb]{\smash{$\Omega_{3,1}$}}}
\put(4000,-3286){\makebox(0,0)[lb]{\smash{$\Omega_{3,2}$}}}
\put(4600,-3286){\makebox(0,0)[lb]{\smash{$\Omega_{3,3}$}}}
\put(5200,-3286){\makebox(0,0)[lb]{\smash{$\Omega_{3,4}$}}}
\put(5800,-3286){\makebox(0,0)[lb]{\smash{$\Omega_{3,5}$}}}
\put(6400,-3286){\makebox(0,0)[lb]{\smash{$\Omega_{3,6}$}}}
\put(7000,-3286){\makebox(0,0)[lb]{\smash{$\Omega_{3,7}$}}}
\put(2776,-2836){\makebox(0,0)[lb]{\smash{$H$}}}
\put(3526,-2836){\makebox(0,0)[lb]{\smash{$G$}}}
\put(4051,-2836){\makebox(0,0)[lb]{\smash{$H$}}}
\put(4726,-2836){\makebox(0,0)[lb]{\smash{$G$}}}
\put(5200,-2836){\makebox(0,0)[lb]{\smash{$H$}}}
\put(5926,-2836){\makebox(0,0)[lb]{\smash{$G$}}}
\put(6451,-2836){\makebox(0,0)[lb]{\smash{$H$}}}
\put(7126,-2836){\makebox(0,0)[lb]{\smash{$G$}}}
\put(6676,-1861){\makebox(0,0)[lb]{\smash{$G$}}}
\put(5701,-1861){\makebox(0,0)[lb]{\smash{$H$}}}
\put(4276,-1861){\makebox(0,0)[lb]{\smash{$G$}}}
\put(3301,-1861){\makebox(0,0)[lb]{\smash{$H$}}}
\put(4201,-961){\makebox(0,0)[lb]{\smash{$H$}}}
\put(5776,-961){\makebox(0,0)[lb]{\smash{$G$}}}
\put(7000,-3950){\makebox(0,0)[lb]{\smash{Frequency}}}
\end{picture}

%% file: thesisfig4.latex
\setlength{\unitlength}{0.00060000in}%
\begingroup\makeatletter\ifx\SetFigFont\undefined
\def\x#1#2#3#4#5#6#7\relax{\def\x{#1#2#3#4#5#6}}%
\expandafter\x\fmtname xxxxxx\relax \def\y{splain}%
\ifx\x\y   
\gdef\SetFigFont#1#2#3{%
  \ifnum #1<17\tiny\else \ifnum #1<20\small\else
  \ifnum #1<24\normalsize\else \ifnum #1<29\large\else
  \ifnum #1<34\Large\else \ifnum #1<41\LARGE\else
     \huge\fi\fi\fi\fi\fi\fi
  \csname #3\endcsname}%
\else
\gdef\SetFigFont#1#2#3{\begingroup
  \count@#1\relax \ifnum 25<\count@\count@25\fi
  \def\x{\endgroup\@setsize\SetFigFont{#2pt}}%
  \expandafter\x
    \csname \romannumeral\the\count@ pt\expandafter\endcsname
    \csname @\romannumeral\the\count@ pt\endcsname
  \csname #3\endcsname}%
\fi
\fi\endgroup
\begin{picture}(7662,2019)(1489,-4897)
\thicklines
\put(2101,-3961){\line( 1, 1){900}}
\put(3001,-3061){\line( 1, 0){1800}}
\put(4801,-3061){\line( 1,-1){900}}
\put(4801,-3961){\line( 1, 1){900}}
\put(5701,-3061){\line( 1, 0){1800}}
\put(7501,-3061){\line( 1,-1){900}}
\put(7501,-3961){\line( 1, 1){900}}
\put(8401,-3061){\line( 1, 0){600}}
\put(1501,-3961){\vector( 1, 0){7500}}
\put(2101,-3886){\line( 0,-1){150}}
\put(3001,-3886){\line( 0,-1){150}}
\put(2551,-3886){\line( 0,-1){150}}
\put(4801,-3886){\line( 0,-1){150}}
\put(5251,-3886){\line( 0,-1){150}}
\put(5701,-3886){\line( 0,-1){150}}
\put(7501,-3886){\line( 0,-1){150}}
\put(7951,-3886){\line( 0,-1){150}}
\put(8401,-3886){\line( 0,-1){150}}
\put(1501,-3061){\line( 1, 0){600}}
\put(2101,-3061){\line( 1,-1){900}}
\put(3901,-2986){\makebox(0,0)[lb]{\smash{$1$}}}
\put(6601,-2986){\makebox(0,0)[lb]{\smash{$1$}}}
\put(3601,-3586){\makebox(0,0)[lb]{\smash{$b_{k}$}}}
\put(6301,-3586){\makebox(0,0)[lb]{\smash{$b_{k+1}$}}}
\put(2400,-4250){\makebox(0,0)[lb]{\smash{$\alpha_{k}$}}}
\put(5000,-4250){\makebox(0,0)[lb]{\smash{$\alpha_{k+1}$}}}
\put(7700,-4250){\makebox(0,0)[lb]{\smash{$\alpha_{k+2}$}}}
\put(9151,-4036){\makebox(0,0)[lb]{\smash{$\bR$}}}
\end{picture}

%% file: thesisfig5.latex
\setlength{\unitlength}{0.00063300in}%
\begingroup\makeatletter\ifx\SetFigFont\undefined
\def\x#1#2#3#4#5#6#7\relax{\def\x{#1#2#3#4#5#6}}%
\expandafter\x\fmtname xxxxxx\relax \def\y{splain}%
\ifx\x\y   
\gdef\SetFigFont#1#2#3{%
  \ifnum #1<17\tiny\else \ifnum #1<20\small\else
  \ifnum #1<24\normalsize\else \ifnum #1<29\large\else
  \ifnum #1<34\Large\else \ifnum #1<41\LARGE\else
     \huge\fi\fi\fi\fi\fi\fi
  \csname #3\endcsname}%
\else
\gdef\SetFigFont#1#2#3{\begingroup
  \count@#1\relax \ifnum 25<\count@\count@25\fi
  \def\x{\endgroup\@setsize\SetFigFont{#2pt}}%
  \expandafter\x
    \csname \romannumeral\the\count@ pt\expandafter\endcsname
    \csname @\romannumeral\the\count@ pt\endcsname
  \csname #3\endcsname}%
\fi
\fi\endgroup
\begin{picture}(6024,3351)(2989,-4300)
\thicklines
\put(3001,-1561){\line( 1, 1){600}}
\put(3601,-961){\line( 1, 0){4800}}
\put(8401,-961){\line( 1,-1){600}}
\put(3001,-2761){\line( 1, 0){6000}}
\put(3001,-2761){\line( 1, 1){600}}
\put(3601,-2161){\line( 1, 0){2100}}
\put(5701,-2161){\line( 1,-1){600}}
\put(5701,-2761){\line( 1, 1){600}}
\put(6301,-2161){\line( 1, 0){2100}}
\put(8401,-2161){\line( 1,-1){600}}
\put(3001,-3961){\line( 1, 0){6000}}
\put(3001,-3961){\line( 1, 1){600}}
\put(6301,-3961){\line(-1, 1){600}}
\put(3601,-3361){\line( 1, 0){750}}
\put(4351,-3361){\line( 1,-1){600}}
\put(5701,-3361){\line(-1, 0){750}}
\put(4951,-3361){\line(-1,-1){600}}
\put(9001,-3961){\line(-1, 1){600}}
\put(8401,-3361){\line(-1, 0){750}}
\put(7651,-3361){\line(-1,-1){600}}
\put(7651,-3961){\line(-1, 1){600}}
\put(7051,-3361){\line(-1, 0){750}}
\put(6301,-3361){\line(-1,-1){600}}
\put(6001,-1261){\vector( 4,-3){1536}}
\put(6001,-1261){\vector(-4,-3){1536}}
\put(4501,-2461){\vector(-1,-2){600}}
\put(4501,-2461){\vector( 3,-4){900}}
\put(3001,-1561){\line( 1, 0){6000}}
\put(7501,-2461){\vector(-3,-4){900}}
\put(7800,-3850){\makebox(0,0)[lb]{\smash{$\Omega_{2,3}$}}}
\put(7501,-2461){\vector( 1,-2){600}}
\put(6001,-2686){\line( 0,-1){150}}
\put(4651,-3886){\line( 0,-1){150}}
\put(6001,-3886){\line( 0,-1){150}}
\put(7351,-3886){\line( 0,-1){150}}
\put(5900,-3100){\makebox(0,0)[lb]{\smash{$\alpha_{k}$}}}
\put(4500,-4300){\makebox(0,0)[lb]{\smash{$\alpha_{k-1}$}}}
\put(5900,-4300){\makebox(0,0)[lb]{\smash{$\alpha_{k}$}}}
\put(7200,-4300){\makebox(0,0)[lb]{\smash{$\alpha_{k+1}$}}}
\put(5850,-1200){\makebox(0,0)[lb]{\smash{$\Omega_{0,0}$}}}
\put(3900,-2450){\makebox(0,0)[lb]{\smash{$\Omega_{1,0}$}}}
\put(7651,-2450){\makebox(0,0)[lb]{\smash{$\Omega_{1,1}$}}}
\put(3700,-3850){\makebox(0,0)[lb]{\smash{$\Omega_{2,0}$}}}
\put(5200,-3850){\makebox(0,0)[lb]{\smash{$\Omega_{2,1}$}}}
\put(6300,-3850){\makebox(0,0)[lb]{\smash{$\Omega_{2,2}$}}}
\end{picture}

%% file: algorithms.tex
\chapter[Factorization of The Discrete Wavelet Transform]
{Factorization of the Discrete Wavelet Transform in Dimension 1 Using
  Elements in $SO_{2}(\bR), SL_{m}(\bR)$}

\section{Introduction}

\subsection{Notation.}

If $T$ is a linear transformation, then $T^{t},T^{-t}$ will be the 
transpose of $T,T^{-1}$, respectively. If ${\bf x} \in \bR^{n}$, then
$S(k){\bf x}(j) = {\bf x}(j-k)$ and 
$\sigma {\bf x}(j) = {\bf x}(-j)$. If $a: \bZ^{n}\rightarrow \bR^{n}$,
$b: \bZ^{n}\rightarrow \bR^{n}$, then 
$a \ast b(k) = \sum_{j\in \bZ}a(j)b(k-j)$. For convenience we will 
occasionally denote Daubechies' shortest filters of length $L$, 
corresponding to an orthonormal compactly
supported wavelet possessing a number of $L/2$
vanishing moments, by $H^{d}_{L}$ for lowpass and $G^{d}_{L}$ for
highpass. We will denote the operation of 
``convolution followed by downsampling by 2'' by $\ast_{2}$.
Using similar notation, in the biorthogonal case 
we will denote the analysis lowpass- and highpass filters by 
$H^{b}_{L}$, $G^{b}_{\tilde{L}}$ respectively, and the synthesis 
lowpass and highpass filters by $\tilde{H}^{b}_{\tilde{L}}$, 
$\tilde{G}^{b}_{L}$, where $L$ and $\tilde{L}$ are the lengths
of the lowpass analysis- and lowpass synthesis filter, respectively. 

\subsection{The idea.}

We can present the main idea in our construction in just a few words.
Consider a conjugate pair of FIR analysis filters $H,G$. 
The filtering of a vector in $\bR^{n}$, $n$ even, into lowpass and highpass
coefficients using this pair of filters for the operation $\ast_{2}$, 
can be arranged as a mapping induced by a linear operator 
$\Theta_{n}(H,G):\bR^{n} \mapsto \bR^{n}$ defined by 

\begin{equation}
  \Theta_{n}(H,G): {\bf x} \in \bR^{n} \mapsto
  \left(c^{1}_{0},d^{1}_{0},c^{1}_{1},d^{1}_{1},\cdots
    ,c^{1}_{n/2-1},d^{1}_{n/2-1}\right). 
  \label{filtering operator}
\end{equation} 

Given a vector ${\bf x} \in \bR^{n}$, we
consider the following linear operators from $\bR^{n}$ to $\bR^{n}$,
defined by

\begin{eqnarray}
  && F_{2}(\alpha): {\bf x} \rightarrow 
  \left\{M_{2}(\alpha)\left(\begin{array}{c}
        x_{2j} \\ x_{2j+1}
      \end{array}\right)\right\}_{j=0}^{n/2-1}, \nonumber \\
  &&  SO_{2}(\bR) \ni M_{2}(\alpha) = \frac{1}{\sqrt{1+\alpha^{2}}}
  \left(\begin{array}{cc}
      1 & -\alpha \\ \alpha & 1
    \end{array}\right), 
  \label{orthogonal 22matrix mapping}
\end{eqnarray}

\begin{eqnarray}
  && \hat{F}_{m}(\alpha): {\bf x} \rightarrow 
  \left\{\hat{M}_{m}(\alpha)\left(\begin{array}{c}
        x_{mj} \\ x_{mj+1} \\ \vdots \\ x_{mj+m-1}
      \end{array}\right)\right\}_{j=0}^{n/m-1}, m \geq 2, \nonumber \\
  && SL_{m}(\bR) \ni \hat{M}_{m}(\alpha)(i,j) = 
  \left\{\begin{array}{ll}
      1 & \mbox{if $i=j$} \\ 
      -\alpha & \mbox{if $i=1,j=m$} \\
      0 & \mbox{otherwise.} 
    \end{array}\right.
  \label{biorthogonal mmmatrix mapping}
\end{eqnarray}

\begin{eqnarray}
  && \tilde{F}_{3}(\alpha): {\bf x} \rightarrow 
  \left\{\tilde{M}_{3}(\alpha)\left(\begin{array}{c}
        x_{2j} \\ x_{2j+1} \\ x_{2j+2}
      \end{array}\right)\right\}_{j=0}^{n/2-1}, \nonumber \\
  && SL_{3}(\bR) \ni \tilde{M}_{3}(\alpha) = 
  \left(\begin{array}{ccc}
      1 & -\alpha & 0 \\
      0 & 1 & 0 \\
      0 & -\alpha & 1 \\
      \end{array}\right),
  \label{symmetric biorthogonal 33matrix mapping}
\end{eqnarray}

\begin{eqnarray}
  && \bar{F}_{2}(\alpha,\beta): {\bf x} \rightarrow 
  \left\{\bar{M}_{2}(\alpha,\beta)\left(\begin{array}{c}
        x_{2j} \\ x_{2j+1}
      \end{array}\right)\right\}_{j=0}^{n/2-1}, \nonumber \\
  &&  SL_{2}(\bR) \ni \bar{M}_{2}(\alpha) =  \frac{1}{1+\alpha\beta}
  \left(\begin{array}{cc}
      1 & -\alpha \\
      \beta & 1
    \end{array}\right),
  \label{biorthogonal special 22matrix mapping}
\end{eqnarray}

\noindent {\bf Remark}. In case the number $n/m, m > 2$ is 
non-integer, just extend ${\bf x} \in \bR^{n}$ by 
periodizing or mirroring or by zeros.

We will show how the operator $\Theta_{n}(H,G)$ can be factored into:

\begin{itemize}

\item A composition of operators $F_{2}(\alpha_{j})$ and $S(k_{j})$, 
  $j$ in some finite index interval, in the
  case where $H,G$ is a conjugate orthogonal pair of filters.
  
\item A composition of operators $\hat{F}_{m}^{-1}(\alpha_{j})$, 
  $\hat{F}_{m}^{-t}(\alpha_{j})$ and $S(k_{j})$, with $m$, 
  $j$ in some finite index intervals, 
  in the case where $H,G$ is a pair of biorthogonal conjugate 
  analysis filters of odd-odd lengths or odd-even/even-odd lengths.

\item A composition of operators $\tilde{F}_{3}^{-1}(\alpha_{j})$, 
  $\hat{F}_{3}^{-t}(\alpha_{j})$ and $S(k_{j})$, with
  $j$ in some finite index interval, 
  in the case where $H,G$ is a pair of biorthogonal conjugate 
  {\em symmetric} analysis filters of odd-odd lenghts.

\item A composition of operators $\hat{F}^{-1}_{m}(\alpha_{j})$, 
  $\hat{F}_{m}^{-t}(\alpha_{j})$,  $\bar{F}_{2}^{-1}(\alpha_{j})$, 
  $\bar{F}_{2}^{-t}(\alpha_{j})$ and $S(k_{j})$, with $m$, 
  $j$ in some finite index intervals, 
  in the case where $H,G$ is a conjugate pair of biorthogonal
  filters of even-even lengths. 

\end{itemize}

We will start considering Daubechies/Coiflet filters, and then
extend our results to biorthogonal filters.

\section{Factorization of the orthonormal wavelet transform.}

Let $H^{d}_{L}, G^{d}_{L}$ be an orthonormal pair of FIR
filters. A linear orthogonal map on ${\bR}^{2}$ is some matrix $M(\alpha)$, 
$\alpha \in \bR$ in the one-parameter family $SO_{2}({\bR})$. 
If we can represent $\Theta_{n}(H^{d}_{L},G^{d}_{L})$ 
by some composition of the 
orthogonal maps $F_{2}(\alpha_{j})$ as defined in 
(\ref{orthogonal 22matrix mapping}),
we have an alternative way of computing the discrete wavelet transform
in the orthonormal case. 

Defining the normalization factor $R_{m}$ by

\begin{equation} 
  R_{m} = \prod_{j=1}^{m}(1+\alpha_{j}^{2})^{- \frac{1}{2}}, 
  \label{normalization factor} 
\end{equation}

\noindent then because of the linearity of the maps $F(\alpha_{j})$, 
we can avoid repeated multiplication by square-root factors, 
instead normalizing in the last step by scalar-multiplication
by some $R_{m}$. 

From the orthogonality relation (\ref{double shift orthogonality}) 
we see that the operation of lowpass filtering when applied to the 
lowpass filter itself, reduces the filter to 1 point. 
Obviously, we have to construct a stepwise procedure to reduce the 
length of the lowpass filter, using some set
of matrices $M_{2}(\alpha_{j})$. We define the first step to be the operation 

\begin{equation}
  H^{d}_{L} \mapsto F_{2}(\alpha_{1})H^{d}_{L},
  \label{orthonormal first step}
\end{equation}

\noindent where we claim 

\begin{equation}
  H^{d}_{L}(0) \mapsto 0, \mbox{ which implies } \alpha_{1} =
  \frac{H^{d}_{L}(0)}{H^{d}_{L}(1)}. 
  \label{orthonormal claim}
\end{equation}

\noindent Then by (\ref{double shift orthogonality}) we have

\begin{eqnarray}
  0 &=& H^{d}_{L}(0)\cdot H^{d}_{L}(L-2) + 
  H^{d}_{L}(1)\cdot H^{d}_{L}(L-1) \nonumber \\ 
  & \Longrightarrow & \frac{H^{d}_{L}(0)}{H^{d}_{L}(1)} = 
  - \frac{H^{d}_{L}(L-1)}{H^{d}_{L}(L-2)} \nonumber \\
    & \Longrightarrow & H^{d}_{L}(L-1) \mapsto 0 \mbox{ under }
    F_{2}(\alpha_{1}).
\label{double shift orthogonality zero map}
\end{eqnarray}

\noindent We see that the number of coefficients in the filter 
$F_{2}(\alpha_{1})H^{d}_{L}$ is 2 less than in $H^{d}_{L}$.
It is now clear how to proceed: To simplify notation we write 

\begin{equation}
  W_{2}(\alpha_{k}) \equiv F_{2}(\alpha_{k})S(-(k-1)).
  \label{definition of W}
\end{equation}

\noindent $W_{2}(\cdot)$ is an orthogonal operator. 
Then we define the $k$'th step in our
procedure, $1 \leq k \leq L/2$ as the operation

\begin{equation}
  W_{2}(\alpha_{k-1})\cdots W_{2}(\alpha_{1})H^{d}_{L} \mapsto 
  W_{2}(\alpha_{k})W_{2}(\alpha_{k-1})\cdots W_{2}(\alpha_{1})H^{d}_{L}. 
  \label{matrix filtering} 
\end{equation}

\noindent where we claim $W_{2}(\alpha_{k})\circ \cdots \circ
W_{2}(\alpha_{1})H^{d}_{L}(0) \mapsto 0$.

\begin{figure}[h]
  \begin{center}
    \input{thesisfig9.latex}
  \end{center}
  \caption{Orthogonal mapping of the lowpass filter $H^{d}_{6}$ to 1 point.}
  \label{reduce lowpass}
\end{figure}
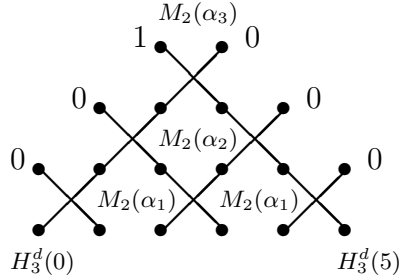

Because every $W_{2}(\alpha_{j})$ is a linear orthogonal operator, 
the orthogonality relation 
(\ref{double shift orthogonality}) is preserved in each
step. Thus each step but the last will reduce the length of the filter
at the previous step by 2, the last step reducing a filter of length 
2 to a filter of length 1. Our construction is illustrated in 
Figure \ref{reduce lowpass} below for the filter $H^{d}_{6}$. Each
cross in Figure \ref{reduce lowpass} represents the mapping induced
by $M_{2}(\alpha)$ on the lower pair of points resulting in the upper
pair of points.
\vspace{5mm}

Since the  highpass filter $G^{d}_{L}$ is the alternating flip of
$H^{d}_{L}$  (see relation (\ref{orthonormal highpass from lowpass})), 
it is easy to see that applying the operation defined in 
(\ref{matrix filtering}) to $G^{d}_{L}$ will reduce it to a
1-point-filter. Because of the orthogonality of lowpass channel to
highpass channel, we see that

\[ H^{d}_{L} \mapsto (1,0) \Longrightarrow G^{d}_{L} \mapsto (0,1).
\]

Thus we get the following result:
  
\begin{theorem}
  Given a pair of orthonormal FIR filters $H^{d}_{L},G^{d}_{L}$ 
  of length $L$, the operator $\Theta_{n}(H^{d}_{L},G^{d}_{L})$ 
  defined in (\ref{filtering operator}) 
  may be decomposed as 
  $\Theta_{n}(H^{d}_{L},G^{d}_{L}) = 
  S(1-L/2)W_{2}(\alpha_{L/2})W_{2}(\alpha_{L/2-1}) 
  \cdots W_{2}(\alpha_{1})$, with $W_{2}(\alpha_{j})$, 
  $1 \leq j \leq L/2$ defined as shown above.
  \label{orthonormal result}
\end{theorem}

\proof

The theorem was proved by construction. The argumentation below is
just a formalization of this construction.
 
\begin{eqnarray}
  c^{1}_{k} & = & \left({\bf x}\ast_{2}\sigma H^{d}_{L}\right)(k)
  = \sum_{j \in \bZ}{\bf x}(j)H^{d}_{L}(j-2k) 
  = \IP{S(2k)H^{d}_{L}}{{\bf x}} \nonumber \\
  & = & \IP{W^{-1}_{2}(\alpha_{1})\cdots W^{-1}_{2}(\alpha_{L/2})
    W_{2}(\alpha_{L/2})\cdots W_{2}(\alpha_{1})
    S(2k)H^{d}_{L}}{{\bf x}} \nonumber \\
  & = & \IP{W_{2}(\alpha_{L/2})\cdots W_{2}(\alpha_{1})S(2k)H^{d}_{L}}
  {W^{-t}_{2}(\alpha_{L/2})\cdots W^{-t}_{2}(\alpha_{1}){\bf x}}
  \nonumber \\
  & = & \IP{\delta(j,2k-(L/2-1))}{W_{2}(\alpha_{L/2})\cdots  
    W_{2}(\alpha_{1}){\bf x}} \nonumber \\ 
  & = & W_{2}(\alpha_{L/2})\cdots W_{2}(\alpha_{1}){\bf x}(2k-(L/2-1)).
  \label{proof of orthonormal lowpass result}
\end{eqnarray}

\begin{eqnarray}
  d^{1}_{k} & = & \left({\bf x}\ast_{2}\sigma G^{d}_{L}\right)(k)
  = \sum_{j \in \bZ}{\bf x}(j)G(j-2k) 
  = \IP{S(2k)G^{d}_{L}}{{\bf x}} \nonumber \\
  & = & \IP{W^{-1}_{2}(\alpha_{1})\cdots W^{-1}_{2}(\alpha_{L/2})
    W_{2}(\alpha_{L/2})\cdots W_{2}(\alpha_{1})
    S(2k)G^{d}_{L}}{{\bf x}} \nonumber \\
  & = & \IP{W_{2}(\alpha_{L/2})\cdots W_{2}(\alpha_{1})S(2k)G^{d}_{L}}
  {W^{-t}_{2}(\alpha_{L/2})\cdots W^{-t}_{2}(\alpha_{1}){\bf x}}
  \nonumber \\
  & = & \IP{\delta(j,2k+1-(L/2-1))}{W_{2}(\alpha_{L/2})\cdots  
    W_{2}(\alpha_{1}){\bf x}} \nonumber \\ 
  & = & W_{2}(\alpha_{L/2})\cdots W_{2}(\alpha_{1}){\bf x}(2k+1-(L/2-1)).
  \label{proof of orthonormal highpass result}
\end{eqnarray}

\endproof

A schematic picture of the algorithm that results from the
factorization of $\Theta_{n}(H^{d}_{L},G^{d}_{L})$ 
is shown in Figure \ref{filteringladder1} below.

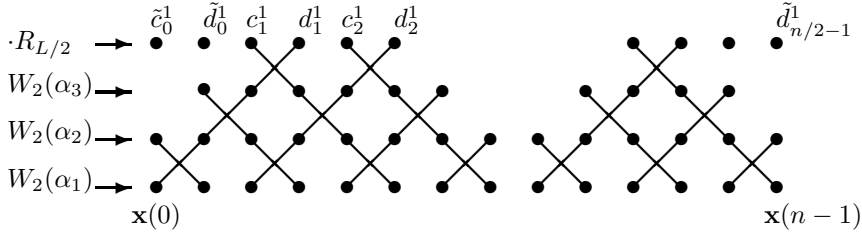
\begin{figure}[h]
  \begin{center}
    \input{figur2.latex}
  \end{center}
  \caption{Filtering-diagram for $H^{d}_{6},G^{d}_{6}$.}
  \label{filteringladder1}
\end{figure}

The ``tilde'' c's and d's at the edges of the filtering-diagram above can
be obtained by mirroring ${\bf x}$ about its endpoints or periodizing.
Counting operations, we see that we have to do 2
multiplications and 2 additions pr. pair of points for each
$W_{2}(\alpha)$. Writing

\begin{equation} 
  \Pi_{n}(H^{d}_{L}, G^{d}_{L}) \equiv \prod_{j=L/2}^{1}W_{2}(\alpha_{j}),
  \label{factored orthogonal wavelet transform}
\end{equation} 

\noindent then by construction $S(1-L/2)\Pi_{n}=\Theta_{n}$, 
but it has a more efficient implementation as shown in 
Table \ref{orthonormal operation cost}. 
\footnote{The additition of 1 to $L/2$ results from
  normalization by scalar-multiplication by $R_{L/2}$.}

\begin{table}[h]
  \setlength{\extrarowheight}{6pt}
  \begin{center}
    \begin{tabular}{|c|c|c|} \hline
      Operation & $\Pi_{n}$ & $\Theta_{n}$  \\ \hline
      $\sharp$ mult's & $(L/2+1)\cdot n$ & $L\cdot n$ \\ \hline
      $\sharp$ add's & $L/2\cdot n$ & $(L-1)\cdot n$ \\ \hline
    \end{tabular}
  \end{center}  
  \caption{The operation cost of filtering a set of $n$ points
    using the pair of filters $H^{d}_{L},G^{d}_{L}$ 
    by 2 different techniques.}
  \label{orthonormal operation cost} 
\end{table}

We note that by Theorem \ref{orthonormal result} 
the inverse wavelet transform has a decomposition given by reversing 
the order of the operators $W_{2}(\alpha_{j})$ and replacing each by
its inverse.
  
\section{Factorization of the biorthogonal wavelet transform.}

Let $H^{b}_{L},G^{b}_{\tilde{L}}$ and 
$\tilde{H}^{b}_{\tilde{L}},\tilde{G}^{b}_{L}$ 
be pairs of biorthogonal FIR analysis filters and synthesis filters, 
respectively. We make no restrictions on them, except that they 
satisfy the relations given in  
(\ref{biorthogonal highpass from lowpass}).

We want to find a factoriztion of $\Theta_{n}(H^{b}_{L},G^{b}_{\tilde{L}})$ 
like we did in the case where the filters were orthogonal. This 
may seem slightly more complicated because of two major differences:
 
\begin{itemize}
 
\item The filter lengths $L$, $\tilde{L}$ of the lowpass and highpass 
  filter need not be even numbers, moreover they will 
  in general not be equal: $\tilde{L}-L$ may be any number.

\item The highpass filter $G^{b}_{\tilde{L}}$ is {\em not} 
  the alternating flip of the lowpass filter $H^{b}_{L}$, 
  it is the alternating flip of the synthesis lowpass filter 
 $\tilde{H}^{b}_{\tilde{L}}$.

\end{itemize}
 
But as we will see below, these differences are not crucial to the
main idea in the method developed in the orthogonal case, and the relations 
(\ref{double shift biorthogonality}), (\ref{biorthogonal highpass from
  lowpass}) will provide us with all that we need to make 
a modified version of the method work in this case. We simply
have to remember that when dealing with the biorthogonal wavelet transform,
we have to work with two ``Multiresolution Analysises'' (MRA's) which
are {\em duals} of each other: One for analysis, the other for
synthesis. Thus, to each linear operation/mapping we choose to perform
on the analysis MRA, there corresponds a linear {\em dual} mapping 
on the synthesis MRA and vice versa. 
Having this in mind, we proceed to solve our problem.

Now that we have the procedure from the orthonormal case to guide us, 
we begin with the definition of a linear map that reduces the
length of the analysis lowpass filter $H^{b}_{L}$. Since this will
uniquely define some dual map on the synthesis lowpass filter, 
we see that we could equally well define our mapping on 
$\tilde{H}^{b}_{\tilde{L}}$. It will be more practical to start
with the shortest filter. We assume $\tilde{L} \geq L$.
Since the filter ${H}^{b}_{L}$ is not double-shift
orthonormal to itself,  $F_{2}(\alpha)$ will only reduce this filter
in one end for some $\alpha$. Therefore, we may as well replace 
$F_{2}(\alpha)$ by $\hat{F}_{2}(\alpha)$ to save useless operations.
We define the first step analogous to (\ref{orthonormal first step}) 

\begin{equation}
  H^{b}_{L}\mapsto 
  \hat{F}_{2}(\alpha_{1})H^{b}_{L},
  \label{biorthogonal first step}
\end{equation}

\noindent where we claim 

\begin{equation}
  H^{b}_{L}(0) \mapsto 0, \mbox{ which implies } 
  \alpha_{1} = \frac{H^{b}_{L}(0)}
  {H^{b}_{L}(1)}. 
  \label{biorthogonal claim}
\end{equation}

\noindent From the ``double shift biorthogonality relation'' 
(\ref{double shift biorthogonality}), we get

\begin{eqnarray}
  \delta_{0,k} & = & \IP{H^{b}_{L}}{S(2k)\tilde{H}^{b}_{\tilde{L}}} 
  \nonumber \\
  & = &\IP{\hat{F}^{-1}_{2}(\alpha_{1})\hat{F}_{2}(\alpha_{1})
    H^{b}_{L}}{S(2k)\tilde{H}^{b}_{\tilde{L}}} \nonumber \\
  & = & \IP{\hat{F}_{2}(\alpha_{1})H^{b}_{L}}
  {\hat{F}^{-t}_{2}(\alpha_{1})S(2k)\tilde{H}^{b}_{\tilde{L}}}.
  \label{definining the dual map}
\end{eqnarray}

Using relation (\ref{double shift biorthogonality}) once again, we 
observe that

\begin{eqnarray}
  0 & = & H^{b}_{L}(0)\tilde{H}^{b}_{\tilde{L}}(\tilde{L}-2) +
  H^{b}_{L}(1)\tilde{H}^{b}_{\tilde{L}}(\tilde{L}-1) \nonumber \\
  & \Longrightarrow & \frac{\tilde{H}^{b}_{\tilde{L}}(\tilde{L}-1)}
  {\tilde{H}^{b}_{\tilde{L}}(\tilde{L}-2)} = 
  - \frac{H^{b}_{L}(0)}{H^{b}_{L}(1)}
  = -\alpha_{1} \nonumber \\
  & \Longrightarrow & \hat{F}^{-t}_{2}(\alpha_{1}): 
  \tilde{H}^{b}_{\tilde{L}}(\tilde{L}-1) \rightarrow 0. 
  \label{biorthogonal reduction of lowpass filter}
\end{eqnarray}

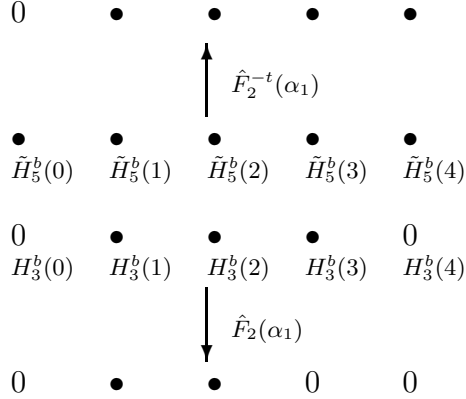
\begin{figure}[h]
  \begin{center}
    \input{thesisfig10.latex}
  \end{center}
  \caption{Illustration of $\hat{F}_{2}(\alpha_{1})H^{b}_{3}$ and
    $\hat{F}^{-t}_{2}(\alpha_{1})\tilde{H}^{b}_{5}$.}
  \label{figure of biorthogonal first step}
\end{figure}

Figure \ref{figure of biorthogonal first step}
illustrates this fact for the case $L = 3$, $\tilde{L} = 5$. 

Then, using relation (\ref{biorthogonal highpass from lowpass}) and the
duality of $\hat{F}_{2}(\alpha_{1})$ to $\hat{F}^{-t}_{2}(\alpha_{1})$
we get

\begin{equation}
  \hat{F}_{2}(\alpha_{1})G^{b}_{\tilde{L}}(0) =
  (-1)^{1}\hat{F}^{-t}_{2}(\alpha_{1})
  \tilde{H}^{b}_{\tilde{L}}(\tilde{L}-1) = 0.
  \label{biorthogonal reduction of highpass filter}
\end{equation}

Now we have to be careful in defining the next steps. 
We look for a sequence of linear non-singular operators of type 
$\hat{F}_{2}(\cdot), \hat{F}_{2}^{t}(\cdot)$, the composition of
which reduces the pair $H^{b}_{L}, G^{b}_{L}$ to a pair of 
1-point filters.
By the relations (\ref{definining the dual map}), 
(\ref{biorthogonal reduction of lowpass filter}), 
(\ref{biorthogonal reduction of highpass filter}), we see that this
goal is reached once we have a sequence of pairs of non-singular 
dual linear maps that reduces the pair of filters 
$H^{b}_{L}, \tilde{H}^{b}_{\tilde{L}}$
to a pair of {\em biorthogonal} 1-point lowpass filters. We will show
that such a sequence exists by construction.

Using the operators $\hat{F}_{2}(\alpha)$, each step in this 
construction will reduce the filter $H^{b}_{L}$ 
in one end only. 
We have to decide which end to reduce at step $k$ such that at 
every step $j$, $1\leq k\leq j < max(\tilde{L},L),$   

\begin{equation}
  supp(\tilde{H}^{b}_{\tilde{L}-j}) \cap supp(H^{b}_{L(j)})
  \neq \emptyset,
  \label{non-empty support intersection}  
\end{equation}

\noindent and such that after the last step we are left with two
dual 1-point filters. 

Since dual filters always have common center, 
we see that condition (\ref{non-empty support intersection}) 
forces a symmetric reduction of the filter
lengths around the common center. As is easily checked out, 
this is achieved by alternatingly making use of suitable 
operators $\hat{F}_{m}(\alpha_{k}),
\hat{F}^{t}_{m}(\alpha_{k+1}), \mbox{ }m \geq 2,
\mbox{ } 1\leq k < max(\tilde{L},L)-1$, together with suitable 
shifts. By the definition of the highpass filter $G^{b}_{\tilde{L}}$
given in (\ref{biorthogonal highpass from lowpass}), we see that the 
centers of the lowpass and highpass filters do not share support, 
the highpass center is delayed 1 time step with regard to the 
lowpass center. Thus, the symmetric reduction of the highpass
filter is delayed 1 time step with regard to the symmetric reduction
of the lowpass filter. 

We note one possible problem: If at some step 
$j$, $1\leq j <  max(\tilde{L},L)$ the filter 
$\tilde{H}^{b}_{\tilde{L}-j}$ has internal zero 
coefficients, we will not be able to map the outer (with regard to the center
of the filter) neighbour coefficient to zero using any
$\hat{F}_{2}(\alpha)$ or $\hat{F}^{t}_{2}(\alpha)$. 
But it is easy to see that by replacing $\hat{F}_{2}(\alpha)$ 
by $\hat{F}_{3}(\alpha)$ or $\hat{F}^{t}_{2}(\alpha)$ 
by $\hat{F}^{t}_{3}(\alpha)$ , as defined in 
(\ref{biorthogonal mmmatrix mapping}), the desired
zero-mapping of the neighbour coefficient can be achieved in both 
$\tilde{H}^{b}_{\tilde{L}-j}$ and $H^{b}_{L(j)}$. 
In general, if there are $m$ consecutive internal zero
coefficients, one should use $\hat{F}_{m+2}(\alpha), 
\hat{F}^{t}_{m+2}(\alpha)$. 

There is really not much left to prove, except to clear up some
details. We separate the biorthogonal filters into 3 classes, 
and give the form of our factorization of the wavelet transform in
each of these classes.

\subsection{The general odd-odd case.}

Here $L\neq \tilde{L}$ are both odd numbers. 
To simplify notation we define

\begin{eqnarray}
  && J = \frac{L+\tilde{L}}{2}, \nonumber \\
  && \hat{W}_{m(k)}(\alpha_{k}) \equiv \left\{\begin{array}{ll}
      \hat{F}_{m(k)}(\alpha_{k})S(-\lfloor \frac{k}{2}\rfloor) & 
      \mbox{if $k, 1\leq k \leq J,$ is odd.} \\
      \hat{F}^{t}_{m(k)}(\alpha_{k})S(-\frac{k}{2}) & 
      \mbox{if $k, 1\leq k \leq J,$ is even.} \nonumber \\
    \end{array}\right. \\
  \label{definition of J and W hat}
\end{eqnarray}

\noindent Then we can state the following result:

\begin{theorem}
  Given a pair of biorthogonal FIR filters 
  $H^{b}_{L},G^{b}_{\tilde{L}}$ of odd lengths $L,\tilde{L}$, there 
  exists a sequence of integers $\left\{m(j)\right\}_{j=1}^{J} 
  \subset \bN -\{1\}$
  such that the operator $\Theta_{n}(H^{b}_{L},G^{b}_{\tilde{L}})$ 
  defined in (\ref{filtering operator}) may be decomposed as   
  $\Theta_{n}(H^{b}_{L},G^{b}_{\tilde{L}}) = 
  S(-\lfloor J/2\rfloor)\hat{W}^{-t}_{m(J)}(\alpha_{J})
  \hat{W}^{-t}_{m(J-1)}(\alpha_{J-1})\cdots
  \hat{W}^{-t}_{m(1)}(\alpha_{1})$,
  with $\hat{W}_{m(j)}(\alpha_{j})$, $1\leq j \leq J$ 
  and $J$ defined as in (\ref{definition of J and W hat}).
  \label{general odd-odd result}
\end{theorem}

\proof

The theorem was proved by construction. The argumentation below is
just a formalization of this construction.
 
\begin{eqnarray}
  c^{1}_{k} & = & \left({\bf x}\ast_{2}\sigma H^{b}_{L}\right)(k)
  = \sum_{j \in \bZ}{\bf x}(j)H(j-2k) 
  = \IP{S(2k)H^{b}_{L}}{{\bf x}} \nonumber \\
  & = & \IP{\hat{W}^{-1}_{m(1)}(\alpha_{1})
    \cdots \hat{W}^{-1}_{m(J)}(\alpha_{J})\hat{W}_{m(J)}(\alpha_{J})
    \cdots \hat{W}_{m(1)}(\alpha_{1})
    S(2k)H^{d}_{L}}{{\bf x}} \nonumber \\
  & = & \IP{\hat{W}_{m(J)}(\alpha_{J})\cdots 
    \hat{W}_{m(1)}(\alpha_{1})S(2k)H^{b}_{L}}
  {\hat{W}^{-t}_{m(J)}(\alpha_{J})\cdots 
    \hat{W}^{-t}_{m(1)}(\alpha_{1}){\bf x}}
  \nonumber \\
  & = & \IP{\delta(j,2k-J/2)}{\hat{W}^{-t}_{m(J)}(\alpha_{J})\cdots  
    \hat{W}^{-t}_{m(1)}(\alpha_{1}){\bf x}} \nonumber \\ 
  & = & \hat{W}^{-t}_{m(J)}(\alpha_{J})\cdots 
  \hat{W}^{-t}_{m(1)}(\alpha_{1})
  {\bf x}(2k-\lfloor J/2\rfloor).
  \label{proof of general odd-odd lowpass result}
\end{eqnarray}

\begin{eqnarray}
  d^{1}_{k} & = & \left({\bf x}\ast_{2}\sigma G^{b}_{\tilde{L}}\right)(k)
  = \sum_{j \in \bZ}{\bf x}(j)G^{b}_{\tilde{L}}(j-2k) 
  = \IP{S(2k)G^{b}_{\tilde{L}}}{{\bf x}} \nonumber \\
  & = & \IP{\hat{W}^{-1}_{m(1)}(\alpha_{1})
    \cdots \hat{W}^{-1}_{m(J)}(\alpha_{J})\hat{W}_{m(J)}(\alpha_{J})
    \cdots \hat{W}_{m(1)}(\alpha_{1})
    S(2k)G^{d}_{\tilde{L}}}{{\bf x}} \nonumber \\
  & = & \IP{\hat{W}_{m(J)}(\alpha_{J})\cdots 
    \hat{W}_{m(1)}(\alpha_{1})S(2k)G^{b}_{\tilde{L}}}
  {\hat{W}^{-t}_{m(J)}(\alpha_{J})\cdots 
    \hat{W}^{-t}_{m(1)}(\alpha_{1}){\bf x}}
  \nonumber \\
  & = & \IP{\delta(j,2k+1-J/2)}{\hat{W}^{-t}_{m(J)}(\alpha_{J})\cdots  
    \hat{W}^{-t}_{m(1)}(\alpha_{1}){\bf x}} \nonumber \\ 
  & = & \hat{W}^{-t}_{m(J)}(\alpha_{J})\cdots 
  \hat{W}^{-t}_{m(1)}(\alpha_{1})
  {\bf x}(2k+1-\lfloor J/2\rfloor).
  \label{proof of general odd-odd highpass result}
\end{eqnarray}

\noindent It remains to prove that the orthogonal length-reduction of
the filters $H^{b}_{L}, \tilde{H}^{b}_{\tilde{L}}$ takes
$\frac{L+\tilde{L}}{2}$ steps. But this is easy to see: We have

\[ H^{b}_{L} \mapsto \delta_{j,0} \Longrightarrow
\tilde{H}^{b}_{\tilde L} \mapsto \tilde{H}^{b}_{\tilde{L}-L+1}. 
\] 

\noindent by relation (\ref{double shift biorthogonality}). 
If $\tilde{L}-L = 2(2k+1), k\geq 0$, then by considering relation 
(\ref{double shift biorthogonality}) we deduce that $2k$ of the
$\tilde{L}-L+1$ coefficients in $\tilde{H}^{b}_{\tilde{L}-L+1}$ are 
$0$, and that  these $0$-coefficients are distributed symmetrically
around the center-coefficient. Thus the total number of steps will be 

\[ L-1 + 2k+2 = L-1 +\frac{\tilde{L}-L}{2} + 1 =
\frac{L+\tilde{L}}{2}. 
\] 

\noindent By a similar argument, if $\tilde{L}-L= 4k, k\geq 1$, then
the total number of steps will be $\frac{L+\tilde{L}}{2}-1$.

Thus, the number $J$ as defined in (\ref{definition of J and W hat})
is at least an upper bound for the total number of steps.

\endproof

A schematic picture of the algorithm that results from the
factorization of $\Theta_{n}(H^{b}_{L},G^{b}_{\tilde{L}})$ 
in this case of odd length filters is shown in 
Figure \ref{filteringladder2} below. At level $k$
in the figure, each slanting line $A$
connected to a vertical line $B$ means multiplication of the number in the
point of origin of $A$ by the factor $\alpha_{k}$, 
followed by addition to the number in the point of origin of $B$,
while single vertical lines just mean the identity. This is easily
seen to be the pairwise mapping induced by either 
$\hat{F}^{-t}_{2}(\alpha_{k})$ or $\hat{F}^{-1}_{2}(\alpha_{k})$.
\vspace{2mm}

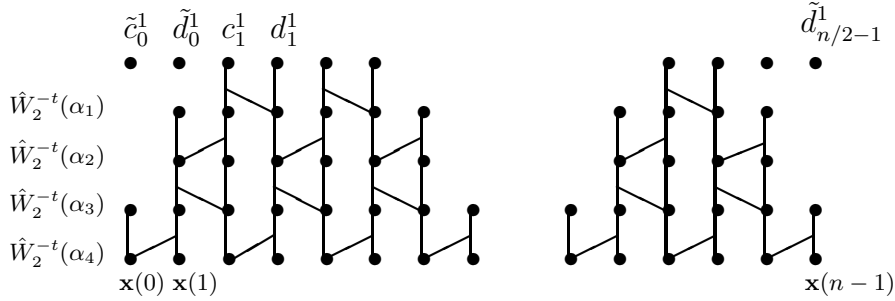
\begin{figure}[h]
  \begin{center}
    \input{thesisfig11.latex}
  \end{center}
  \caption{Filtering-diagram for a conjugate pair of biorthogonal
    filters of length 5 and 3.}
  \label{filteringladder2}
\end{figure}

The ``tilde'' c's and d's at the edges of the filtering-diagram above can
be obtained by mirroring ${\bf x}$ about its endpoints or periodizing.
Counting operations, we see that we have to do 1
multiplication and 1 addition pr. pair of points for each
$\hat{W}_{m(j)}(\alpha_{j})$. Writing 

\begin{equation} 
  \Pi_{n}(H^{b}_{L},G^{b}_{\tilde{L}}) 
  \equiv \prod_{j=J}^{1}
  \hat{W}^{-t}_{m(j)}(\alpha_{j}), 
  \label{general odd-odd factored wavelet transform}
\end{equation}

\noindent then by construction $\Pi_{n}= 
S(\lfloor \frac{J}{2} \rfloor)\Theta_{n}$, 
but it has a more efficient 
implementation as shown
in Table \ref{odd-odd biorthogonal operation cost}. 
\footnote{The cost result shown in Table \ref{odd-odd biorthogonal operation cost} 
  is only an upper bound if the $H^{b}_{L},G^{b}_{\tilde{L}}$ 
  have internal zero coefficients.}

\begin{table}[h]
  \setlength{\extrarowheight}{6pt}
  \begin{center}
    \begin{tabular}{|c|c|c|} \hline
      Operation 
      & $\Pi_{n}$ & $\Theta_{n}$  \\ \hline
      $\sharp$ mult's & $\frac{\tilde{L}+L}{4}n$ 
      & $\frac{\tilde{L}+L}{2}n$ \\ \hline
      $\sharp$ add's & $\frac{\tilde{L}+L}{4}n$  
      & $\frac{\tilde{L}+L-2}{2}n$ \\ \hline
    \end{tabular}
\end{center}
\caption{The operation cost of filtering a set of $n$ points
  using a pair of biorthogonal odd-odd filters 
  $H^{b}_{L},G^{b}_{\tilde{L}}$ by 2 different techniques.}
    \label{odd-odd biorthogonal operation cost} 
\end{table}

We note that by Theorem \ref{general odd-odd result} 
the inverse wavelet transform has a decomposition given by reversing 
the order of the operators $\hat{W}^{-t}_{2}(\alpha_{j})$ and
replacing each by its inverse.

\subsection{The symmetric odd-odd case.}

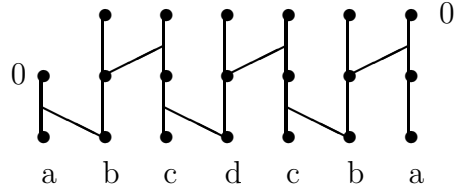
\begin{figure}[h]
  \begin{center}
    \input{thesisfig12.latex}
  \end{center}
  \caption{2-step reduction of a symmetric filter of odd length using some
    $\hat{F}_{2}(\alpha_{1}),\hat{F}^{t}_{2}(\alpha_{2})S(-1)$.}
  \label{reduction of a symmetric odd filter} 
\end{figure}

There is only one crucial observation to be made here, then the 
results follow immediately from our work in the previous section. 
This is best illustrated by an example. 
Figure \ref{reduction of a symmetric odd filter} shows the first 2
steps in reducing a symmetric filter of length 7 to a 1-point filter.

By the definition of our length reducing maps in the section
above, we have $\alpha_{2k-1} = \alpha_{2k}, 
1\leq k \leq max(\tilde{L},L)/2$, because the maps \newline 
$\hat{F}_{2}(\alpha_{2k-1})S(-(k-1))$, 
$\hat{F}^{t}_{2}(\alpha_{2k})S(-k)$
preserve symmetry in pairs. Thus we see that replacing the operator 
$\hat{F}^{t}_{2}(\alpha_{2k})S(-k)\hat{F}^{t}_{2}(\alpha_{2k-1})
S(-(k-1)), 1 \leq k \leq max(\tilde{L},L)/2$
by the operator $\tilde{F}_{3}(\alpha_{k})S(-(k-1)), 1\leq k 
\leq max(\tilde{L},L)/2$ as defined in 
(\ref{symmetric biorthogonal 33matrix mapping}), 
will improve efficiency: $\tilde{F}_{3}(\alpha_{k}): 
H^{b}_{L-2k} \mapsto H^{b}_{L-2k-2}$. 

Using similar notation to the previous section, we define

\begin{eqnarray}
  && J = \lfloor \frac{L+\tilde{L}}{4}\rfloor , \nonumber \\
  && \tilde{W}_{3}(\alpha_{k}) \equiv 
  \tilde{F}_{3}(\alpha_{k})S(-(k-1)).
  \label{definition of J and W tilde}
\end{eqnarray}  

\noindent Then we have the following refinement of 
Theorem \ref{general odd-odd result}:

\begin{corollary}
  Given a pair of biorthogonal symmetric FIR filters 
  $H^{b}_{L},G^{b}_{\tilde{L}}$ of odd lengths $L,\tilde{L}$,  
  $\Theta_{n}(H^{b}_{L},G^{b}_{\tilde{L}})$ 
  defined in (\ref{filtering operator}) may be decomposed as \newline  
  $\Theta_{n}(H^{b}_{L},G^{b}_{\tilde{L}}) = 
  S(-\lceil \frac{J}{2}\rceil)\tilde{W}^{-t}_{3}(\alpha_{J})
  \tilde{W}^{-t}_{3}(\alpha_{J-1})\cdots
  \tilde{W}^{-t}_{3}(\alpha_{1})$,
  with $\tilde{W}_{3}(\alpha_{j})$, $1\leq j \leq J$ 
  and $J$ defined as in (\ref{definition of J and W tilde}).
  \label{symmetric odd-odd result}
\end{corollary}

\proof

The only thing left to prove is that the total number of
steps equals $J$ as defined in (\ref{definition of J and W tilde}).
But this follows easily by the same argumentation as in the proof of 
Theorem \ref{general odd-odd result}.

\endproof 

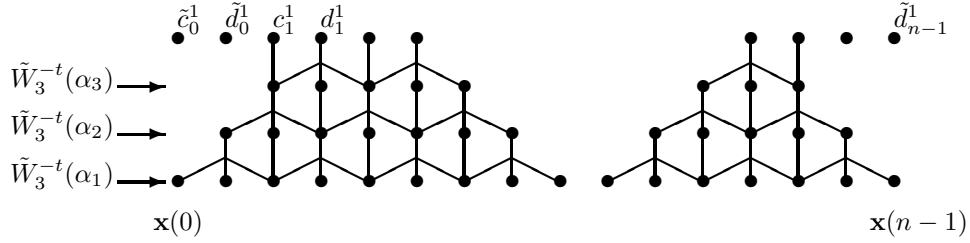
\begin{figure}[h]
  \begin{center}
    \input{figur9.latex}
  \end{center}
  \caption{Filtering diagram for a symmetric pair of conjugate filters of length $7$ and $5$.}
  \label{filteringladder3}
\end{figure}

A schematic picture of the algorithm that results from the
factorization of $\Theta_{n}(H^{b}_{L},G^{b}_{\tilde{L}})$ 
in this case of symmetric odd length filters is shown in 
Figure \ref{filteringladder3} below with the same use of symbols
as in Figure  \ref{filteringladder2}.

The ``tilde'' c's and d's at the edges of the filtering-diagram above can
be obtained by mirroring ${\bf x}$ about its endpoints or periodizing.
Counting operations, we see that we have to do 1
multiplication and 2 additions pr. pair of points for each
$\tilde{W}_{m(j)}(\alpha_{j})$. We write

\begin{equation}
  \Pi_{n}(H^{b}_{L},G^{b}_{\tilde{L}}) 
  \equiv \prod_{j=J}^{1}
  \tilde{W}^{-t}_{m(j)}(\alpha_{j}).
  \label{symmetric odd-odd factored wavelet transform}
\end{equation}
 
\noindent By construction 
$\Pi_{n}=S(\lceil\frac{J}{2}\rceil-1)\Theta_{n}$. 
$\Pi_{n}$ has a more efficient implementation as shown 
in Table \ref{symmetric odd-odd biorthogonal operation cost}. 
We note that if the filters $H,G$ have inner zero-coefficients, we
simply replace $\tilde{F}_{3}(\alpha)$ by the composition
$\hat{F}^{t}(\alpha)\hat{F}_{m}(\alpha)$, some $m\geq 3$, and it is
easy to see that this does not affect the total number of additions
and multiplications. 
\footnote{The cost result shown 
in Table \ref{symmetric odd-odd biorthogonal operation cost} 
is only an upper bound if $H^{b}_{L},G^{b}_{\tilde{L}}$ 
have internal zero coefficients.}

\begin{table}[h]
  \setlength{\extrarowheight}{6pt}
  \begin{center}
    { \begin{tabular}{|c|c|c|}\hline
        Operation 
        & $\Pi_{n}$ 
        & $\Theta_{n}$  \\ \hline 
        $\sharp$ mult's 
        &  $\frac{1}{2}\lfloor \frac{L+\tilde{L}}{4}\rfloor n$ 
        & $\frac{\tilde{L}+L}{2}n$ \\ \hline
        $\sharp$ add's
        & $\lfloor \frac{L+\tilde{L}}{4}\rfloor n$
        & $\frac{\tilde{L}+L-2}{2}n$ \\ \hline
      \end{tabular}}
  \end{center}
  \caption{The cost of filtering a set of $n$ points
    using a pair of biorthogonal symmetric odd-odd filters 
    $H^{b}_{L},G^{b}_{\tilde{L}}$ by 2 different techniques.}
  \label{symmetric odd-odd biorthogonal operation cost} 
\end{table}

\subsection{The even-odd case.}

We assume without loss of generality that $L$ is odd and $\tilde{L}$
is even. We observe that there is only one $n$ such that 
$H^{b}_{L}(n)\tilde{H}^{b}_{\tilde{L}}(n) \neq 0$. 

Now, define 

\begin{eqnarray}
  && J = \frac{L+\tilde{L}-1}{2} \nonumber \\
  && \hat{W}_{m(j)}(\alpha_{j}) \mbox{  as in (\ref{definition of J and
      W hat})} \\
  \label{second definition of J and W hat}
\end{eqnarray}

\noindent Then we have the result:

\begin{corollary}
  Given a pair of biorthogonal FIR filters 
  $H^{b}_{L},G^{b}_{\tilde{L}}$ of odd-even lengths $L,\tilde{L}$, 
  there exists a sequence of integers 
  $\left\{m(j)\right\}_{j=1}^{J} 
  \subset \bN -\{1\}$
  such that the operator $\Theta_{n}(H^{b}_{L},G^{b}_{\tilde{L}})$ 
  defined in (\ref{filtering operator}) may be decomposed as \newline   
  $\Theta_{n}(H^{b}_{L},G^{b}_{\tilde{L}}) = 
  S(1-\lfloor \frac{J}{2}\rfloor)\hat{W}^{-t}_{2}(\alpha_{J})
  \hat{W}^{-t}_{m(J-1)}(\alpha_{J-1})\cdots
  \hat{W}^{-t}_{m(1)}(\alpha_{1})$,
  with  $\hat{W}_{m(j)}(\alpha_{j})$, $1\leq j \leq J$ 
  and $J$ defined as in (\ref{second definition of J and W hat}).
  \label{even-odd result}
\end{corollary}

\proof

We only need to show that the total number of steps equals $J$ as
defined in (\ref{second definition of J and W hat}). We have

\[ H^{b}_{L} \mapsto \delta_{j,0} \Longrightarrow 
\tilde{H}^{b}_{\tilde{L}} \mapsto \tilde{H}^{b}_{\tilde{L}-L+1}
\]

\noindent Now, setting $\tilde{L}-L = 2k+1, k\geq 0$, we observe that 
$\tilde{H}^{b}_{\tilde{L}-L+1}$ has only $k+1$ nonzero coefficients
because of relation (\ref{double shift biorthogonality}). This yields

\[ J = L-1+k+1 = L-1 +\frac{\tilde{L}-L+1}{2} =
\frac{L+\tilde{L}-1}{2}.
\]

\endproof

Counting operations, we see that we have to do 1
multiplication and 1 addition pr. pair of points for each
$\hat{W}_{m(j)}(\alpha_{j}), 1\leq j \leq J$.

\begin{equation}
  \Pi_{n}(H^{b}_{L},G^{b}_{\tilde{L}}) 
  \equiv \prod_{j=J}^{1}\hat{W}^{-t}_{m(j)}(\alpha_{j}),
  \label{even-odd factored wavelet transform}
\end{equation}
 
\noindent then by construction $S(1-\lfloor \frac{J}{2}\rfloor)
\Pi_{n}= \Theta_{n}$. 
$\Pi_{n}$ has a more efficient implementation as shown in 
Table \ref{even-odd biorthogonal operation cost}. 
\footnote{The cost result shown 
  in Table \ref{even-odd biorthogonal operation cost} 
  is only an upper bound if $H^{b}_{L},G^{b}_{\tilde{L}}$ 
  have internal zero coefficients.}

\begin{table}[h]
  \setlength{\extrarowheight}{6pt}
  \begin{center}
    {\begin{tabular}{|c|c|c|}\hline
        Operation 
        & $\Pi_{n}$ & $\Theta_{n}$ \\ \hline 
        $\sharp$ mult's
        &  $\frac{L+\tilde{L}-1}{4}n$ 
        & $\frac{\tilde{L}+L}{2}n$ \\ \hline
        $\sharp$ add's
        & $\frac{L+\tilde{L}-1}{4}n$ 
        & $\frac{\tilde{L}+L-2}{2}n$\\ \hline
      \end{tabular}}
  \end{center}
  \caption{The cost of filtering a set of $n$ points
    using a pair of biorthogonal even-odd filters 
    $H^{b}_{L},G^{b}_{\tilde{L}}$ by 2 different techniques.}
  \label{even-odd biorthogonal operation cost} 
\end{table}

\subsection{The even-even case.} 

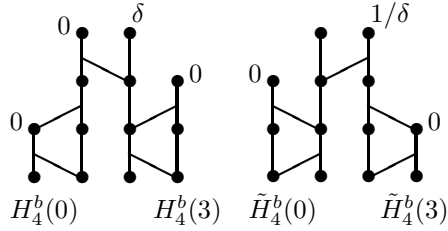
\begin{figure}[h]
  \begin{center}
    \input{figur8.latex}
  \end{center}
  \caption{The result of reducing a pair of dual biorthogonal filters of length
    4 to 1 point.} 
  \label{reduce even length common center filters} 
\end{figure}

Here, both $\tilde{L},L$ are even numbers.
There is only one important difference in this case:
The last step will {\em not} reduce the dual filter from a 2-point filter
to a 1-point filter, as illustrated in 
Figure \ref{reduce even length common center filters}.    

This flaw is easily repaired by replacing $\hat{F}_{2}(\alpha_{k})$ by
$\bar{F}_{2}(\alpha_{k},\beta)$ in the last step $k$, where $\beta$
is chosen such that $\bar{F}^{-t}_{2}(\alpha_{k},\beta)$ 
yields a 1-point dual filter. 
Using similar notation to the previous sections, we define

\begin{eqnarray}
  && J = \frac{L+\tilde{L}}{2}, \nonumber \\
  && \hat{W}_{m}(\alpha_{k}) \equiv \left\{\begin{array}{ll}
      \bar{F}_{2}(\alpha_{J},\beta)S(-(J-1)/2) 
      &  \mbox{if $k=J$.} \\
      \hat{F}_{m}(\alpha_{k})S(-\lfloor \frac{k}{2}\rfloor) & 
      \mbox{if $k, 1\leq k \leq J-1,$ is odd.} \\
      \hat{F}^{t}_{m}(\alpha_{k})S(-\lfloor \frac{k-1}{2}\rfloor) & 
      \mbox{if $k, 1\leq k \leq J,$ is even.} \nonumber \\
    \end{array}\right. \\
  \label{redefinition of J and W hat}
\end{eqnarray}  

\noindent Then we have the following result:

\begin{corollary}
  Given a pair of biorthogonal FIR filters 
  $H^{b}_{L},G^{b}_{\tilde{L}}$ of even lengths $L,\tilde{L}$, 
  there exists a sequence of integers 
  $\left\{m(j)\right\}_{j=1}^{J} 
  \subset \bN -\{1\}$
  such that the operator $\Theta_{n}(H^{b}_{L},G^{b}_{\tilde{L}})$ 
  defined in (\ref{filtering operator}) may be decomposed as \newline   
  $\Theta_{n}(H^{b}_{L},G^{b}_{\tilde{L}}) = 
  S(1-\lfloor \frac{J}{2}\rfloor)\hat{W}^{-t}_{2}(\alpha_{J})
  \hat{W}^{-t}_{m(J-1)}(\alpha_{J-1})\cdots
  \hat{W}^{-t}_{m(1)}(\alpha_{1})$,
  with  $\hat{W}_{m(j)}(\alpha_{j})$, $1\leq j \leq J$ 
  and $J$ defined as in (\ref{redefinition of J and W hat}).
  \label{even-even result}
\end{corollary}

Counting operations, we see that we have to do 1
multiplication and 1 addition pr. pair of points for each
$\hat{W}_{m(j)}(\alpha_{j}), 1\leq j < J$, and 
2 multiplications and 2 additions pr. pair of points
when implementing $\hat{W}_{J}(\alpha_{J})$. Writing

\begin{equation}
  \Pi_{n}(H^{b}_{L},G^{b}_{\tilde{L}}) 
  \equiv \prod_{j=J}^{1}\hat{W}^{-t}_{m(j)}(\alpha_{j}),
  \label{even-even factored wavelet transform}
\end{equation}
 
\noindent then by construction $S(1-\lfloor \frac{J}{2}\rfloor)
\Pi_{n}= \Theta_{n}$. 
$\Pi_{n}$ has a more efficient implementation as shown in 
Table \ref{even-even biorthogonal operation cost}. 
\footnote{The cost result shown 
  in Table \ref{even-even biorthogonal operation cost} 
  is only an upper bound if $H^{b}_{L},G^{b}_{\tilde{L}}$ 
  have internal zero coefficients.}

\begin{table}[h]
  \setlength{\extrarowheight}{6pt}
  \begin{center}
    {\begin{tabular}{|c|c|c|}\hline
        Operation 
        & $\Pi_{n}$ & $\Theta_{n}$ \\ \hline 
        $\sharp$ mult's
        &  $\frac{L+\tilde{L}+6}{4}n$ 
        & $\frac{\tilde{L}+L}{2}n$ \\ \hline
        {$\sharp$ add's}
        & $\frac{L+\tilde{L}+2}{4}n$ 
        & $\frac{\tilde{L}+L-2}{2}n$\\ \hline
      \end{tabular}}
  \end{center}
  \caption{The cost of filtering a set of $n$ points
    using a pair of biorthogonal even-even filters 
    $H^{b}_{L},G^{b}_{\tilde{L}}$ by 2 different techniques.}
  \label{even-even biorthogonal operation cost} 
\end{table}

\noindent {\bf Remark}. Since orthonormal FIR filters are a special case of
the filters discussed in this section, we see that it is possible 
to reformulate the theory for orthonormal filters by 
exclusively using elements in $SL_{2}(\bR)$. 
Furthermore, it is possible to achieve the same operation cost 
result as before. 

We note that in \cite{strang} is shown fast algorithms 
for computing wavelet coefficients with similar operation-counts to
ours using a different approach.

\chapter[Extension of Results From Dimension 1]
{Extension of Results From Dimension 1 to Higher Dimensions.}

\section{Introduction.}

When considering tensor wavelet bases in dimension 2 or 3, one will 
face the problem of filtering a set of $n \times n$ or 
$n \times n \times n$ points, respectively, in
an effective way, where $n$ is some power of 2. Assuming
identical MRA's in all dimensions and using notation as in the 
previous chapter, we look for efficient 
implementations of $\Theta_{n}(H,G)\otimes \Theta_{n}(H,G)$,  
$\Theta_{n}(H,G)\otimes \Theta_{n}(H,G)\otimes \Theta_{n}(H,G)$. Proceeding  
straightforward, we can use the factorization of the wavelet transform
in dimension 1 separately in each of the 2 or 3 dimensions, 
and thus achieve the same operation cost result as before: Reducing 
the operation count roughly by half. However, in both the orthonormal and 
biorthogonal case it is possible to obtain substantially better results.  

\section{Dimension 2.}

\subsection{The orthonormal case.}

Since our factored wavelet transform in dimension 1 only works with
2 points at a time, and tensor products commute,
we can restrict to a square consisting of a set of
$2 \times 2$ points at a time, as illustrated in 
Figure \ref{2dim-unitcell1} below. 

\begin{figure}[h]
  \begin{center}
    \input{figur12.latex}
  \end{center}
  \caption{Filtering-diagram for the discrete wavelet transform 
    in dimension 2.} 
  \label{2dim-unitcell1}  
\end{figure}
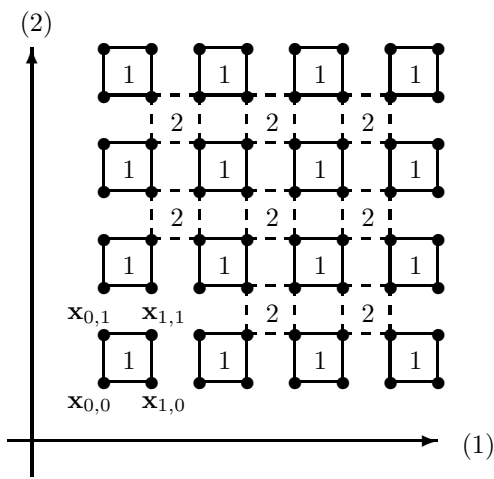

\noindent The Figure \ref{2dim-unitcell1} illustrates the division of 
the point-set into point-squares. 
Each point-square represents the map induced by 
restricting some $F_{2}(\alpha)\otimes F_{2}(\alpha)$ to this point
set, more specificly each line connecting a pair of points  represents the 
map induced by restricting some $F_{2}(\alpha)$ to this pair of points. 
The squares will alternatingly be in the position marked $1$
and $2$ as $k$ in $F_{2}(\alpha_{k})\otimes F_{2}(\alpha_{k})$ varies. 
It will turn out favourably to use the matrix $M_{2}(\alpha)$ on
its trigonometric form, that is

\begin{equation}
  M_{2}(\alpha) = \left(\begin{array}{cc}
      \cos\alpha & -\sin\alpha \\
      \sin\alpha & \cos\alpha
    \end{array} \right).
  \label{SO2 matrix on trigonometric form} 
\end{equation}

\noindent Using this form and restricting to a point-square, we write

\begin{equation}
  \left. F_{2}(\alpha)\otimes F_{2}(\alpha)\right|_{2\times 2}: 
  \left\{ \begin{array}{cc}
      {\bf x}_{0,0} & {\bf x}_{1,0} \\
      {\bf x}_{0,1} & {\bf x}_{1,1}
    \end{array} \right\} \mapsto  \left\{ \begin{array}{cc}
      {\bf y}_{0,0} & {\bf y}_{1,0} \\
      {\bf y}_{0,1} & {\bf y}_{1,1}
    \end{array} \right\}. 
  \label{F tensor F on a square}
\end{equation}

\noindent Computing on a square yields

\begin{eqnarray*}
  \left( \begin{array}{c}
      {\bf y}_{0,0} \\ 
      {\bf y}_{0,1}
    \end{array} \right) =  
  \left( \begin{array}{c}
      \cos^{2}\alpha(
      {\bf x}_{0,0} - {\bf x}_{1,1}) - 
      \sin\alpha\cos\alpha({\bf x}_{0,1} + {\bf x}_{1,0}) +
      {\bf x}_{1,1} \\
      - \sin^{2}\alpha({\bf x}_{0,1} + {\bf x}_{1,0}) + \sin\alpha \cos\alpha
      ({\bf x}_{0,0} - {\bf x}_{1,1}) + {\bf x}_{0,1} \end{array} \right), \\
  \\
  \left( \begin{array}{c}
      {\bf y}_{1,0} \\
      {\bf y}_{1,1}
    \end{array} \right) = \left( \begin{array}{c}
      -\sin^{2}\alpha({\bf x}_{0,1} + {\bf x}_{1,0}) 
      + \sin\alpha\cos\alpha ({\bf x}_{0,0} - {\bf x}_{1,1}) + {\bf x}_{1,0} \\
      - \cos^{2}\alpha({\bf x}_{0,0} - {\bf x}_{1,1}) + \sin\alpha \cos\alpha
      ({\bf x}_{0,1} + {\bf x}_{1,0}) + {\bf x}_{0,0} \end{array} \right). 
\end{eqnarray*} \newline 

\noindent Writing 

\begin{equation*}
  d = \cos^{2}\alpha({\bf x}_{0,0} - {\bf x}_{1,1}) -
  \sin\alpha\cos\alpha({\bf x}_{0,1} + {\bf x}_{1,0}),
\end{equation*}

\noindent we get 

\begin{eqnarray*}
  {\bf y}_{0,0} &= &d + {\bf x}_{1,1}  \\
  {\bf y}_{0,1} &=& \cot \alpha \cdot d + {\bf x}_{0,1}  \\
  {\bf y}_{1,0} &=& \cot \alpha \cdot d + {\bf x}_{1,0}  \\
  {\bf y}_{1,1} &=& - d + {\bf x}_{0,0}. 
\end{eqnarray*}

\noindent Combining these relations, we see that 
$\left. F_{2}(\alpha)\otimes F_{2}(\alpha)\right|_{2\times 2}$ 
can be implemented by 3 multiplications and 7 additions. 
Using notation as in dimension 1, we write 

\begin{equation}
  \Xi_{n}(H^{d}_{L},G^{d}_{L}) \equiv  
  \prod_{j=L/2}^{1}\left.(W_{2}(\alpha_{j})\otimes W_{2}(\alpha_{j}))
  \right|_{2\times 2},
  \label{square factored orthonormal 2tensor wavelet transform}
\end{equation}
 
\noindent Then by construction 
$\Xi_{n}= S(L/2-1)\Theta_{n}\otimes S(L/2-1)\Theta_{n}$, but $\Xi_{n}$
has a more efficient implementation, as shown in 
Table \ref{2tensor orthonormal results}.
\footnote{The addition of 1 to $L$ in the multiply-cost for 
  $\Pi_{n}\otimes \Pi_{n}$ results from scalar multiplication by 
  $R^{2}_{L/2}$.}

\begin{table}[h]
  \setlength{\extrarowheight}{6pt}
  \begin{center}
    \begin{tabular}{|l|c|c|c|}\hline
     Operation & $\Xi_{n}$ 
     & $\Pi_{n}\otimes \Pi_{n}$
     & $\Theta_{n}\otimes \Theta_{n}$ \\ \hline
     $\sharp$ mult's & $\frac{3}{8}Ln^{2}$ 
     & $(L+1)n^{2}$ & $2Ln^{2}$ \\ \hline 
     $\sharp$ add's &  $\frac{7}{8}Ln^{2}$ 
     & $Ln^{2}$ & $(2L-2)n^{2}$ \\ \hline
   \end{tabular}
 \end{center}
 \caption{The operation-cost of filtering a set of $n\times n$ points
   using the pair of filters $H^{d}_{L},G^{d}_{L}$ 
   by 3 different techniques.}
 \label{2tensor orthonormal results} 
\end{table}

\subsection{The biorthogonal case }
 
Arguing as in the last section and using the same notation as in 
dimension 1, we write 

\begin{equation} 
  \left. \hat{F}^{-t}_{2}(\alpha)\otimes
    \hat{F}^{-t}_{2}(\alpha)\right|_{2\times 2}:
  \left\{ \begin{array}{cc}
      {\bf x}_{0,0} & {\bf x}_{1,0}  \\
      {\bf x}_{0,1} & {\bf x}_{1,1}  \\
    \end{array} \right\} \mapsto  \left\{ \begin{array}{cc}
      {\bf y}_{0,0} & {\bf y}_{1,0}  \\
      {\bf x}_{0,1} & {\bf y}_{1,1}  \\
    \end{array} \right\} 
  \label{F hat tensor F hat}
\end{equation}

\begin{eqnarray}
  && \left. \tilde{F}^{-t}_{3}(\alpha)\otimes
    \tilde{F}^{-t}_{3}(\alpha)\right|_{3\times 3}: 
  \left\{ \begin{array}{ccc}
      {\bf x}_{0,0} & {\bf x}_{1,0} & {\bf x}_{2,0} \\
      {\bf x}_{0,1} & {\bf x}_{1,1} & {\bf x}_{2,1} \\
      {\bf x}_{0,2} & {\bf x}_{1,2} & {\bf x}_{2,2}
    \end{array} \right\} \mapsto  \nonumber \\
  && \left\{ \begin{array}{ccc}
      {\bf x}_{0,0} & {\bf y}_{1,0} & {\bf x}_{2,0} \\
      {\bf y}_{0,1} & {\bf y}_{1,1} & {\bf y}_{2,1} \\
      {\bf x}_{0,2} & {\bf y}_{1,2} & {\bf x}_{2,2}
    \end{array} \right\}. 
  \label{F tilde tensor F tilde}
\end{eqnarray}

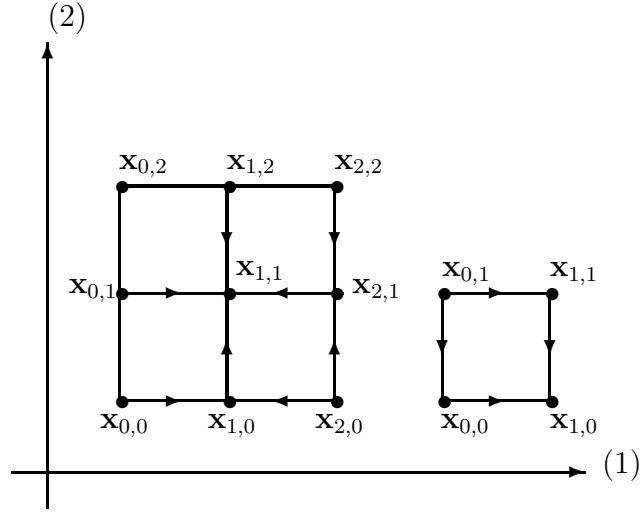
\begin{figure}[h]
  \begin{center}
    \input{thesisfig13.latex}
  \end{center}
  \caption{The point-squares for symmetric and non-symmetric 
    biorthogonal filters, respectively.}
  \label{2dim-unitcell2} 
\end{figure}

We see that for even-even filters and general odd-odd filters we 
can restrict to a set of $2\times 2$ points at a time, while for
symmetric odd-odd filters we restrict to a set of $3\times 3$ points at
a time. This ``factorization'' of operations allows us to implement 
the biorthogonal wavelet transform more efficiently in dimension 2, as
illustrated in Figure \ref{2dim-unitcell2}. Each arrow-line in the
figure means multiplication of the point of origin of the arrow by the
number $\alpha$, followed by addition of the product to the point 
of termination of the arrow line. 

To implement $\hat{F}^{-t}_{2}(\alpha)\otimes 
\hat{F}^{-t}_{2}(\alpha)$ efficiently, we compute on a 
square of $2\times 2$ points as shown in Figure \ref{2dim-unitcell2}.
Writing 

\begin{eqnarray*}
  {\bf y}_{0,0} & = & {\bf x}_{0,0} +\alpha {\bf x}_{0,1} \\
  {\bf y}_{0,1} & = & {\bf x}_{0,1} \\
  {\bf y}_{1,1} & = & {\bf x}_{1,1} + \alpha {\bf x}_{0,1} \\
  {\bf y}_{1,1} & = & {\bf x}_{1,0} + 
  \alpha({\bf y}_{1,1}+{\bf x}_{0,0}),
\end{eqnarray*}
  
\noindent we see that we can implement 
$\left. \hat{F}^{-t}_{2}(\alpha)\otimes 
\hat{F}^{-t}_{2}(\alpha)\right|_{2\times 2}$ by 2 multiplications and
4 additions. 

Similarly, to implement $\tilde{F}^{-t}_{3}(\alpha)\otimes 
\tilde{F}^{-t}_{3}(\alpha)$ efficiently, we write

\begin{eqnarray*}
  {\bf y}_{0,0} & = & {\bf x}_{0,0} \\
  {\bf y}_{0,1} & = & {\bf x}_{0,1} \\
  {\bf y}_{0,2} & = & {\bf x}_{0,2}  \\
  {\bf y}_{1,2} & = & {\bf x}_{1,2}  \\
  {\bf y}_{2,2} & = & {\bf x}_{2,2}  \\
  {\bf y}_{2,0} & = & {\bf x}_{2,0}  \\
  {\bf y}_{1,0} & = & {\bf x}_{1,0} + 
  \alpha{\bf x}_{0,0}+\alpha{\bf x}_{2,0} \\
  {\bf y}_{2,1} & = & {\bf x}_{2,1} + 
  \alpha{\bf x}_{2,2}+\alpha{\bf x}_{2,0} \\
  {\bf y}_{1,1} & = & {\bf x}_{1,1} + 
  \alpha({\bf y}_{1,0}+\alpha{\bf x}_{2,2}+{\bf x}_{2,1}+
  {\bf x}_{0,1}+{\bf x}_{1,2}) 
\end{eqnarray*}
  
\noindent Thus, we see that 
$\left. \tilde{F}^{-t}_{3}(\alpha)
\otimes \tilde{F}^{-t}_{3}(\alpha)\right|_{3\times 3}$ can be
implemented by 2 multiplications and 8 additions, assuming that we
have stored the numbers $\alpha{\bf x}_{2,2},\alpha{\bf x}_{0,0},
\alpha{\bf x}_{2,2}+{\bf x}_{1,2}$ from previous steps in the
implementation of the algorithm.   

Using notation as in (\ref{definition of J and W hat}),
in the general odd-odd case we define 

\begin{equation}
  \Xi_{n}(H^{b}_{L},G^{b}_{\tilde{L}}) \equiv  
  \prod_{j=J}^{1}\left. (\hat{W}_{2}(\alpha_{j})\otimes 
    \hat{W}_{2}(\alpha_{j})) \right|_{2\times 2},
  \label{square factored biorthogonal odd-odd 2tensor wavelet transform}
\end{equation}
 
\noindent and thus get $\Xi_{n}=S(\lfloor J/2\rfloor)\Theta_{n}
\otimes S(\lfloor J/2\rfloor)\Theta_{n}$.

In the odd-even case we define 

\begin{equation}
  \Xi_{n}(H^{b}_{L},G^{b}_{\tilde{L}}) \equiv  
  \prod_{j=J}^{1}\left. (\hat{W}_{2}(\alpha_{j})\otimes 
    \hat{W}_{2}(\alpha_{j})) \right|_{2\times 2},
  \label{square factored biorthogonal odd-even 2tensor wavelet transform}
\end{equation}

\noindent where we use the notation defined in 
(\ref{second definition of J and W hat}). We get 
$\Xi_{n} = S(\lfloor \frac{J}{2}\rfloor-1)
\Theta_{n}\otimes S(\lfloor \frac{J}{2}\rfloor-1)\Theta_{n}$.

In the even-even case we define 

\begin{equation}
  \Xi_{n}(H^{b}_{L},G^{b}_{\tilde{L}}) \equiv  
  \prod_{j=J}^{1}\left. (\hat{W}_{2}(\alpha_{j})\otimes 
    \hat{W}_{2}(\alpha_{j})) \right|_{2\times 2},
  \label{square factored biorthogonal even-even 2tensor wavelet transform}
\end{equation}

\noindent where we use the notation defined in 
(\ref{redefinition of J and W hat}). We get 
$\Xi_{n} = S(\lfloor \frac{J}{2}\rfloor-1)
\Theta_{n}\otimes S(\lfloor \frac{J}{2}\rfloor-1)\Theta_{n}$.

Finally, in the symmetric odd-odd case we define 

\begin{equation}
  \Xi_{n}(H^{b}_{L},G^{b}_{\tilde{L}}) \equiv  
  \prod_{j=J}^{1}\left. (\tilde{W}_{3}(\alpha_{j})\otimes 
    \tilde{W}_{3}(\alpha_{j})) \right|_{3\times 3},
  \label{square factored biorthogonal odd-odd symmetric 
    2tensor wavelet transform}
\end{equation}

\noindent where we use the notation in (\ref{definition of J and W
  tilde}). This yields
$\Xi_{n}= S(\lceil \frac{J}{2}\rceil)\Theta_{n}\otimes 
S(\lceil \frac{J}{2}\rceil)\Theta_{n}$.

Thus, we get the operation-cost results shown in Table 
\ref{2tensor odd-odd biorthogonal results}, Table 
\ref{2tensor even-even biorthogonal results}, Table 
\ref{2tensor symmetric odd-odd biorthogonal results}. 
\footnote{The operation-counts shown in 
  Table \ref{2tensor odd-odd biorthogonal results},
  Table \ref{2tensor odd-even biorthogonal results},
  Table \ref{2tensor even-even biorthogonal results}, 
  Table \ref{2tensor symmetric odd-odd biorthogonal results}, are only
  upper bounds if $H^{b}_{L},G^{b}_{\tilde{L}}$ have inner zero
  coefficients.}

\begin{table}[h]
  \setlength{\extrarowheight}{6pt}
  \begin{center}
    {\begin{tabular}{|l|c|c|c|}\hline
        Operation & $\Xi_{n}$ & $\Pi_{n}\otimes \Pi_{n}$ 
        & $\Theta_{n}\otimes \Theta_{n}$  \\ \hline
        $\sharp$ mult's 
        & $\frac{L+\tilde{L}}{4}n^{2}$ 
        & $\frac{L+\tilde{L}}{2}n^{2}$ 
        & $(L+\tilde{L})n^{2}$ \\ \hline  
        $\sharp$ add's
        & $\frac{L+\tilde{L}}{2}n^{2}$ 
        & $\frac{L+\tilde{L}}{2}n^{2}$ 
        & $(L+\tilde{L}-2)n^{2}$ \\ \hline
      \end{tabular}}
  \end{center}
  \caption{The operation cost of filtering a set of $n \times n$ points
    using a pair of biorthogonal odd-odd filters
    $H^{b}_{L},G^{b}_{\tilde{L}}$ by 3 different techniques.}
  \label{2tensor odd-odd biorthogonal results} 
\end{table}

\begin{table}[h]
  \setlength{\extrarowheight}{6pt}
  \begin{center}
    {\begin{tabular}{|l|c|c|c|}\hline
        Operation & $\Xi_{n}$ & $\Pi_{n}\otimes \Pi_{n}$ 
        & $\Theta_{n}\otimes \Theta_{n}$  \\ \hline
        $\sharp$ mult's 
        & $\frac{L+\tilde{L}-1}{4}n^{2}$ 
        & $\frac{L+\tilde{L}-1}{2}n^{2}$ 
        & $(L+\tilde{L})n^{2}$ \\ \hline  
        $\sharp$ add's
        & $\frac{L+\tilde{L}-1}{2}n^{2}$ 
        & $\frac{L+\tilde{L}-1}{2}n^{2}$ 
        & $(L+\tilde{L}-2)n^{2}$ \\ \hline
      \end{tabular}}
  \end{center}
  \caption{The operation cost of filtering a set of $n \times n$ points
    using a pair of biorthogonal odd-even filters
    $H^{b}_{L},G^{b}_{\tilde{L}}$ by 3 different techniques.}
  \label{2tensor odd-even biorthogonal results} 
\end{table}

\begin{table}[h]
  \setlength{\extrarowheight}{6pt}
  \begin{center}
    {\begin{tabular}{|l|c|c|c|}\hline
        Operation & $\Xi_{n}$ & $\Pi_{n}\otimes \Pi_{n}$ 
        & $\Theta_{n}\otimes \Theta_{n}$  \\ \hline
        $\sharp$ mult's 
        & $\frac{L+\tilde{L}+10}{4}n^{2}$ 
        & $\frac{L+\tilde{L}+4}{2}n^{2}$ 
        & $(L+\tilde{L})n^{2}$ \\ \hline  
        $\sharp$ add's
        & $\frac{L+\tilde{L}+2}{2}n^{2}$ 
        & $\frac{L+\tilde{L}+2}{2}n^{2}$ 
        & $(L+\tilde{L}-2)n^{2}$ \\ \hline
      \end{tabular}}
  \end{center}
  \caption{The operation cost of filtering a set of $n \times n$ points
    using a pair of biorthogonal even-even filters
    $H^{b}_{L},G^{b}_{\tilde{L}}$ by 3 different techniques.}
  \label{2tensor even-even biorthogonal results} 
\end{table}

\begin{table}[h]
  \setlength{\extrarowheight}{6pt}
  \begin{center}
    {\begin{tabular}{|l|c|c|c|}\hline
        Operation & $\Xi_{n}$ & $\Pi_{n}\otimes \Pi_{n}$ 
        & $\Theta_{n}\otimes \Theta_{n}$  \\ \hline
        $\sharp$ mult's 
        & $\frac{1}{2}\lfloor \frac{L+\tilde{L}}{4}\rfloor n^{2}$ 
        & $\lfloor \frac{L+\tilde{L}}{4}\rfloor n^{2}$ 
        & $(L+\tilde{L})n^{2}$ \\ \hline  
        $\sharp$ add's
        & $\frac{9}{4}\lfloor \frac{L+\tilde{L}}{4}\rfloor n^{2}$ 
        & $2\lfloor \frac{L+\tilde{L}}{4}\rfloor n^{2}$ 
        & $(L+\tilde{L}-2)n^{2}$ \\ \hline
      \end{tabular}}
  \end{center}
  \caption{The operation cost of filtering a set of $n \times n$ points
    using a pair of biorthogonal symmetric odd-odd filters
    $H^{b}_{L},G^{b}_{\tilde{L}}$ by 3 different techniques.}
  \label{2tensor symmetric odd-odd biorthogonal results} 
\end{table}

\section{Dimension 3.}

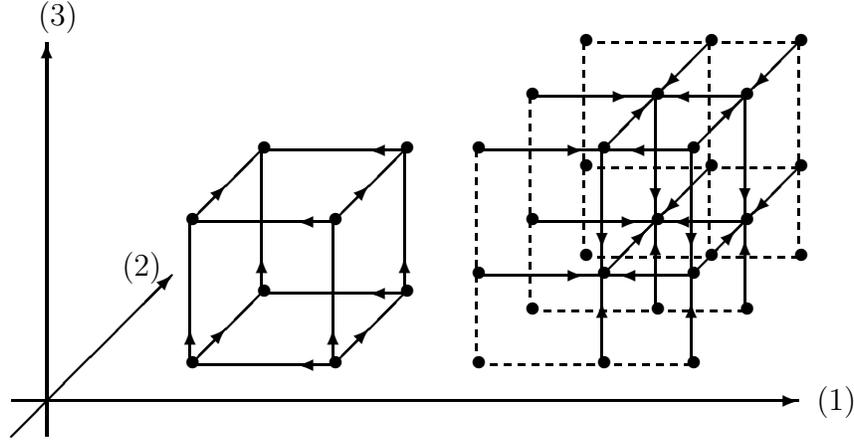
\begin{figure}[h]
  \begin{center}
    \input{figur15.latex}
  \end{center}
  \caption{The cube for biorthogonal odd-odd/even-even filters on the
    left and for symmetric odd-odd filters on the right.}
  \label{3dim-unitcell} 
\end{figure}

It is straightforward to see that we can group the operations of 
$F_{2}(\alpha)\otimes F_{2}(\alpha)\otimes F_{2}(\alpha)$,
$\hat{F}_{2}(\alpha)\otimes \hat{F}_{2}(\alpha)\otimes 
\hat{F}_{2}(\alpha)$ into cubes of $2\times 2\times 2$ points, 
and $\tilde{F}_{3}(\alpha)\otimes \tilde{F}_{3}(\alpha)\otimes 
\tilde{F}_{3}(\alpha)$ into cubes of $3\times 3\times 3$ points. This 
is illustrated in Figure \ref{3dim-unitcell}. To make things easy, we will use 
some ``loose'' geometrical terminology. 

\subsection{The orthonormal case.}

We did not succeed in finding any symmetries in the 
trigonometric expressions in the 3-dimensional case. The cheapest 
arrangement of operations will therefore be to apply
$\left. W_{2}(\alpha)\otimes W_{2}(\alpha)\right|_{2\times 2}$
to 2 ``opposing planes'' of $2\times 2$ points in the 
$2\times 2\times 2$ cube, and then $W_{2}(\alpha)$ to the 4 ``lines''
that connect these 2 planes. 
Thus, one has to do 14 multiplications and 22 additions
pr. cube, not counting normalization. We denote this ``factored'' 
wavelet transform in 3 dimensions on $n\times n\times n$ 
points by $\Omega_{n}(H^{d}_{L},G^{d}_{\tilde{L}})$. Like before, 
we have $\Omega_{n}= S(L/2-1)\Theta_{n}\otimes S(L/2-1)\Theta_{n}
\otimes S(L/2-1)\Theta_{n}$. 
The operation cost-result is shown in 
Table \ref{3tensor orthonormal results}. \footnote{The addition of 1
  to $\frac{7}{8}L$ and $\frac{3}{2}L$ is because of normalization 
  by $R_{L/2}^{3}$.}

\begin{table}[h]
  \setlength{\extrarowheight}{6pt}
  \begin{center}
    {\begin{tabular}{|l|c|c|c|}\hline
        Operation & $\Omega_{n}$ & $\Pi_{n}\otimes \Pi_{n} \otimes \Pi_{n}$ 
        & $\Theta_{n}\otimes \Theta_{n}\otimes \Theta_{n}$ \\ \hline
        $\sharp$ mult's
        & $(\frac{7}{8}L+1)n^{3}$ & $(\frac{3}{2}L+1)n^{3}$ 
        & $3Ln^{3}$   \\ \hline      
        $\sharp$ add's  
        & $\frac{11}{8}Ln^{3}$ & $\frac{3}{2}Ln^{3}$ &
        $3(L-1)n^{3}$ \\  \hline 
      \end{tabular}}
  \end{center}
  \caption{The cost of filtering a set of $n\times
    n\times n$ points using the pair of filters $H^{d}_{L},G^{d}_{L}$,
    by 3 different techniques.}
  \label{3tensor orthonormal results}
\end{table}

\subsection{The biorthogonal case.}

By rearranging the arithmetic operations,
we find we have to do 6 multiplications and 12 additions pr. cube in the
even-even/odd-even and non-symmetric odd-odd case.
In the symmetric odd-odd case we found an operation arrangement
yielding 6 multiplications and 24 additions pr. cube. 
All these operation counts are easily verified by considering 
Figure \ref{3dim-unitcell}. Denoting in each case the factored
biorthogonal discrete wavelet transform on $n\times n\times n$ points 
by $\Omega_{n}(H^{b}_{L},G^{b}_{\tilde{L}})$, we get that up to shifts,
$\Omega_{n}$ equals $\Theta_{n}\otimes \Theta_{n}\otimes \Theta_{n}$.
The cost-results are shown in 
Table \ref{3tensor odd-odd biorthogonal results},
Table \ref{3tensor odd-even biorthogonal results},
Table \ref{3tensor even-even biorthogonal results} and
Table \ref{3tensor symmetric odd-odd biorthogonal results}.
\footnote{If $H^{b}_{L},G^{b}_{\tilde{L}}$ has inner
  zero-coefficients, the operation counts in 
  Table \ref{3tensor odd-odd biorthogonal results},
  Table \ref{3tensor odd-even biorthogonal results},
  Table \ref{3tensor even-even biorthogonal results} and
  Table \ref{3tensor symmetric odd-odd biorthogonal results} are only
  upper bounds.}

\begin{table}[h]
  \setlength{\extrarowheight}{6pt}
  \begin{center}
    {\begin{tabular}{|l|c|c|c|}\hline
        Operation & $\Omega_{n}$ & $\Pi_{n}\otimes \Pi_{n}\otimes \Pi_{n}$
        & $\Theta_{n}\otimes \Theta_{n}\otimes \Theta_{n}$ \\ \hline
        $\sharp$ mult's
        & $\frac{3}{8}(L+\tilde{L})n^{3}$ 
        & $\frac{3}{4}(L+\tilde{L})n^{3}$ 
        & $\frac{3}{2}(\tilde{L}+L)n^{3}$   \\ \hline     
        $\sharp$ add's
        & $\frac{3}{4}(L+\tilde{L})n^{3}$ 
        & $\frac{3}{4}(L+\tilde{L})n^{3}$
        & $\frac{3}{2}(L+\tilde{L}-2)n^{3}$ \\ \hline 
      \end{tabular}}
  \end{center}
\caption{The cost of filtering a set of $n \times
    n\times n$ points using a pair of odd-odd filters 
    $H^{b}_{L},G^{b}_{\tilde{L}}$, by 3 different techniques.}
  \label{3tensor odd-odd biorthogonal results}
\end{table}

\begin{table}[h]
  \setlength{\extrarowheight}{6pt}
  \begin{center}
    {\begin{tabular}{|l|c|c|c|}\hline
        Operation & $\Omega_{n}$ & $\Pi_{n}\otimes \Pi_{n}\otimes \Pi_{n}$
        & $\Theta_{n}\otimes \Theta_{n}\otimes \Theta_{n}$ \\ \hline
        $\sharp$ mult's
        & $\frac{3}{8}(L+\tilde{L}-1)n^{3}$ 
        & $\frac{3}{4}(L+\tilde{L}-1)n^{3}$ 
        & $\frac{3}{2}(L+\tilde{L})n^{3}$   \\ \hline     
        $\sharp$ add's
        & $\frac{3}{4}(L+\tilde{L}-1)n^{3}$ 
        & $\frac{3}{4}(L+\tilde{L}-1)n^{3}$
        & $\frac{3}{2}(L+\tilde{L}-2)n^{3}$ \\ \hline 
      \end{tabular}}
  \end{center}
\caption{The cost of filtering a set of $n \times
    n\times n$ points using a pair of odd-even filters 
    $H^{b}_{L},G^{b}_{\tilde{L}}$, by 3 different techniques.}
  \label{3tensor odd-even biorthogonal results}
\end{table}

\begin{table}[h]
  \setlength{\extrarowheight}{6pt}
  \begin{center}
    {\begin{tabular}{|l|c|c|c|}\hline
        Operation & $\Omega_{n}$ & $\Pi_{n}\otimes \Pi_{n}\otimes \Pi_{n}$
        & $\Theta_{n}\otimes \Theta_{n}\otimes \Theta_{n}$ \\ \hline
        $\sharp$ mult's
        & $\frac{3}{8}(L+\tilde{L}+\frac{14}{3})n^{3}$ 
        & $\frac{3}{4}(L+\tilde{L}+\frac{10}{3})n^{3}$ 
        & $\frac{3}{2}(L+\tilde{L})n^{3}$   \\ \hline     
        $\sharp$ add's
        & $\frac{3}{4}(L+\tilde{L})n^{3}$ 
        & $\frac{3}{4}(L+\tilde{L}+2)n^{3}$
        & $\frac{3}{2}(L+\tilde{L}-2)n^{3}$ \\ \hline 
      \end{tabular}}
  \end{center}
\caption{The cost of filtering a set of $n \times
    n\times n$ points using a pair of even-even filters 
    $H^{b}_{L},G^{b}_{\tilde{L}}$, by 3 different techniques.}
  \label{3tensor even-even biorthogonal results}
\end{table}

\begin{table}[h]
  \setlength{\extrarowheight}{6pt}
  \begin{center}
    {\begin{tabular}{|l|c|c|c|}\hline
        Operation & $\Omega_{n}$ & $\Pi_{n}\otimes \Pi_{n}\otimes \Pi_{n}$
        & $\Theta_{n}\otimes \Theta_{n}\otimes \Theta_{n}$ \\ \hline
        $\sharp$ mult's
        & $\frac{3}{4}\lfloor\frac{L+\tilde{L}}{4}\rfloor n^{3}$ 
        & $\frac{3}{2}\lfloor\frac{L+\tilde{L}}{4}\rfloor n^{3}$ 
        & $\frac{3}{2}(\tilde{L}+L)n^{3}$   \\ \hline     
        $\sharp$ add's
        & $3\lfloor\frac{L+\tilde{L}}{4}\rfloor n^{3}$ 
        & $3\lfloor\frac{L+\tilde{L}}{4}\rfloor n^{3}$
        & $\frac{3}{2}(\tilde{L}+L-2)n^{3}$ \\ \hline 
      \end{tabular}}
  \end{center}
\caption{The cost of filtering a set of $n \times
    n\times n$ points using a the pair of symmetric odd-odd filters 
    $H^{b}_{L},G^{b}_{\tilde{L}}$, by 3 different techniques.}
  \label{3tensor symmetric odd-odd biorthogonal results}
\end{table}


%% file: thesisfig9.latex
\setlength{\unitlength}{0.00053300in}%
\begingroup\makeatletter\ifx\SetFigFont\undefined
\def\x#1#2#3#4#5#6#7\relax{\def\x{#1#2#3#4#5#6}}%
\expandafter\x\fmtname xxxxxx\relax \def\y{splain}%
\ifx\x\y   
\gdef\SetFigFont#1#2#3{%
  \ifnum #1<17\tiny\else \ifnum #1<20\small\else
  \ifnum #1<24\normalsize\else \ifnum #1<29\large\else
  \ifnum #1<34\Large\else \ifnum #1<41\LARGE\else
     \huge\fi\fi\fi\fi\fi\fi
  \csname #3\endcsname}%
\else
\gdef\SetFigFont#1#2#3{\begingroup
  \count@#1\relax \ifnum 25<\count@\count@25\fi
  \def\x{\endgroup\@setsize\SetFigFont{#2pt}}%
  \expandafter\x
    \csname \romannumeral\the\count@ pt\expandafter\endcsname
    \csname @\romannumeral\the\count@ pt\endcsname
  \csname #3\endcsname}%
\fi
\fi\endgroup
\begin{picture}(3600,2394)(3001,-3472)
\thicklines
\put(3301,-3061){\line( 1, 1){600}}
\put(4501,-3061){\line( 1, 1){600}}
\put(4501,-2461){\line( 1,-1){600}}
\put(5701,-3061){\line( 1, 1){600}}
\put(5701,-2461){\line( 1,-1){600}}
\put(3901,-1861){\line( 1,-1){600}}
\put(3901,-2461){\line( 1, 1){600}}
\put(5101,-2461){\line( 1, 1){600}}
\put(5101,-1861){\line( 1,-1){600}}
\put(4501,-1261){\line( 1,-1){600}}
\put(4501,-1861){\line( 1, 1){600}}
\put(3201,-3136){\makebox(0,0)[lb]{\smash{$\bullet$}}}
\put(3801,-3136){\makebox(0,0)[lb]{\smash{$\bullet$}}}
\put(4401,-3136){\makebox(0,0)[lb]{\smash{$\bullet$}}}
\put(5001,-3136){\makebox(0,0)[lb]{\smash{$\bullet$}}}
\put(5601,-3136){\makebox(0,0)[lb]{\smash{$\bullet$}}}
\put(6201,-3136){\makebox(0,0)[lb]{\smash{$\bullet$}}}
\put(3201,-2536){\makebox(0,0)[lb]{\smash{$\bullet$}}}
\put(3801,-2536){\makebox(0,0)[lb]{\smash{$\bullet$}}}
\put(4401,-2536){\makebox(0,0)[lb]{\smash{$\bullet$}}}
\put(5001,-2536){\makebox(0,0)[lb]{\smash{$\bullet$}}}
\put(5601,-2536){\makebox(0,0)[lb]{\smash{$\bullet$}}}
\put(3301,-2461){\line( 1,-1){600}}
\put(6201,-2536){\makebox(0,0)[lb]{\smash{$\bullet$}}}
\put(4450,-1000){\makebox(0,0)[lb]{\smash{\scriptsize{$M_{2}(\alpha_{3})$}}}}
\put(3801,-1936){\makebox(0,0)[lb]{\smash{$\bullet$}}}
\put(4401,-1936){\makebox(0,0)[lb]{\smash{$\bullet$}}}
\put(5001,-1936){\makebox(0,0)[lb]{\smash{$\bullet$}}}
\put(5601,-1936){\makebox(0,0)[lb]{\smash{$\bullet$}}}
\put(4401,-1336){\makebox(0,0)[lb]{\smash{$\bullet$}}}
\put(5001,-1336){\makebox(0,0)[lb]{\smash{$\bullet$}}}
\put(4201,-1261){\makebox(0,0)[lb]{\smash{1}}}
\put(5301,-1261){\makebox(0,0)[lb]{\smash{0}}}
\put(3601,-1861){\makebox(0,0)[lb]{\smash{0}}}
\put(5901,-1861){\makebox(0,0)[lb]{\smash{0}}}
\put(3001,-2461){\makebox(0,0)[lb]{\smash{0}}}
\put(6501,-2461){\makebox(0,0)[lb]{\smash{0}}}
\put(3001,-3436){\makebox(0,0)[lb]{\smash{\scriptsize{$H^{d}_{3}(0)$}}}}
\put(6201,-3436){\makebox(0,0)[lb]{\smash{\scriptsize{$H^{d}_{3}(5)$}}}}
\put(3850,-2836){\makebox(0,0)[lb]{\smash{\scriptsize{$M_{2}(\alpha_{1})$}}}}
\put(5050,-2836){\makebox(0,0)[lb]{\smash{\scriptsize{$M_{2}(\alpha_{1})$}}}}
\put(4450,-2236){\makebox(0,0)[lb]{\smash{\scriptsize{$M_{2}(\alpha_{2})$}}}}
\end{picture}

%% file: figur2.latex
\setlength{\unitlength}{0.0125in}
\begin{picture}(280,89)(105,465)
\thicklines
\put(320,540){\line( 1,-1){ 20}}
\put(320,520){\line( 1, 1){ 20}}
\put(360,520){\line(-1,-1){ 20}}
\put(340,520){\line( 1,-1){ 20}}
\put(300,520){\line( 1,-1){ 20}}
\put(300,500){\line( 1, 1){ 20}}
\put(360,500){\line( 1,-1){ 20}}
\put(360,480){\line( 1, 1){ 20}}
\put(377,477){\makebox(0,0)[lb]{\raisebox{0pt}[0pt][0pt]{$\bullet$}}}
\put(340,500){\line(-1,-1){ 20}}
\put(320,500){\line( 1,-1){ 20}}
\put(280,500){\line( 1,-1){ 20}}
\put(280,480){\line( 1, 1){ 20}}
\put(357,477){\makebox(0,0)[lb]{\raisebox{0pt}[0pt][0pt]{$\bullet$}}}
\put(337,477){\makebox(0,0)[lb]{\raisebox{0pt}[0pt][0pt]{$\bullet$}}}
\put(317,477){\makebox(0,0)[lb]{\raisebox{0pt}[0pt][0pt]{$\bullet$}}}
\put(297,477){\makebox(0,0)[lb]{\raisebox{0pt}[0pt][0pt]{$\bullet$}}}
\put(277,477){\makebox(0,0)[lb]{\raisebox{0pt}[0pt][0pt]{$\bullet$}}}
\put(200,540){\line( 1,-1){ 20}}
\put(200,520){\line( 1, 1){ 20}}
\put(220,520){\line( 1,-1){ 20}}
\put(220,500){\line( 1, 1){ 20}}
\put(240,500){\line( 1,-1){ 20}}
\put(240,480){\line( 1, 1){ 20}}
\put(257,477){\makebox(0,0)[lb]{\raisebox{0pt}[0pt][0pt]{$\bullet$}}}
\put(237,477){\makebox(0,0)[lb]{\raisebox{0pt}[0pt][0pt]{$\bullet$}}}
\put(160,540){\line( 1,-1){ 20}}
\put(160,520){\line( 1, 1){ 20}}
\put(180,520){\line( 1,-1){ 20}}
\put(180,500){\line( 1, 1){ 20}}
\put(140,520){\line( 1,-1){ 20}}
\put(140,500){\line( 1, 1){ 20}}
\put(200,500){\line( 1,-1){ 20}}
\put(200,480){\line( 1, 1){ 20}}
\put(217,477){\makebox(0,0)[lb]{\raisebox{0pt}[0pt][0pt]{$\bullet$}}}
\put(160,500){\line( 1,-1){ 20}}
\put(160,480){\line( 1, 1){ 20}}
\put(120,500){\line( 1,-1){ 20}}
\put(120,480){\line( 1, 1){ 20}}
\put(197,477){\makebox(0,0)[lb]{\raisebox{0pt}[0pt][0pt]{$\bullet$}}}
\put(177,477){\makebox(0,0)[lb]{\raisebox{0pt}[0pt][0pt]{$\bullet$}}}
\put(157,477){\makebox(0,0)[lb]{\raisebox{0pt}[0pt][0pt]{$\bullet$}}}
\put(137,477){\makebox(0,0)[lb]{\raisebox{0pt}[0pt][0pt]{$\bullet$}}}
\put(117,477){\makebox(0,0)[lb]{\raisebox{0pt}[0pt][0pt]{$\bullet$}}}
\put(137,518){\makebox(0,0)[lb]{\raisebox{0pt}[0pt][0pt]{$\bullet$}}}
\put(380,547){\makebox(0,0)[lb]{\raisebox{0pt}[0pt][0pt]{{\footnotesize $\tilde{d}^1_{n/2-1}$}}}} 
\put(377,537){\makebox(0,0)[lb]{\raisebox{0pt}[0pt][0pt]{$\bullet$}}}
\put(357,537){\makebox(0,0)[lb]{\raisebox{0pt}[0pt][0pt]{$\bullet$}}}
\put(337,537){\makebox(0,0)[lb]{\raisebox{0pt}[0pt][0pt]{$\bullet$}}}
\put(317,537){\makebox(0,0)[lb]{\raisebox{0pt}[0pt][0pt]{$\bullet$}}}
\put(297,517){\makebox(0,0)[lb]{\raisebox{0pt}[0pt][0pt]{$\bullet$}}}
\put(317,517){\makebox(0,0)[lb]{\raisebox{0pt}[0pt][0pt]{$\bullet$}}}
\put(337,517){\makebox(0,0)[lb]{\raisebox{0pt}[0pt][0pt]{$\bullet$}}}
\put(357,517){\makebox(0,0)[lb]{\raisebox{0pt}[0pt][0pt]{$\bullet$}}}
\put(377,497){\makebox(0,0)[lb]{\raisebox{0pt}[0pt][0pt]{$\bullet$}}}
\put(357,497){\makebox(0,0)[lb]{\raisebox{0pt}[0pt][0pt]{$\bullet$}}}
\put(337,497){\makebox(0,0)[lb]{\raisebox{0pt}[0pt][0pt]{$\bullet$}}}
\put(317,497){\makebox(0,0)[lb]{\raisebox{0pt}[0pt][0pt]{$\bullet$}}}
\put(297,497){\makebox(0,0)[lb]{\raisebox{0pt}[0pt][0pt]{$\bullet$}}}
\put(277,497){\makebox(0,0)[lb]{\raisebox{0pt}[0pt][0pt]{$\bullet$}}}
\put(375,465){\makebox(0,0)[lb]{\raisebox{0pt}[0pt][0pt]{{\footnotesize
        ${\bf x}(n-1)$}}}}
\put(117,537){\makebox(0,0)[lb]{\raisebox{0pt}[0pt][0pt]{$\bullet$}}}
\put(140,547){\makebox(0,0)[lb]{\raisebox{0pt}[0pt][0pt]{{\footnotesize $\tilde{d}^1_0$}}}}
\put(118,547){\makebox(0,0)[lb]{\raisebox{0pt}[0pt][0pt]{{\footnotesize $\tilde{c}^1_0$}}}}
\put(137,537){\makebox(0,0)[lb]{\raisebox{0pt}[0pt][0pt]{$\bullet$}}}
\put(110,465){\makebox(0,0)[lb]{\raisebox{0pt}[0pt][0pt]{{\footnotesize
        ${\bf x}(0)$}}}}
\put(220,547){\makebox(0,0)[lb]{\raisebox{0pt}[0pt][0pt]{{\footnotesize $d^1_2$}}}}
\put(198,547){\makebox(0,0)[lb]{\raisebox{0pt}[0pt][0pt]{{\footnotesize $c^1_2$}}}}
\put(180,547){\makebox(0,0)[lb]{\raisebox{0pt}[0pt][0pt]{{\footnotesize $d^1_1$}}}}
\put(158,547){\makebox(0,0)[lb]{\raisebox{0pt}[0pt][0pt]{{\footnotesize $c^1_1$}}}}
\put(157,537){\makebox(0,0)[lb]{\raisebox{0pt}[0pt][0pt]{$\bullet$}}}
\put(177,537){\makebox(0,0)[lb]{\raisebox{0pt}[0pt][0pt]{$\bullet$}}}
\put(197,537){\makebox(0,0)[lb]{\raisebox{0pt}[0pt][0pt]{$\bullet$}}}
\put(217,537){\makebox(0,0)[lb]{\raisebox{0pt}[0pt][0pt]{$\bullet$}}}
\put(157,517){\makebox(0,0)[lb]{\raisebox{0pt}[0pt][0pt]{$\bullet$}}}
\put(177,497){\makebox(0,0)[lb]{\raisebox{0pt}[0pt][0pt]{$\bullet$}}}
\put(197,497){\makebox(0,0)[lb]{\raisebox{0pt}[0pt][0pt]{$\bullet$}}}
\put(217,497){\makebox(0,0)[lb]{\raisebox{0pt}[0pt][0pt]{$\bullet$}}}
\put(237,497){\makebox(0,0)[lb]{\raisebox{0pt}[0pt][0pt]{$\bullet$}}}
\put(257,497){\makebox(0,0)[lb]{\raisebox{0pt}[0pt][0pt]{$\bullet$}}}
\put(237,517){\makebox(0,0)[lb]{\raisebox{0pt}[0pt][0pt]{$\bullet$}}}
\put(217,517){\makebox(0,0)[lb]{\raisebox{0pt}[0pt][0pt]{$\bullet$}}}
\put(197,517){\makebox(0,0)[lb]{\raisebox{0pt}[0pt][0pt]{$\bullet$}}}
\put(177,517){\makebox(0,0)[lb]{\raisebox{0pt}[0pt][0pt]{$\bullet$}}}
\put(137,497){\makebox(0,0)[lb]{\raisebox{0pt}[0pt][0pt]{$\bullet$}}}
\put(157,497){\makebox(0,0)[lb]{\raisebox{0pt}[0pt][0pt]{$\bullet$}}}
\put(117,497){\makebox(0,0)[lb]{\raisebox{0pt}[0pt][0pt]{$\bullet$}}}
\put(58,520){\makebox(0,0)[lb]{\raisebox{0pt}[0pt][0pt]{{\footnotesize $W_{2}(\alpha_3)$}}}}
\put(58,500){\makebox(0,0)[lb]{\raisebox{0pt}[0pt][0pt]{{\footnotesize $W_{2}(\alpha_2)$}}}}
\put(58,480){\makebox(0,0)[lb]{\raisebox{0pt}[0pt][0pt]{{\footnotesize $W_{2}(\alpha_1)$}}}} 
\put(58,538){\makebox(0,0)[lb]{\raisebox{0pt}[0pt][0pt]{{\footnotesize
$\cdot R_{L/2}$}}}}
\put(95,520){\vector( 1, 0){ 15}}
\put(95,500){\vector( 1, 0){ 15}}
\put(95,480){\vector( 1, 0){ 15}}
\put(95,540){\vector( 1, 0){ 15}}
\end{picture}


%% file: thesisfig10.latex
\setlength{\unitlength}{0.00043300in}%
\begingroup\makeatletter\ifx\SetFigFont\undefined
\def\x#1#2#3#4#5#6#7\relax{\def\x{#1#2#3#4#5#6}}%
\expandafter\x\fmtname xxxxxx\relax \def\y{splain}%
\ifx\x\y   
\gdef\SetFigFont#1#2#3{%
  \ifnum #1<17\tiny\else \ifnum #1<20\small\else
  \ifnum #1<24\normalsize\else \ifnum #1<29\large\else
  \ifnum #1<34\Large\else \ifnum #1<41\LARGE\else
     \huge\fi\fi\fi\fi\fi\fi
  \csname #3\endcsname}%
\else
\gdef\SetFigFont#1#2#3{\begingroup
  \count@#1\relax \ifnum 25<\count@\count@25\fi
  \def\x{\endgroup\@setsize\SetFigFont{#2pt}}%
  \expandafter\x
    \csname \romannumeral\the\count@ pt\expandafter\endcsname
    \csname @\romannumeral\the\count@ pt\endcsname
  \csname #3\endcsname}%
\fi
\fi\endgroup
\begin{picture}(4800,5310)(3001,-6238)
\thicklines
\put(5401,-4561){\vector( 0,-1){900}}
\put(3001,-2836){\makebox(0,0)[lb]{\smash{$\bullet$}}}
\put(4201,-2836){\makebox(0,0)[lb]{\smash{$\bullet$}}}
\put(5401,-2836){\makebox(0,0)[lb]{\smash{$\bullet$}}}
\put(6601,-2836){\makebox(0,0)[lb]{\smash{$\bullet$}}}
\put(7801,-2836){\makebox(0,0)[lb]{\smash{$\bullet$}}}
\put(4201,-1300){\makebox(0,0)[lb]{\smash{$\bullet$}}}
\put(5401,-1300){\makebox(0,0)[lb]{\smash{$\bullet$}}}
\put(6601,-1300){\makebox(0,0)[lb]{\smash{$\bullet$}}}
\put(7801,-1300){\makebox(0,0)[lb]{\smash{$\bullet$}}}
\put(3001,-1300){\makebox(0,0)[lb]{\smash{0}}}
\put(3001,-4036){\makebox(0,0)[lb]{\smash{0}}}
\put(4201,-4036){\makebox(0,0)[lb]{\smash{$\bullet$}}}
\put(5401,-4036){\makebox(0,0)[lb]{\smash{$\bullet$}}}
\put(6601,-4036){\makebox(0,0)[lb]{\smash{$\bullet$}}}
\put(7801,-4036){\makebox(0,0)[lb]{\smash{0}}}
\put(3001,-5836){\makebox(0,0)[lb]{\smash{0}}}
\put(4201,-5836){\makebox(0,0)[lb]{\smash{$\bullet$}}}
\put(5401,-5836){\makebox(0,0)[lb]{\smash{$\bullet$}}}
\put(6601,-5836){\makebox(0,0)[lb]{\smash{0}}}
\put(7801,-5836){\makebox(0,0)[lb]{\smash{0}}}
\put(5401,-2461){\vector( 0, 1){900}}
\put(3001,-3211){\makebox(0,0)[lb]{\smash{\scriptsize{$\tilde{H}^{b}_{5}(0)$}}}}
\put(5701,-5161){\makebox(0,0)[lb]{\smash{\scriptsize{$\hat{F}_{2}(\alpha_{1})$}}}}
\put(4201,-3211){\makebox(0,0)[lb]{\smash{\scriptsize{$\tilde{H}^{b}_{5}(1)$}}}}
\put(5401,-3211){\makebox(0,0)[lb]{\smash{\scriptsize{$\tilde{H}^{b}_{5}(2)$}}}}
\put(6601,-3211){\makebox(0,0)[lb]{\smash{\scriptsize{$\tilde{H}^{b}_{5}(3)$}}}}
\put(7801,-3211){\makebox(0,0)[lb]{\smash{\scriptsize{$\tilde{H}^{b}_{5}(4)$}}}}
\put(3001,-4411){\makebox(0,0)[lb]{\smash{\scriptsize{$H^{b}_{3}(0)$}}}}
\put(4201,-4411){\makebox(0,0)[lb]{\smash{\scriptsize{$H^{b}_{3}(1)$}}}}
\put(5401,-4411){\makebox(0,0)[lb]{\smash{\scriptsize{$H^{b}_{3}(2)$}}}}
\put(6601,-4411){\makebox(0,0)[lb]{\smash{\scriptsize{$H^{b}_{3}(3)$}}}}
\put(7801,-4411){\makebox(0,0)[lb]{\smash{\scriptsize{$H^{b}_{3}(4)$}}}}
\put(5701,-2161){\makebox(0,0)[lb]{\smash{\scriptsize{$\hat{F}^{-t}_{2}(\alpha_{1})$}}}}
\end{picture}

%% file: thesisfig11.latex
\setlength{\unitlength}{0.00043300in}%
\begingroup\makeatletter\ifx\SetFigFont\undefined
\def\x#1#2#3#4#5#6#7\relax{\def\x{#1#2#3#4#5#6}}%
\expandafter\x\fmtname xxxxxx\relax \def\y{splain}%
\ifx\x\y   
\gdef\SetFigFont#1#2#3{%
  \ifnum #1<17\tiny\else \ifnum #1<20\small\else
  \ifnum #1<24\normalsize\else \ifnum #1<29\large\else
  \ifnum #1<34\Large\else \ifnum #1<41\LARGE\else
     \huge\fi\fi\fi\fi\fi\fi
  \csname #3\endcsname}%
\else
\gdef\SetFigFont#1#2#3{\begingroup
  \count@#1\relax \ifnum 25<\count@\count@25\fi
  \def\x{\endgroup\@setsize\SetFigFont{#2pt}}%
  \expandafter\x
    \csname \romannumeral\the\count@ pt\expandafter\endcsname
    \csname @\romannumeral\the\count@ pt\endcsname
  \csname #3\endcsname}%
\fi
\fi\endgroup
\begin{picture}(9612,3180)(301,-4672)
\thicklines
\put(2776,-4261){\line( 5, 3){518.382}}
\put(4501,-3661){\line( 0,-1){600}}
\put(3901,-4261){\line( 2, 1){600}}
\put(5701,-3661){\line( 0,-1){600}}
\put(5101,-4261){\line( 2, 1){600}}
\put(7501,-3661){\line( 0,-1){600}}
\put(6901,-4261){\line( 2, 1){600}}
\put(8701,-3661){\line( 0,-1){600}}
\put(8101,-4261){\line( 2, 1){600}}
\put(2701,-3661){\line( 0,-1){600}}
\put(3901,-3661){\line( 0,-1){600}}
\put(5101,-3661){\line( 0,-1){600}}
\put(6901,-3661){\line( 0,-1){600}}
\put(8101,-3661){\line( 0,-1){600}}
\put(3301,-3061){\line( 0,-1){600}}
\put(3901,-3661){\line(-2, 1){600}}
\put(4501,-3061){\line( 0,-1){600}}
\put(4501,-3361){\line( 2,-1){600}}
\put(3901,-3061){\line( 0,-1){600}}
\put(5101,-3061){\line( 0,-1){600}}
\put(7501,-3061){\line( 0,-1){600}}
\put(8101,-3061){\line( 0,-1){600}}
\put(7501,-3361){\line( 2,-1){600}}
\put(1501,-3661){\line( 0,-1){600}}
\put(2101,-3661){\line( 0,-1){600}}
\put(1501,-4261){\line( 2, 1){600}}
\put(2101,-3061){\line( 0,-1){600}}
\put(2701,-3061){\line( 0,-1){600}}
\put(2101,-3361){\line( 2,-1){600}}
\put(2701,-2461){\line( 0,-1){600}}
\put(2101,-2461){\line( 0,-1){600}}
\put(3301,-2461){\line( 0,-1){600}}
\put(3901,-2461){\line( 0,-1){600}}
\put(4501,-2461){\line( 0,-1){600}}
\put(5101,-2461){\line( 0,-1){600}}
\put(9901,-3661){\line( 0,-1){600}}
\put(9301,-3661){\line( 0,-1){600}}
\put(9301,-4261){\line( 2, 1){600}}
\put(8701,-3061){\line( 0,-1){600}}
\put(9301,-3061){\line( 0,-1){600}}
\put(8701,-3361){\line( 2,-1){600}}
\put(2101,-3061){\line( 2, 1){600}}
\put(3301,-3061){\line( 2, 1){600}}
\put(4501,-3061){\line( 2, 1){600}}
\put(7501,-2461){\line( 0,-1){600}}
\put(8101,-2461){\line( 0,-1){600}}
\put(8701,-2461){\line( 0,-1){600}}
\put(9301,-2461){\line( 0,-1){600}}
\put(7501,-3061){\line( 2, 1){600}}
\put(8701,-3061){\line( 5, 2){600}}
\put(2701,-1861){\line( 0,-1){600}}
\put(3301,-1861){\line( 0,-1){600}}
\put(3901,-1861){\line( 0,-1){600}}
\put(4501,-1861){\line( 0,-1){600}}
\put(2701,-2161){\line( 2,-1){600}}
\put(3901,-2161){\line( 2,-1){600}}
\put(8101,-1861){\line( 0,-1){600}}
\put(8701,-1861){\line( 0,-1){600}}
\put(8101,-2161){\line( 2,-1){600}}
\put(2650,-4336){\makebox(0,0)[lb]{$\bullet$}}
\put(3250,-4336){\makebox(0,0)[lb]{$\bullet$}}
\put(3301,-3661){\line( 0,-1){600}}
\put(3850,-4336){\makebox(0,0)[lb]{$\bullet$}}
\put(50,-4336){\makebox(0,0)[lb]{\scriptsize{$\hat{W}^{-t}_{2}(\alpha_{4})$}}}
\put(4440,-4336){\makebox(0,0)[lb]{$\bullet$}}
\put(5040,-4336){\makebox(0,0)[lb]{$\bullet$}}
\put(5640,-4336){\makebox(0,0)[lb]{$\bullet$}}
\put(6840,-4336){\makebox(0,0)[lb]{$\bullet$}}
\put(7440,-4336){\makebox(0,0)[lb]{$\bullet$}}
\put(8040,-4336){\makebox(0,0)[lb]{$\bullet$}}
\put(8640,-4336){\makebox(0,0)[lb]{$\bullet$}}
\put(2640,-3736){\makebox(0,0)[lb]{$\bullet$}}
\put(3840,-3736){\makebox(0,0)[lb]{$\bullet$}}
\put(4440,-3736){\makebox(0,0)[lb]{$\bullet$}}
\put(5040,-3736){\makebox(0,0)[lb]{$\bullet$}}
\put(5640,-3736){\makebox(0,0)[lb]{$\bullet$}}
\put(6840,-3736){\makebox(0,0)[lb]{$\bullet$}}
\put(7440,-3736){\makebox(0,0)[lb]{$\bullet$}}
\put(8040,-3736){\makebox(0,0)[lb]{$\bullet$}}
\put(8640,-3736){\makebox(0,0)[lb]{$\bullet$}}
\put(3240,-3736){\makebox(0,0)[lb]{$\bullet$}}
\put(2040,-4336){\makebox(0,0)[lb]{$\bullet$}}
\put(1440,-4336){\makebox(0,0)[lb]{$\bullet$}}
\put(1440,-3736){\makebox(0,0)[lb]{$\bullet$}}
\put(2040,-3736){\makebox(0,0)[lb]{$\bullet$}}
\put(2040,-3136){\makebox(0,0)[lb]{$\bullet$}}
\put(2640,-3136){\makebox(0,0)[lb]{$\bullet$}}
\put(3240,-3136){\makebox(0,0)[lb]{$\bullet$}}
\put(3840,-3136){\makebox(0,0)[lb]{$\bullet$}}
\put(4440,-3136){\makebox(0,0)[lb]{$\bullet$}}
\put(5040,-3136){\makebox(0,0)[lb]{$\bullet$}}
\put(7440,-3136){\makebox(0,0)[lb]{$\bullet$}}
\put(8040,-3136){\makebox(0,0)[lb]{$\bullet$}}
\put(9240,-4336){\makebox(0,0)[lb]{$\bullet$}}
\put(9840,-4336){\makebox(0,0)[lb]{$\bullet$}}
\put(9240,-3736){\makebox(0,0)[lb]{$\bullet$}}
\put(9840,-3736){\makebox(0,0)[lb]{$\bullet$}}
\put(8640,-3136){\makebox(0,0)[lb]{$\bullet$}}
\put(9240,-3136){\makebox(0,0)[lb]{$\bullet$}}
\put(2040,-2536){\makebox(0,0)[lb]{$\bullet$}}
\put(2640,-2536){\makebox(0,0)[lb]{$\bullet$}}
\put(3240,-2536){\makebox(0,0)[lb]{$\bullet$}}
\put(3840,-2536){\makebox(0,0)[lb]{$\bullet$}}
\put(4440,-2536){\makebox(0,0)[lb]{$\bullet$}}
\put(5040,-2536){\makebox(0,0)[lb]{$\bullet$}}
\put(7440,-2536){\makebox(0,0)[lb]{$\bullet$}}
\put(8040,-2536){\makebox(0,0)[lb]{$\bullet$}}
\put(8640,-2536){\makebox(0,0)[lb]{$\bullet$}}
\put(9240,-2536){\makebox(0,0)[lb]{$\bullet$}}
\put(2640,-1936){\makebox(0,0)[lb]{$\bullet$}}
\put(3240,-1936){\makebox(0,0)[lb]{$\bullet$}}
\put(3840,-1936){\makebox(0,0)[lb]{$\bullet$}}
\put(4440,-1936){\makebox(0,0)[lb]{$\bullet$}}
\put(8040,-1936){\makebox(0,0)[lb]{$\bullet$}}
\put(8640,-1936){\makebox(0,0)[lb]{$\bullet$}}
\put(1400,-4700){\makebox(0,0)[lb]{\scriptsize{${\bf x}(0)$}}}
\put(2050,-4700){\makebox(0,0)[lb]{\scriptsize{${\bf x}(1)$}}}
\put(9800,-4700){\makebox(0,0)[lb]{\scriptsize{${\bf x}(n-1)$}}}
\put(1440,-1936){\makebox(0,0)[lb]{$\bullet$}}
\put(2040,-1936){\makebox(0,0)[lb]{$\bullet$}}
\put(9240,-1936){\makebox(0,0)[lb]{$\bullet$}}
\put(9840,-1936){\makebox(0,0)[lb]{$\bullet$}}
\put(1450,-1636){\makebox(0,0)[lb]{$\tilde{c}^{1}_{0}$}}
\put(2050,-1636){\makebox(0,0)[lb]{$\tilde{d}^{1}_{0}$}}
\put(2650,-1636){\makebox(0,0)[lb]{$c^{1}_{1}$}}
\put(3250,-1636){\makebox(0,0)[lb]{$d^{1}_{1}$}}
\put(9751,-1636){\makebox(0,0)[lb]{$\tilde{d}^{1}_{n/2-1}$}}
\put(50,-2536){\makebox(0,0)[lb]{\scriptsize{$\hat{W}^{-t}_{2}(\alpha_{1})$}}}
\put(50,-3136){\makebox(0,0)[lb]{\scriptsize{$\hat{W}^{-t}_{2}(\alpha_{2})$}}}
\put(50,-3736){\makebox(0,0)[lb]{\scriptsize{$\hat{W}^{-t}_{2}(\alpha_{3})$}}}
\end{picture}

%% file: thesisfig12.latex
\setlength{\unitlength}{0.00053300in}%
\begingroup\makeatletter\ifx\SetFigFont\undefined
\def\x#1#2#3#4#5#6#7\relax{\def\x{#1#2#3#4#5#6}}%
\expandafter\x\fmtname xxxxxx\relax \def\y{splain}%
\ifx\x\y   
\gdef\SetFigFont#1#2#3{%
  \ifnum #1<17\tiny\else \ifnum #1<20\small\else
  \ifnum #1<24\normalsize\else \ifnum #1<29\large\else
  \ifnum #1<34\Large\else \ifnum #1<41\LARGE\else
     \huge\fi\fi\fi\fi\fi\fi
  \csname #3\endcsname}%
\else
\gdef\SetFigFont#1#2#3{\begingroup
  \count@#1\relax \ifnum 25<\count@\count@25\fi
  \def\x{\endgroup\@setsize\SetFigFont{#2pt}}%
  \expandafter\x
    \csname \romannumeral\the\count@ pt\expandafter\endcsname
    \csname @\romannumeral\the\count@ pt\endcsname
  \csname #3\endcsname}%
\fi
\fi\endgroup
\begin{picture}(4200,1710)(2401,-3538)
\thicklines
\put(2701,-2761){\line( 2,-1){600}}
\put(3901,-2461){\line( 0,-1){600}}
\put(3901,-2761){\line( 2,-1){600}}
\put(5101,-2461){\line( 0,-1){600}}
\put(5101,-2761){\line( 2,-1){600}}
\put(6301,-3061){\line( 0, 1){600}}
\put(3301,-2461){\line( 0,-1){600}}
\put(4501,-2461){\line( 0,-1){600}}
\put(5701,-2461){\line( 0,-1){600}}
\put(6301,-1861){\line( 0,-1){600}}
\put(5701,-2461){\line( 2, 1){600}}
\put(5701,-2461){\line( 0, 1){600}}
\put(5101,-1861){\line( 0,-1){600}}
\put(4501,-2461){\line( 2, 1){600}}
\put(4501,-1861){\line( 0,-1){600}}
\put(3901,-1861){\line( 0,-1){600}}
\put(3301,-1861){\line( 0,-1){600}}
\put(3301,-2461){\line( 2, 1){600}}
\put(2650,-3136){\makebox(0,0)[lb]{$\bullet$}}
\put(3250,-3136){\makebox(0,0)[lb]{$\bullet$}}
\put(3850,-3136){\makebox(0,0)[lb]{$\bullet$}}
\put(4450,-3136){\makebox(0,0)[lb]{$\bullet$}}
\put(5050,-3136){\makebox(0,0)[lb]{$\bullet$}}
\put(2701,-2461){\line( 0,-1){600}}
\put(5650,-3136){\makebox(0,0)[lb]{$\bullet$}}
\put(6601,-1936){\makebox(0,0)[lb]{0}}
\put(6250,-3136){\makebox(0,0)[lb]{$\bullet$}}
\put(2701,-3511){\makebox(0,0)[lb]{a}}
\put(3301,-3511){\makebox(0,0)[lb]{b}}
\put(3901,-3511){\makebox(0,0)[lb]{c}}
\put(4501,-3511){\makebox(0,0)[lb]{d}}
\put(5101,-3511){\makebox(0,0)[lb]{c}}
\put(5701,-3511){\makebox(0,0)[lb]{b}}
\put(6301,-3511){\makebox(0,0)[lb]{a}}
\put(2650,-2536){\makebox(0,0)[lb]{$\bullet$}}
\put(3850,-2536){\makebox(0,0)[lb]{$\bullet$}}
\put(5050,-2536){\makebox(0,0)[lb]{$\bullet$}}
\put(6250,-2536){\makebox(0,0)[lb]{$\bullet$}}
\put(3250,-2536){\makebox(0,0)[lb]{$\bullet$}}
\put(4450,-2536){\makebox(0,0)[lb]{$\bullet$}}
\put(5650,-2536){\makebox(0,0)[lb]{$\bullet$}}
\put(2401,-2536){\makebox(0,0)[lb]{0}}
\put(3250,-1936){\makebox(0,0)[lb]{$\bullet$}}
\put(3850,-1936){\makebox(0,0)[lb]{$\bullet$}}
\put(4450,-1936){\makebox(0,0)[lb]{$\bullet$}}
\put(5050,-1936){\makebox(0,0)[lb]{$\bullet$}}
\put(5650,-1936){\makebox(0,0)[lb]{$\bullet$}}
\put(6250,-1936){\makebox(0,0)[lb]{$\bullet$}}
\end{picture}

%% file: figur9.latex
\setlength{\unitlength}{0.0125in}%
\begin{picture}(345,94)(75,340)
\thicklines
\put( 95,400){\vector( 1, 0){ 20}}
\put( 95,380){\vector( 1, 0){ 20}}
\put( 95,360){\vector( 1, 0){ 20}}
\put(360,410){\line( 2,-1){ 20}}
\put(340,400){\line( 2, 1){ 20}}
\put(360,420){\line( 0,-1){ 20}}
\put(380,390){\line( 2,-1){ 20}}
\put(360,380){\line( 2, 1){ 20}}
\put(340,390){\line( 2,-1){ 20}}
\put(320,380){\line( 2, 1){ 20}}
\put(380,400){\line( 0,-1){ 20}}
\put(360,400){\line( 0,-1){ 20}}
\put(340,400){\line( 0,-1){ 20}}
\put(200,420){\line( 0,-1){ 20}}
\put(220,410){\line( 2,-1){ 20}}
\put(200,400){\line( 2, 1){ 20}}
\put(220,420){\line( 0,-1){ 20}}
\put(400,370){\line( 2,-1){ 20}}
\put(380,360){\line( 2, 1){ 20}}
\put(360,370){\line( 2,-1){ 20}}
\put(340,360){\line( 2, 1){ 20}}
\put(320,370){\line( 2,-1){ 20}}
\put(300,360){\line( 2, 1){ 20}}
\put(400,380){\line( 0,-1){ 20}}
\put(380,380){\line( 0,-1){ 20}}
\put(360,380){\line( 0,-1){ 20}}
\put(340,380){\line( 0,-1){ 20}}
\put(320,380){\line( 0,-1){ 20}}
\put(240,390){\line( 2,-1){ 20}}
\put(220,380){\line( 2, 1){ 20}}
\put(240,400){\line( 0,-1){ 20}}
\put(220,400){\line( 0,-1){ 20}}
\put(240,380){\line( 0,-1){ 20}}
\put(180,410){\line( 2,-1){ 20}}
\put(160,400){\line( 2, 1){ 20}}
\put(180,420){\line( 0,-1){ 20}}
\put(260,370){\line( 2,-1){ 20}}
\put(240,360){\line( 2, 1){ 20}}
\put(260,380){\line( 0,-1){ 20}}
\put(200,390){\line( 2,-1){ 20}}
\put(180,380){\line( 2, 1){ 20}}
\put(160,390){\line( 2,-1){ 20}}
\put(140,380){\line( 2, 1){ 20}}
\put(200,400){\line( 0,-1){ 20}}
\put(180,400){\line( 0,-1){ 20}}
\put(160,400){\line( 0,-1){ 20}}
\put(220,370){\line( 2,-1){ 20}}
\put(200,360){\line( 2, 1){ 20}}
\put(180,370){\line( 2,-1){ 20}}
\put(160,360){\line( 2, 1){ 20}}
\put(140,370){\line( 2,-1){ 20}}
\put(120,360){\line( 2, 1){ 20}}
\put(220,380){\line( 0,-1){ 20}}
\put(200,380){\line( 0,-1){ 20}}
\put(180,380){\line( 0,-1){ 20}}
\put(160,380){\line( 0,-1){ 20}}
\put(140,380){\line( 0,-1){ 20}}
\put(160,420){\line( 0,-1){ 20}}
\put(380,420){\line( 0,-1){ 20}}
\put( 50,400){\makebox(0,0)[lb]{\raisebox{0pt}[0pt][0pt]{{\footnotesize
$\tilde{W}^{-t}_{3}(\alpha_{3})$}}}} 
\put( 50,380){\makebox(0,0)[lb]{\raisebox{0pt}[0pt][0pt]{{\footnotesize
$\tilde{W}^{-t}_{3}(\alpha_{2})$}}}} 
\put( 50,360){\makebox(0,0)[lb]{\raisebox{0pt}[0pt][0pt]{{\footnotesize
$\tilde{W}^{-t}_{3}(\alpha_{1})$}}}}
\put(420,425){\makebox(0,0)[lb]{\raisebox{0pt}[0pt][0pt]{{\footnotesize
$\tilde{d}^{1}_{n-1}$}}}}
\put(180,425){\makebox(0,0)[lb]{\raisebox{0pt}[0pt][0pt]{{\footnotesize $d^{1}_{1}$}}}}
\put(160,425){\makebox(0,0)[lb]{\raisebox{0pt}[0pt][0pt]{{\footnotesize $c^{1}_{1}$}}}}
\put(140,425){\makebox(0,0)[lb]{\raisebox{0pt}[0pt][0pt]{{\footnotesize $\tilde{d}^{1}_{0}$}}}}
\put(120,425){\makebox(0,0)[lb]{\raisebox{0pt}[0pt][0pt]{{\footnotesize $\tilde{c}^{1}_{0}$}}}}
\put(417,417){\makebox(0,0)[lb]{\raisebox{0pt}[0pt][0pt]{$\bullet$}}}
\put(397,417){\makebox(0,0)[lb]{\raisebox{0pt}[0pt][0pt]{$\bullet$}}}
\put(377,417){\makebox(0,0)[lb]{\raisebox{0pt}[0pt][0pt]{$\bullet$}}}
\put(157,417){\makebox(0,0)[lb]{\raisebox{0pt}[0pt][0pt]{$\bullet$}}}
\put(137,417){\makebox(0,0)[lb]{\raisebox{0pt}[0pt][0pt]{$\bullet$}}}
\put(117,417){\makebox(0,0)[lb]{\raisebox{0pt}[0pt][0pt]{$\bullet$}}}
\put(410,340){\makebox(0,0)[lb]{\raisebox{0pt}[0pt][0pt]{{\footnotesize
        ${\bf x}(n-1)$}}}}
\put(357,417){\makebox(0,0)[lb]{\raisebox{0pt}[0pt][0pt]{$\bullet$}}}
\put(377,397){\makebox(0,0)[lb]{\raisebox{0pt}[0pt][0pt]{$\bullet$}}}
\put(357,397){\makebox(0,0)[lb]{\raisebox{0pt}[0pt][0pt]{$\bullet$}}}
\put(337,397){\makebox(0,0)[lb]{\raisebox{0pt}[0pt][0pt]{$\bullet$}}}
\put(217,417){\makebox(0,0)[lb]{\raisebox{0pt}[0pt][0pt]{$\bullet$}}}
\put(197,417){\makebox(0,0)[lb]{\raisebox{0pt}[0pt][0pt]{$\bullet$}}}
\put(397,377){\makebox(0,0)[lb]{\raisebox{0pt}[0pt][0pt]{$\bullet$}}}
\put(377,377){\makebox(0,0)[lb]{\raisebox{0pt}[0pt][0pt]{$\bullet$}}}
\put(357,377){\makebox(0,0)[lb]{\raisebox{0pt}[0pt][0pt]{$\bullet$}}}
\put(337,377){\makebox(0,0)[lb]{\raisebox{0pt}[0pt][0pt]{$\bullet$}}}
\put(317,377){\makebox(0,0)[lb]{\raisebox{0pt}[0pt][0pt]{$\bullet$}}}
\put(417,357){\makebox(0,0)[lb]{\raisebox{0pt}[0pt][0pt]{$\bullet$}}}
\put(397,357){\makebox(0,0)[lb]{\raisebox{0pt}[0pt][0pt]{$\bullet$}}}
\put(377,357){\makebox(0,0)[lb]{\raisebox{0pt}[0pt][0pt]{$\bullet$}}}
\put(357,357){\makebox(0,0)[lb]{\raisebox{0pt}[0pt][0pt]{$\bullet$}}}
\put(337,357){\makebox(0,0)[lb]{\raisebox{0pt}[0pt][0pt]{$\bullet$}}}
\put(317,357){\makebox(0,0)[lb]{\raisebox{0pt}[0pt][0pt]{$\bullet$}}}
\put(297,357){\makebox(0,0)[lb]{\raisebox{0pt}[0pt][0pt]{$\bullet$}}}
\put(177,417){\makebox(0,0)[lb]{\raisebox{0pt}[0pt][0pt]{$\bullet$}}}
\put(217,397){\makebox(0,0)[lb]{\raisebox{0pt}[0pt][0pt]{$\bullet$}}}
\put(237,377){\makebox(0,0)[lb]{\raisebox{0pt}[0pt][0pt]{$\bullet$}}}
\put(237,397){\makebox(0,0)[lb]{\raisebox{0pt}[0pt][0pt]{$\bullet$}}}
\put(257,377){\makebox(0,0)[lb]{\raisebox{0pt}[0pt][0pt]{$\bullet$}}}
\put(110,340){\makebox(0,0)[lb]{\raisebox{0pt}[0pt][0pt]{{\footnotesize
        ${\bf x}(0)$}}}}
\put(277,357){\makebox(0,0)[lb]{\raisebox{0pt}[0pt][0pt]{$\bullet$}}}
\put(257,357){\makebox(0,0)[lb]{\raisebox{0pt}[0pt][0pt]{$\bullet$}}}
\put(237,357){\makebox(0,0)[lb]{\raisebox{0pt}[0pt][0pt]{$\bullet$}}}
\put(217,357){\makebox(0,0)[lb]{\raisebox{0pt}[0pt][0pt]{$\bullet$}}}
\put(197,357){\makebox(0,0)[lb]{\raisebox{0pt}[0pt][0pt]{$\bullet$}}}
\put(177,357){\makebox(0,0)[lb]{\raisebox{0pt}[0pt][0pt]{$\bullet$}}}
\put(157,357){\makebox(0,0)[lb]{\raisebox{0pt}[0pt][0pt]{$\bullet$}}}
\put(137,357){\makebox(0,0)[lb]{\raisebox{0pt}[0pt][0pt]{$\bullet$}}}
\put(117,357){\makebox(0,0)[lb]{\raisebox{0pt}[0pt][0pt]{$\bullet$}}}
\put(197,397){\makebox(0,0)[lb]{\raisebox{0pt}[0pt][0pt]{$\bullet$}}}
\put(177,397){\makebox(0,0)[lb]{\raisebox{0pt}[0pt][0pt]{$\bullet$}}}
\put(157,397){\makebox(0,0)[lb]{\raisebox{0pt}[0pt][0pt]{$\bullet$}}}
\put(217,377){\makebox(0,0)[lb]{\raisebox{0pt}[0pt][0pt]{$\bullet$}}}
\put(197,377){\makebox(0,0)[lb]{\raisebox{0pt}[0pt][0pt]{$\bullet$}}}
\put(177,377){\makebox(0,0)[lb]{\raisebox{0pt}[0pt][0pt]{$\bullet$}}}
\put(157,377){\makebox(0,0)[lb]{\raisebox{0pt}[0pt][0pt]{$\bullet$}}}
\put(137,377){\makebox(0,0)[lb]{\raisebox{0pt}[0pt][0pt]{$\bullet$}}}
\end{picture}


%% file: figur8.latex
\setlength{\unitlength}{0.0125in}%
\begin{picture}(180,94)(90,380)
\thicklines
\put(220,440){\line( 2, 1){ 20}}
\put(220,460){\line( 0,-1){ 20}}
\put(240,460){\line( 0,-1){ 20}}
\put(240,430){\line( 2,-1){ 20}}
\put(240,440){\line( 0,-1){ 20}}
\put(220,440){\line( 0,-1){ 20}}
\put(200,430){\line( 2,-1){ 20}}
\put(200,440){\line( 0,-1){ 20}}
\put(200,400){\line( 2, 1){ 20}}
\put(240,400){\line( 2, 1){ 20}}
\put(260,420){\line( 0,-1){ 20}}
\put(240,420){\line( 0,-1){ 20}}
\put(220,420){\line( 0,-1){ 20}}
\put(200,420){\line( 0,-1){ 20}}
\put(120,450){\line( 2,-1){ 20}}
\put(140,460){\line( 0,-1){ 20}}
\put(120,460){\line( 0,-1){ 20}}
\put(100,420){\line( 2, 1){ 20}}
\put(120,440){\line( 0,-1){ 20}}
\put(140,440){\line( 0,-1){ 20}}
\put(140,420){\line( 2, 1){ 20}}
\put(160,440){\line( 0,-1){ 20}}
\put(160,420){\line( 0,-1){ 20}}
\put(140,410){\line( 2,-1){ 20}}
\put(140,420){\line( 0,-1){ 20}}
\put(120,420){\line( 0,-1){ 20}}
\put(100,410){\line( 2,-1){ 20}}
\put(100,420){\line( 0,-1){ 20}}
\put(245,383){\makebox(0,0)[lb]{\raisebox{0pt}[0pt][0pt]{{\footnotesize $\tilde{H}^{b}_{4}(3)$}}}}
\put(190,383){\makebox(0,0)[lb]{\raisebox{0pt}[0pt][0pt]{{\footnotesize $\tilde{H}^{b}_{4}(0)$}}}}
\put(241,465){\makebox(0,0)[lb]{\raisebox{0pt}[0pt][0pt]{{\footnotesize $1/\delta$}}}}
\put(237,457){\makebox(0,0)[lb]{\raisebox{0pt}[0pt][0pt]{$\bullet$}}}
\put(217,457){\makebox(0,0)[lb]{\raisebox{0pt}[0pt][0pt]{$\bullet$}}}
\put(190,440){\makebox(0,0)[lb]{\raisebox{0pt}[0pt][0pt]{{\footnotesize $0$}}}}
\put(237,437){\makebox(0,0)[lb]{\raisebox{0pt}[0pt][0pt]{$\bullet$}}}
\put(217,437){\makebox(0,0)[lb]{\raisebox{0pt}[0pt][0pt]{$\bullet$}}}
\put(197,437){\makebox(0,0)[lb]{\raisebox{0pt}[0pt][0pt]{$\bullet$}}}
\put(265,420){\makebox(0,0)[lb]{\raisebox{0pt}[0pt][0pt]{{\footnotesize $0$}}}}
\put(257,397){\makebox(0,0)[lb]{\raisebox{0pt}[0pt][0pt]{$\bullet$}}}
\put(237,397){\makebox(0,0)[lb]{\raisebox{0pt}[0pt][0pt]{$\bullet$}}}
\put(217,397){\makebox(0,0)[lb]{\raisebox{0pt}[0pt][0pt]{$\bullet$}}}
\put(197,397){\makebox(0,0)[lb]{\raisebox{0pt}[0pt][0pt]{$\bullet$}}}
\put(257,417){\makebox(0,0)[lb]{\raisebox{0pt}[0pt][0pt]{$\bullet$}}}
\put(237,417){\makebox(0,0)[lb]{\raisebox{0pt}[0pt][0pt]{$\bullet$}}}
\put(217,417){\makebox(0,0)[lb]{\raisebox{0pt}[0pt][0pt]{$\bullet$}}}
\put(197,417){\makebox(0,0)[lb]{\raisebox{0pt}[0pt][0pt]{$\bullet$}}}
\put(150,383){\makebox(0,0)[lb]{\raisebox{0pt}[0pt][0pt]{{\footnotesize $H^{b}_{4}(3)$}}}}
\put(90,383){\makebox(0,0)[lb]{\raisebox{0pt}[0pt][0pt]{{\footnotesize $H^{b}_{4}(0)$}}}}
\put(141,465){\makebox(0,0)[lb]{\raisebox{0pt}[0pt][0pt]{{\footnotesize $\delta$}}}}
\put(110,460){\makebox(0,0)[lb]{\raisebox{0pt}[0pt][0pt]{{\footnotesize $0$}}}}
\put(137,457){\makebox(0,0)[lb]{\raisebox{0pt}[0pt][0pt]{$\bullet$}}}
\put(117,457){\makebox(0,0)[lb]{\raisebox{0pt}[0pt][0pt]{$\bullet$}}}
\put(165,440){\makebox(0,0)[lb]{\raisebox{0pt}[0pt][0pt]{{\footnotesize $0$}}}}
\put(137,437){\makebox(0,0)[lb]{\raisebox{0pt}[0pt][0pt]{$\bullet$}}}
\put(157,437){\makebox(0,0)[lb]{\raisebox{0pt}[0pt][0pt]{$\bullet$}}}
\put(117,437){\makebox(0,0)[lb]{\raisebox{0pt}[0pt][0pt]{$\bullet$}}}
\put( 90,420){\makebox(0,0)[lb]{\raisebox{0pt}[0pt][0pt]{{\footnotesize $0$}}}}
\put(157,417){\makebox(0,0)[lb]{\raisebox{0pt}[0pt][0pt]{$\bullet$}}}
\put(137,417){\makebox(0,0)[lb]{\raisebox{0pt}[0pt][0pt]{$\bullet$}}}
\put(117,417){\makebox(0,0)[lb]{\raisebox{0pt}[0pt][0pt]{$\bullet$}}}
\put(97,417){\makebox(0,0)[lb]{\raisebox{0pt}[0pt][0pt]{$\bullet$}}}
\put(157,397){\makebox(0,0)[lb]{\raisebox{0pt}[0pt][0pt]{$\bullet$}}}
\put(137,397){\makebox(0,0)[lb]{\raisebox{0pt}[0pt][0pt]{$\bullet$}}}
\put(117,397){\makebox(0,0)[lb]{\raisebox{0pt}[0pt][0pt]{$\bullet$}}}
\put(97,397){\makebox(0,0)[lb]{\raisebox{0pt}[0pt][0pt]{$\bullet$}}}
\end{picture}


%% file: figur12.latex
\setlength{\unitlength}{0.0125in}%
\begin{picture}(190,197)(100,380)
\thicklines
\put(100,395){\vector( 1, 0){180}}
\put(110,380){\vector( 0, 1){180}}
\multiput(240,460)(0.00000,-8.00000){3}{\line( 0,-1){  4.000}}
\multiput(240,440)(8.00000,0.00000){3}{\line( 1, 0){  4.000}}
\multiput(260,440)(0.00000,8.00000){3}{\line( 0, 1){  4.000}}
\multiput(260,460)(-8.00000,0.00000){3}{\line(-1, 0){  4.000}}
\multiput(200,460)(0.00000,-8.00000){3}{\line( 0,-1){  4.000}}
\multiput(200,440)(8.00000,0.00000){3}{\line( 1, 0){  4.000}}
\multiput(220,440)(0.00000,8.00000){3}{\line( 0, 1){  4.000}}
\multiput(220,460)(-8.00000,0.00000){3}{\line(-1, 0){  4.000}}
\multiput(240,500)(0.00000,-8.00000){3}{\line( 0,-1){  4.000}}
\multiput(240,480)(8.00000,0.00000){3}{\line( 1, 0){  4.000}}
\multiput(260,480)(0.00000,8.00000){3}{\line( 0, 1){  4.000}}
\multiput(260,500)(-8.00000,0.00000){3}{\line(-1, 0){  4.000}}
\multiput(200,500)(0.00000,-8.00000){3}{\line( 0,-1){  4.000}}
\multiput(200,480)(8.00000,0.00000){3}{\line( 1, 0){  4.000}}
\multiput(220,480)(0.00000,8.00000){3}{\line( 0, 1){  4.000}}
\multiput(220,500)(-8.00000,0.00000){3}{\line(-1, 0){  4.000}}
\multiput(160,500)(0.00000,-8.00000){3}{\line( 0,-1){  4.000}}
\multiput(160,480)(8.00000,0.00000){3}{\line( 1, 0){  4.000}}
\multiput(180,480)(0.00000,8.00000){3}{\line( 0, 1){  4.000}}
\multiput(180,500)(-8.00000,0.00000){3}{\line(-1, 0){  4.000}}
\multiput(240,540)(8.00000,0.00000){3}{\line( 1, 0){  4.000}}
\multiput(260,540)(0.00000,-8.00000){3}{\line( 0,-1){  4.000}}
\multiput(240,520)(8.00000,0.00000){3}{\line( 1, 0){  4.000}}
\multiput(240,540)(0.00000,-8.00000){3}{\line( 0,-1){  4.000}}
\multiput(200,540)(0.00000,-8.00000){3}{\line( 0,-1){  4.000}}
\multiput(200,520)(8.00000,0.00000){3}{\line( 1, 0){  4.000}}
\multiput(220,520)(0.00000,8.00000){3}{\line( 0, 1){  4.000}}
\multiput(220,540)(-8.00000,0.00000){3}{\line(-1, 0){  4.000}}
\multiput(160,540)(8.00000,0.00000){3}{\line( 1, 0){  4.000}}
\multiput(180,520)(0.00000,8.00000){3}{\line( 0, 1){  4.000}}
\multiput(160,520)(8.00000,0.00000){3}{\line( 1, 0){  4.000}}
\multiput(160,540)(0.00000,-8.00000){3}{\line( 0,-1){  4.000}}
\put(160,440){\line( 0,-1){ 20}}
\put(140,440){\line( 1, 0){ 20}}
\put(140,440){\line( 0,-1){ 20}}
\put(260,440){\line( 0,-1){ 20}}
\put(260,420){\line( 1, 0){ 20}}
\put(280,420){\line( 0, 1){ 20}}
\put(280,440){\line(-1, 0){ 20}}
\put(220,440){\line( 1, 0){ 20}}
\put(240,440){\line( 0,-1){ 20}}
\put(220,440){\line( 0,-1){ 20}}
\put(220,420){\line( 1, 0){ 20}}
\put(180,440){\line( 0,-1){ 20}}
\put(180,420){\line( 1, 0){ 20}}
\put(200,420){\line( 0, 1){ 20}}
\put(200,440){\line(-1, 0){ 20}}
\put(260,480){\line( 1, 0){ 20}}
\put(280,460){\line( 0, 1){ 20}}
\put(260,480){\line( 0,-1){ 20}}
\put(260,460){\line( 1, 0){ 20}}
\put(220,480){\line( 1, 0){ 20}}
\put(240,460){\line( 0, 1){ 20}}
\put(220,460){\line( 1, 0){ 20}}
\put(220,480){\line( 0,-1){ 20}}
\put(180,480){\line( 1, 0){ 20}}
\put(200,460){\line( 0, 1){ 20}}
\put(180,460){\line( 1, 0){ 20}}
\put(180,480){\line( 0,-1){ 20}}
\put(140,480){\line( 1, 0){ 20}}
\put(160,460){\line( 0, 1){ 20}}
\put(140,460){\line( 1, 0){ 20}}
\put(140,480){\line( 0,-1){ 20}}
\put(260,520){\line( 1, 0){ 20}}
\put(280,500){\line( 0, 1){ 20}}
\put(260,500){\line( 1, 0){ 20}}
\put(260,520){\line( 0,-1){ 20}}
\put(220,520){\line( 1, 0){ 20}}
\put(240,500){\line( 0, 1){ 20}}
\put(220,500){\line( 1, 0){ 20}}
\put(220,520){\line( 0,-1){ 20}}
\put(180,520){\line( 1, 0){ 20}}
\put(200,500){\line( 0, 1){ 20}}
\put(180,500){\line( 1, 0){ 20}}
\put(180,520){\line( 0,-1){ 20}}
\put(140,520){\line( 1, 0){ 20}}
\put(160,500){\line( 0, 1){ 20}}
\put(140,500){\line( 1, 0){ 20}}
\put(140,520){\line( 0,-1){ 20}}
\put(260,560){\line( 1, 0){ 20}}
\put(280,540){\line( 0, 1){ 20}}
\put(260,540){\line( 1, 0){ 20}}
\put(260,560){\line( 0,-1){ 20}}
\put(220,560){\line( 1, 0){ 20}}
\put(240,540){\line( 0, 1){ 20}}
\put(220,540){\line( 1, 0){ 20}}
\put(220,560){\line( 0,-1){ 20}}
\put(180,560){\line( 1, 0){ 20}}
\put(200,540){\line( 0, 1){ 20}}
\put(180,540){\line( 1, 0){ 20}}
\put(180,560){\line( 0,-1){ 20}}
\put(160,540){\line( 0, 1){ 20}}
\put(140,540){\line( 1, 0){ 20}}
\put(140,560){\line( 0,-1){ 20}}
\put(140,560){\line( 1, 0){ 20}}
\put(140,420){\line( 1, 0){ 20}}
\put(248,445){\makebox(0,0)[lb]{\raisebox{0pt}[0pt][0pt]{{\footnotesize $ 2$}}}}
\put(208,445){\makebox(0,0)[lb]{\raisebox{0pt}[0pt][0pt]{{\footnotesize $ 2$}}}}
\put(248,485){\makebox(0,0)[lb]{\raisebox{0pt}[0pt][0pt]{{\footnotesize $ 2$}}}}
\put(208,485){\makebox(0,0)[lb]{\raisebox{0pt}[0pt][0pt]{{\footnotesize $ 2$}}}}
\put(168,485){\makebox(0,0)[lb]{\raisebox{0pt}[0pt][0pt]{{\footnotesize $ 2$}}}}
\put(248,525){\makebox(0,0)[lb]{\raisebox{0pt}[0pt][0pt]{{\footnotesize $ 2$}}}}
\put(208,525){\makebox(0,0)[lb]{\raisebox{0pt}[0pt][0pt]{{\footnotesize $ 2$}}}}
\put(168,525){\makebox(0,0)[lb]{\raisebox{0pt}[0pt][0pt]{{\footnotesize $ 2$}}}}
\put(268,425){\makebox(0,0)[lb]{\raisebox{0pt}[0pt][0pt]{{\footnotesize $ 1$}}}}
\put(228,425){\makebox(0,0)[lb]{\raisebox{0pt}[0pt][0pt]{{\footnotesize $ 1$}}}}
\put(188,425){\makebox(0,0)[lb]{\raisebox{0pt}[0pt][0pt]{{\footnotesize $ 1$}}}}
\put(148,425){\makebox(0,0)[lb]{\raisebox{0pt}[0pt][0pt]{{\footnotesize $ 1$}}}}
\put(268,465){\makebox(0,0)[lb]{\raisebox{0pt}[0pt][0pt]{{\footnotesize $ 1$}}}}
\put(228,465){\makebox(0,0)[lb]{\raisebox{0pt}[0pt][0pt]{{\footnotesize $ 1$}}}}
\put(188,465){\makebox(0,0)[lb]{\raisebox{0pt}[0pt][0pt]{{\footnotesize $ 1$}}}}
\put(148,465){\makebox(0,0)[lb]{\raisebox{0pt}[0pt][0pt]{{\footnotesize $ 1$}}}}
\put(268,505){\makebox(0,0)[lb]{\raisebox{0pt}[0pt][0pt]{{\footnotesize $ 1$}}}}
\put(228,505){\makebox(0,0)[lb]{\raisebox{0pt}[0pt][0pt]{{\footnotesize $ 1$}}}}
\put(188,505){\makebox(0,0)[lb]{\raisebox{0pt}[0pt][0pt]{{\footnotesize $ 1$}}}}
\put(148,505){\makebox(0,0)[lb]{\raisebox{0pt}[0pt][0pt]{{\footnotesize $ 1$}}}}
\put(268,545){\makebox(0,0)[lb]{\raisebox{0pt}[0pt][0pt]{{\footnotesize $ 1$}}}}
\put(228,545){\makebox(0,0)[lb]{\raisebox{0pt}[0pt][0pt]{{\footnotesize $ 1$}}}}
\put(188,545){\makebox(0,0)[lb]{\raisebox{0pt}[0pt][0pt]{{\footnotesize $ 1$}}}}
\put(148,545){\makebox(0,0)[lb]{\raisebox{0pt}[0pt][0pt]{{\footnotesize $ 1$}}}}
\put(290,390){\makebox(0,0)[lb]{\raisebox{0pt}[0pt][0pt]{{\footnotesize $ (1)$}}}}
\put(105,567){\makebox(0,0)[lb]{\raisebox{0pt}[0pt][0pt]{{\footnotesize $ (2)$}}}}
\put(156,446){\makebox(0,0)[lb]{\raisebox{0pt}[0pt][0pt]{{\footnotesize
        ${\bf x}_{1,1}$}}}}
\put(125,446){\makebox(0,0)[lb]{\raisebox{0pt}[0pt][0pt]{{\footnotesize
        ${\bf x}_{0,1}$}}}}
\put(156,409){\makebox(0,0)[lb]{\raisebox{0pt}[0pt][0pt]{{\footnotesize
        ${\bf x}_{1,0}$}}}}
\put(125,409){\makebox(0,0)[lb]{\raisebox{0pt}[0pt][0pt]{{\footnotesize
        ${\bf x}_{0,0}$}}}}
\put(277,416){\makebox(0,0)[lb]{\raisebox{0pt}[0pt][0pt]{$\bullet$}}}
\put(257,416){\makebox(0,0)[lb]{\raisebox{0pt}[0pt][0pt]{$\bullet$}}}
\put(237,416){\makebox(0,0)[lb]{\raisebox{0pt}[0pt][0pt]{$\bullet$}}}
\put(217,416){\makebox(0,0)[lb]{\raisebox{0pt}[0pt][0pt]{$\bullet$}}}
\put(197,416){\makebox(0,0)[lb]{\raisebox{0pt}[0pt][0pt]{$\bullet$}}}
\put(177,416){\makebox(0,0)[lb]{\raisebox{0pt}[0pt][0pt]{$\bullet$}}}
\put(157,416){\makebox(0,0)[lb]{\raisebox{0pt}[0pt][0pt]{$\bullet$}}}
\put(137,416){\makebox(0,0)[lb]{\raisebox{0pt}[0pt][0pt]{$\bullet$}}}
\put(277,436){\makebox(0,0)[lb]{\raisebox{0pt}[0pt][0pt]{$\bullet$}}}
\put(257,436){\makebox(0,0)[lb]{\raisebox{0pt}[0pt][0pt]{$\bullet$}}}
\put(237,436){\makebox(0,0)[lb]{\raisebox{0pt}[0pt][0pt]{$\bullet$}}}
\put(217,436){\makebox(0,0)[lb]{\raisebox{0pt}[0pt][0pt]{$\bullet$}}}
\put(197,436){\makebox(0,0)[lb]{\raisebox{0pt}[0pt][0pt]{$\bullet$}}}
\put(177,436){\makebox(0,0)[lb]{\raisebox{0pt}[0pt][0pt]{$\bullet$}}}
\put(157,436){\makebox(0,0)[lb]{\raisebox{0pt}[0pt][0pt]{$\bullet$}}}
\put(137,436){\makebox(0,0)[lb]{\raisebox{0pt}[0pt][0pt]{$\bullet$}}}
\put(277,456){\makebox(0,0)[lb]{\raisebox{0pt}[0pt][0pt]{$\bullet$}}}
\put(257,456){\makebox(0,0)[lb]{\raisebox{0pt}[0pt][0pt]{$\bullet$}}}
\put(237,456){\makebox(0,0)[lb]{\raisebox{0pt}[0pt][0pt]{$\bullet$}}}
\put(217,456){\makebox(0,0)[lb]{\raisebox{0pt}[0pt][0pt]{$\bullet$}}}
\put(197,456){\makebox(0,0)[lb]{\raisebox{0pt}[0pt][0pt]{$\bullet$}}}
\put(177,456){\makebox(0,0)[lb]{\raisebox{0pt}[0pt][0pt]{$\bullet$}}}
\put(157,456){\makebox(0,0)[lb]{\raisebox{0pt}[0pt][0pt]{$\bullet$}}}
\put(137,456){\makebox(0,0)[lb]{\raisebox{0pt}[0pt][0pt]{$\bullet$}}}
\put(277,476){\makebox(0,0)[lb]{\raisebox{0pt}[0pt][0pt]{$\bullet$}}}
\put(257,476){\makebox(0,0)[lb]{\raisebox{0pt}[0pt][0pt]{$\bullet$}}}
\put(237,476){\makebox(0,0)[lb]{\raisebox{0pt}[0pt][0pt]{$\bullet$}}}
\put(217,476){\makebox(0,0)[lb]{\raisebox{0pt}[0pt][0pt]{$\bullet$}}}
\put(197,476){\makebox(0,0)[lb]{\raisebox{0pt}[0pt][0pt]{$\bullet$}}}
\put(177,476){\makebox(0,0)[lb]{\raisebox{0pt}[0pt][0pt]{$\bullet$}}}
\put(157,476){\makebox(0,0)[lb]{\raisebox{0pt}[0pt][0pt]{$\bullet$}}}
\put(137,476){\makebox(0,0)[lb]{\raisebox{0pt}[0pt][0pt]{$\bullet$}}}
\put(277,496){\makebox(0,0)[lb]{\raisebox{0pt}[0pt][0pt]{$\bullet$}}}
\put(257,496){\makebox(0,0)[lb]{\raisebox{0pt}[0pt][0pt]{$\bullet$}}}
\put(237,496){\makebox(0,0)[lb]{\raisebox{0pt}[0pt][0pt]{$\bullet$}}}
\put(217,496){\makebox(0,0)[lb]{\raisebox{0pt}[0pt][0pt]{$\bullet$}}}
\put(197,496){\makebox(0,0)[lb]{\raisebox{0pt}[0pt][0pt]{$\bullet$}}}
\put(177,496){\makebox(0,0)[lb]{\raisebox{0pt}[0pt][0pt]{$\bullet$}}}
\put(157,496){\makebox(0,0)[lb]{\raisebox{0pt}[0pt][0pt]{$\bullet$}}}
\put(137,496){\makebox(0,0)[lb]{\raisebox{0pt}[0pt][0pt]{$\bullet$}}}
\put(277,516){\makebox(0,0)[lb]{\raisebox{0pt}[0pt][0pt]{$\bullet$}}}
\put(257,516){\makebox(0,0)[lb]{\raisebox{0pt}[0pt][0pt]{$\bullet$}}}
\put(237,516){\makebox(0,0)[lb]{\raisebox{0pt}[0pt][0pt]{$\bullet$}}}
\put(217,516){\makebox(0,0)[lb]{\raisebox{0pt}[0pt][0pt]{$\bullet$}}}
\put(197,516){\makebox(0,0)[lb]{\raisebox{0pt}[0pt][0pt]{$\bullet$}}}
\put(177,516){\makebox(0,0)[lb]{\raisebox{0pt}[0pt][0pt]{$\bullet$}}}
\put(157,516){\makebox(0,0)[lb]{\raisebox{0pt}[0pt][0pt]{$\bullet$}}}
\put(137,516){\makebox(0,0)[lb]{\raisebox{0pt}[0pt][0pt]{$\bullet$}}}
\put(277,536){\makebox(0,0)[lb]{\raisebox{0pt}[0pt][0pt]{$\bullet$}}}
\put(257,536){\makebox(0,0)[lb]{\raisebox{0pt}[0pt][0pt]{$\bullet$}}}
\put(237,536){\makebox(0,0)[lb]{\raisebox{0pt}[0pt][0pt]{$\bullet$}}}
\put(217,536){\makebox(0,0)[lb]{\raisebox{0pt}[0pt][0pt]{$\bullet$}}}
\put(197,536){\makebox(0,0)[lb]{\raisebox{0pt}[0pt][0pt]{$\bullet$}}}
\put(177,536){\makebox(0,0)[lb]{\raisebox{0pt}[0pt][0pt]{$\bullet$}}}
\put(157,536){\makebox(0,0)[lb]{\raisebox{0pt}[0pt][0pt]{$\bullet$}}}
\put(137,536){\makebox(0,0)[lb]{\raisebox{0pt}[0pt][0pt]{$\bullet$}}}
\put(277,556){\makebox(0,0)[lb]{\raisebox{0pt}[0pt][0pt]{$\bullet$}}}
\put(257,556){\makebox(0,0)[lb]{\raisebox{0pt}[0pt][0pt]{$\bullet$}}}
\put(237,556){\makebox(0,0)[lb]{\raisebox{0pt}[0pt][0pt]{$\bullet$}}}
\put(217,556){\makebox(0,0)[lb]{\raisebox{0pt}[0pt][0pt]{$\bullet$}}}
\put(197,556){\makebox(0,0)[lb]{\raisebox{0pt}[0pt][0pt]{$\bullet$}}}
\put(177,556){\makebox(0,0)[lb]{\raisebox{0pt}[0pt][0pt]{$\bullet$}}}
\put(157,556){\makebox(0,0)[lb]{\raisebox{0pt}[0pt][0pt]{$\bullet$}}}
\put(137,556){\makebox(0,0)[lb]{\raisebox{0pt}[0pt][0pt]{$\bullet$}}}
\end{picture}


%% file: thesisfig13.latex
\setlength{\unitlength}{0.00063300in}%
\begingroup\makeatletter\ifx\SetFigFont\undefined
\def\x#1#2#3#4#5#6#7\relax{\def\x{#1#2#3#4#5#6}}%
\expandafter\x\fmtname xxxxxx\relax \def\y{splain}%
\ifx\x\y   
\gdef\SetFigFont#1#2#3{%
  \ifnum #1<17\tiny\else \ifnum #1<20\small\else
  \ifnum #1<24\normalsize\else \ifnum #1<29\large\else
  \ifnum #1<34\Large\else \ifnum #1<41\LARGE\else
     \huge\fi\fi\fi\fi\fi\fi
  \csname #3\endcsname}%
\else
\gdef\SetFigFont#1#2#3{\begingroup
  \count@#1\relax \ifnum 25<\count@\count@25\fi
  \def\x{\endgroup\@setsize\SetFigFont{#2pt}}%
  \expandafter\x
    \csname \romannumeral\the\count@ pt\expandafter\endcsname
    \csname @\romannumeral\the\count@ pt\endcsname
  \csname #3\endcsname}%
\fi
\fi\endgroup
\begin{picture}(4962,4131)(1789,-3973)
\thicklines
\put(1801,-3661){\vector( 1, 0){4800}}
\put(2701,-3061){\line( 0, 1){1800}}
\put(2701,-1261){\line( 1, 0){1800}}
\put(4501,-1261){\vector( 0,-1){525}}
\put(4501,-3061){\vector( 0, 1){525}}
\put(3601,-1261){\vector( 0,-1){525}}
\put(3601,-3061){\vector( 0, 1){525}}
\put(4501,-3061){\vector(-1, 0){525}}
\put(2701,-3061){\vector( 1, 0){525}}
\put(4576,-2161){\vector(-1, 0){600}}
\put(2701,-2161){\vector( 1, 0){525}}
\put(5401,-2161){\vector( 0,-1){525}}
\put(6301,-2161){\vector( 0,-1){525}}
\put(5401,-3061){\vector( 1, 0){525}}
\put(5401,-2161){\vector( 1, 0){525}}
\put(5401,-2161){\line( 0,-1){900}}
\put(5401,-3061){\line( 1, 0){900}}
\put(6301,-3061){\line( 0, 1){900}}
\put(6301,-2161){\line(-1, 0){900}}
\put(2701,-3061){\line( 1, 0){1800}}
\put(4501,-3061){\line( 0, 1){1800}}
\put(3601,-1261){\line( 0,-1){1800}}
\put(2701,-2161){\line( 1, 0){1800}}
\put(2660,-1336){\makebox(0,0)[lb]{$\bullet$}}
\put(2660,-2236){\makebox(0,0)[lb]{$\bullet$}}
\put(2660,-3136){\makebox(0,0)[lb]{$\bullet$}}
\put(3560,-3136){\makebox(0,0)[lb]{$\bullet$}}
\put(2101,-3961){\vector( 0, 1){3900}}
\put(4460,-3136){\makebox(0,0)[lb]{$\bullet$}}
\put(2101, 14){\makebox(0,0)[lb]{(2)}}
\put(4460,-2236){\makebox(0,0)[lb]{$\bullet$}}
\put(4460,-1336){\makebox(0,0)[lb]{$\bullet$}}
\put(3560,-1336){\makebox(0,0)[lb]{$\bullet$}}
\put(3560,-2236){\makebox(0,0)[lb]{$\bullet$}}
\put(5360,-2236){\makebox(0,0)[lb]{$\bullet$}}
\put(6260,-2236){\makebox(0,0)[lb]{$\bullet$}}
\put(5360,-3136){\makebox(0,0)[lb]{$\bullet$}}
\put(6260,-3136){\makebox(0,0)[lb]{$\bullet$}}
\put(2541,-3361){\makebox(0,0)[lb]{${\bf x}_{0,0}$}}
\put(3441,-3361){\makebox(0,0)[lb]{${\bf x}_{1,0}$}}
\put(4341,-3361){\makebox(0,0)[lb]{${\bf x}_{2,0}$}}
\put(2280,-2236){\makebox(0,0)[lb]{${\bf x}_{0,1}$}}
\put(3676,-2086){\makebox(0,0)[lb]{${\bf x}_{1,1}$}}
\put(4651,-2236){\makebox(0,0)[lb]{${\bf x}_{2,1}$}}
\put(2701,-1190){\makebox(0,0)[lb]{${\bf x}_{0,2}$}}
\put(3601,-1190){\makebox(0,0)[lb]{${\bf x}_{1,2}$}}
\put(4501,-1190){\makebox(0,0)[lb]{${\bf x}_{2,2}$}}
\put(5391,-3361){\makebox(0,0)[lb]{${\bf x}_{0,0}$}}
\put(5401,-2090){\makebox(0,0)[lb]{${\bf x}_{0,1}$}}
\put(6301,-2090){\makebox(0,0)[lb]{${\bf x}_{1,1}$}}
\put(6301,-3361){\makebox(0,0)[lb]{${\bf x}_{1,0}$}}
\put(6741,-3736){\makebox(0,0)[lb]{(1)}}
\end{picture}

%% file: figur15.latex
\setlength{\unitlength}{0.009375in}%
\begin{picture}(450,237)(20,300)
\thicklines
\put(440,500){\line(-1,-1){ 25}}
\put(390,500){\line(-1,-1){ 20}}
\put(350,460){\vector( 1, 1){ 25}}
\put(400,460){\vector( 1, 1){ 25}}
\put(460,520){\vector(-1,-1){ 25}}
\multiput(340,450)(7.74194,0.00000){16}{\line( 1, 0){  3.871}}
\put(330,460){\line( 1, 0){ 40}}
\put(400,460){\vector(-1, 0){ 35}}
\put( 20,300){\vector( 1, 1){ 90}}
\put( 40,300){\vector( 0, 1){220}}
\put( 20,320){\vector( 1, 0){440}}
\multiput(340,400)(7.74194,0.00000){16}{\line( 1, 0){  3.871}}
\put(390,430){\line(-1,-1){ 20}}
\put(350,390){\vector( 1, 1){ 25}}
\put(410,450){\vector(-1,-1){ 25}}
\put(280,460){\vector( 1, 0){ 60}}
\multiput(310,490)(0.00000,-7.74194){16}{\line( 0,-1){  3.871}}
\put(410,520){\vector(-1,-1){ 25}}
\put(430,370){\vector( 0, 1){ 40}}
\put(220,460){\line(-1, 0){ 60}}
\put(240,460){\vector(-1, 0){ 20}}
\put(120,420){\line( 1, 0){ 65}}
\put(160,400){\line( 0, 1){ 60}}
\put(240,400){\line( 0, 1){ 60}}
\put(220,440){\line( 1, 1){ 20}}
\put(140,440){\line( 1, 1){ 20}}
\put(200,420){\vector(-1, 0){ 20}}
\put(240,380){\vector( 0, 1){ 20}}
\put(160,380){\vector( 0, 1){ 20}}
\put(120,360){\line( 0, 1){ 60}}
\put(140,360){\line( 1, 1){ 20}}
\put(220,380){\line(-1, 0){ 60}}
\put(200,360){\line( 0, 1){ 60}}
\put(180,340){\line(-1, 0){ 60}}
\put(120,340){\vector( 0, 1){ 20}}
\put(200,340){\vector( 0, 1){ 20}}
\put(200,340){\vector(-1, 0){ 20}}
\put(240,380){\vector(-1, 0){ 20}}
\put(120,420){\vector( 1, 1){ 20}}
\put(200,420){\vector( 1, 1){ 20}}
\put(120,340){\vector( 1, 1){ 20}}
\put(220,360){\line( 1, 1){ 20}}
\put(200,340){\vector( 1, 1){ 20}}
\multiput(410,520)(0.00000,-7.74194){16}{\line( 0,-1){  3.871}}
\multiput(340,520)(0.00000,-7.74194){16}{\line( 0,-1){  3.871}}
\multiput(310,370)(7.74194,0.00000){16}{\line( 1, 0){  3.871}}
\multiput(280,340)(24.89312,24.89312){3}{\line( 1, 1){ 10.214}}
\multiput(280,390)(24.89312,24.89312){3}{\line( 1, 1){ 10.214}}
\multiput(280,460)(0.00000,-7.74194){16}{\line( 0,-1){  3.871}}
\multiput(340,520)(-24.89312,-24.89312){3}{\line(-1,-1){ 10.214}}
\multiput(280,340)(7.74194,0.00000){16}{\line( 1, 0){  3.871}}
\multiput(460,400)(-24.89312,-24.89312){3}{\line(-1,-1){ 10.214}}
\multiput(460,520)(0.00000,-7.74194){16}{\line( 0,-1){  3.871}}
\multiput(340,520)(7.74194,0.00000){16}{\line( 1, 0){  3.871}}
\put(430,490){\vector(-1, 0){ 40}}
\put(430,490){\vector( 0,-1){ 60}}
\put(350,460){\vector( 0,-1){ 55}}
\put(350,415){\line( 0,-1){ 45}}
\put(350,340){\vector( 0, 1){ 35}}
\put(430,440){\line( 0,-1){ 45}}
\put(380,435){\line( 0,-1){ 30}}
\put(380,490){\vector( 0,-1){ 60}}
\put(310,420){\vector( 1, 0){ 60}}
\put(355,420){\line( 1, 0){ 40}}
\put(430,420){\vector(-1, 0){ 40}}
\put(380,370){\vector( 0, 1){ 40}}
\put(360,490){\line( 1, 0){ 40}}
\put(310,490){\vector( 1, 0){ 55}}
\put(440,430){\line(-1,-1){ 20}}
\put(460,450){\vector(-1,-1){ 25}}
\put(400,390){\vector( 1, 1){ 25}}
\put(335,390){\line( 1, 0){ 30}}
\put(400,390){\vector(-1, 0){ 40}}
\put(280,390){\vector( 1, 0){ 60}}
\put(400,410){\line( 0,-1){ 40}}
\put(400,340){\vector( 0, 1){ 35}}
\put(400,460){\vector( 0,-1){ 55}}
\put(470,315){\makebox(0,0)[lb]{\raisebox{0pt}[0pt][0pt]{$(1)$}}}
\put( 82,390){\makebox(0,0)[lb]{\raisebox{0pt}[0pt][0pt]{$(2)$}}}
\put( 35,530){\makebox(0,0)[lb]{\raisebox{0pt}[0pt][0pt]{$(3)$}}}
\put(197,417){\makebox(0,0)[lb]{\raisebox{0pt}[0pt][0pt]{$\bullet$}}}
\put(117,417){\makebox(0,0)[lb]{\raisebox{0pt}[0pt][0pt]{$\bullet$}}}
\put(237,457){\makebox(0,0)[lb]{\raisebox{0pt}[0pt][0pt]{$\bullet$}}}
\put(157,457){\makebox(0,0)[lb]{\raisebox{0pt}[0pt][0pt]{$\bullet$}}}
\put(157,377){\makebox(0,0)[lb]{\raisebox{0pt}[0pt][0pt]{$\bullet$}}}
\put(237,377){\makebox(0,0)[lb]{\raisebox{0pt}[0pt][0pt]{$\bullet$}}}
\put(197,337){\makebox(0,0)[lb]{\raisebox{0pt}[0pt][0pt]{$\bullet$}}}
\put(117,337){\makebox(0,0)[lb]{\raisebox{0pt}[0pt][0pt]{$\bullet$}}}
\put(407,397){\makebox(0,0)[lb]{\raisebox{0pt}[0pt][0pt]{$\bullet$}}}
\put(407,447){\makebox(0,0)[lb]{\raisebox{0pt}[0pt][0pt]{$\bullet$}}}
\put(407,517){\makebox(0,0)[lb]{\raisebox{0pt}[0pt][0pt]{$\bullet$}}}
\put(377,417){\makebox(0,0)[lb]{\raisebox{0pt}[0pt][0pt]{$\bullet$}}}
\put(307,417){\makebox(0,0)[lb]{\raisebox{0pt}[0pt][0pt]{$\bullet$}}}
\put(377,367){\makebox(0,0)[lb]{\raisebox{0pt}[0pt][0pt]{$\bullet$}}}
\put(427,417){\makebox(0,0)[lb]{\raisebox{0pt}[0pt][0pt]{$\bullet$}}}
\put(347,387){\makebox(0,0)[lb]{\raisebox{0pt}[0pt][0pt]{$\bullet$}}}
\put(347,337){\makebox(0,0)[lb]{\raisebox{0pt}[0pt][0pt]{$\bullet$}}}
\put(347,457){\makebox(0,0)[lb]{\raisebox{0pt}[0pt][0pt]{$\bullet$}}}
\put(377,487){\makebox(0,0)[lb]{\raisebox{0pt}[0pt][0pt]{$\bullet$}}}
\put(397,387){\makebox(0,0)[lb]{\raisebox{0pt}[0pt][0pt]{$\bullet$}}}
\put(277,387){\makebox(0,0)[lb]{\raisebox{0pt}[0pt][0pt]{$\bullet$}}}
\put(457,447){\makebox(0,0)[lb]{\raisebox{0pt}[0pt][0pt]{$\bullet$}}}
\put(337,447){\makebox(0,0)[lb]{\raisebox{0pt}[0pt][0pt]{$\bullet$}}}
\put(427,487){\makebox(0,0)[lb]{\raisebox{0pt}[0pt][0pt]{$\bullet$}}}
\put(397,457){\makebox(0,0)[lb]{\raisebox{0pt}[0pt][0pt]{$\bullet$}}}
\put(307,487){\makebox(0,0)[lb]{\raisebox{0pt}[0pt][0pt]{$\bullet$}}}
\put(307,367){\makebox(0,0)[lb]{\raisebox{0pt}[0pt][0pt]{$\bullet$}}}
\put(427,367){\makebox(0,0)[lb]{\raisebox{0pt}[0pt][0pt]{$\bullet$}}}
\put(277,457){\makebox(0,0)[lb]{\raisebox{0pt}[0pt][0pt]{$\bullet$}}}
\put(457,517){\makebox(0,0)[lb]{\raisebox{0pt}[0pt][0pt]{$\bullet$}}}
\put(337,517){\makebox(0,0)[lb]{\raisebox{0pt}[0pt][0pt]{$\bullet$}}}
\put(457,397){\makebox(0,0)[lb]{\raisebox{0pt}[0pt][0pt]{$\bullet$}}}
\put(337,397){\makebox(0,0)[lb]{\raisebox{0pt}[0pt][0pt]{$\bullet$}}}
\put(397,337){\makebox(0,0)[lb]{\raisebox{0pt}[0pt][0pt]{$\bullet$}}}
\put(277,337){\makebox(0,0)[lb]{\raisebox{0pt}[0pt][0pt]{$\bullet$}}}
\end{picture}

%% file: edgemap.tex
\chapter{The $SO_{2}(\bR)$-Method in Computing Filters and Edge Maps}

In this chapter we exploit our 2-point factorization of the
orthonormal discrete wavelet transform in the last chapter to compute filters
by solving non-linear equations in several variables numerically. We
also try to improve on the preservation of regularity in the discrete
wavelet transform by constructing some (non-orthogonal) edge maps. 

\section{Computation of orthonormal filters}

We start with the construction of an orthonormal filter of length 4
with a maximum number of vanishing moments compatible with its support
width. That is, we want $\alpha_{1}$, $\alpha_{2}$ so that 

\begin{eqnarray}
  && F_{2}(\alpha_{2})S(1)F_{2}(\alpha_{1}){\bf 1} \mapsto
  \delta_{1,k} \nonumber \\
  && F_{2}(\alpha_{2})S(1)F_{2}(\alpha_{1}){\bf n} \mapsto
  \delta_{1,k} \nonumber \\
  \label{vanishing moments}
\end{eqnarray}

This is simple enough to do by handcalculation.
 
\begin{equation}
  c^{1}_{1} = \frac{\alpha_{1}{\bf x}(0) + {\bf x}(1) - \alpha_{2}{\bf
      x}(2) + \alpha_{1}\alpha_{2}{\bf x}(3)}
  {(1 + \alpha_{1}^2)^{1/2}(1 + \alpha_{2}^2)^{1/2}} 
  \label{computed c_{1}}
\end{equation}

\begin{equation}
  d^{1}_{1} = \frac{\alpha_{1}\alpha_{2}{\bf x}(0) + \alpha_{2}{\bf x}(1) 
    + {\bf x}(2) - \alpha_{1}{\bf x}(3)}
  {(1 + \alpha_{1}^2)^{1/2}(1 + \alpha_{2}^2)^{1/2}} 
  \label{computed d_{1}}
\end{equation}

\noindent We insert the vanishing moment conditions from 
(\ref{vanishing moments}) and get the equation-set

\[ \left[ \begin{array}{l}
    \alpha_{1}\alpha_{2}+ \alpha_{2} - \alpha_{1} + 1 = 0 \\
    \alpha_{2} - 3\alpha_{1} + 2 = 0
  \end{array}\right] 
\]

\noindent Solving yields

\[ (\alpha_{1},\alpha_{2}) \in \{(1/\sqrt{3},\sqrt{3} - 2),(-1/\sqrt{3},2 +
\sqrt{3})\}
\]

\noindent We find the lowpass filter by inserting the solutions in 
(\ref{computed c_{1}}).

\begin{table}[h]
  \begin{center}
    {\scriptsize \begin{tabular}{|c|c|c|}\hline
        n & $H_{4}(n)$      & $\alpha_{n}$ \\ \hline
        0 &  0.482962913    &  \\ \hline
        1 &  0.836516304    & $1/\sqrt{3}$ \\ \hline
        2 &  0.224143868    & $\sqrt{3} - 2$  \\ \hline
        3 & -0.129409523    & \\ \hline
      \end{tabular}}
  \end{center}
  \caption{The filter $H_{4}$ and the entries in the
    corresponding  $M_{2}(\alpha_{n})$.} 
  \label{L=4} 
\end{table}

The two solution-sets are mirror-images of each other, it is the filter
shown in Table \ref{L=4} that corresponds to our choice of rotation-matrix
$M_{2}(\alpha_{k})$. 

To find orthonormal filters of length $L$, 
the above method leads to the wearying task of solving $\frac{L}{2}$ 
nonlinear equations in $\frac{L}{2}$
variables. However, by use of some Maple-routines we succeeded in
generating $H^{d}_{6}$ and {\em some} orthogonal filter $H_{8}$
corresponding to a wavelet with 4 vanishing moments.

\begin{table}[h]
  \begin{center}
    {\scriptsize \begin{tabular}{|c|c|c|}\hline
        n &         ${\bf H}^{d}_{8}(n)$  & $\alpha_{n}$ \\ \hline
        0 &   .03222310057  &                  \\ \hline
        1 & - .0126039675   & - 2.556583915    \\ \hline
        2 & - .09921954341  & - .1434214911    \\ \hline
        3 &   .2978577953   &  .7958755204     \\ \hline
        4 &   .8037387510   & - 2.351285662    \\ \hline
        5 &   .4976186668   &                  \\ \hline
        6 & - .02963552763  &                  \\ \hline
        7 & - .07576571472  &                  \\ \hline
      \end{tabular}}
  \end{center}
  \caption{The computed filter $H_{8}$ and the entries in the
    corresponding  $M_{2}(\alpha_{k})$.}
  \label{case4} 
\end{table}

\noindent {\bf Remark:} The filter $H_{8}$ in Table \ref{case4}
is not the one given in \cite{ten}, but these filters are not uniquely
determined by claiming a maximum number of vanishing moments only.

Now we turn to coiflets. Claiming vanishing moments on the scaling
function $\phi$ leads to the following observation: 

\begin{eqnarray}
  && \mbox{If } \sum_{n=0}^{L-1}{H(n)} \cdot n^{p} = 
  0 \ \mbox{ for } \forall \ p
  \mbox{, } 1 \leq p \leq M \leq \frac{L}{2}-1, \nonumber \\
  && \mbox{where $M$ is odd, then} \nonumber \\
  && \sum_{n=0}^{L-1}H(n) \cdot n^{M+1} = 0.
  \label{free vanishing moments}
\end{eqnarray}

This means that it suffices to impose odd vanishing moments on
$\phi$. Counting degrees of freedom in the filtering-diagram we find
that for $H^{coiflet}_{L}$, if the highpass-filter has $M$ vanishing
moments, then the lowpassfilter has $2(\frac{L}{2}-M)$ vanishing
moments. The proof of this is given below. 
Maple were able to solve the nonlinear 
equations to give $H^{coiflet}_{6}$ and  {\em some} $H^{coiflet}_{8}$,
where we used 3 out 4 degrees of freedom to provide the highpass filter
with 3 vanishing moments, yielding 2 vanishing moments on the scaling
function. 

\begin{table}[h]
  \begin{center}
    {\scriptsize \begin{tabular}{|c|c|c|}\hline
        n &  $H^{coiflet}_{8}(n)$  & $\alpha_{n}$ \\ \hline
        -3 & - .07342587213  &                    \\ \hline
        -2 & - .03314563036  &                    \\ \hline
        -1 &   .4854426593   &                    \\ \hline
        0 &   .8065436723    &                    \\ \hline
        1 &   .3100524696    &    2.215250436     \\ \hline
        2 & - .09943689110   &     .1504720765    \\ \hline
        3 & - .01496247550   &   - .7384168123    \\ \hline
        4 &   .03314563037   &   - .451416230     \\ \hline
      \end{tabular}}
    \end{center}
    \caption{The computed filter $H^{coiflet}_{8}$ and the entries in the
      corresponding  $M_{2}(\alpha_{n})$. The highpass filter has 3
      vanishing moments, the lowpass filter has 2 vanishing moments.} 
    \label{lowpass coiflet of length 8}
  \end{table}

\vspace{5mm}

\noindent {\bf Proof of} (\ref{free vanishing moments}): \newline

\noindent We have 

\begin{eqnarray*}
  & & 0 = \int x^{l} \phi(x) dx \mbox{ , } \forall \ p \mbox{, }
  1 \leq p \leq M \leq \frac{L}{2}-1  \\
  \Longleftrightarrow & & \left. \frac{d^{p}}{d\xi^{p}}
    \widehat{\phi} \right|_{\xi=0} = 0  \mbox{, for } \forall \ p \mbox{, } 1
  \leq p \leq M \leq \frac{L}{2}-1. 
\end{eqnarray*}

\noindent Fourier transforming equation (\ref{scaling equation}) we
get 

\[ \widehat{\phi}(\xi) = m_{0}(\xi/2) \widehat{\phi}(\xi/2) 
\]  

\noindent where $m_{0}$ is the trigonometric polynomial that is the
Fourier transform of the filter $H^{coif}_{L}$. Differentiation yields

\[ 0 = \widehat{\phi}^{\prime}(0) = \frac{1}{2}m^{\prime}_{0}(0) 
\widehat{\phi}(0) + \frac{1}{2} m_{0}(0) \widehat{\phi}^{\prime}(0) = 
\frac{1}{2 \sqrt{\pi}}m^{\prime}_{0}(0).
\]

\noindent Repeating this argument $M$ times we get

\[ \left. \frac{d^{p}}{d\xi^{p}} m_{0} \right|_{\xi=0} = 0,  \mbox{  for
  } \forall \ p \mbox{, } 1 \leq p \leq M, 
\]

\noindent thus we conclude that $m_{0}$ must be on the form

\begin{equation}
  m_0(\xi) = 1 + \left( 1- e^{-i\xi} \right)^{M+1} P(\xi) \label{P}
\end{equation} 

\noindent for some trigonometric polynomial P. Assuming that the coefficients
$H^{coif}_{L}(n)$ of $m_{0}$ are real, we have 

\[ m_0(-\xi) = \Re \ m_0(\xi) - i \cdot \Im \ m_{0}(\xi) =
\overline{m_{0}(\xi)},
\] 

\noindent which implies

\[ \left| m_{0}(\xi) \right|^{2} = m_{0}(\xi) \cdot m_{0}(-\xi) 
\]

\noindent Expanding $m_{0}$ in a Taylorseries around the origin
\[ m_{0}(\xi) = 1 + a_{1}\xi + a_{2}\xi^{2} + \cdots + a_{n}\xi^{n} +
\cdots \mbox{ ,where } a_{k} = \frac{m_{0}^{(k)}(0)}{k!} 
\] 

\noindent we get

\begin{eqnarray*}
  \left| m_{0}(\xi) \right|^{2}  &=& ( 1 + a_{1}\xi + a_{2}\xi^{2} +
  a_{3}\xi^{3} + \cdots )( 1 - a_{1}\xi + a_{2}\xi^{2} -
  a_{3}\xi^{3} + \cdots )  \\
  &= &  1 + c_{M+1}\xi^{M+1} + c_{M+2}\xi^{M+2} + \cdots + c_{M+n}\xi^{M+n}
  + \cdots 
\end{eqnarray*}

\noindent for some sequence $\{c_{M+i}\}_{i=1}^{\infty}$ because of
(\ref{P}). By computing the above product of the two series, we
get the relations

\[ c_{0} = 1 \mbox{ , }  c_{2k+1} = 0, 
\]

\begin{equation}
  c_{2k} = (-1)^k a_{k}^{2} + 2 \sum_{i=0}^{k-1}(-1)^i a_{2k-i}a_{i}, 
  \mbox{ where $a_{0}$ = 1.} 
  \label{c}
\end{equation}

\noindent Moreover
\
\begin{eqnarray}
  \left| a_{k}k! \right| & = &\left| m_{0}^{(k)}(0) \right| = 
  \left| \frac{1}{\sqrt{2}}
    \sum_{n}H^{coif}_{L}(n) (-in)^{k} e^{-in\cdot 0 } \right| \nonumber \\
  & = & \frac{1}{\sqrt{2}} \sum_{n}H^{coif}_{L}(n)n^{k}. \nonumber \\ 
  \label{a}
\end{eqnarray}

\noindent Now, claiming the $c_{k} = 0 \mbox{ for } 1 \leq k \leq M$
and using (\ref{c}), (\ref{a}), the result follows. 
Indeed, one has

\begin{eqnarray*}
  & &\sum_{n=0}^{L-1}H^{coif}_{L}(n) n^{p} = 0 \ \mbox{ for } \forall \ p
  \mbox{, } 1 \leq p \leq M\leq \frac{L}{2}-1, \\
  && \mbox{where $M$ is odd,} \\
  \Longrightarrow & & \sum_{n=0}^{L-1}H^{coif}_{L}(n)n^{M+1} =
  \sum_{n=0}^{L-1}H^{coif}_{L}(n)n^{M+3} = 0.  
\end{eqnarray*}

\endproof

\section{Ways of dealing with the edge problem}

We look for ways of ``smoothing'' the edges of a vector in $\bR^{n}$.
That is, we look for operators acting on the edge coefficients
that to some degree possess the following properties:

\begin{itemize}
\item {\bf P1}: Linearity and orthogonality.

\item {\bf P2}: In some strict sense, ``nice functions'' should map to ``nice
  functions''.

\item {\bf P3}: The map should ensure some vanishing moments on the $\tilde{d}$'s.
\end{itemize} 

We have to make clear the meaning of {\bf P2}: Considering the coefficients
$c_{k}$ at some level $j$, we see from

\begin{eqnarray}
  \int x^{M} \phi_{j,k}(x) dx & = & 2^{j/2} \int x^{M}
  \phi(2^{j}x-k)dx \nonumber \\
  &=& 2^{-j/2} \int (2^{-j}(x+k))^{M} \phi(x)dx \nonumber \\
  &=& 2^{-j(M+\frac{1}{2})} \int (x+k)^{M} \phi(x) dx \nonumber \\
  &=& P_{M}(k) \mbox{ , $P_{M}$ a polynomial of degree M}, \nonumber
  \\
  \label{polynomial mapsto polynomial}
\end{eqnarray}

\noindent that a polynomial maps to a polynomial of the same degree under
$\Theta_{n}(H_{L},G_{L})$. Depending on the number of vanishing moments on
the corresponding $\psi$, the $d_{k}$'s will be zero.

\subsection{The method of periodizing}

This method simply periodizes the vector around its edges. This 
yields orthogonality of the transform, but only preserves regularity
up to constant functions, and we get only 1 vanishing moment on the
$\tilde{d}_{k}$'s. 

\subsection{The method of mirroring}
 
One way to overcome the edge problem is to mirror the
input vector about each of its endpoints and then apply the filter to 
the mirrored points.

The method of mirroring does not lead to an orthogonal transform, but
it preserves regularity up to continuity, and yields
1 vanishing moment on the $\tilde{d}_{k}$'s.

\subsection{The method of edge matrices}

Another approach is to try to construct a matrix map acting directly on
the edgepoints such that the properties {\bf P1}, {\bf P2} and {\bf P3}
are satisfied to some extent. We will do this construction by claiming
some vanishing moments on the $\tilde{d}_{k}$'s and claiming some
degree of preservation of regularity of the transform at edges by
claiming monomials up to some degree $M$ mapping continuously to
polynomials of degree $M$. That is, we compute the polynomials 
$P_{j}$ given in (\ref{polynomial mapsto polynomial}) 
for $0\leq j \leq M$, and impose $x^{j} \rightarrow P_{j}\mbox{ ,}
0\leq j \leq k(L)< M$ at edges, $L$ is the length of lowpass filter. 
For a orthonormal FIR filter $H_{L}$ we will construct two 
$(\frac{L}{2}-1) \times (\frac{L}{2}-1)$ matrices 
${\bf E}_{H_{L},l}$, ${\bf E}_{H_{L},r}$, each acting on the ordered
set ${\bf e}_{l}$, ${\bf e}_{r}$  
of $\frac{L}{2}-1$ edgepoints on the lefthand and righthandside, 
respectively, to give the ``tilde'' coefficients. For example 

\[{\bf E}_{H_{10},l} : {\bf e}_{l} \mapsto 
(\tilde{c}_{0}, \ \tilde{d}_{0},
\ \tilde{c}_{1}, \ \tilde{d}_{1}) 
\]

\[ {\bf E}_{H_{10},r} : {\bf e}_{r} \mapsto 
(\tilde{c}_{n/2-2}, \ \tilde{d}_{n/2-2},
\ \tilde{c}_{n/2-1}, \ \tilde{d}_{n/2-1}) 
\]    

\noindent This is illustrated in Figure \ref{edgematrix1}.
\newline

\begin{figure}[h]
  \begin{center}
    \input{figur6.latex}
  \end{center}
  \caption{}
  \label{edgematrix1}
\end{figure}

\noindent As for fulfilling the above listed properties, this edge map is of
course linear, but not orthogonal. We have sacrificed orthogonality 
to be able to achieve {\bf P2} and {\bf P3}. 
Some edge-matrices are shown in the appendix.

\subsection{Comparing the methods}

The profits of the edge matrices are obvious: 

\begin{itemize}

\item {\bf Good}: Preservation of regularity to a higher degree than
  by simply mirroring at edges. 
\end{itemize}

The drawbacks are equally obvious: 

\begin{itemize}
\item {\bf Bad}: This edgemap is very far from orthogonal in
  the sense that the operator norm of the map turns out to be large, 
  and it seems to grow with increasing degree of preservation of regularity
  as can be seen in Table \ref{daubechiesbounds} and Table \ref{coifbounds}.

\end{itemize} 

To be able to compare the methods of mirroring and edge matrices in this
non-orthogonality respect, we proceed as follows:

Given the vector ${\bf x} \in {\bR}^{n}$ and some orthogonal FIR filter
$H_{L}$, we compute the $\tilde{c}_{k}$'s and the  
$\tilde{d}_{k}$'s, getting

\begin{equation} 
  \tilde{c}_{k} = \sum_{i=0}^{K(k)}\gamma(i){\bf x}(i)
  \label{edge lowpass}
\end{equation} 

\noindent when we use edge matrices, and   

\begin{equation} 
  \tilde{c}_{k} = \sum_{i=0}^{J(k)}\epsilon(i){\bf x}(i)
  \label{mirror lowpass} 
\end{equation} 

\noindent when we mirror ${\bf x}$. The expressions for the 
$\tilde{d}_{k}$'s are similar. We use Cauchy-Schwartz on 
(\ref{edge lowpass}) and (\ref{mirror lowpass}). On the left edge we set 

\begin{eqnarray*}
  Q_{l}(c_{k}) &=&
  \left(\sum_{i=0}^{K(k)}\left|\gamma(i)\right|^{2}\right)^{1/2} \\
  P_{l}(c_{k}) &=& \
  \left(\sum_{i=0}^{J(k)}\left|\epsilon(i)\right|^{2}\right)^{1/2}. \\
\end{eqnarray*} 

\noindent On the right edge, the bounds $Q_{r}$, $P_{r}$ are defined
similarly. Now, we may compare the two methods in the non-orthogonality
respect. Results corresponding to the edge matrices shown in the
appendix are given in Table \ref{daubechiesbounds} and 
Table \ref{coifbounds}.

\begin{table}[h]
  \begin{center}
    {\scriptsize \begin{tabular}{|c|l|l|l|l|}\hline
        \multicolumn{1}{|p{20mm}|}{$l^{2}$ bounds on the 
          $|\tilde{c}_{k}|$, $|\tilde{d}_{k}|$.} &
          \multicolumn{4}{|c|}{$L$ in $H^{d}_{L}$.} \\ 
          \cline{2-5}
          & 6 & 8 & 10 & 12 \\ \hline

          $P_{l}(\tilde{c}_{0})$ &.852  &.935  &1.05  &1.04    \\ \hline 
          $Q_{l}(\tilde{c}_{0})$ &9.50  &21.8  &310   &757   \\ \hline
          $P_{l}(\tilde{d}_{0})$ &.989  &.907  &.973  &.963   \\ \hline
          $Q_{l}(\tilde{d}_{0})$ &1.39  &2.14  &11.0  &2.02   \\ \hline
          $P_{l}(\tilde{c}_{1})$ &      &      &1.04  &1.12   \\ \hline
          $Q_{l}(\tilde{c}_{1})$ &      &      &48.0  &103   \\ \hline
          $P_{l}(\tilde{d}_{1})$ &      &.994  &.999  &.996    \\ \hline
          $Q_{l}(\tilde{d}_{1})$ &      &2.59  &1.58  &7.98    \\ \hline
          $P_{l}(\tilde{d}_{2})$ &      &      &      &1.00      \\ \hline
          $Q_{l}(\tilde{d}_{2})$ &      &      &      &1.10       \\ \hline
          $P_{r}(\tilde{c}_{0})$ &.852  &1.01  &1.05  &1.01   \\ \hline 
          $Q_{r}(\tilde{c}_{0})$ &1.01  &1.03  &1.01  &1.03   \\ \hline
          $P_{r}(\tilde{d}_{0})$ &.989  &.994  &.973  &.963   \\ \hline
          $Q_{r}(\tilde{d}_{0})$ &.512  &.402  &.973  &3.13   \\ \hline
          $P_{r}(\tilde{c}_{1})$ &      &.935  &1.04  &1.04   \\ \hline
          $Q_{r}(\tilde{c}_{1})$ &      &1.00  &1.00  &1.00   \\ \hline
          $P_{r}(\tilde{d}_{1})$ &      &      &.999  &.996   \\ \hline
          $Q_{r}(\tilde{d}_{1})$ &      &      &2.58  &5.03   \\ \hline
          $P_{r}(\tilde{c}_{2})$ &      &      &      &1.12   \\ \hline
          $Q_{r}(\tilde{c}_{2})$ &      &      &      &1.00   \\ \hline
          
        \end{tabular}}
    \end{center}
    \caption{The $l^{2}$ bounds for the edge operators 
      ${\bf E}_{H^{d}_{L},\cdot}$ using Daubechies shortest filters.}
    \label{daubechiesbounds}
\end{table}

\begin{table}[h]
  \begin{center}
    {\scriptsize \begin{tabular}{|c|l|l|l|}\hline
        \multicolumn{1}{|p{20mm}|}{$l^{2}$ bounds on the 
          $|\tilde{c}_{k}|$, $|\tilde{d}_{k}|$.}&
        \multicolumn{3}{|c|}{$L$ in $H^{coif}_{L}$.} \\ 
        \cline{2-4}
        & 6 & 8 & 12 \\ \hline
        
        $P_{l}(\tilde{c}_{0})$ &1.13  &.949  &1.16 \\ \hline 
        $Q_{l}(\tilde{c}_{0})$ &4.75  &2.43  &25.0 \\ \hline
        $P_{l}(\tilde{d}_{0})$ &.930  &1.15  &.909 \\ \hline
        $Q_{l}(\tilde{d}_{0})$ &1.97  &10.4  &2.31 \\ \hline
        $P_{l}(\tilde{c}_{1})$ &      &      &.997  \\ \hline
        $Q_{l}(\tilde{c}_{1})$ &      &      &2.50 \\ \hline
        $P_{l}(\tilde{d}_{1})$ &      &1.02  &.999  \\ \hline
        $Q_{l}(\tilde{d}_{1})$ &      &1.54  &2.35 \\ \hline
        $P_{l}(\tilde{d}_{2})$ &      &      &1.00  \\ \hline
        $Q_{l}(\tilde{d}_{2})$ &      &      &2.31  \\ \hline
        $P_{r}(\tilde{c}_{0})$ &1.13  &.907  &.997  \\ \hline 
        $Q_{r}(\tilde{c}_{0})$ &1.02  &1.97  &1.45  \\ \hline
        $P_{r}(\tilde{d}_{0})$ &.930  &1.02  &.999  \\ \hline
        $Q_{r}(\tilde{d}_{0})$ &.705  &.908  &3.33 \\ \hline
        $P_{r}(\tilde{c}_{1})$ &      &.949  &1.16 \\ \hline
        $Q_{r}(\tilde{c}_{1})$ &      &1.10  &1.03\\ \hline
        $P_{r}(\tilde{d}_{1})$ &      &      &1.00  \\ \hline
        $Q_{r}(\tilde{d}_{1})$ &      &      &19.5  \\ \hline
        $P_{r}(\tilde{c}_{2})$ &      &      &.997  \\ \hline
        $Q_{r}(\tilde{c}_{2})$ &      &      &1.16  \\ \hline

      \end{tabular}}
  \end{center}
  \caption{The $l^{2}$ bounds for the different edge operators 
    ${\bf E}_{H^{coif}_{L},\cdot}$ using Coiflet filters.}
  \label{coifbounds}
\end{table}

Table \ref{daubechiesbounds} and Table \ref{coifbounds} clearly show
the blowup of some of the edge-coefficients that results from using edge
matrices. However, for the Daubechies filters the blowup effect almost
exclusively affects the lefthandside, while for the Coiflet filters the blowup
effect does not seem to favour any side, and is on the whole more
moderate. Anyway, this blowup will affect the numerical stability of
the filtering-reconstruction procedure.


%% file: figur6.latex
\setlength{\unitlength}{0.0125in}%
\begin{picture}(165,129)(195,365)
\thicklines
\put(225,430){\vector( 0, 1){ 40}}
\put(300,480){\line(-1, 0){105}}
\multiput(320,380)(7.27273,0.00000){6}{\line( 1, 0){  3.636}}
\put(200,380){\line( 1, 0){120}}
\put(240,420){\line( 1,-1){ 20}}
\put(240,400){\line( 1, 1){ 20}}
\put(280,385){\line( 0,-1){ 10}}
\put(260,400){\line( 1,-1){ 20}}
\put(260,380){\line( 1, 1){ 20}}
\put(240,400){\line(-1,-1){ 20}}
\put(220,400){\line( 1,-1){ 20}}
\put(260,385){\line( 0,-1){ 10}}
\put(240,385){\line( 0,-1){ 10}}
\put(220,385){\line( 0,-1){ 10}}
\put(200,380){\line( 1, 1){100}}
\put(175,445){\makebox(0,0)[lb]{\raisebox{0pt}[0pt][0pt]{${\bf E}_{H_{L},l}$}}}
\put(245,460){\makebox(0,0)[lb]{\raisebox{0pt}[0pt][0pt]{${\bf e}_{l}(3)$}}}
\put(205,420){\makebox(0,0)[lb]{\raisebox{0pt}[0pt][0pt]{${\bf e}_{l}(1)$}}}
\put(225,440){\makebox(0,0)[lb]{\raisebox{0pt}[0pt][0pt]{${\bf e}_{l}(2)$}}}
\put(185,400){\makebox(0,0)[lb]{\raisebox{0pt}[0pt][0pt]{${\bf e}_{l}(0)$}}}
\put(298,495){\makebox(0,0)[lb]{\raisebox{0pt}[0pt][0pt]{$c^{1}_{2}$}}}
\put(274,495){\makebox(0,0)[lb]{\raisebox{0pt}[0pt][0pt]{$\tilde{d}_1$}}}
\put(254,495){\makebox(0,0)[lb]{\raisebox{0pt}[0pt][0pt]{$\tilde{c}_1$}}}
\put(234,495){\makebox(0,0)[lb]{\raisebox{0pt}[0pt][0pt]{$\tilde{d}_0$}}}
\put(214,495){\makebox(0,0)[lb]{\raisebox{0pt}[0pt][0pt]{$\tilde{c}_0$}}}
\put(298,477){\makebox(0,0)[lb]{\raisebox{0pt}[0pt][0pt]{$\bullet$}}}
\put(278,477){\makebox(0,0)[lb]{\raisebox{0pt}[0pt][0pt]{$\bullet$}}}
\put(258,477){\makebox(0,0)[lb]{\raisebox{0pt}[0pt][0pt]{$\bullet$}}}
\put(238,477){\makebox(0,0)[lb]{\raisebox{0pt}[0pt][0pt]{$\bullet$}}}
\put(218,477){\makebox(0,0)[lb]{\raisebox{0pt}[0pt][0pt]{$\bullet$}}}
\put(276,365){\makebox(0,0)[lb]{\raisebox{0pt}[0pt][0pt]{${\bf x}_{3}$}}}
\put(258,417){\makebox(0,0)[lb]{\raisebox{0pt}[0pt][0pt]{$\bullet$}}}
\put(238,397){\makebox(0,0)[lb]{\raisebox{0pt}[0pt][0pt]{$\bullet$}}}
\put(258,397){\makebox(0,0)[lb]{\raisebox{0pt}[0pt][0pt]{$\bullet$}}}
\put(278,397){\makebox(0,0)[lb]{\raisebox{0pt}[0pt][0pt]{$\bullet$}}}
\put(256,365){\makebox(0,0)[lb]{\raisebox{0pt}[0pt][0pt]{${\bf x}_{2}$}}}
\put(236,365){\makebox(0,0)[lb]{\raisebox{0pt}[0pt][0pt]{${\bf x}_{1}$}}}
\put(216,365){\makebox(0,0)[lb]{\raisebox{0pt}[0pt][0pt]{${\bf x}_{0}$}}}
\put(278,457){\makebox(0,0)[lb]{\raisebox{0pt}[0pt][0pt]{$\bullet$}}}
\put(258,437){\makebox(0,0)[lb]{\raisebox{0pt}[0pt][0pt]{$\bullet$}}}
\put(238,417){\makebox(0,0)[lb]{\raisebox{0pt}[0pt][0pt]{$\bullet$}}}
\put(218,397){\makebox(0,0)[lb]{\raisebox{0pt}[0pt][0pt]{$\bullet$}}}
\end{picture}


%% file: features.tex
\chapter[Classification of Radar Signals]{Classification of Radar
  Signals  Using Local Feature Extraction in the Space-Frequency Plane}

\section{Formulation of the problem}

We consider the problem of separating two different distributions 
(classes) of electromagnetic radar signal sources from one another by doing a
space-frequency analysis on the signals. Each distribution $D_{n}(\alpha)$
of signal sources will consist of a number of $n$ coherent signal
transmitters distributed randomly over $n$ regularly spaced plane 
domains $\{\Omega_{n,i}(\alpha)\}_{i=1}^{n}$ with one and only one 
transmitter in each domain $\Omega(n,i)$. The domains may overlap.

\[ \Omega_{n,i}(\alpha) = \{(r\cos\theta,r\sin\theta): 
1\leq r \leq 10, \mbox{ } 2\pi\frac{i}{n} \leq \theta \leq 
2\pi\frac{i}{n} + \alpha\}.
\]
 
\begin{figure}[h]
  \begin{center}
    \includegraphics[scale = 0.6]{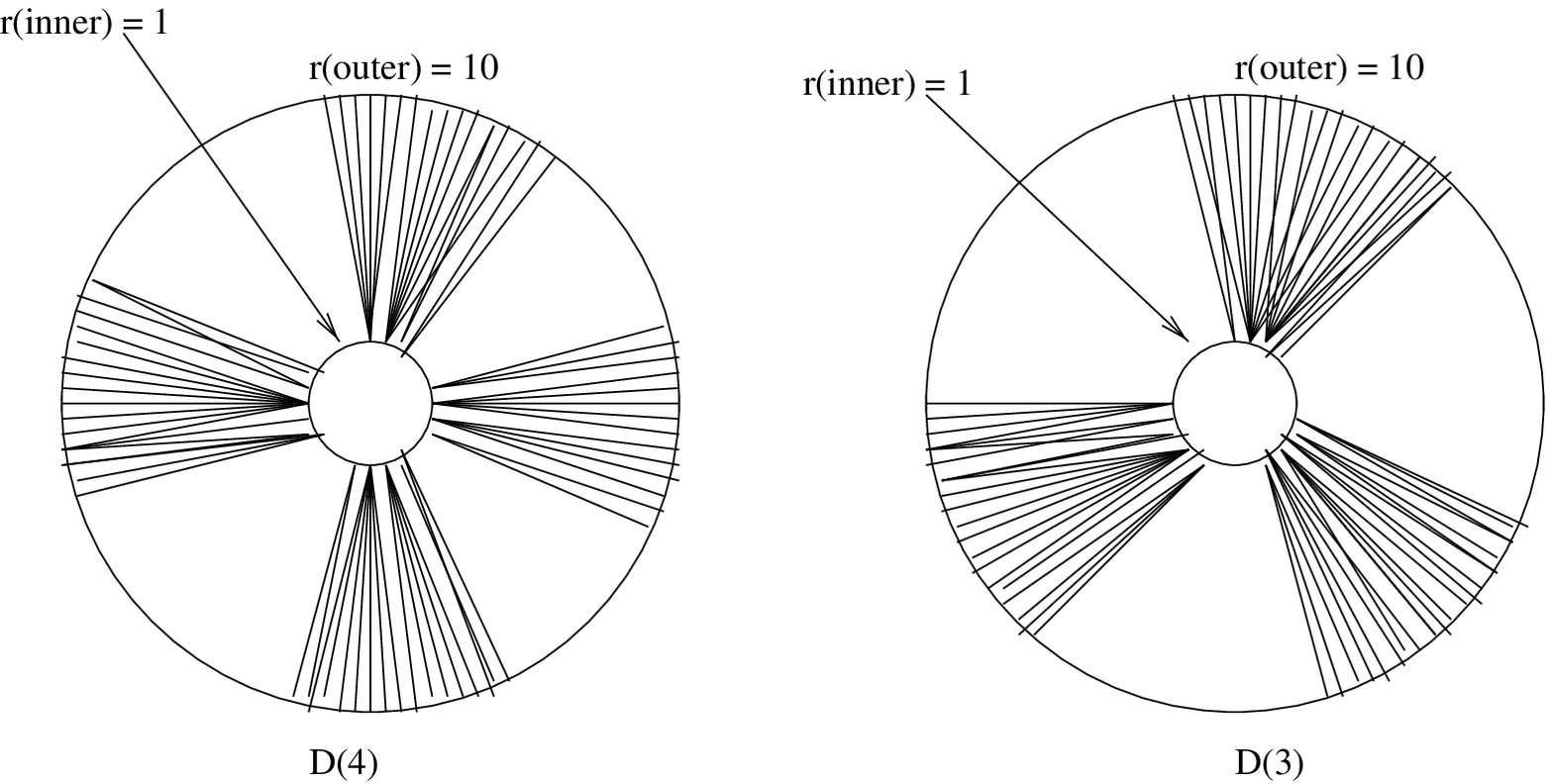}
  \end{center}
  \caption{The domains of $D_{4}(\cdot)$ and $D_{3}(\cdot)$.}
  \label{signalsources}
\end{figure}

\noindent A picture of $D_{3}(\cdot)$ and $D_{4}(\cdot)$ is shown in 
Figure \ref{signalsources}, where the shaded areas are the 
$\{\Omega_{3,i}\}_{i=1}^{3}$ and the $\{\Omega_{4,i}\}_{i=1}^{4}$.

The analytical forms of the signals are solutions of the wave equation
in three dimensions:

\begin{equation}
  \frac{\partial^{2}\psi({\bf x},t)}{\partial^{2}t} -
  c^{2}\nabla^{2}\psi({\bf x},t) = 0,
\end{equation}

\noindent where $\nabla^{2} = \frac{\partial^{2}}{\partial^{2}x_{1}}
+  \frac{\partial^{2}}{\partial^{2}x_{2}} + 
 \frac{\partial^{2}}{\partial^{2}x_{3}}$.

Making the ansatz $\psi({\bf x},t) = \phi({\bf x})e^{-i\omega t}$, 
we reduce the problem to the Helmholtz equation

\begin{equation} 
  \nabla^{2}\phi({\bf x}) +k^{2}\phi({\bf x}) = 0, \mbox{ }
  k = \frac{\omega}{c}.
\end{equation}

\noindent The fundamental solution of this equation may be
expressed in polar coordinates as

\[ \phi(r,\theta) = \phi(r) = A\frac{e^{ikr}}{r}, \mbox{ } 
A \mbox{ is a constant}. \] 

Writing $\{{\bf p}_{i}\}_{i=1}^{n}$ for the locations of the
signal sources in a signal source distribution $D_{n}(\cdot)$, we get
the following analytical expression for the signal $s_{n}({\bf x},t)$
corresponding to this distribution:

\begin{equation}
  s_{n}({\bf x},t) = e^{-ikt}\sum_{i=1}^{n}\frac{A_{n}(i)}
  {\left|{\bf x}-{\bf p}_{i}\right|}
  e^{ik\left|{\bf x}-{\bf p}_{i}\right|}.
  \label{signal}
\end{equation}

We write 

\[ {\bf x} = (R\cos\theta, R\sin\theta), \mbox{ } {\bf p}_{i} = 
(r_{i}\cos\theta_{i},r_{i}\sin\theta_{i}). \]

\noindent Now, we assume $R$ is large compared to the $r_{i}$'s
and make an expansion:

\begin{eqnarray*}
  \left|{\bf x}-{\bf p}_{i}\right| &=& R\left[1+\frac{r_{i}^{2}-
      2Rr_{i}(\cos\theta\cos\theta_{i}+\sin\theta\sin\theta_{i})}
    {R^{2}}\right]^{1/2} \\
  &\approx& R + \frac{r_{i}^{2}}{2R}-r_{i}
  (\cos\theta\cos\theta_{i}+\sin\theta\sin\theta_{i}) \\
  &=& R + \frac{r_{i}^{2}}{2R}-r_{i}
  \cos(\theta-\theta_{i}),
\end{eqnarray*}

\noindent then we can write 

\begin{eqnarray}
  s_{n}(R,\theta,t) &\approx& e^{-ikt} \sum_{i=1}^{n}A_{n}(i)
  \frac{e^{ik(R+\frac{r_{i}^{2}}{2R}-r_{i}\cos(\theta-\theta_{i}))}}
  {R+\frac{r_{i}^{2}}{2R}-r_{i}\cos(\theta-\theta_{i})} \nonumber \\
  &\approx& \frac{e^{-ik(t-R)}}{R} \sum_{i=1}^{n}A_{n}(i)
  e^{ik(\frac{r_{i}^{2}}{2R}-r_{i}\cos(\theta-\theta_{i}))},
  \label{approximated signal}
\end{eqnarray}

\noindent which is highly accurate if $R$ is large enough.

\begin{figure}[h]
  \begin{center}
    \includegraphics[scale = 0.7]{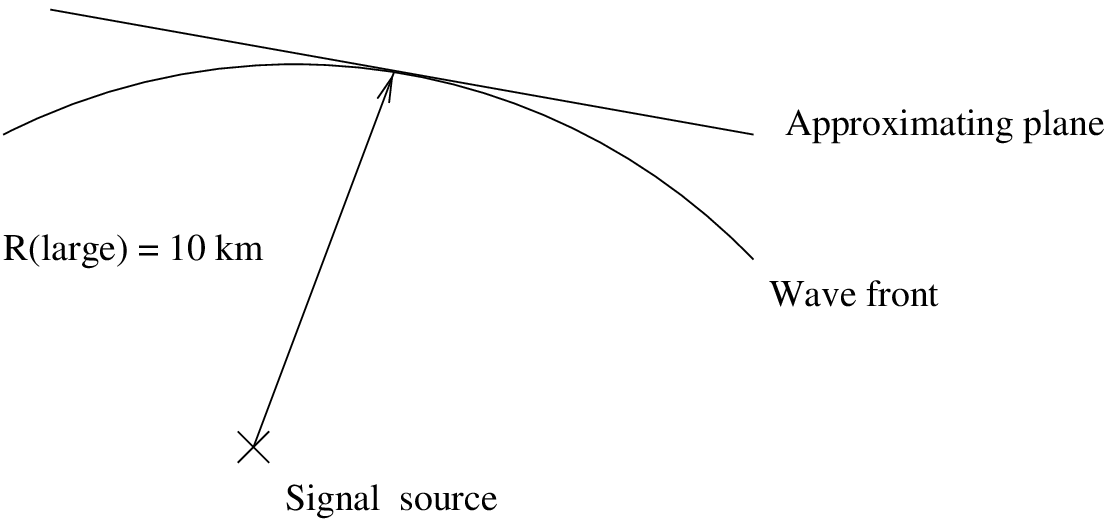}
  \end{center}
  \caption{Local approximation of wavefront by a plane.}
  \label{measurement of signal}
\end{figure}

For fixed $t = t^{\prime}$, a plot of the magnitude of this 
function will be a smooth closed surface in $\bR^{3}$ with a more or 
less spherical shape. This 
surface is called the wavefront at time $t^{\prime}$, and may be
approximated locally ``quite accurately'' by a plane as illustrated in Figure 
\ref{measurement of signal}, if the wavenumber $k$ is not ``very large''.

We now formulate our problem precisely: Is it possible, within 
some reasonable degree of accuracy, to separate $D_{n}(\alpha)$
from $D_{m}(\alpha), \mbox{ }n \neq m$, by simply doing constant phase
measurements of the signals $s_{n},s_{m}$ in some small space 
interval far away from the signal source?

We will use the concepts and algorithms introduced in chapter 1
to give an answer to this question.

\section{Experimental results}

The results given in this section are very ``experimental'' in nature 
and were generated by implementing the Local Discriminant Basis
Selection Algorithm described in Chapter 1. 

By plotting the two most discriminating coordinates in the 
LDB for a collection of signals from two different distributions,
we observe that one class is more or less centered around the origin
in a relatively small area whereas the other class is spread over a 
much larger area around the origin. This phenomenon
is to some degree documented in the scatter plots shown in appendix B.   
This indicates that it should be 
possible to achieve a reasonable degree of discrimination between
the two classes by using suitable ``hypersphere'' surfaces 
in constructing the classifier. Define the set $B_{n,r}$ by

\[ B_{n,r} = \{{\bf x}: x_{1}^{2}+x_{2}^{2}+ \cdots + x_{n}^{2} \leq r
\}.
\]
 
\noindent Let $\widehat{{\bf s}}\in \bR^{m}$ be the point defined by the 
$m$ most discriminating coordinates of some test signal ${\bf s} \in
\bR^{n}$. Given two distributions 
$D_{x}(\alpha) \neq D_{y}(\alpha)$ and $m$ and $r$,
we define the classifier $c_{m,r}$ by

\[ c_{m,r}(\widehat{{\bf s}}) \in 
\left\{\begin{array}{l} D_{x}(\alpha) \mbox{ if }
\widehat{{\bf s}}\in B_{m,r}, \\
D_{y}(\alpha) \mbox{ otherwise. }
\end{array} \right.
\]  

We will concentrate on the problem of separating $D_{n}$ from $D_{m}$ 
when $|m-n| = 1$, since the results obtained in this case should get
even better when $|m-n| >1$.
We compute the misclassification rate  
on both the training dataset and the test dataset for each of four 
different real orthonormal bases: The ``standard euclidean basis'' (STB), 
a real Fourier-basis of type ``discrete cosine basis'' (DCB), 
the ``best local cosine packet basis'' 
(BLCPB) and the ``best coiflet packet basis'' (BCPB) 
using the coiflet of length 18. The discrete cosine basis
will be of type IV, that is cosines evaluated at half integers
in both time and frequency. 

We note that the time-frequency
localization properties of these four bases are very different:
The elements of the standard euclidean basis are perfectly 
localized in time but possess no localization in frequency, 
the elements of the discrete cosine basis are perfectly
localized in frequency and possess no localization in time, while 
the elements of BCPB and BLCPB possess localization in 
both time and frequency.

We use a training dataset of 250 signals from each class 
generated by randomly choosing points 
${\bf p}_{i}\in \Omega_{n,i}(\alpha) \mbox{ } 
1 \leq n \leq 250$ in (\ref{signal}), 
and a test dataset of 2500 randomly generated signals of the same type
from each class.                  
We set the wavenumber $k$ to 100 and the distance of observation $R$
to $10^{4}$. To not make the problem of extracting relevant features
too easy for the LDB-algorithm, we equalize the maximum amplitude of the   
resultant signals $s_{n}$ and $s_{n+1}$ by setting 
$A_{n}(i)=\frac{1.0}{n}$. The real parts of the 
signals $s_{n}$, $s_{n+1}$ were sampled $2^{11} = 2048$ times with period 
$\frac{2\pi}{16\cdot k} \approx 4\cdot 10^{-3}$ units, yielding 
a sampling interval at the point of observation of length 
$\frac{2\pi}{16\cdot k}\cdot 2^{11} \approx
8.04$ units containing about $2^{7} = 128$ oscillations of the signals
$s_{n}$, $s_{n+1}$. The expansion depth in the packet bases is fixed to 8.
The performance of the classifier $c_{m,r}$ is 
maximized subject to the conditions: $2\leq m \leq 10$, 
$0.0 < r \leq 5.0$, where $r$ is an integer multiplum of $2^{-7}$. 

Given $D_{n}$, $D_{n+1}$, we will study the performance of the 
LDB-algorithm and our particular 
classifier $c_{m,r}$ for the cases 
$\alpha=30^{\circ}$, $\alpha=45^{\circ}$, $\alpha=60^{\circ}$.

\subsection{The case $D_{2}(\cdot),D_{3}(\cdot)$}

The numbers given in boldface behind the error rates
are the best $m$ in the upper righthand corner and the best
$r$ in the lower righthand corner. The same $m$ and $r$ are
applied to the test dataset, of course. The best test results
for each $\alpha$ are written in bold face.
\vspace{5mm}

\begin{table}[h]
\begin{tabular}{|c|c|c|c|c|c|c|} \hline
  Method   & \multicolumn{6}{c|}{Error rate (\%)} \\ \hline
  $c_{m,r}$ on   
  &  \multicolumn{3}{c|}{Training dataset} & 
  \multicolumn{3}{c|}{Test dataset} \\ \cline{2-7}
  & $\alpha=30^{\circ}$ & $\alpha=45^{\circ}$ & $\alpha=60^{\circ}$ 
  & $\alpha=30^{\circ}$ & $\alpha=45^{\circ}$ &
  $\alpha=60^{\circ}$ \\ \hline 

  STB    & $35.80^{{\bf 9}}_{{\bf 1.27}}$ & $38.40^{{\bf 10}}_{{\bf 1.39}}$
  & $21.40^{{\bf 10}}_{{\bf 1.43}}$ & 32.78  & 33.60 & 29.80 \\ \hline
  DCB    & $37.60^{{\bf 5}}_{{\bf 1.07}}$ & $39.60^{{\bf 3}}_{{\bf 0.77}}$ 
  & $48.60^{{\bf 4}}_{{\bf 0.28}}$ & 44.52 & 45.46 & 49.68 \\ \hline
  BCPB   & $15.00^{{\bf 10}}_{{\bf 2.55}}$  & $17.80^{{\bf 8}}_{{\bf 2.12}}$ 
  & $20.20^{{\bf 10}}_{{\bf 2.31}}$ & 16.62 & 22.84 & 25.20 \\ \hline
  BLCPB  & $13.40^{{\bf 10}}_{{\bf 2.67}}$ & $17.60^{{\bf 10}}_{{\bf 2.09}}$  
  & $18.80^{{\bf 10}}_{{\bf 1.98}}$ & {\bf 15.94} & {\bf 17.78} & 
  {\bf 19.80} \\ \hline
 
\end{tabular}
\label{table23} 
\caption{Numerical results from testing the performance of  
  $c_{m,r}$ on $D_{2}(\cdot)$, $D_{3}(\cdot)$.}

\end{table}

\subsection{The case $D_{3}(\cdot),D_{4}(\cdot)$}

The numbers given in boldface behind the error rates
are the best $m$ in the upper righthand corner and the best
$r$ in the lower righthand corner. The same $m$ and $r$ are
applied to the test dataset, of course. The best test results
for each $\alpha$ are written in bold face.
\vspace{5mm}

\begin{table}[h]
\begin{tabular}{|c|c|c|c|c|c|c|} \hline
  Method   & \multicolumn{6}{c|}{Error rate (\%)} \\ \hline
  $c_{m,r}$ on   
  &  \multicolumn{3}{c|}{Training dataset} & 
  \multicolumn{3}{c|}{Test dataset} \\ \cline{2-7}
  & $\alpha=30^{\circ}$ & $\alpha=45^{\circ}$ & $\alpha=60^{\circ}$ 
  & $\alpha=30^{\circ}$ & $\alpha=45^{\circ}$ &
  $\alpha=60^{\circ}$ \\ \hline 

  STB    & $31.20^{{\bf 7}}_{{\bf 0.89}}$ & $31.40^{{\bf 9}}_{{\bf 1.16}}$
  & $40.00^{{\bf 9}}_{{\bf 1.09}}$ & 41.18  & 40.68 & 39.94 \\ \hline
  DCB    & $38.40^{{\bf 3}}_{{\bf 0.92}}$ & $45.00^{{\bf 2}}_{{\bf 0.95}}$ 
  & $40.40^{{\bf 10}}_{{\bf 2.31}}$ & 38.14 & 49.14 & 41.68 \\ \hline
  BCPB   & $7.60^{{\bf 7}}_{{\bf 0.42}}$  & $26.20^{{\bf 2}}_{{\bf 0.14}}$ 
  & $21.60^{{\bf 7}}_{{\bf 0.19}}$ & {\bf 7.72} & {\bf 24.94} 
  & {\bf 21.92} \\ \hline
  BLCPB  & $6.80^{{\bf 8}}_{{\bf 0.78}}$ & $23.80^{{\bf 6}}_{{\bf 0.48}}$  
  & $30.80^{{\bf 9}}_{{\bf 1.12}}$ & 10.56 & 25.20 & 31.74 \\ \hline
 
\end{tabular}
\label{table34} 
\caption{Numerical results from testing the performance of  
  $c_{m,r}$ on $D_{3}(\cdot)$, $D_{4}(\cdot)$.}

\end{table}

\subsection{The case $D_{4}(\cdot),D_{5}(\cdot)$}

The numbers given in boldface behind the error rates
are the best $m$ in the upper righthand corner and the best
$r$ in the lower righthand corner. The same $m$ and $r$ are
applied to the test dataset, of course. The best test results
for each $\alpha$ are written in boldface. 
\vspace{5mm}

\begin{table}[h]
\begin{tabular}{|c|c|c|c|c|c|c|} \hline
  Method   & \multicolumn{6}{c|}{Error rate (\%)} \\ \hline
  $c_{m,r}$ on   
  &  \multicolumn{3}{c|}{Training dataset} & 
  \multicolumn{3}{c|}{Test dataset} \\ \cline{2-7}
  & $\alpha=30^{\circ}$ & $\alpha=45^{\circ}$ & $\alpha=60^{\circ}$ 
  & $\alpha=30^{\circ}$ & $\alpha=45^{\circ}$ &
  $\alpha=60^{\circ}$ \\ \hline 

  STB    & $34.40^{{\bf 4}}_{{\bf 0.63}}$  & $35.80^{{\bf 6}}_{{\bf 0.76}}$
  & $43.00^{{\bf 6}}_{{\bf 0.63}}$ & 44.78 & 44.86 & 44.54 \\ \hline
  DCB    & $35.80^{{\bf 2}}_{{\bf 0.61}}$  & $42.00^{{\bf 2}}_{{\bf 1.01}}$
  & $41.20^{{\bf 9}}_{{\bf 1.48}}$ & 42.32  & 43.08 & 45.10 \\ \hline
  BCPB   & $21.60^{{\bf 10}}_{{\bf 1.62}}$ & $22.80^{{\bf 2}}_{{\bf 0.10}}$
  & $40.80^{{\bf 2}}_{{\bf 0.66}}$ & 23.62 & 25.32 & 43.96 \\ \hline
  BLCPB  & $15.60^{{\bf 10}}_{{\bf 1.42}}$ & $15.60^{{\bf 5}}_{{\bf 0.32}}$
  & $39.60^{{\bf 4}}_{{\bf 0.73}}$  & {\bf 16.74} & {\bf 18.12}
  & {\bf 43.86} \\ \hline
 
\end{tabular}
\label{table45} 
\caption{Numerical results from testing the performance of 
  $c_{m,r}$ on $D_{4}(\cdot)$, $D_{5}(\cdot)$.}

\end{table}

\subsection{The case $D_{5}(\cdot),D_{6}(\cdot)$.}

The numbers given in boldface behind the error rates
are the best $m$ in the upper righthand corner and the best
$r$ in the lower righthand corner. The same $m$ and $r$ are
applied to the test dataset, of course. The best test results
for each $\alpha$ are written in boldface. 
\vspace{5mm}

\begin{table}[h]
\begin{tabular}{|c|c|c|c|c|c|c|} \hline
  Method   & \multicolumn{6}{c|}{Error rate (\%)} \\ \hline
  $c_{m,r}$ on   
  &  \multicolumn{3}{c|}{Training dataset} & 
  \multicolumn{3}{c|}{Test dataset} \\ \cline{2-7}
  & $\alpha=30^{\circ}$ & $\alpha=45^{\circ}$ & $\alpha=60^{\circ}$ 
  & $\alpha=30^{\circ}$ & $\alpha=45^{\circ}$ &
  $\alpha=60^{\circ}$ \\ \hline 

  STB    & $32.60^{{\bf 7}}_{{\bf 0.65}}$  & $43.00^{{\bf 3}}_{{\bf 0.58}}$  
  & $32.20^{{\bf 9}}_{{\bf 0.82}}$ & 46.32 & 46.06 & 44.50 \\ \hline
  DCB    & $41.20^{{\bf 8}}_{{\bf 1.30}}$  & $35.80^{{\bf 10}}_{{\bf 1.74}}$  
  & $38.40^{{\bf 9}}_{{\bf 1.69}}$ & 44.74 & 43.50 & {\bf 44.22} \\ \hline
  BCPB   & $29.60^{{\bf 4}}_{{\bf 0.45}}$  & $31.80^{{\bf 5}}_{{\bf 0.61}}$
  & $41.60^{{\bf 10}}_{{\bf 0.10}}$ & 35.46& 33.80 & 49.18 \\ \hline
  BLCPB  & $27.80^{{\bf 6}}_{{\bf 0.81}}$  & $22.80^{{\bf 5}}_{{\bf 0.52}}$ 
  & $35.20^{{\bf 10}}_{{\bf 0.47}}$  & {\bf 32.48} & {\bf 28.62} 
  & 46.72 \\ \hline
 
\end{tabular}
\label{table56} 
\caption{Numerical results from testing the performance of  
  $c_{m,r}$ on $D_{5}(\cdot)$, $D_{6}(\cdot)$.}

\end{table}

\subsection{The case $D_{10}(\cdot),D_{20}(\cdot)$}

The numbers given in boldface behind the error rates
are the best $m$ in the upper righthand corner and the best
$r$ in the lower righthand corner. The same $m$ and $r$ are
applied to the test dataset, of course. The best test results
for each $\alpha$ are written in boldface. 
\vspace{5mm}

\begin{table}[h]
\begin{tabular}{|c|c|c|c|c|c|c|} \hline
  Method   & \multicolumn{6}{c|}{Error rate (\%)} \\ \hline
  $c_{m,r}$ on   
  &  \multicolumn{3}{c|}{Training dataset} & 
  \multicolumn{3}{c|}{Test dataset} \\ \cline{2-7}
  & $\alpha=30^{\circ}$ & $\alpha=45^{\circ}$ & $\alpha=60^{\circ}$ 
  & $\alpha=30^{\circ}$ & $\alpha=45^{\circ}$ &
  $\alpha=60^{\circ}$ \\ \hline 

  STB    & $21.00^{{\bf 8}}_{{\bf 0.48}}$  & $26.60^{{\bf 10}}_{{\bf 0.59}}$ 
  & $32.00^{{\bf 10}}_{{\bf 0.52}}$        & 26.62  & 29.22 & 29.08 \\ \hline
  DCB    & $24.00^{{\bf 10}}_{{\bf 1.01}}$ & $24.80^{{\bf 10}}_{{\bf 1.04}}$ 
  & $26.20^{{\bf 9}}_{{\bf 0.96}}$         & 25.90  & 25.16 
  & {\bf 26.16} \\ \hline
  BCPB   & $20.00^{{\bf 9}}_{{\bf 0.67}}$  & $28.40^{{\bf 6}}_{{\bf 0.56}}$ 
  & $39.60^{{\bf 10}}_{{\bf 0.71}}$        & 27.56 & 35.10 & 34.42 \\ \hline
  BLCPB  & $22.80^{{\bf 10}}_{{\bf 1.06}}$ & $17.80^{{\bf 10}}_{{\bf 1.08}}$  
  & $20.20^{{\bf 9}}_{{\bf 0.97}}$        & {\bf 24.50} & {\bf 24.86} 
  & 26.72 \\ \hline
 
\end{tabular}
\label{table1020} 
\caption{Numerical results from testing the performance of 
  $c_{m,r}$ on $D_{10}(\cdot)$, $D_{20}(\cdot)$.}

\end{table}    

\section{Conclusion}

Table \ref{table23} - Table \ref{table56} show that the
classifier $c_{m,r}$ performs best on the coiflet packet
and local cosine packet bases. Thus, we see that localization in 
{\em both} time/space and frequency is essential to provide the
classifier with relevant features. We note that coiflets seem to be
less ``resistant'' to overtraining than local cosines, they adapt too well to 
the training signals, that is.


%% file: alphas.tex
\chapter{$SO_{2}(\bR)$ elements for some orthogonal FIR filters.}

Below we give the parameters $\alpha_{k}, 1\leq k \leq L$, that
determines the operators $W_{2}(\alpha_{k})$ for 2 types of orthonormal 
filters of length $L$, Daubechies original shortest filters and
Coiflet filters.

\section{Daubechies shortest filters.}

\noindent Daub 4

\begin{eqnarray*}
&& \alpha_{1} =  0.5773502691 \\
&& \alpha_{2} = -0.2679491923 
\end{eqnarray*}

\noindent Daub 6

\begin{eqnarray*}
&& \alpha_{1} =  0.4122865951 \\
&& \alpha_{2} =  1.831178514 \\
&& \alpha_{3} = -0.1058894200
\end{eqnarray*}

\noindent Daub 8

\begin{eqnarray*}
&&\alpha_{1} = 0.3222758836 \\
&&\alpha_{2} = 1.233150027 \\
&&\alpha_{3} = 3.856627874 \\
&&\alpha_{4} = -0.04600009616
\end{eqnarray*}

\noindent Daub 10

\begin{eqnarray*} 
&&\alpha_{1} = 0.2651451339 \\
&&\alpha_{2} = 0.9398995872 \\
&&\alpha_{3} =  2.353886784 \\
&&\alpha_{4} = 7.508378888 \\
&&\alpha_{5} = -0.02083494630
\end{eqnarray*}

\newpage
\noindent Daub 12

\begin{eqnarray*}
&&\alpha_{1} = 0.2255061720 \\
&&\alpha_{2} = 0.7643296306 \\
&&\alpha_{3} =  1.696013010 \\
&&\alpha_{4} =  4.114979257 \\
&&\alpha_{5} = 14.28573961 \\
&&\alpha_{6} = -0.009658362993
\end{eqnarray*}

\noindent Daub 14

\begin{eqnarray*}
&&\alpha_{1} = 0.1963287126 \\
&&\alpha_{2} = 0.6466065217 \\
&&\alpha_{3} = 1.333037518 \\
&&\alpha_{4} = 2.764759661 \\
&&\alpha_{5} = 7.035232916 \\
&&\alpha_{6} = 27.00281769 \\
&&\alpha_{7} = -0.004543409641
\end{eqnarray*}

\noindent Daub 16

\begin{eqnarray*}
&&\alpha_{1} = 0.1739238836 \\
&&\alpha_{2} = 0.5617332940 \\
&&\alpha_{3} = 1.103629937 \\
&&\alpha_{4} = 2.074598026 \\
&&\alpha_{5} = 4.380557848 \\
&&\alpha_{6} = 12.05139151 \\
&&\alpha_{7} = 49.52666172 \\
&&\alpha_{8} = -0.002443028170
\end{eqnarray*}

\newpage
\noindent Daub 18

\begin{eqnarray*}
&&\alpha_{1} = 0.1561629731 \\
&&\alpha_{2} = 0.4973943657 \\
&&\alpha_{3} = 0.9452416623 \\
&&\alpha_{4} = 1.664172294 \\
&&\alpha_{5} =  3.114016860 \\
&&\alpha_{6} =  6.915226655 \\
&&\alpha_{7} = 20.60043019 \\
&&\alpha_{8} = 96.49772819 \\
&&\alpha_{9} = -0.001033336055
\end{eqnarray*}

\noindent Daub 20

\begin{eqnarray*}
&&\alpha_{1} = 0.1417287200 \\ 
&&\alpha_{2} = 0.4467987788 \\
&&\alpha_{3} = 0.8289658876 \\
&&\alpha_{4} = 1.394189716 \\
&&\alpha_{5} = 2.402966640 \\
&&\alpha_{6} = 4.635603726 \\
&&\alpha_{7} = 10.98508401 \\
&&\alpha_{8} = 35.63003753 \\
&&\alpha_{9} = 183.0054911 \\
&&\alpha_{10}= -0.0004973444230
\end{eqnarray*}

\section{Coiflet filters.}

\noindent Coif 6

\begin{eqnarray*}
&&\alpha_{1} = -0.2152504427 \\
&&\alpha_{2} =  0.3779644639 \\
&&\alpha_{3} = -0.2152504427 \\
\end{eqnarray*}

\newpage
\noindent Coif 12

\begin{eqnarray*}
&&\alpha_{1} = -0.3952094767 \\
&&\alpha_{2} =  -0.5625481503 \\
&&\alpha_{3} =  0.1165449040 \\
&&\alpha_{4} = 1.317233974 \\
&&\alpha_{5} = 6.198029576 \\
&&\alpha_{6} = -0.04396989341 \\
\end{eqnarray*}

\noindent Coif 18

\begin{eqnarray*}
&&\alpha_{1} = -0.4874353702 \\
&&\alpha_{2} = -1.119071133 \\
&&\alpha_{3} = -0.2570708497 \\
&&\alpha_{4} =  0.1290348165 \\
&&\alpha_{5} =  0.4411074710 \\
&&\alpha_{6} =  2.215422179 \\
&&\alpha_{7} =  8.338120664 \\
&&\alpha_{8} =  15.03636438 \\
&&\alpha_{9} = -0.009120773147 
\end{eqnarray*}

\noindent Coif 24

\begin{eqnarray*}
&&\alpha_{1}  = -0.5476023581 \\
&&\alpha_{2}  = -1.457533881 \\
&&\alpha_{3}  = -0.7720754411 \\
&&\alpha_{4}  = -0.1309276144 \\
&&\alpha_{5}  =  0.1710353887 \\
&&\alpha_{6}  =  0.2957793746 \\
&&\alpha_{7}  =  0.8070747686 \\
&&\alpha_{8}  =  3.126528296 \\
&&\alpha_{9}  = 11.27596534 \\
&&\alpha_{10} = 12.66598170 \\
&&\alpha_{11} = 53.96686137  \\
&&\alpha_{12} = -0.002000409650
\end{eqnarray*}

\newpage
\noindent Coif 30

\begin{eqnarray*}
&&\alpha_{1} = -0.5914303923 \\
&&\alpha_{2} = -1.718001035 \\
&&\alpha_{3} = -1.195010469 \\
&&\alpha_{4} = -0.4056552189 \\
&&\alpha_{5} = -0.1316532923 \\
&&\alpha_{6} = 0.1205373016 \\
&&\alpha_{7} = 0.3671126852 \\ 
&&\alpha_{8} = 0.4678947012 \\
&&\alpha_{9} = 1.165968370 \\
&&\alpha_{10} = 4.100416655 \\
&&\alpha_{11} = 15.61099604 \\
&&\alpha_{12} = 11.59905847 \\
&&\alpha_{13} = 37.56973541 \\
&&\alpha_{14} = 197.1316159 \\
&&\alpha_{15} = -0.0004543371650
\end{eqnarray*}


%% file: edgematrices.tex
\chapter{Edge matrices for orthonormal filters}

The matrices ${\bf E}_{H_{L},\cdot}$ are computed as follows:
Each row in a  ${\bf E}_{H_{L},\cdot}$ corresponds to some $\tilde{c}$ or
$\tilde{d}$ under the action of  ${\bf E}_{H_{L},\cdot}$ on ${\bf
  e}^{j}_{\cdot}$. The rows corresponding to $\tilde{c}$'s are uniquely
determined by claiming polynomials up to degree $\frac{L}{2}-1$ mapping to
polynomials of the same degree. 
The rows corresponding to $\tilde{d}$'s are not uniquely determined in
this way, since claiming $\frac{L}{2}-1$ vanishing moments on each
of the $\tilde{d}$'s results in 
the matrix ${\bf E}_{H_{L},\cdot}$ becoming singular. This gives some ``degrees
of freedom'' which were used to minimize the quantity 
$\| {\bf E}_{H_{L},\cdot} \|_{2} + \|{\bf E}_{H_{L},\cdot}^{-1} \|_{2}$, 
that is the sum of the operator norms of the matrices, on some coarse grid.

\noindent {\bf Remarks}: 
\begin{itemize}

\item The edge-matrices ${\bf E}_{H_{L},\cdot}$ operates on edge vectors
${\bf e}^{j}_{\cdot}$ resulting from non-normalized inner rotations. 

\item Entries less than $10^{-6}$ were set to zero.

\item Superscripts, $n$, on the matrix entries denote
multiplication by $10^{-n}$.

\item The matrix norm is the maximum column sum.

\end{itemize}

\section{Edge matrices for $H^{d}_{6}$, $H^{coif}_{6}$}

Applying ${\bf E}_{H^{d}_{6},\cdot}$ or ${\bf E}_{H^{coif}_{6},\cdot}$
at edges leads to 

\begin{itemize}

\item 1 vanishing moment on $\tilde{d}_{0}$.

\item 1 vanishing moment on $\tilde{d}_{n/2-1}$.

\item Polynomials up to and including degree $1$ map 
  to polynomials of the same degree.

\end{itemize}

Grid dimension = 1, Gridsize = .1
\newline

\scriptsize

\[ \begin{array}{lll}

{\bf E}_{H^{d}_{6},l}  = & \left( 
\begin{array}{cc}
 0            &   9.549704 \\
 .7719849     &    - 1.35   
\end{array} 
\right) & \vspace{4mm } \\ 

{\bf E}^{-1}_{H^{d}_{6},l} = & \left( 
\begin{array}{cc}
 .1831197     &  1.295362  \\
 .1047153     &  0  
\end{array} 
\right) & \vspace{4mm} \\

{\bf E}_{H^{d}_{6},r} = & \left( 
\begin{array}{cc}
 .7865626     &  - .35 \\
 0            &    1.011213
\end{array} 
\right) & \vspace{4mm}   \\

{\bf E}^{-1}_{H^{d}_{6},r} = & \left( 
\begin{array}{cc}
 1.271355     &  .4400402  \\
 0            &  .9889118  
\end{array} 
\right) & \vspace{4mm}\\

{\bf E}_{H^{coif}_{6},l} = & \left( 
\begin{array}{cc}
 0            &  4.861003 \\
 .5221843     &  - 1.95  
\end{array} 
\right) & \vspace{4mm}   \\

{\bf E}^{-1}_{H^{coif}_{6},l} = & \left( 
\begin{array}{cc}
 .7682189     & 1.915033   \\
 .2057189     &  0  
\end{array} 
\right) & \vspace{4mm} \\

{\bf E}_{H^{d}_{coif},r} = & \left( 
\begin{array}{cc}
 .6742919     &  - .35 \\
 0            &   1.046333
\end{array} 
\right) & \vspace{4mm}   \\

{\bf E}^{-1}_{H^{coif}_{6},r} = & \left( 
\begin{array}{cc}
 1.483037     & .4960784  \\
 0            & .9557189
\end{array} 
\right) & \vspace{4mm} \\

\end{array} \]

\normalsize

\section{Edge matrices for $H^{d}_{8}$, $H^{coif}_{8}$} 
 
Applying ${\bf E}_{H^{d}_{8},\cdot}$, ${\bf E}_{H^{coif}_{8},\cdot}$  
leads to  

\begin{itemize}

\item 1 vanishing moment on $\tilde{d}_{0}$.

\item 2 vanishing moments on $\tilde{d}_{1}$ and 
  $\tilde{d}_{n/2-1}$.

\item Polynomials up to and including degree 2 map to polynomials 
  of the same degree.

\end{itemize}

\noindent {\bf Remark}: The coiflet of length 8 is the one we computed
numerically in Table \ref{lowpass coiflet of length 8}.

\noindent Left case: Grid dimension = 3, Gridsize = 1.\newline
Right case: Grid dimension = 1, Gridsize = .1
\newline

\scriptsize 

\[ \begin{array}{lll}

{\bf E}_{H^{d}_{8},l} = & \left( 
\begin{array}{ccc}
  5.702903     & - 7.5           & - .5   \\
 - .11143^3   &  .1560114^3     &  21.78508 \\
   4.698626    &  - 7.876422     &    1.5 
\end{array} 
\right) & \vspace{4mm} \\

{\bf E}^{-1}_{H^{d}_{8},l} = & \left( 
\begin{array}{ccc}
   .8137842     &  .7203228^1     & - .7748912  \\
  .4854576      &  .5171221^1     &   - .5892172 \\
   0            &   .4590297^1    &    0
\end{array} 
\right) & \vspace{4mm} \\

{\bf E}_{H^{d}_{8},r} = & \left( 
\begin{array}{ccc}
   0            &  4.124631      & -  .1195278^1  \\
 - 1.186296     & 1.289667       &  -  .15          \\
   0            &  0             &   1.002116
\end{array} 
\right) & \vspace{4mm} \\

{\bf E}^{-1}_{H^{d}_{8},r} = & \left( 
\begin{array}{ccc}
  .2635720     &  - .8429597     &  - .1230332   \\
   .2424459    &   0             &   .2891783^2   \\
   0           &   0             &   .9978885 
\end{array} 
\right) & \vspace{4mm} \\

{\bf E}_{H^{coif}_{8},l} = & \left( 
\begin{array}{ccc}
  - .1888304^1 & - 5.5         & 10.5      \\
   0           &  0            &  2.666667  \\
 - .8293807    &  .4881421     &   -  1.5
\end{array} 
\right) & \vspace{4mm} \\

{\bf E}^{-1}_{H^{coif}_{8},l} = & \left( 
\begin{array}{ccc}
  - .1067955    &  - .2563419     &  - 1.203287     \\
  - .1814515    &    .7167892     &   .4131254^2  \\
   0            &    .3750000     &   0
\end{array} 
\right) & \vspace{4mm} \\

{\bf E}_{H^{coif}_{8},r} = & \left( 
\begin{array}{ccc}
 0             &  - 2.519102     &  .7359046     \\
  .7396251     &  .8283321       &   -  .45          \\
 0             &  0              &   1.203777    
\end{array} 
\right) & \vspace{4mm} \\

{\bf E}^{-1}_{H^{coif}_{8},r} = & \left( 
\begin{array}{ccc}
  .4445772      &  1.352037     &   .2336397    \\
  - .3969669    &  0            &   .2426777   \\
   0            &  0            &   .8307189
\end{array} 
\right) & \vspace{4mm} \\

\end{array} \]

\normalsize

\section{Edge matrices for $H^{d}_{10}$}

Applying ${\bf E}_{H^{d}_{8},\cdot}$ leads to  

\begin{itemize}

\item 2 vanishing moments on $\tilde{d}_{0}$ and the 
  $\tilde{d}_{n/2-1}$.

\item 3 vanishing moments on $\tilde{d}_{1}$ and the
  $\tilde{d}_{n/2-2}$.

\item Polynomials up to and including degree $3$ map to 
  polynomials of the same degree.

\end{itemize}

\noindent Left case : Grid dimension = 3, Gridsize = 1.\newline
Right case : Grid dimension = 3, Gridsize = 1.
\newline

\scriptsize 

\[ \begin{array}{ll}

{\bf E}_{H^{d}_{10},l} = & \left( 
\begin{array}{cccc}
   .1889731^1    & -.2458571^1      & - 366.9168     & 306.9367  \\
  -51.61952      &  50.72522        &  .5            & 10.5          \\
   .2652604^2    & -.3472944^2      & .1291323^2     & 48.01625   \\
 - 3.922730      &  5.645330        &  -2.755913     &  1.5 
\end{array} 
\right) \vspace{4mm} \\

{\bf E}^{-1}_{H^{d}_{10},l} = &  \left( 
\begin{array}{cccc}
 -.4205230^2     & -.6107834^1      &  .2309330^1   & .5488057   \\
 -.4252506^2     & - .4244118^1     &  .1901781^1   &  .5584793   \\
 - .2725347^2    & 0                & .1742140^1    &  0            \\
 0               & 0                & .2082591^1    &  .1007584^{4} 
\end{array} 
\right)  \vspace{4mm}  \\

{\bf E}_{H^{d}_{10},r} = & \left( 
\begin{array}{cccc}
   10.45706      &  - 2.394158          &  - 5.5        &  .5         \\
  0              &  -.2181276           &  7.644881     & - .2776091^2 \\
  20.02188       &  -35.08449           &  12.07371     &  - .5         \\
 0               &  .1417283^{5}        &  0            &  1.000434 
\end{array} 
\right) \vspace{4mm} \\

{\bf E}^{-1}_{H^{d}_{10},r} = & \left( 
\begin{array}{cccc}
  .1100016       &  .9099426^1      & -.7506492^2     & -.5847607^1 \\
.6277529^1       & .9694300^1       & - .3278640^1    & -.4749111^1   \\
 0               & .1308065         & 0               &  .3629740^{3}  \\
 0               & 0                & 0               &  .9995661
\end{array} 
\right)  \vspace{4mm} \\
\end{array} \]

\normalsize

\section{Edge matrices for $H^{d}_{12}$, $H^{coif}_{12}$}

Applying ${\bf E}_{H^{d}_{12},\cdot}$, ${\bf E}_{H^{coif}_{12},\cdot}$ 
leads to  

\begin{itemize}

\item 2 vanishing moments on $\tilde{d}_{0}$.

\item 3 vanishing moments on $\tilde{d}_{1}$ and 
 $\tilde{d}_{n/2-1}$.

\item 4 vanishing moments on $\tilde{d}_{2}$ and $\tilde{d}_{n/2-2}$.

\item Polynomials up to and including degree $4$ 
  map to polynomials of the same degree.

\end{itemize}

\noindent Left case : Grid dimension = 6, Gridsize = 3. \newline
Right case : Grid dimension = 3, Gridsize = 1.

\scriptsize 

\[ \begin{array}{ll}

{\bf E}_{H^{d}_{12},l} = & \left( 
\begin{array}{ccccc}
 -9.312776    & 16.23743       & - 10            &  -1           & 2  \\
  -6.91477    & 5.863835       & - 3.247505      & - 1484.263    & 751.2569  \\
 - 47.71849   & 24.84963       &  8.319144       &  - 1          &  8 \\
 - .8985454   & .7639290       & - .4245146      &  .2903221     & 103.6496 \\
  11.10650    &  -12.40799     &  9.908950       & -5.400903     & -   1  \\
\end{array} 
\right) \vspace{4mm} \\

{\bf E}^{-1}_{H^{d}_{12},l} = & \left( 
\begin{array}{ccccc}
  .1264939        & -.5806068^{3}    & -.1332401^{1}    & .4135265^{2}  &  .1388293  \\
  .2045857        & -.8716865^{3}    &  .6821282^{2}    & .3779404^{2}  & .2006150 \\
 .1144054         & -.8074226^{3}    &  .2347018^{1}    &.3729639^{2}   &  .1965661 \\
 -.2361415^{5}    &  -.6727232^{3}   &  .2410276^{5}    & .4875633^{2}  & - .1600175^{4}  \\
  .5729844^{5}    & 0                & -.6966225^{4}    & .9657503^{2}  & .5300424^{3}
\end{array} 
\right)  \vspace{4mm} \\

{\bf E}_{H^{d}_{12},r} = & \left( 
\begin{array}{ccccc}
 .6614876^{4}  & -.2352082^{3}  & 62.56825       & -.2539657    & .2359844^{2} \\
  - 196.6045   &  298.3539      &  - 61.96023    & - 10.5       &  .5 \\
-.1009016^{5}  & -.3181084^{4}  & .3673955^{4}   & 14.35707     & - .6760348^{3}  \\
 - 139.7373    & 332.5374       &- 210.6613      & 30.02657     & - .5  \\
 -.3170654^{4} & .6220284^{4}   & -.3292688^{4}  &.4123728^{5}  & 1.000093
\end{array} 
\right) \vspace{4mm} \\

{\bf E}^{-1}_{H^{d}_{12},r} =  & \left( 
\begin{array}{ccccc}
  .2850594^{1}   &-.1403871^{1}   & - .3610553^{1}  & .1259560^{1}  & .1322425^{1}  \\
 .2210355^{1}    & -.5899283^{2}  & - .2128227^{1}  &.8300054^{2}  &  .7032465^{2} \\
 .1598260^{1}    & 0              &  .2826771^{3}   &  0            &  -.3750916^{4} \\
 0               & 0              & .6965203^{1}    &  0            & .4709945^{4}  \\
 0               & 0              & 0               &  0             & .9999068
\end{array} 
\right)  \vspace{4mm} \\

{\bf E}_{H^{coif}_{12},l} = & \left( 
\begin{array}{ccccc}
  .3964447^1    & -2.236082     &  2           & -  1          & 2  \\
  6.924950      &  24.51832     &  22.46071    & -97.05257     & -16.09946 \\
  .1434265      &  -1.288120    & -.7670091    & 2             &  2 \\
   1.480799     &  5.242960     & 4.802957     &  10.23351     &  1.464548  \\
   -.1388966    &  -.4917635    & -.4504867    & -.9600        & 2   \\
\end{array} 
\right) \vspace{4mm} \\

{\bf E}^{-1}_{H^{coif}_{12},l} = & \left( 
\begin{array}{ccccc}
   -.2325044     &  .6560194^1     &  2.279093      & .1082415^1   &  - 1.526436   \\
   - .1959253    & - .3493090^3    &  -  .2975522   &.7647028^1    &.4346684 \\
    .2855577     & - .5141012^2    & -  .3778586    &.5405817^1    &  .1133166^1 \\
  0               &  - .6900798^2  &  0             &.2532173^1    &  - .7409259^1 \\
 - .1446773^5    & 0               & .6555962^5     &.4388504^1    & .4678566
\end{array} 
\right)  \vspace{4mm} \\

{\bf E}_{H^{coif}_{12},r} = & \left( 
\begin{array}{ccccc}
 - 6.463859    &  12.73118      &  6.045167      &  -.1153122   &  -.4242841^2 \\
   - 17.25484  &   -5.83848     &  24.97822      & - 10.5       &   .5 \\
 1.145926      & -2.257005      & 1.273759       &  6.245740    & - .2633039^3 \\
 - 83.70978    &  164.8742      &  - 93.04824    &  9.198964    &   - .5   \\
 -.4281287     &  .8432407      & -.4758912      &  .4704796^1  & .9993761
\end{array} 
\right) \vspace{4mm} \\

{\bf E}^{-1}_{H^{coif}_{12},r} =  & \left( 
\begin{array}{ccccc}
  .8105815^1   & - .4945805^1   &  - .6847096^1  &  - .9052881^2   & .2054129^1   \\
  .8381187^1   &  - .2511078^1  &  - .3799309^1  & - .1877050^2    &  .1196993^1 \\
  .7558504^1   & 0              &  .9794265^2    &  - .5689511^2   & - .2523067^2  \\
 0              & 0             & .1569448       &  .2142764^2     & .1113393^2  \\
 0              & 0             & 0              &  -.5104579^2    &  .9980704
\end{array} 
\right)  \vspace{4mm} \\
\end{array} \]
\normalsize


%% file: plots.tex
\chapter{Signal and Scatter Plots.}

\section{Plots for  $D_{2}(45^{\circ})$, $D_{3}(45^{\circ})$}

\begin{figure}[h]
  \begin{center}
    \includegraphics[scale = 0.65]{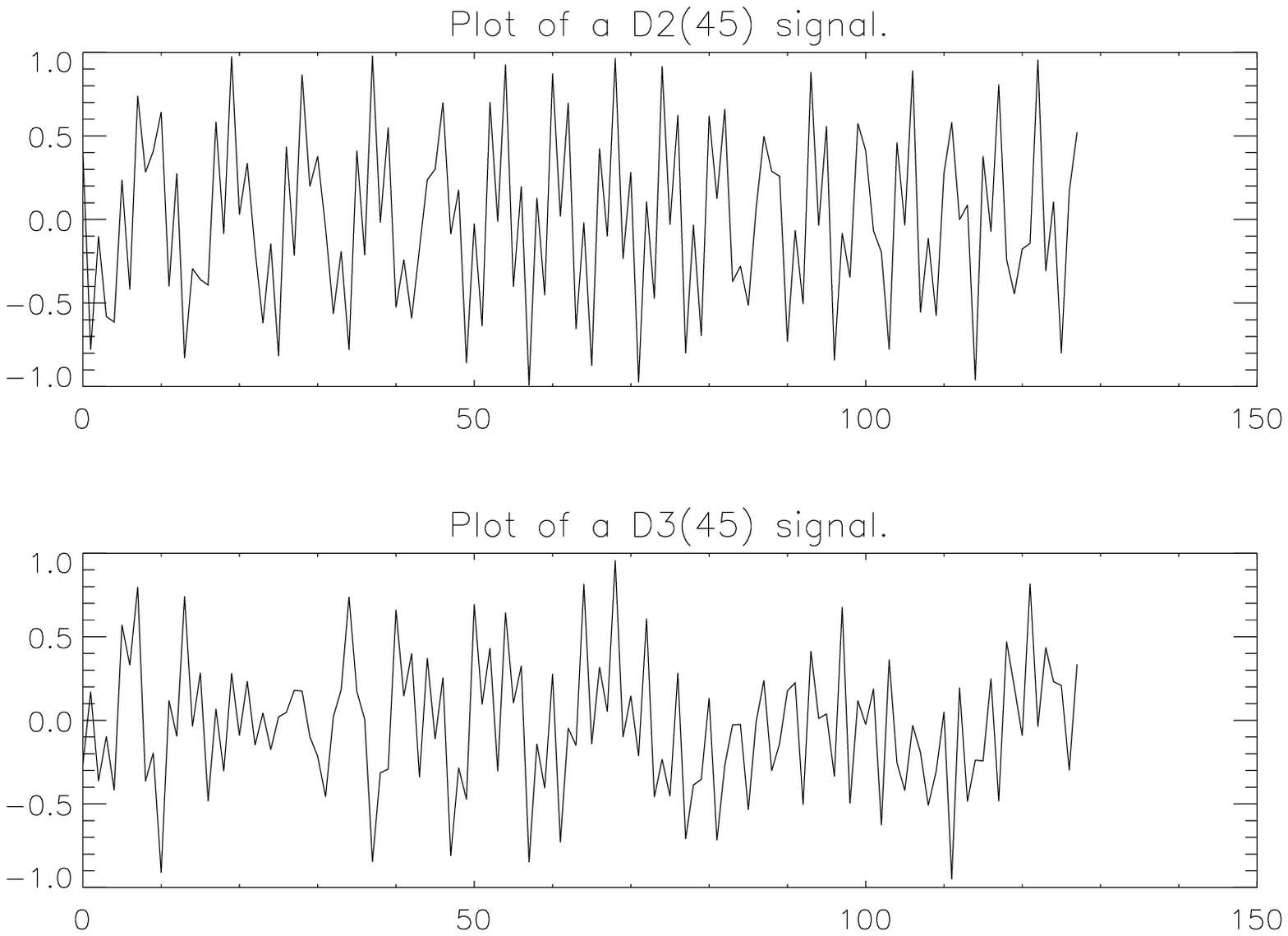}
  \end{center}
  \caption{Plot of the first 128 samples of the signals $s_{2}$ and $s_{3}$.}
  \label{signals23}
\end{figure} 

\begin{figure}[h]
  \begin{center}
    \includegraphics[scale = 0.65]{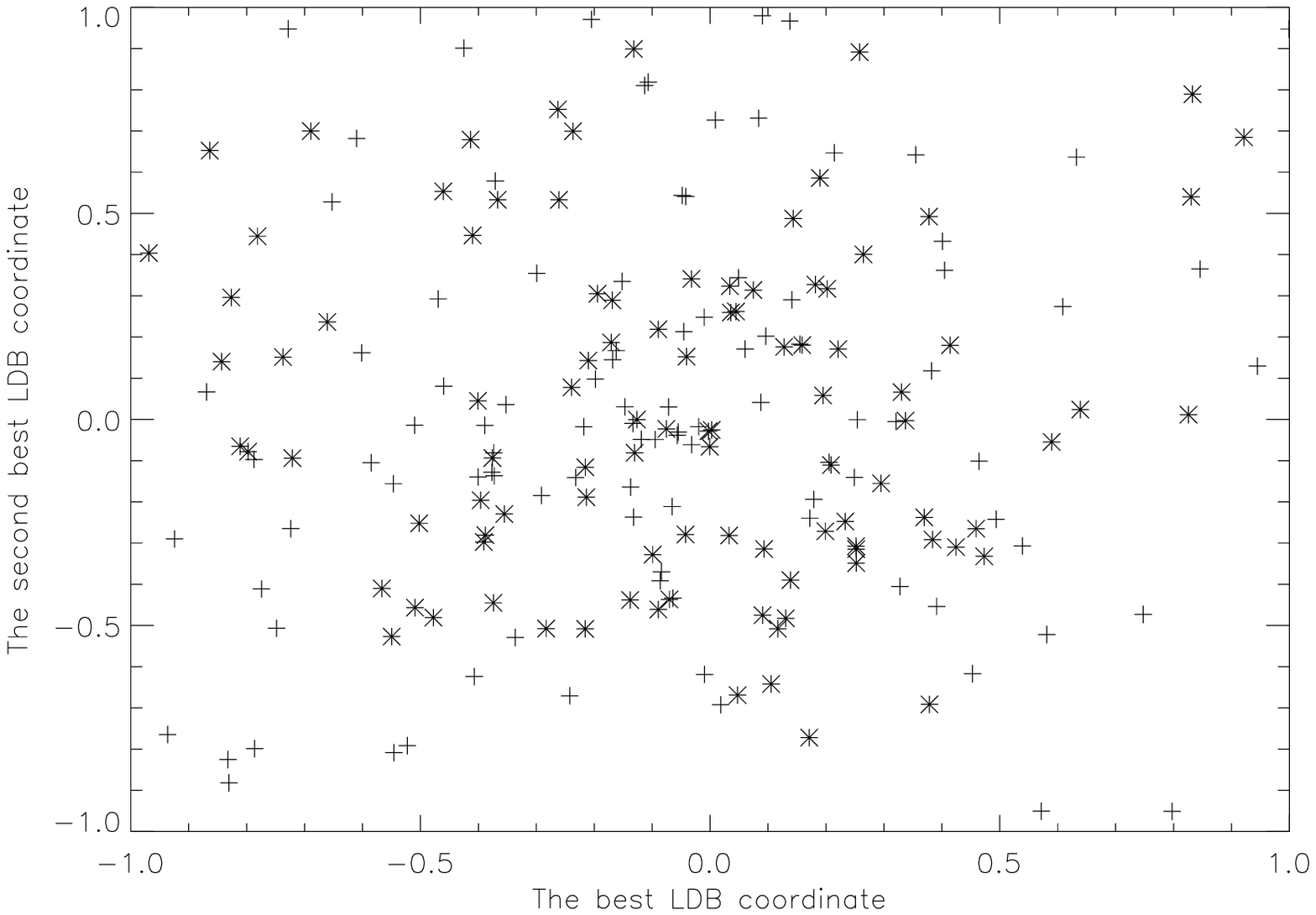}
  \end{center}
  \caption{Scatter plot of the 2 most discriminating coordinates in 
    the standard basis for 100 signals from each of the
    classes $D_{2}(45^{\circ})$ and $D_{3}(45^{\circ})$.}
  \label{stascatter23}
\end{figure} 

\begin{figure}[h]
  \begin{center}
    \includegraphics[scale = 0.65]{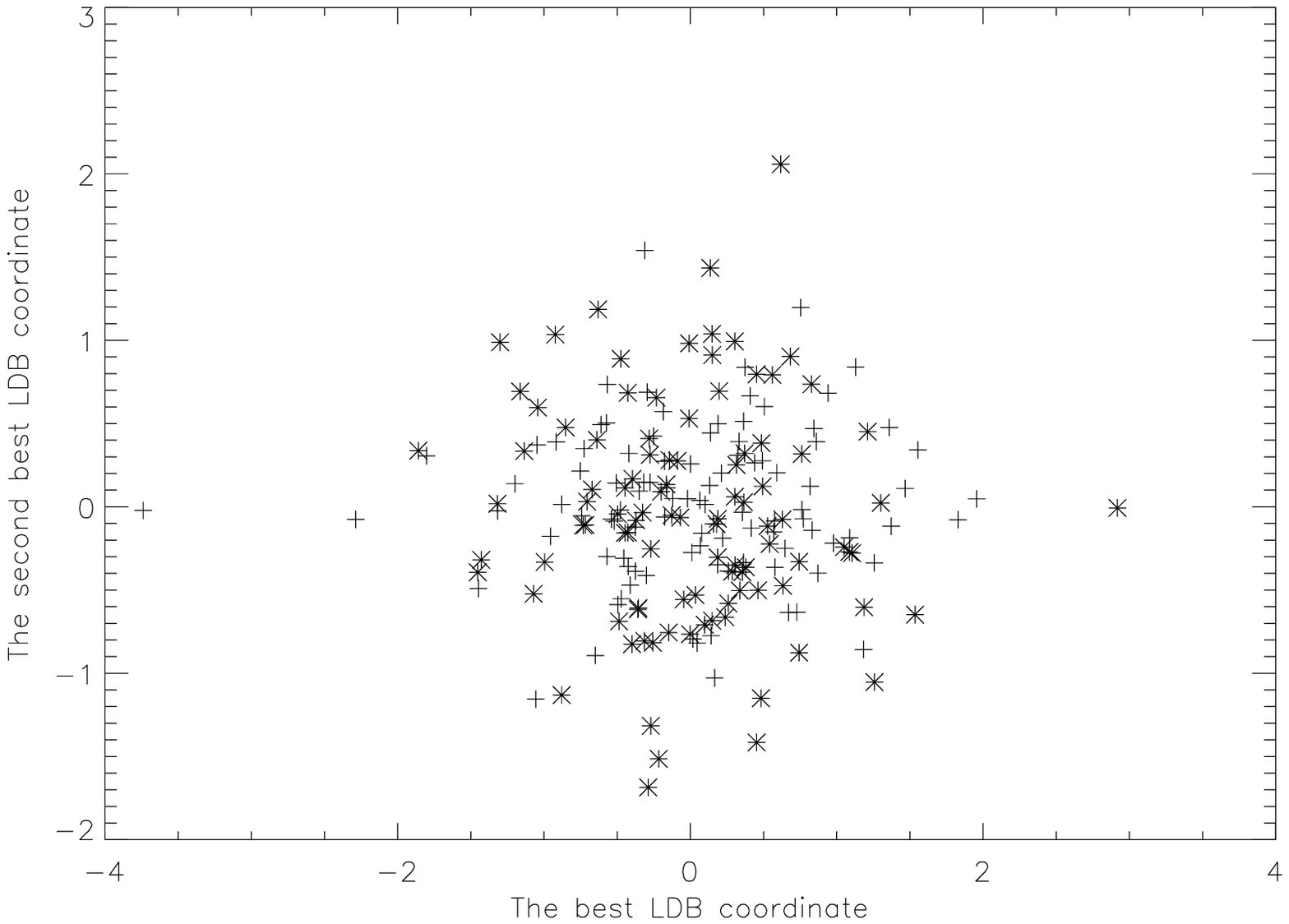}
  \end{center}
  \caption{Scatter plot of the 2 most discriminating
    coordinates in the discrete cosine IV basis for 100 signals from
    each of the classes $D_{2}(45^{\circ})$ 
    and $D_{3}(45^{\circ})$.}
  \label{cosinescatter23}
\end{figure} 

\begin{figure}[h]
  \begin{center}
    \includegraphics[scale = 0.65]{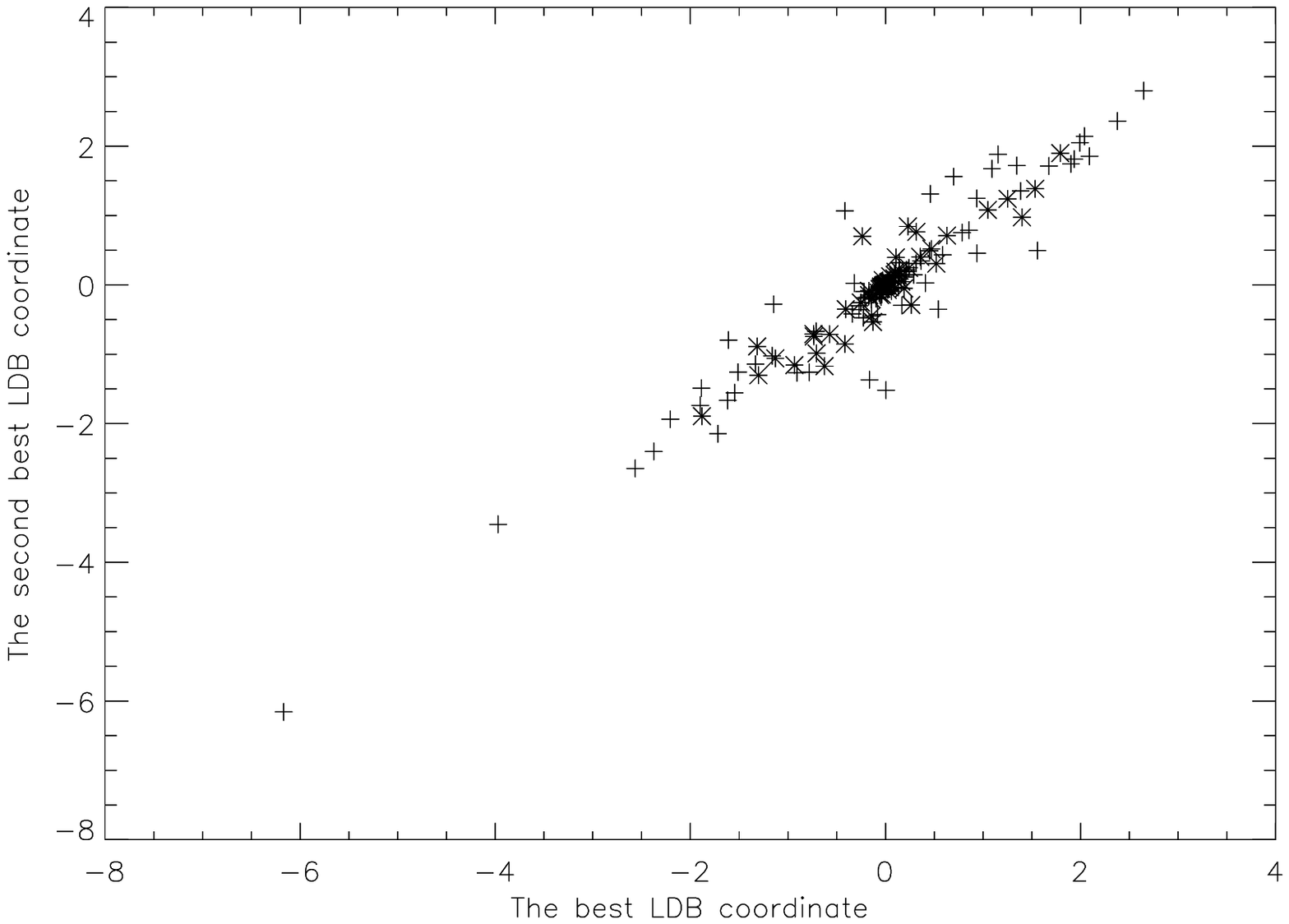}
  \end{center}
  \caption{Scatter plot of the 2 most discriminating coordinates in 
  the coiflet packet basis for 100 signals from each of the 
  classes $D_{2}(45^{\circ})$ and $D_{3}(45^{\circ})$.}
  \label{coifscatter23}
\end{figure} 
 
\begin{figure}[h]
  \begin{center}
    \includegraphics[scale = 0.65]{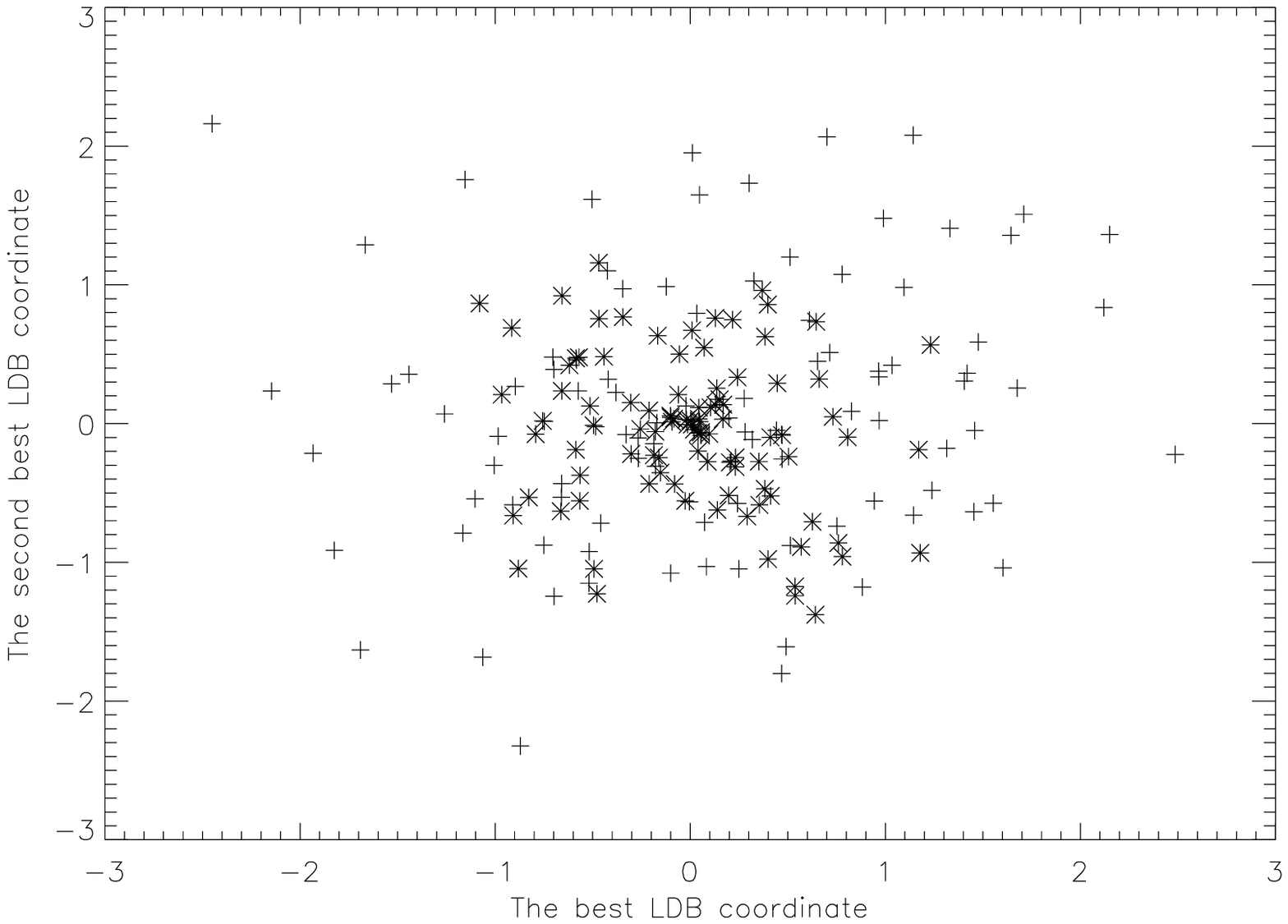}
  \end{center}
  \caption{Scatter plot of the 2 most discriminating coordinates in 
  the local cosine basis for 100 signals from each of the classes 
  $D_{2}(45^{\circ})$ and $D_{3}(45^{\circ})$. }
  \label{localcosinescatter23}
\end{figure}  
\clearpage

\section{Plots for  $D_{3}(45^{\circ})$, $D_{4}(45^{\circ})$}

\begin{figure}[h]
  \begin{center}
    \includegraphics[scale = 0.65]{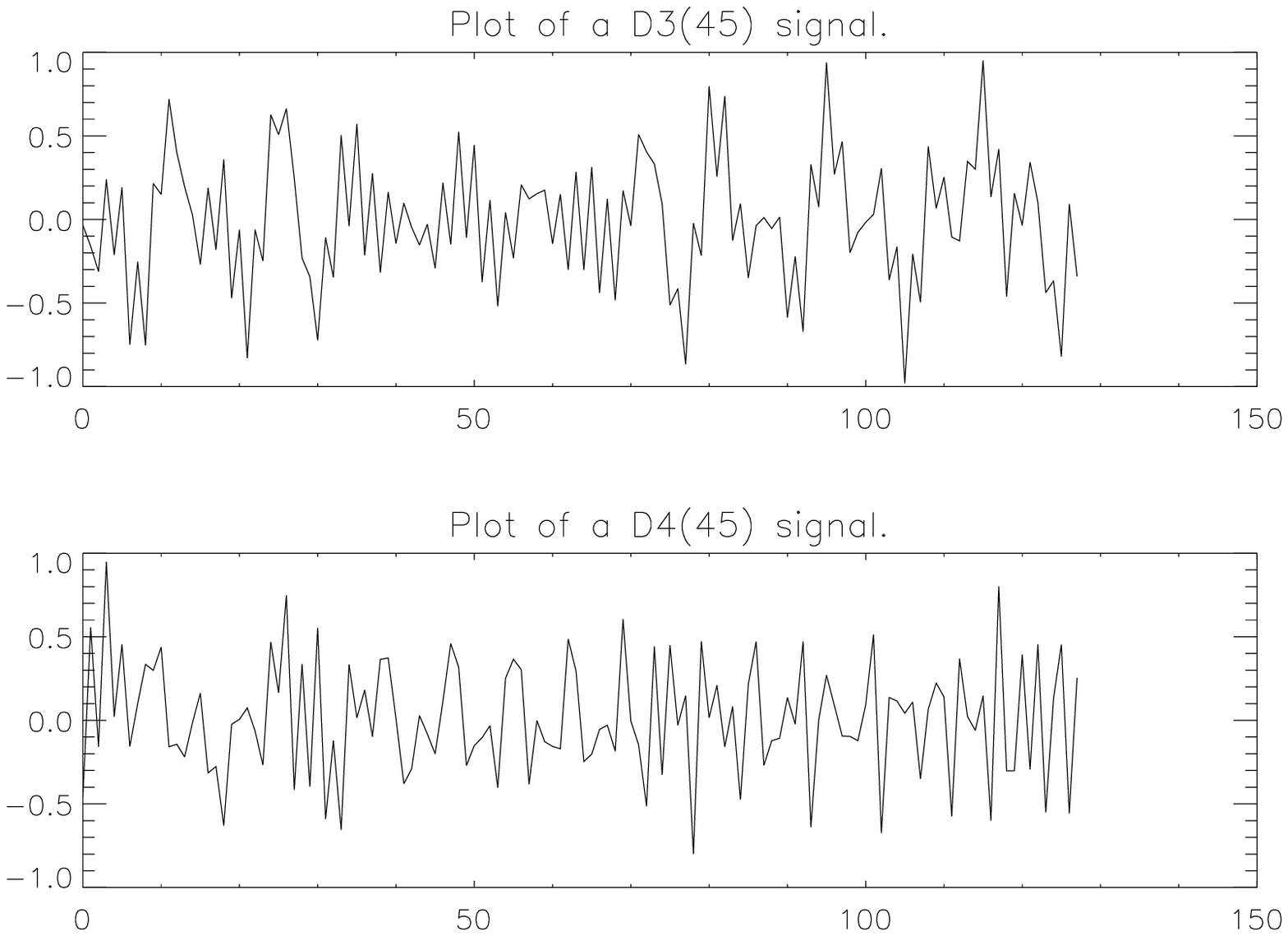}
  \end{center}
  \caption{Plot of the first 128 samples of the signals $s_{3}$ and $s_{4}$.}
  \label{signals34}
\end{figure} 

\begin{figure}[h]
  \begin{center}
    \includegraphics[scale = 0.65]{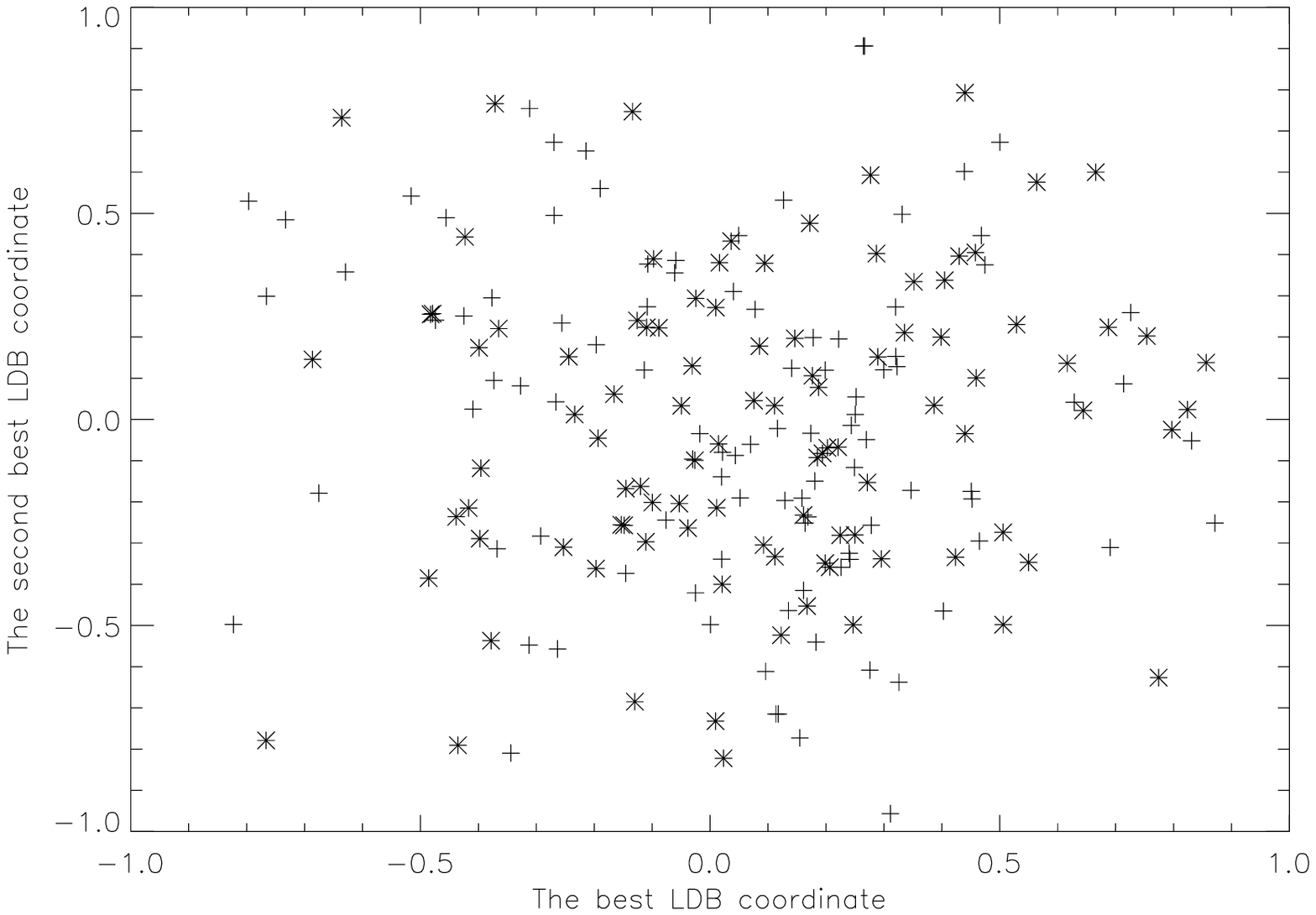}
  \end{center}
  \caption{Scatter plot of the 2 most discriminating coordinates in 
  the standard basis for 100 signals from each of the classes 
  $D_{3}(45^{\circ})$ and $D_{4}(45^{\circ})$. }
  \label{stascatter34}
\end{figure}  

\begin{figure}[h]
  \begin{center}
    \includegraphics[scale = 0.65]{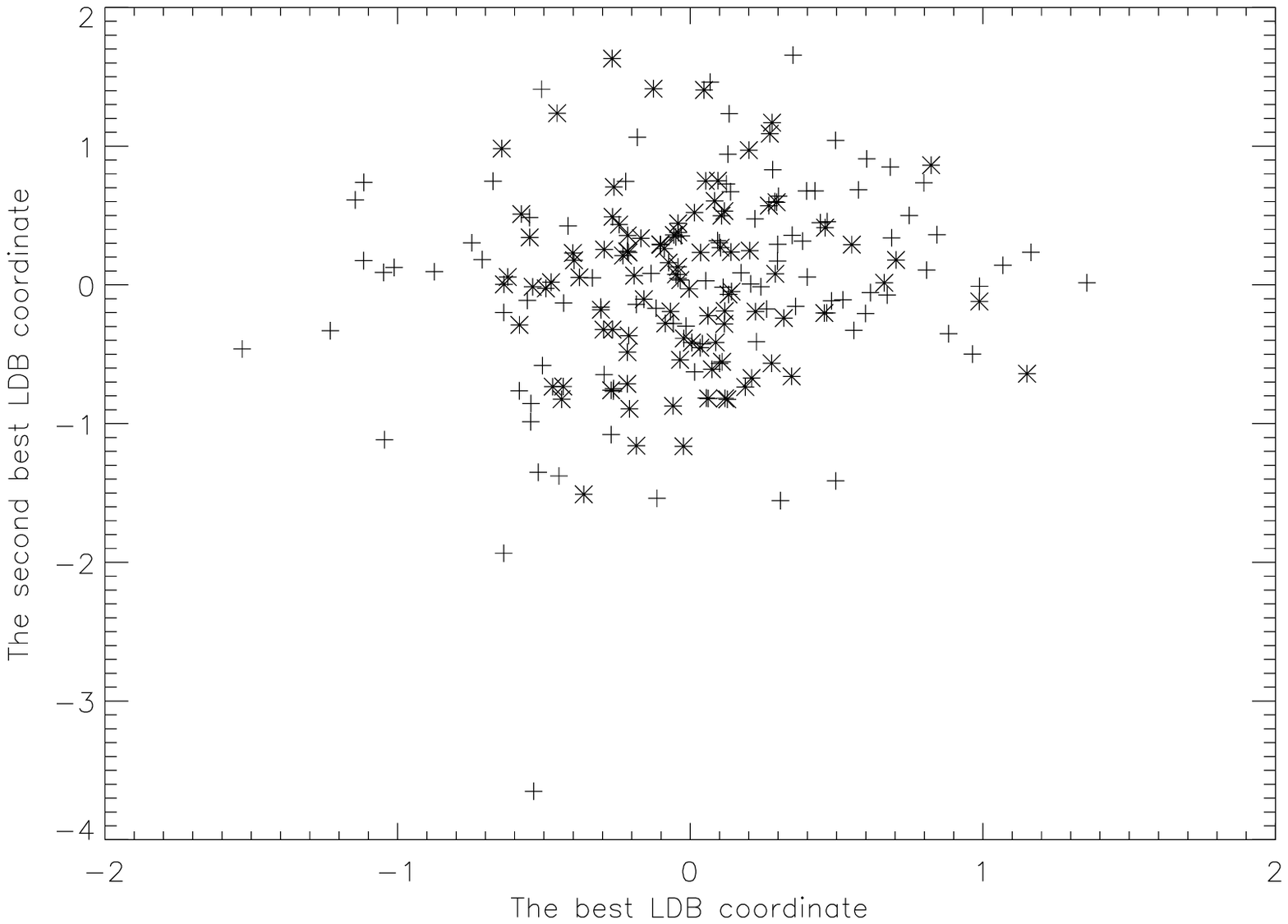}
  \end{center}
  \caption{Scatter plot of the 2 most discriminating coordinates in the 
    discrete cosine IV basis for 100 signals from each of the classes 
    $D_{3}(45^{\circ})$ and $D_{4}(45^{\circ})$. }
  \label{cosinescatter34}
\end{figure}

\begin{figure}[h]
  \begin{center}
    \includegraphics[scale = 0.65]{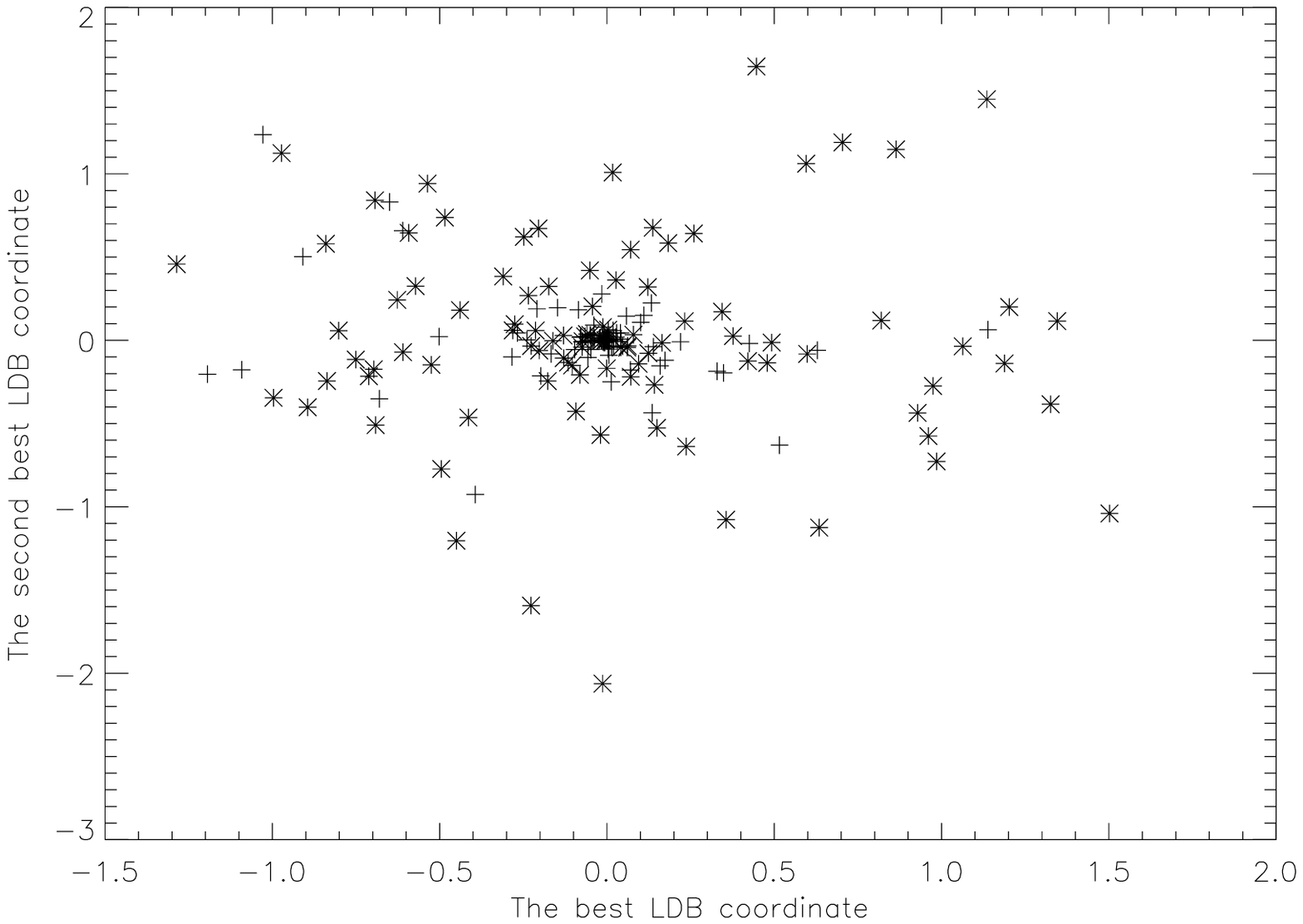}
  \end{center}
  \caption{Scatter plot of the 2 most discriminating
    coordinates in the coiflet packet basis for 100 signals from
    each of the classes $D_{3}(45^{\circ})$ 
    and $D_{4}(45^{\circ})$. }
  \label{coifscatter34}
\end{figure} 

\begin{figure}[h]
  \begin{center}
    \includegraphics[scale = 0.65]{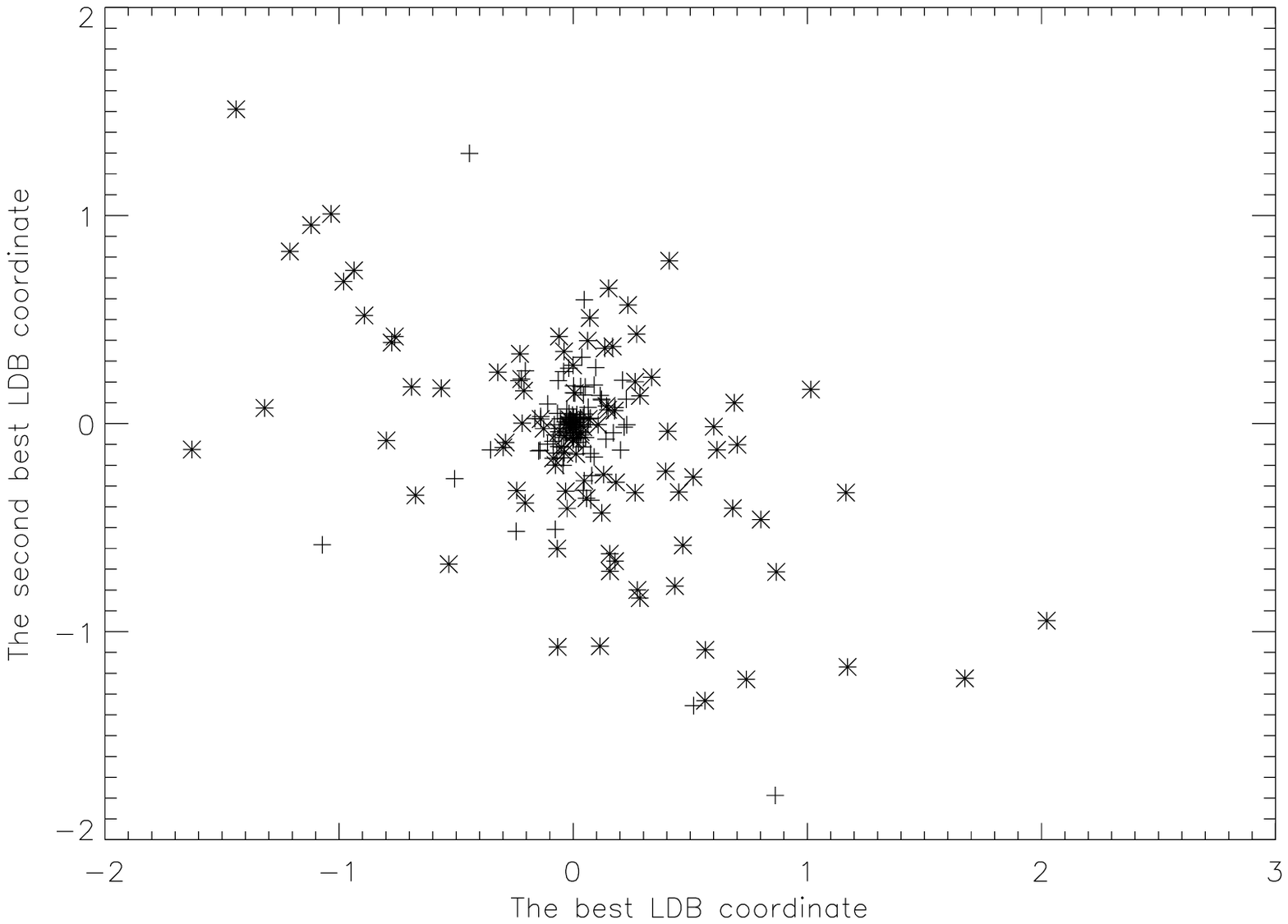}
  \end{center}
  \caption{Scatter plot of the 2 most discriminating
  coordinates in the local cosine basis for 100 signals from
  each of the classes $D_{3}(45^{\circ})$ 
  and $D_{4}(45^{\circ})$. }
\label{localcosinescatter34}
\end{figure}  
\clearpage

\section{Plots for  $D_{4}(45^{\circ})$, $D_{5}(45^{\circ})$}

\begin{figure}[h]
  \begin{center}
    \includegraphics[scale = 0.65]{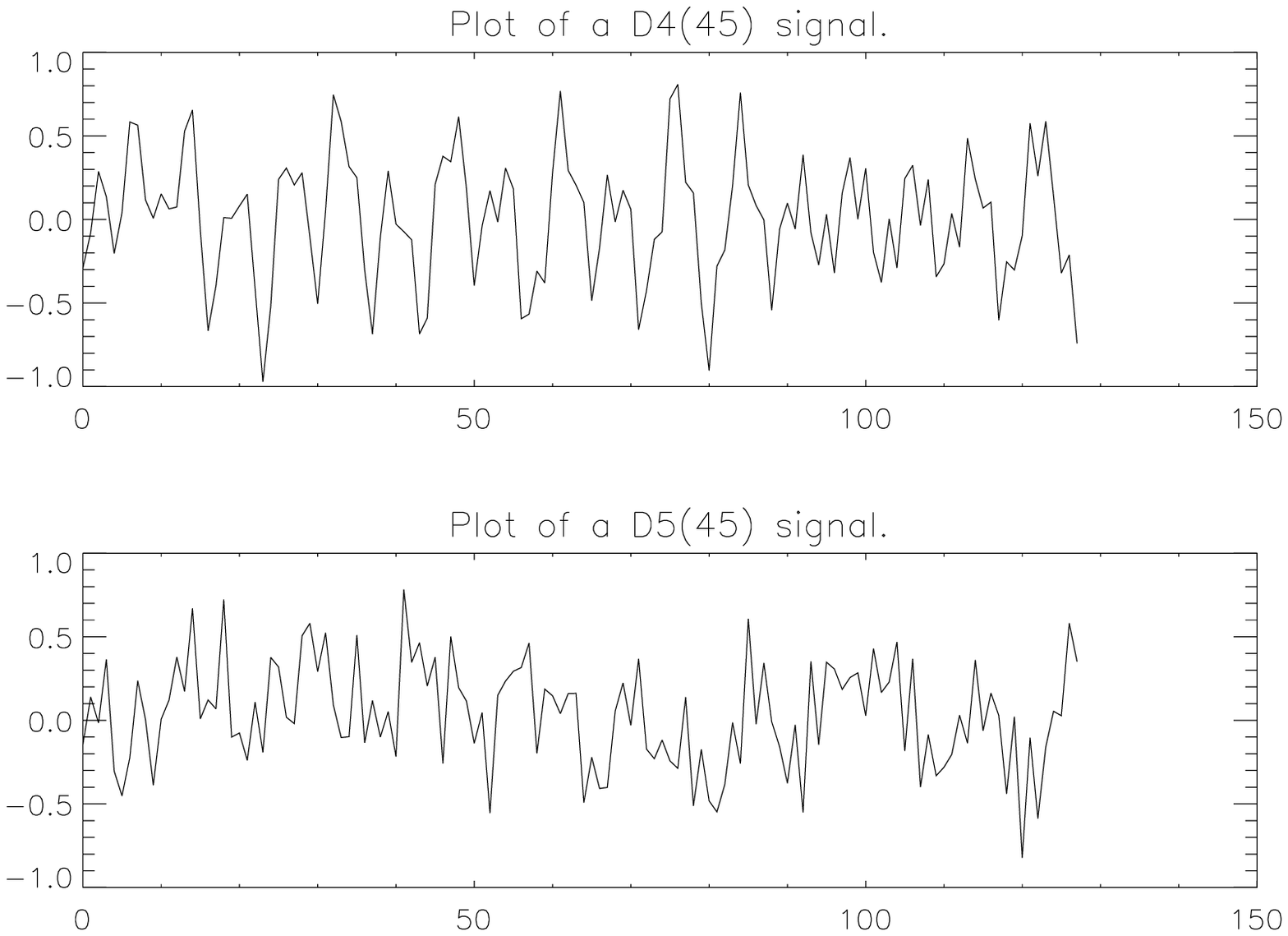}
  \end{center}
  \caption{Plot of the first 128 samples of the signals $s_{4}$ and $s_{5}$.}
  \label{signals45}
\end{figure} 

\begin{figure}[h]
  \begin{center}
    \includegraphics[scale = 0.65]{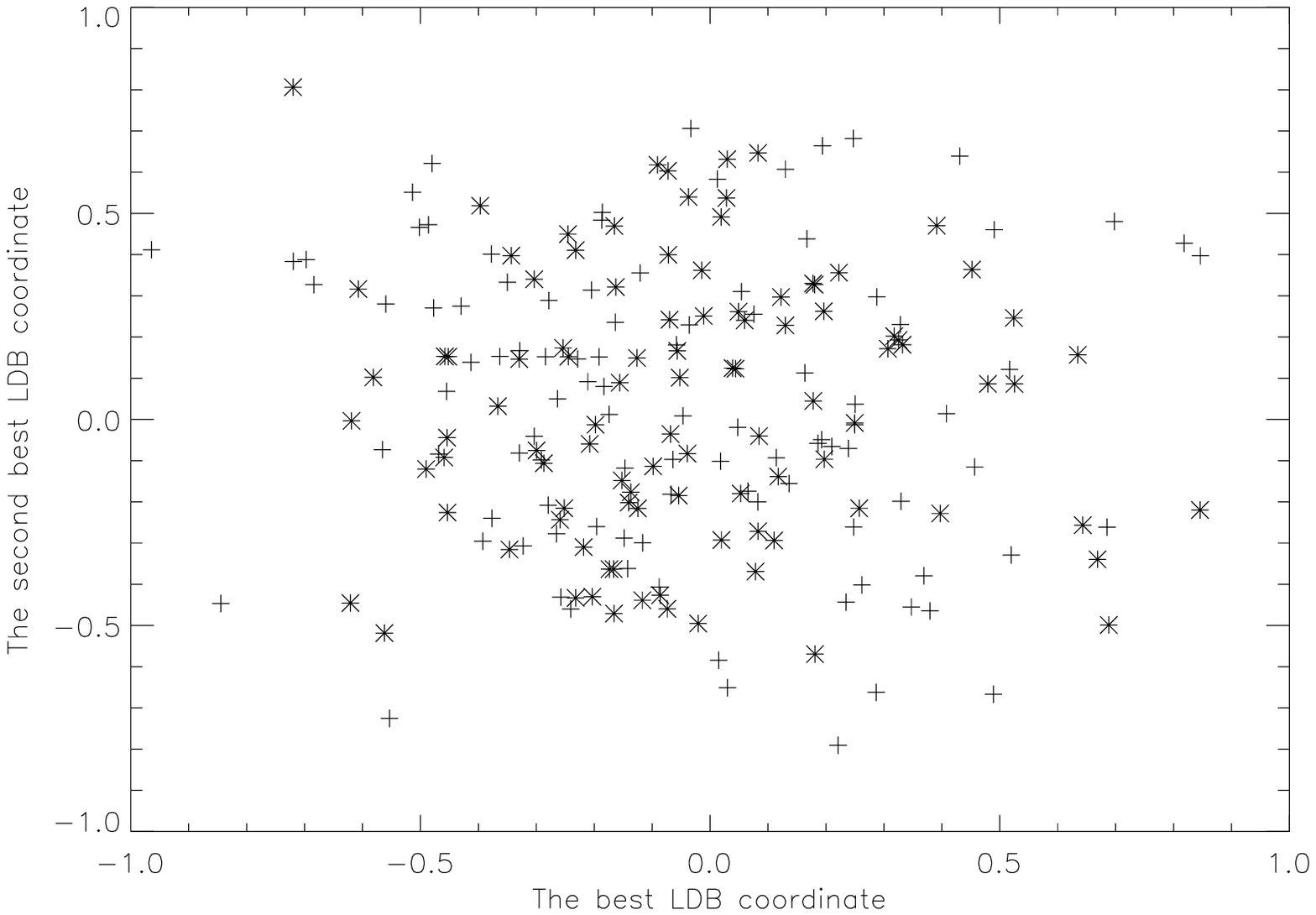}
  \end{center}
  \caption{Scatter plot of the 2 most discriminating
  coordinates in the standard basis for 100 signals from each of the
  classes $D_{4}(45^{\circ})$ and $D_{5}(45^{\circ})$. }
\label{stascatter45}
\end{figure}  

\begin{figure}[h]
  \begin{center}
    \includegraphics[scale = 0.65]{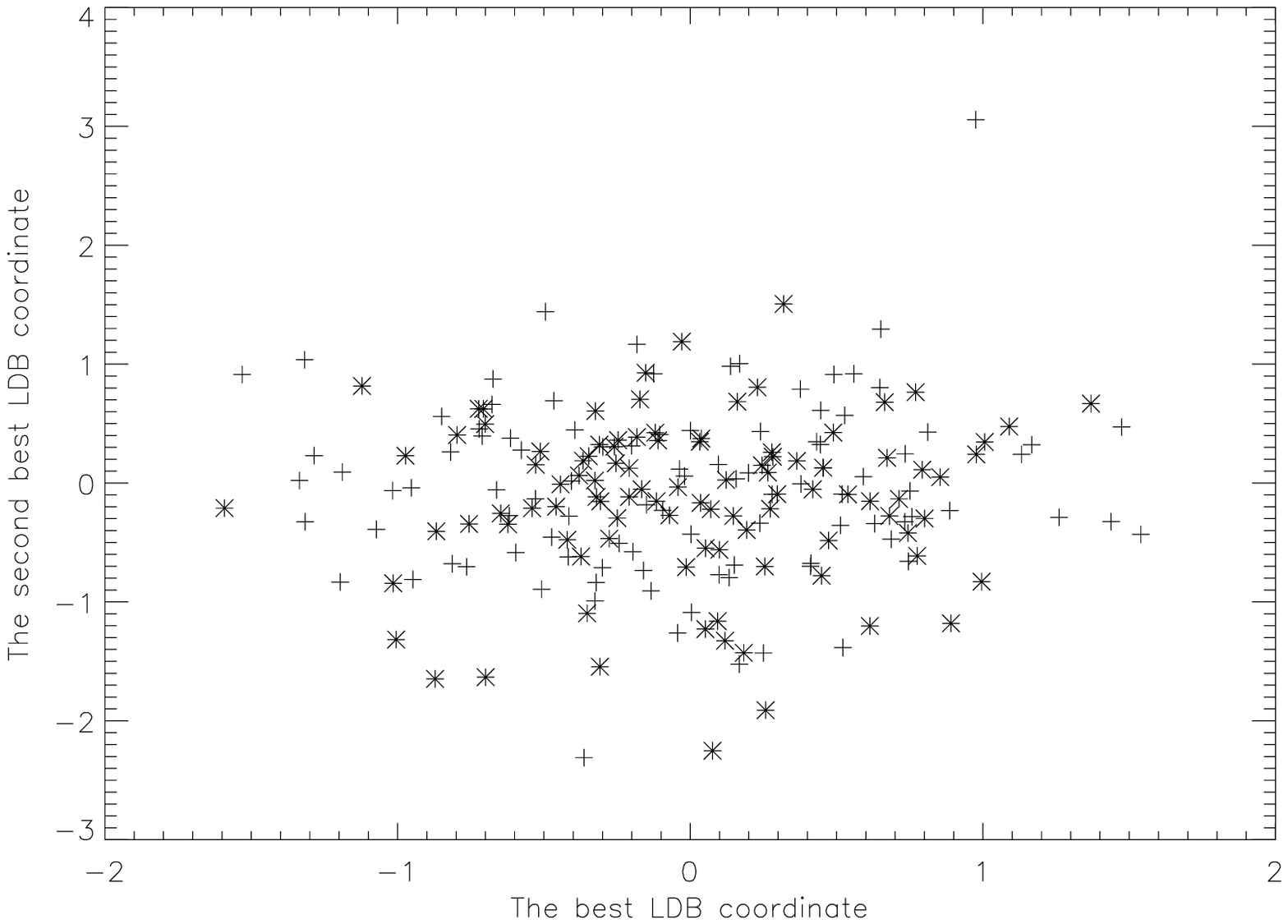}
  \end{center}
  \caption{Scatter plot of the 2 most discriminating coordinates in 
    the discrete cosine IV basis for 100 signals from
    each of the classes $D_{4}(45^{\circ})$ 
    and $D_{5}(45^{\circ})$.}
  \label{cosinescatter45}
\end{figure}  

\begin{figure}[h]
  \begin{center}
    \includegraphics[scale = 0.65]{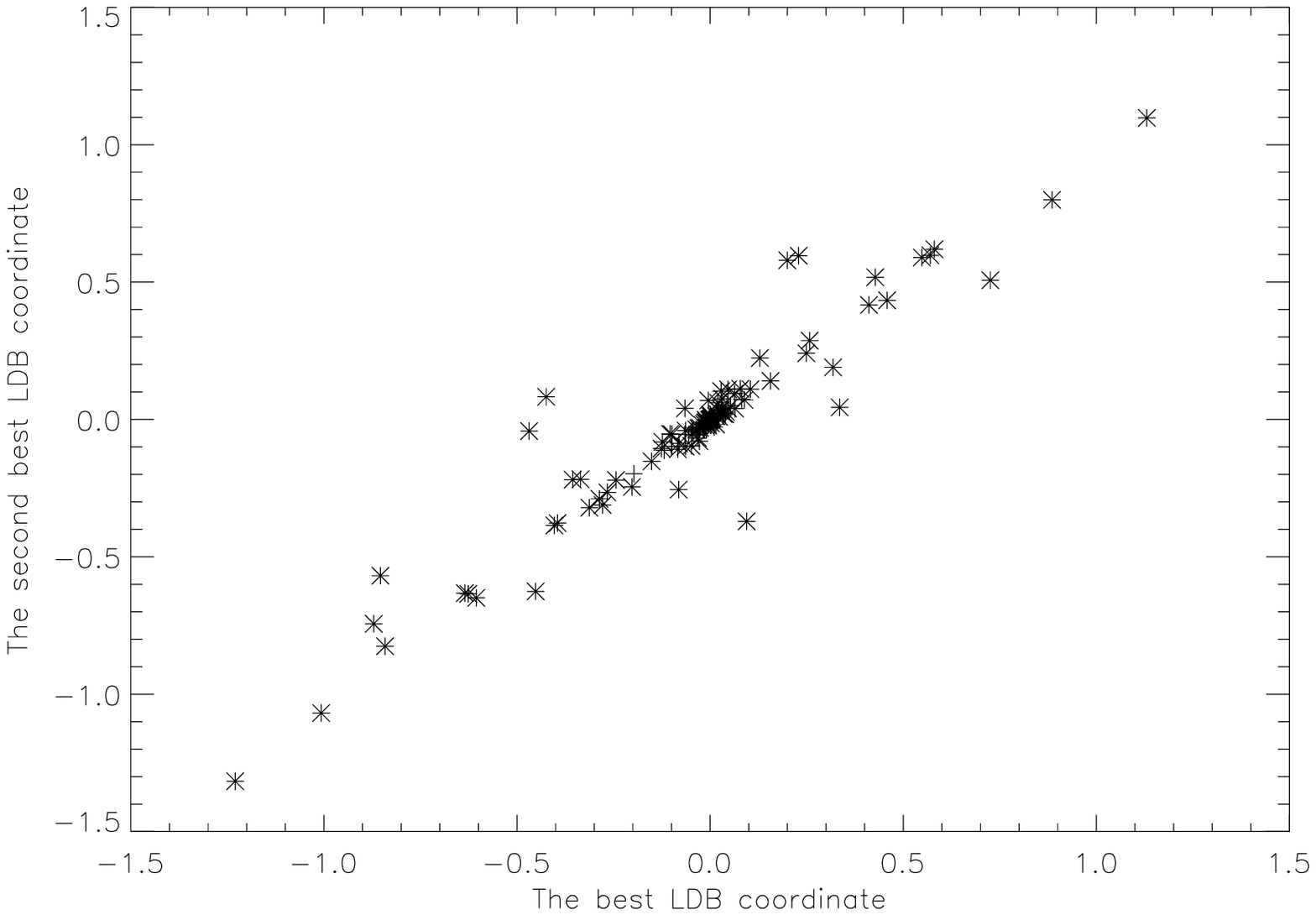}
  \end{center}
  \caption{Scatter plot of the 2 most discriminating
    coordinates in the coiflet packet basis for 100 signals from
    each of the classes $D_{4}(45^{\circ})$ 
    and $D_{5}(45^{\circ})$. }
  \label{coifscatter45}
\end{figure} 

\begin{figure}[h]
  \begin{center}
    \includegraphics[scale = 0.65]{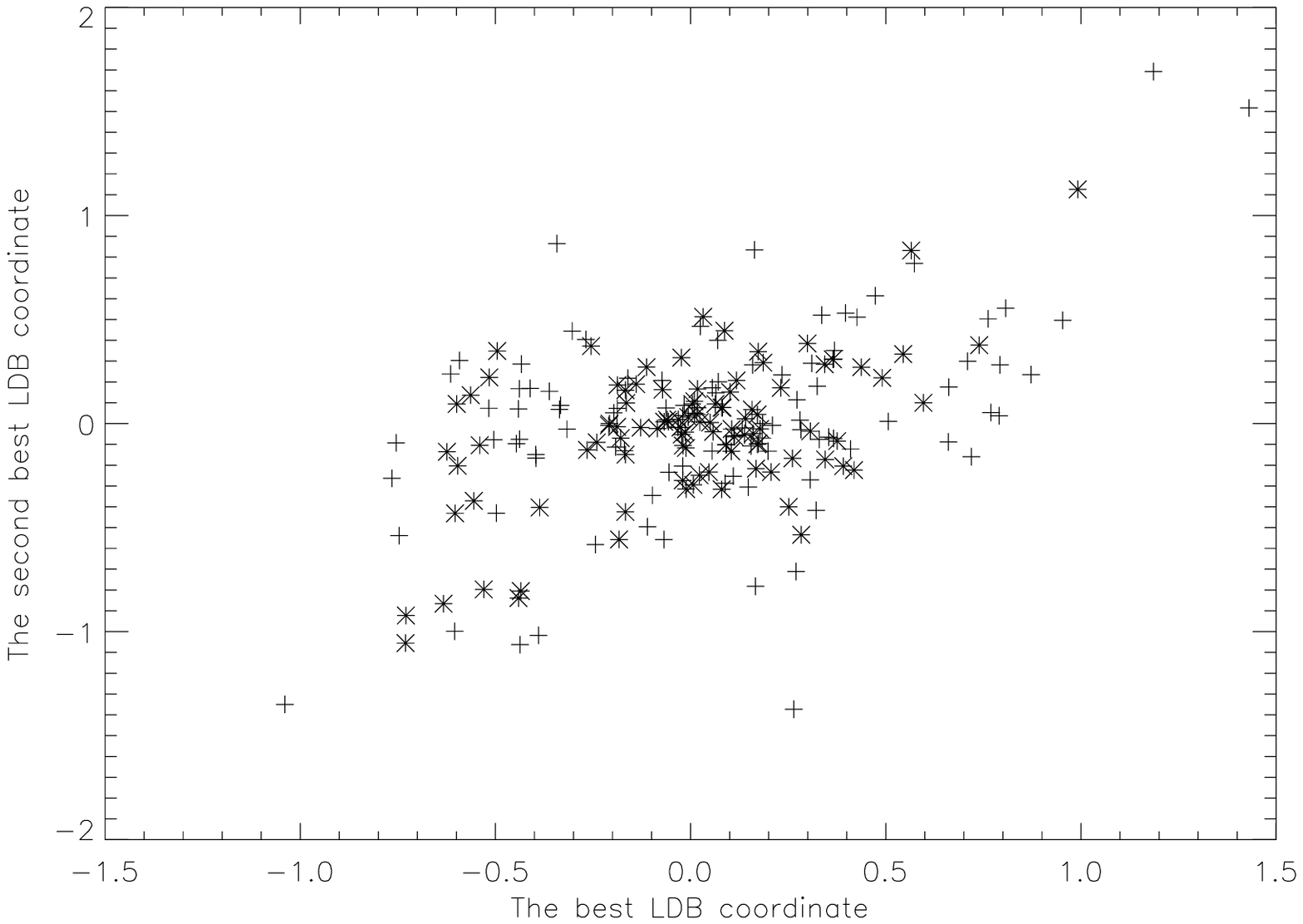}
  \end{center}
  \caption{Scatter plot of the 2 most discriminating
    coordinates in the local cosine basis for 100 signals from
    each of the classes $D_{4}(45^{\circ})$ 
    and $D_{5}(45^{\circ})$. }
  \label{localcosinescatter45}
\end{figure}  
\clearpage

\section{Plots for  $D_{5}(45^{\circ})$, $D_{6}(45^{\circ})$}
  
\begin{figure}[h]
  \begin{center}
    \includegraphics[scale = 0.65]{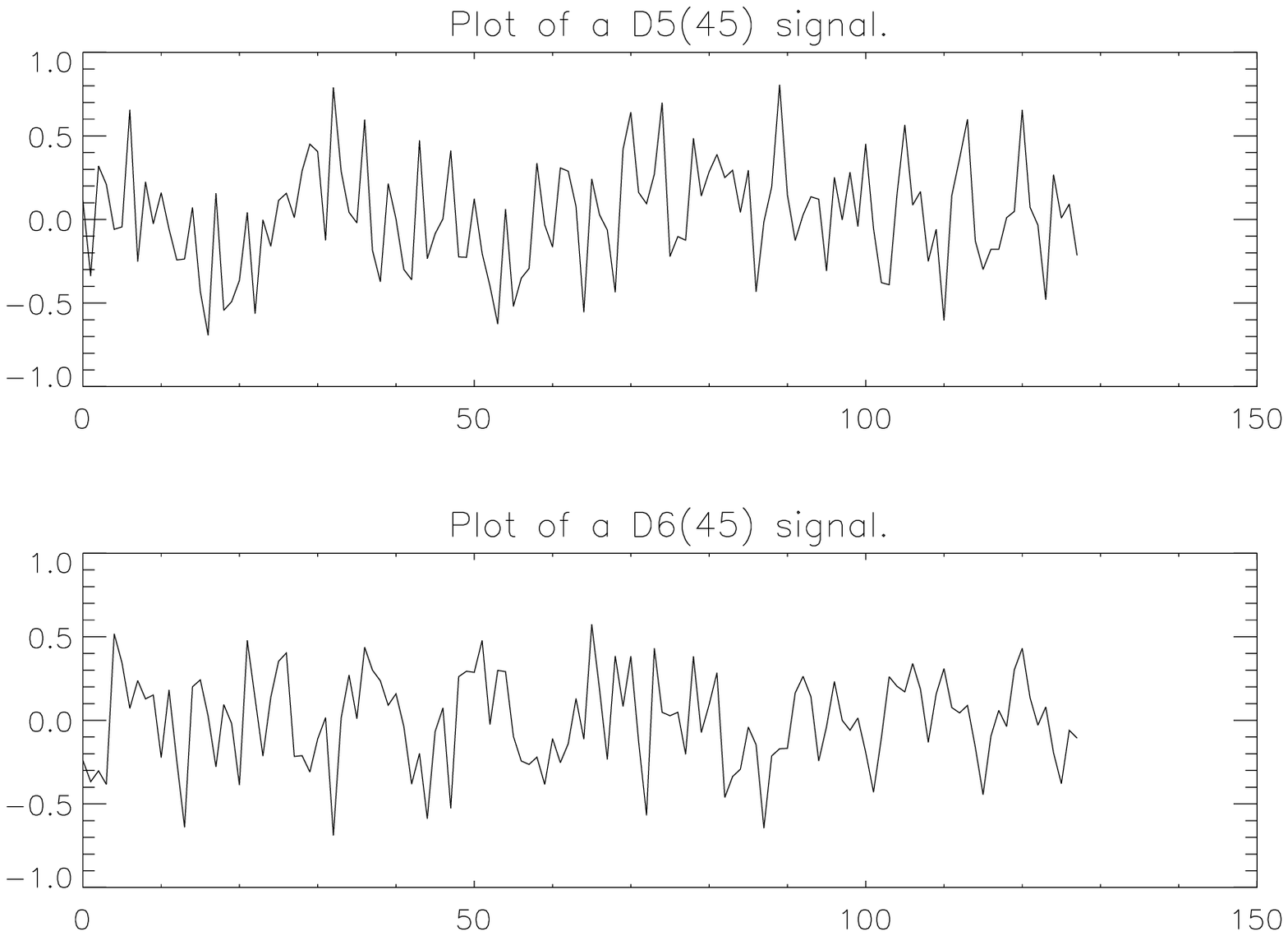}
  \end{center}
  \caption{Plot of the first 128 samples of the signals $s_{5}$ and $s_{6}$.}
  \label{signals56}
\end{figure} 

\begin{figure}[h]
  \begin{center}
    \includegraphics[scale = 0.65]{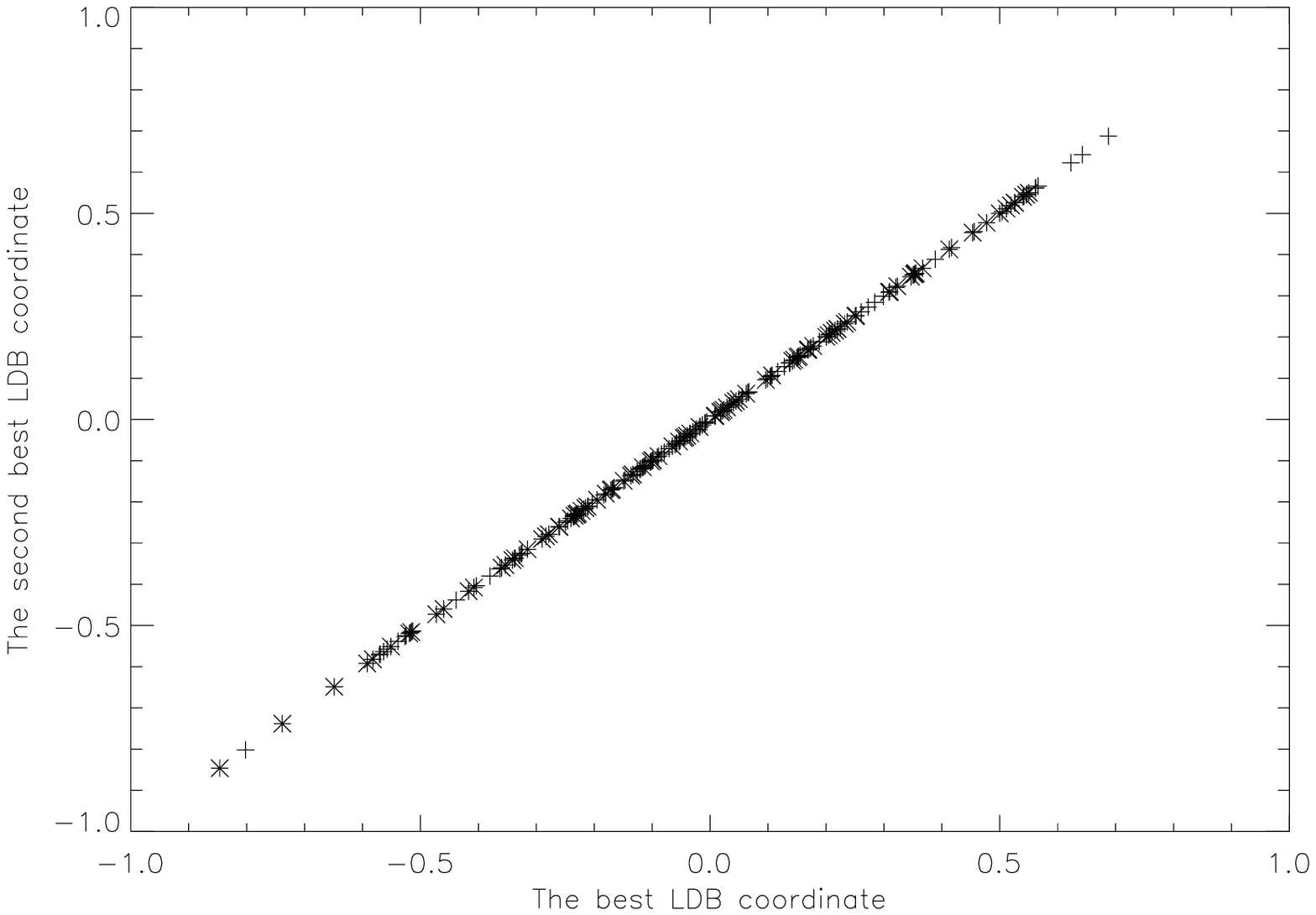}
  \end{center}
  \caption{Scatter plot of the 2 most discriminating
  coordinates in the standard basis for 100 signals from each of the
  classes $D_{5}(45^{\circ})$ and $D_{6}(45^{\circ})$. }
  \label{stascatter56}
\end{figure}  

\begin{figure}[h]
  \begin{center}
    \includegraphics[scale = 0.65]{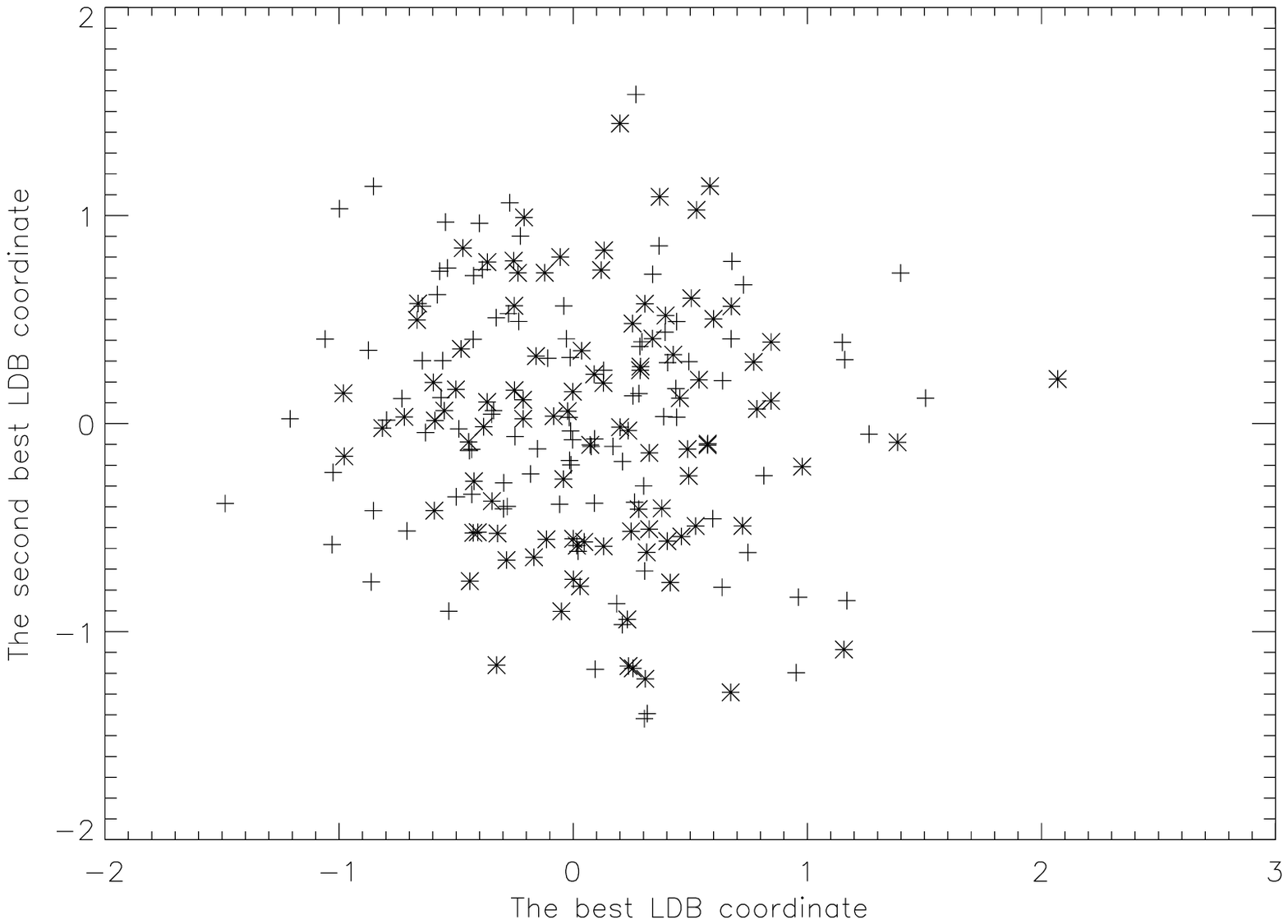}
  \end{center}
  \caption{Scatter plot of the 2 most discriminating
    coordinates in the discrete cosine IV basis for 100 signals from
    each of the classes $D_{5}(45^{\circ})$ 
    and $D_{6}(45^{\circ})$. }
  \label{cosinecatter56}
\end{figure}  

\begin{figure}[h]
  \begin{center}
    \includegraphics[scale = 0.65]{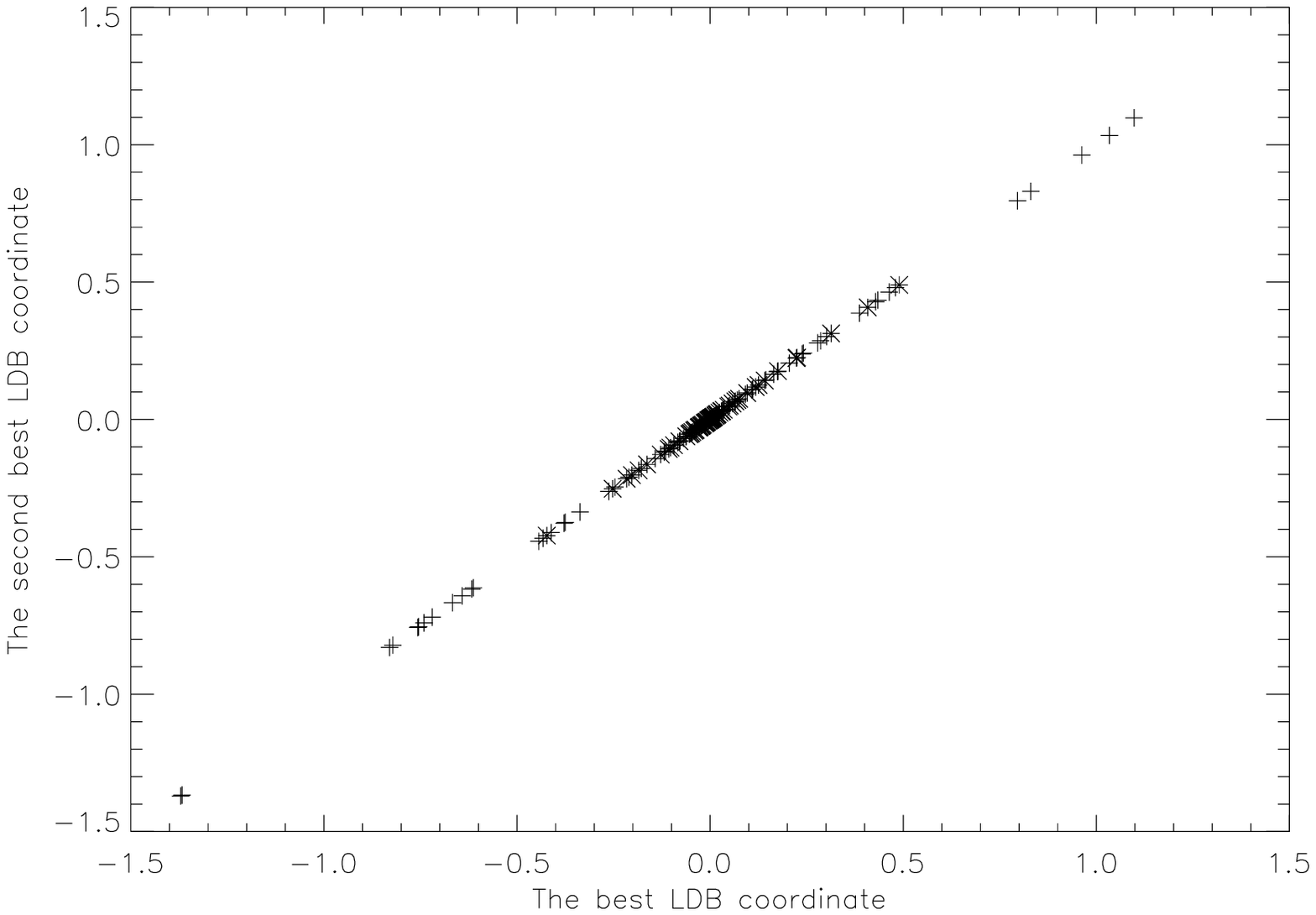}
  \end{center}
  \caption{Scatter plot of the 2 most discriminating
    coordinates in the coiflet packet basis for 100 signals from
    each of the classes $D_{5}(45^{\circ})$ 
    and $D_{6}(45^{\circ})$.}
  \label{coifscatter56}
\end{figure}

\begin{figure}[h]
  \begin{center}
    \includegraphics[scale = 0.65]{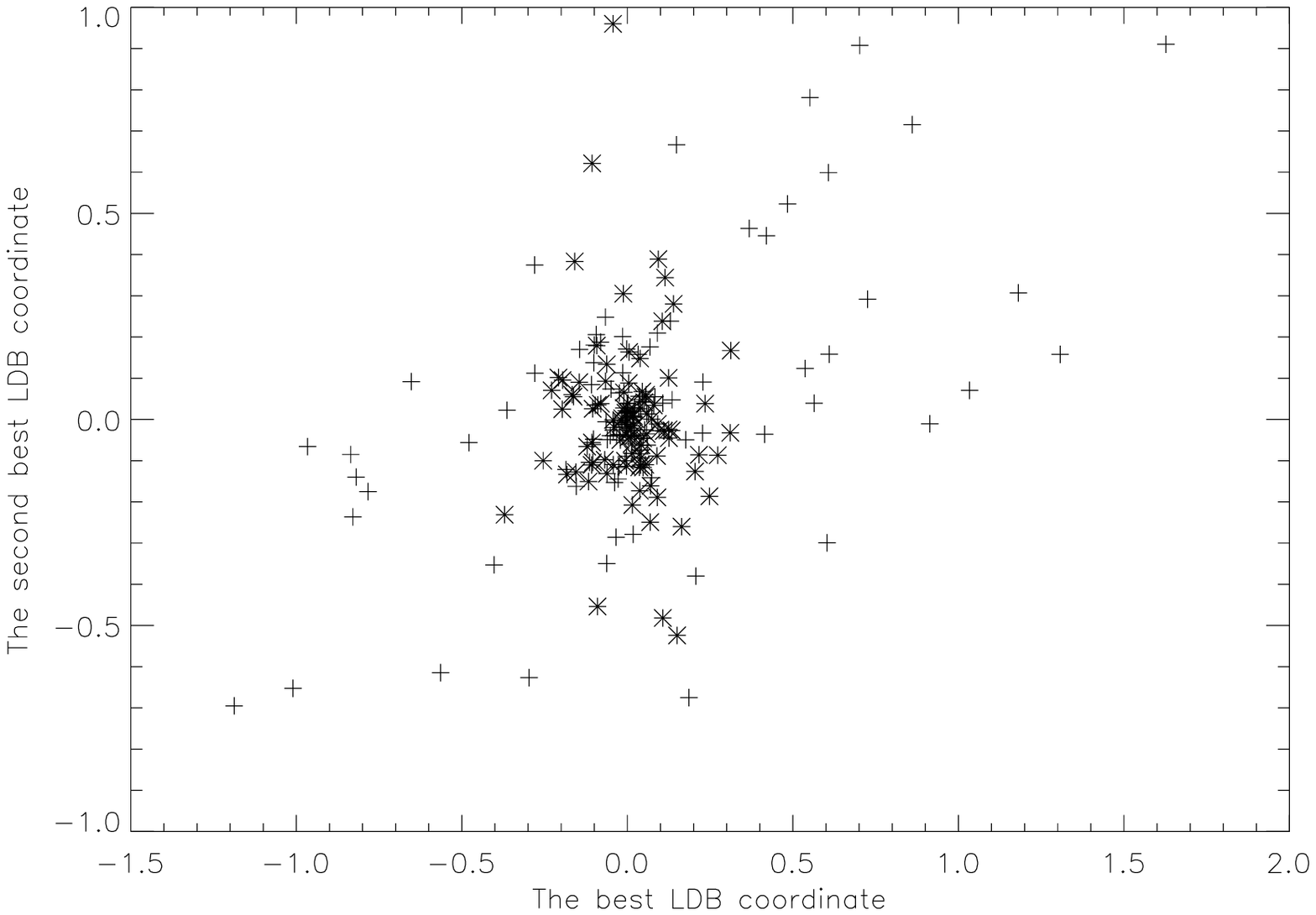}
  \end{center}
  \caption{Scatter plot of the 2 most discriminating
    coordinates in the local cosine basis for 100 signals from
    each of the classes $D_{5}(45^{\circ})$ 
    and $D_{6}(45^{\circ})$. }
  \label{localcosinescatter56}
\end{figure}
\clearpage

\section{Plots for  $D_{10}(45^{\circ})$, $D_{20}(45^{\circ})$}

\begin{figure}[h]
  \begin{center}
    \includegraphics[scale = 0.65]{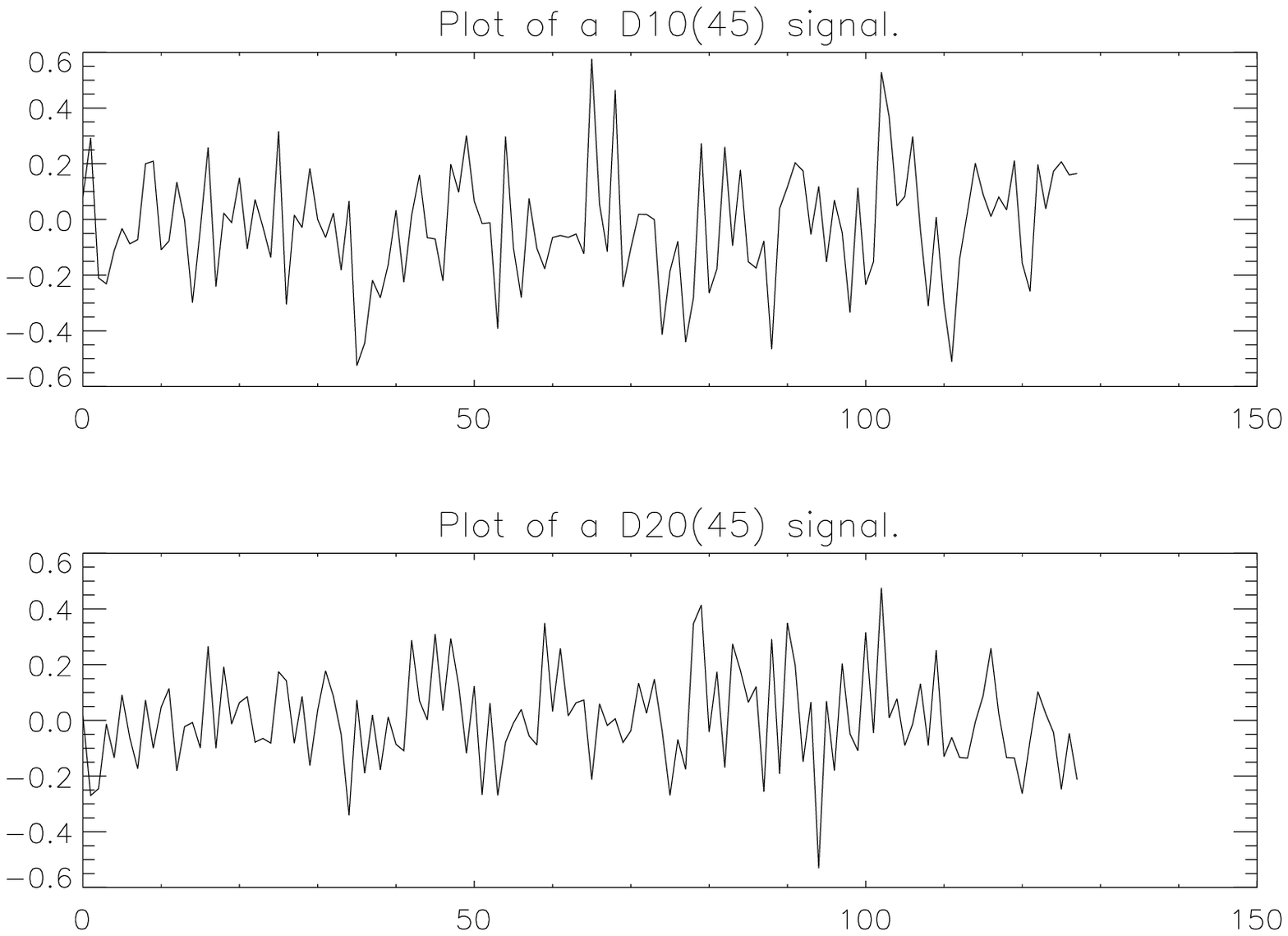}
  \end{center}
  \caption{Plot of the first 128 samples of the signals $s_{10}$ and $s_{20}$.}
  \label{signals1020}
\end{figure} 

\begin{figure}[h]
  \begin{center}
    \includegraphics[scale = 0.65]{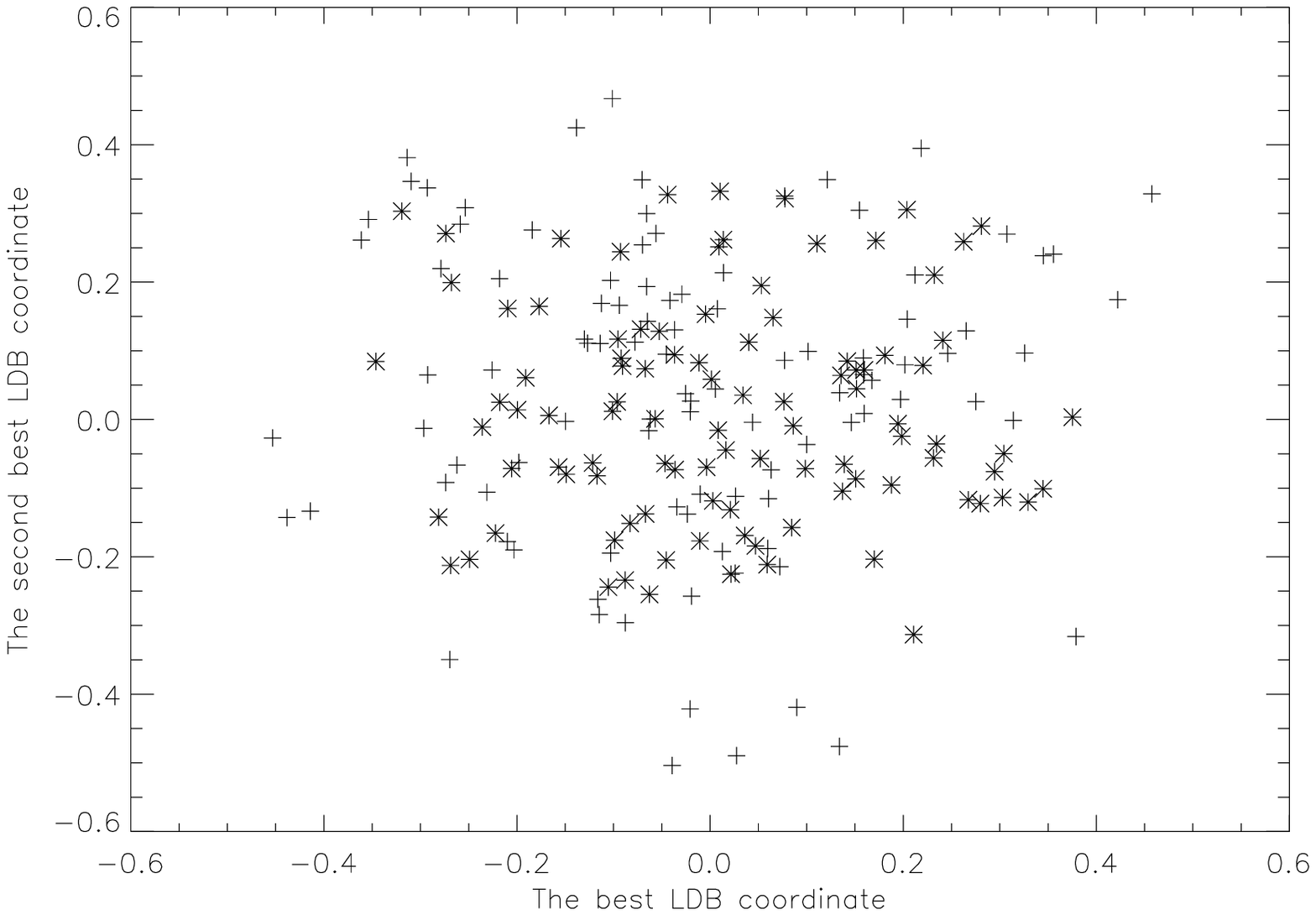}
  \end{center}
  \caption{Scatter plot of the 2 most discriminating
  coordinates in the standard basis for 100 signals from each of the
  classes $D_{10}(45^{\circ})$ and $D_{20}(45^{\circ})$. }
  \label{stascatter1020}
\end{figure} 

\begin{figure}[h]
  \begin{center}
    \includegraphics[scale = 0.65]{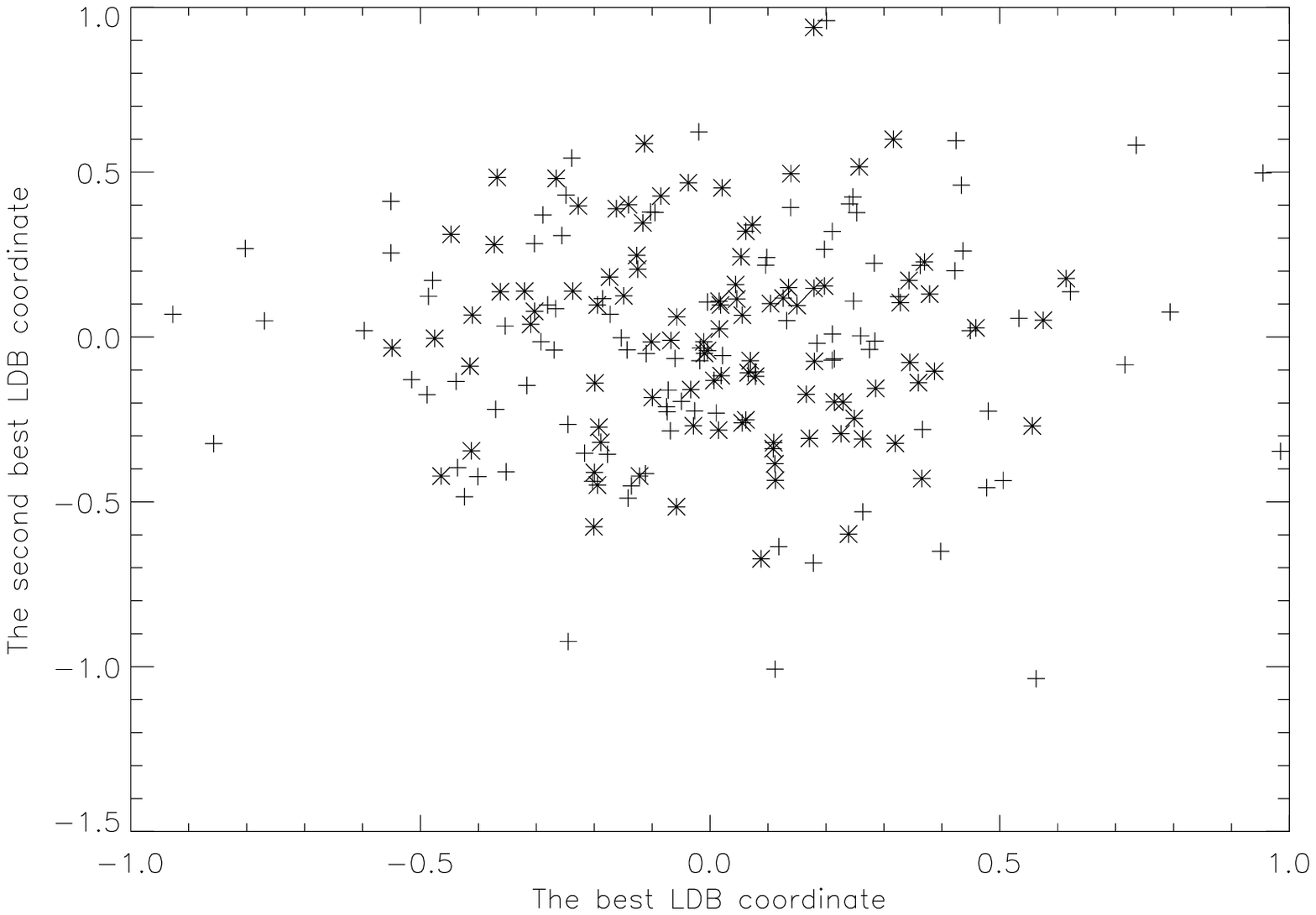}
  \end{center}
  \caption{Scatter plot of the 2 most discriminating
  coordinates in the discrete cosine IV basis for 100 signals from
  each of the classes $D_{10}(45^{\circ})$ 
  and $D_{20}(45^{\circ})$. }
  \label{cosinescatter1020}
\end{figure}

\begin{figure}[h]
  \begin{center}
    \includegraphics[scale = 0.65]{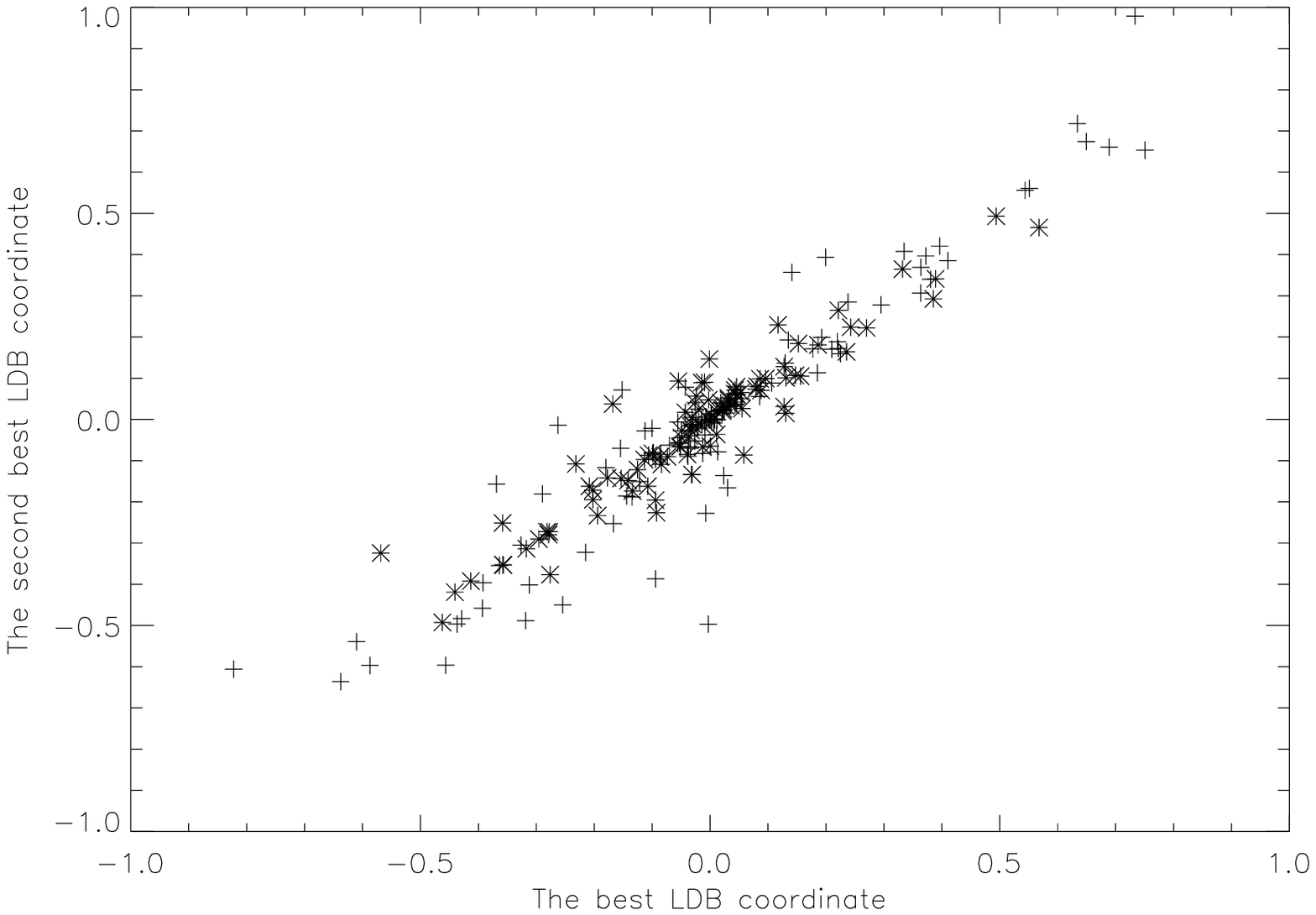}
  \end{center}
  \caption{Scatter plot of the 2 most discriminating
  coordinates in the coiflet packet basis for 100 signals from
  each of the classes $D_{10}(45^{\circ})$ 
  and $D_{20}(45^{\circ})$. }
  \label{coifscatter1020}
\end{figure} 

\begin{figure}[h]
  \begin{center}
    \includegraphics[scale = 0.65]{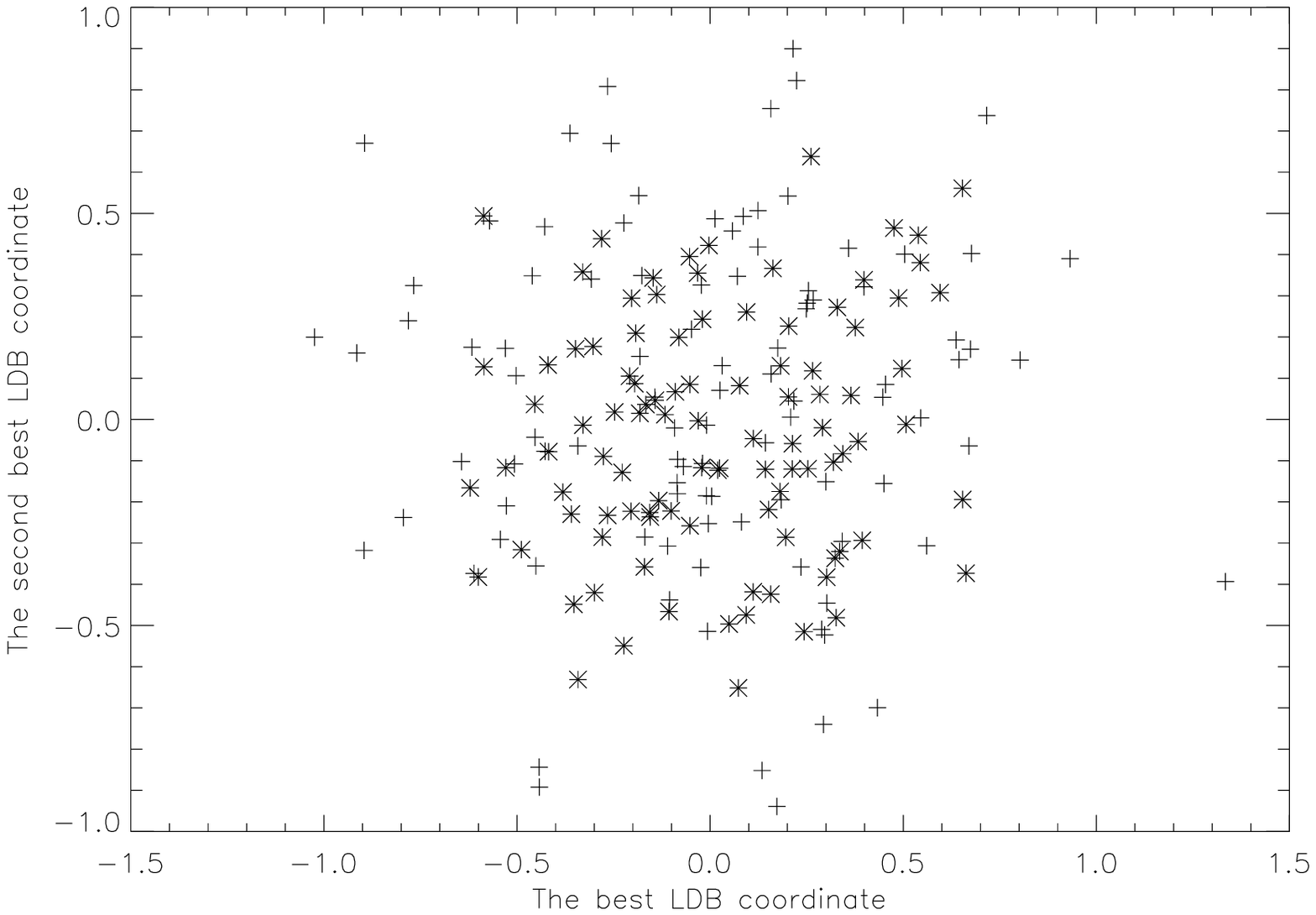}
  \end{center}
  \caption{Scatter plot of the 2 most discriminating
  coordinates in the local cosine basis for 100 signals from
  each of the classes $D_{10}(45^{\circ})$ 
  and $D_{20}(45^{\circ})$. }
  \label{localcosinescatter1020}
\end{figure}


%% file: dummyappendix.tex
\chapter{Computer Programs}

All of the numerical work discussed in this paper was carried
through in Ansi-C or Maple. In the three sections below we give the 
source code of all important transforms and algorithms used in this paper.
The code in the first section was, except some slight
modifications, copied from \cite{wick}. The code in the second
and third section was generated by the author himself.

\section{Borrowed Ansi C source code}

\input wickprograms
\newpage

\section{Author's Ansi C source code} 

\input myprograms
\newpage

\section{Author's Maple code} 

\input mapleprograms


%% file: wickprograms.tex

\File{wickprograms.c}{1997}{Oct}{30}{23:13}{9487}
\L{\LB{\C{}\1\*\*\*\*\*\*\*\*\*\*\*\*\*\*\*\*\*\*\*\*\*\*\*\*\*\*\*\*\*\*\*\*\*\*\*\*\*\*\*\*\*\*\*\*\*\*\*\*\*\*\*\*\*\*\*\*\*\*\*\*\*\*\*\*\*\*\*\*\*\*\1\CE{}}}
\L{\LB{}}
\L{\LB{\C{}\1\*_Some_useful_functions:_The_discrete_fourier,_cosine\1sine,_}}
\L{\LB{___local_cosine\1sine_functions_and_their_companions_are_copied_}}
\L{\LB{___(and_somewhat_adapted_for_our_special_use)_from_M.L.Wickerhauser:_}}
\L{\LB{___\3Adapted_Wavelet_Analysis_from_Theory_to_Software\3._\*\1\CE{}}}
\L{\LB{}}
\L{\LB{\C{}\1\*\*\*\*\*\*\*\*\*\*\*\*\*\*\*\*\*\*\*\*\*\*\*\*\*\*\*\*\*\*\*\*\*\*\*\*\*\*\*\*\*\*\*\*\*\*\*\*\*\*\*\*\*\*\*\*\*\*\*\*\*\*\*\*\*\*\*\*\*\*\1\CE{}}}
\L{\LB{}}
\L{\LB{}}
\L{\LB{\C{}\1\*_typedef_complex:_Define_a_new_data_structure_of_type_\3complex\3_}}
\L{\LB{___containing_two_members:}}
\L{\LB{___}}
\L{\LB{___RE_=_Floating_point_number_describing_the_}}
\L{\LB{___real_part_of_a_complex_number._}}
\L{\LB{___IM_=__Floating_point_number_describing_the_imaginary_part_}}
\L{\LB{___of_a_complex_number._\*\1\CE{}}}
\L{\LB{}}
\L{\LB{\K{typedef}_\K{struct}_\{}}
\L{\LB{__\K{double}_\V{RE};}}
\L{\LB{__\K{double}_\V{IM};}}
\L{\LB{\}_\V{complex};}}
\L{\LB{}}
\L{\LB{}}
\L{\LB{\C{}\1\*_struct_interval:_Define_a_new_data_structure_of_type_\3interval\3,_}}
\L{\LB{___containing_three_members:}}
\L{\LB{___}}
\L{\LB{___ORIGIN_=_Floating_point_pointer_to_origin_of_an_array_of_both_}}
\L{\LB{___positive_and_negative_indexes.}}
\L{\LB{___LEAST_=_Integer_describing_the_least_index_of_the_array.}}
\L{\LB{___FINAL_=_Integer_describing_the_largest_index_of_the_array._\*\1\CE{}}}
\L{\LB{}}
\L{\LB{\K{struct}_\V{interval}_\{}}
\L{\LB{__\K{double}_\*\V{ORIGIN};}}
\L{\LB{__\K{int}_\V{LEAST};}}
\L{\LB{__\K{int}_\V{FINAL};}}
\L{\LB{\};}}
\L{\LB{}}
\L{\LB{\K{\#define}__\V{CCMULRE}(\V{Z1},\V{Z2})__________________(\V{Z1}.\V{RE}\*\V{Z2}.\V{RE}\-\V{Z1}.\V{IM}\*\V{Z2}.\V{IM})}}
\L{\LB{\K{\#define}__\V{CCMULIM}(\V{Z1},\V{Z2})__________________(\V{Z1}.\V{RE}\*\V{Z2}.\V{IM}+\V{Z1}.\V{IM}\*\V{Z2}.\V{RE})}}
\L{\LB{\K{\#define}__\V{CRRMULRE}(\V{Z},\V{YRE},\V{YIM})_____________(\V{Z}.\V{RE}\*\V{YRE}\-\V{Z}.\V{IM}\*\V{YIM})}}
\L{\LB{\K{\#define}__\V{CRRMULIM}(\V{Z},\V{YRE},\V{YIM})_____________(\V{Z}.\V{RE}\*\V{YIM}+\V{Z}.\V{IM}\*\V{YRE})}}
\L{\LB{\K{\#define}__\V{max}(\V{X},\V{Y})________________________((\V{X}_\>_\V{Y})_?_\V{X}:\V{Y})}}
\L{\LB{\K{\#define}__\V{min}(\V{X},\V{Y})________________________((\V{X}_\<_\V{Y})_?_\V{X}:\V{Y})}}
\L{\LB{\K{\#define}__\V{rcf}(\V{T})__________________________\V{rcfis}(\N{1},\V{T})}}
\L{\LB{\K{\#define}__\V{abtblock}(\V{N},\V{L},\V{B})_________________((\V{L})\*(\V{N})+(\V{B})\*((\V{N})\>\>(\V{L})))}}
\L{\LB{\K{\#define}__\V{abtlength}(\V{N},\V{L})__________________((\V{N})\>\>(\V{L}))}}
\L{\LB{}}
\L{\LB{\C{}\1\*_makeinterval:_Allocate_an_interval_data_structure_}}
\L{\LB{___and_assign_its_data_array_\*\1\CE{}}}
\L{\LB{}}
\L{\LB{\Proc{makeinterval}\K{struct}_\V{interval}_\V{makeinterval}(\K{double}_\*\V{DATA},_\K{int}_\V{LEAST},_\K{int}_\V{FINAL})}}
\L{\LB{\{}}
\L{\LB{__\K{int}_\V{LENGTH},_\V{K};}}
\L{\LB{__}}
\L{\LB{__\K{struct}_\V{interval}_\V{SEG};}}
\L{\LB{__}}
\L{\LB{__\V{LENGTH}_=_\N{1}+\V{FINAL}\-\V{LEAST};}}
\L{\LB{__\K{if}_(\V{LENGTH}_\>_\N{0})_\{}}
\L{\LB{____\V{SEG}.\V{ORIGIN}_=_(\K{double}_\*)\V{calloc}(\V{LENGTH},_\K{sizeof}(\K{double}));}}
\L{\LB{____\V{SEG}.\V{ORIGIN}_\-=_\V{LEAST};}}
\L{\LB{____\K{if}_(\V{DATA}_!=_\V{NULL})_\{}}
\L{\LB{______\K{for}_(\V{K}_=_\V{LEAST};_\V{K}_\<=_\V{FINAL};_\V{K}++)_\{}}
\L{\LB{}\Tab{8}{\V{SEG}.\V{ORIGIN}[\V{K}]_=_\V{DATA}[\V{K}\-\V{LEAST}];}}
\L{\LB{______\}}}
\L{\LB{____\}}}
\L{\LB{__\}}}
\L{\LB{__\V{SEG}.\V{LEAST}_=_\V{LEAST};}}
\L{\LB{__\V{SEG}.\V{FINAL}_=_\V{FINAL};}}
\L{\LB{__\K{return}_\V{SEG};}}
\L{\LB{\}}}
\L{\LB{}}
\L{\LB{\C{}\1\*_freeinterval:_Deallocate_an_interval_data_structure_}}
\L{\LB{___and_its_data_array_\*\1\CE{}}}
\L{\LB{}}
\L{\LB{\Proc{freeinterval}\K{double}_\*\V{freeinterval}(\K{struct}_\V{interval}_\*\V{SEG})}}
\L{\LB{\{}}
\L{\LB{__\K{if}_(\V{SEG}_!=_\V{NULL})_\{}}
\L{\LB{____\K{if}_((\V{SEG}\-\!\>\V{ORIGIN})_!=_\V{NULL})_\{}}
\L{\LB{______(\V{SEG}\-\!\>\V{ORIGIN})_+=_\V{SEG}\-\!\>\V{LEAST};}}
\L{\LB{______\V{free}(\V{SEG}\-\!\>\V{ORIGIN});}}
\L{\LB{____\}}}
\L{\LB{____\V{free}(\V{SEG});}}
\L{\LB{__\}}}
\L{\LB{__\K{return}_\V{NULL};}}
\L{\LB{\}}}
\L{\LB{}}
\L{\LB{\C{}\1\*_br:_Return_the_input_integer_bit-reversed_\*\1\CE{}_}}
\L{\LB{}}
\L{\LB{\Proc{br}\K{int}_\V{br}(\K{int}_\V{N},_\K{int}_\V{LOG2LEN})}}
\L{\LB{\{}}
\L{\LB{__\K{int}_\V{U},_\V{J};}}
\L{\LB{__}}
\L{\LB{__\V{U}_=_\V{N}\&\N{1};}}
\L{\LB{__\K{for}_(\V{J}_=_\N{1};_\V{J}_\<_\V{LOG2LEN};_\V{J}++)_\{}}
\L{\LB{____\V{N}_\>\>=_\N{1};}}
\L{\LB{____\V{U}_\<\<=_\N{1};}}
\L{\LB{____\V{U}_+=_\V{N}\&\N{1};}}
\L{\LB{__\}}}
\L{\LB{__\K{return}_\V{U};}}
\L{\LB{\}}}
\L{\LB{}}
\L{\LB{\C{}\1\*_bitrevd:_Permute_to_a_disjoint_array_by_bit-reversing_}}
\L{\LB{___the_indices_\*\1\CE{}}}
\L{\LB{}}
\L{\LB{\Proc{bitrevd}\K{void}_\V{bitrevd}(\V{complex}_\*\V{OUT},_\V{complex}_\*\V{IN},_\K{int}_\V{Q})}}
\L{\LB{\{}}
\L{\LB{__\K{int}_\V{M},_\V{N},_\V{U};}}
\L{\LB{__}}
\L{\LB{__\V{M}_=_\N{1}\<\<\V{Q};}}
\L{\LB{__\K{for}_(\V{N}_=_\N{0};_\V{N}_\<_\V{M};_\V{N}++)_\{}}
\L{\LB{____\V{U}_=_\V{br}(\V{N},_\V{Q});}}
\L{\LB{____\V{OUT}[\V{U}]_=_\V{IN}[\V{N}];}}
\L{\LB{__\}}}
\L{\LB{\}}}
\L{\LB{}}
\L{\LB{\C{}\1\*_bitrevi:_Permute_an_array_in_place_via_index_bit-reversal_\*\1\CE{}}}
\L{\LB{}}
\L{\LB{\Proc{bitrevi}\K{void}_\V{bitrevi}_(\V{complex}_\*\V{X},_\K{int}_\V{Q})}}
\L{\LB{\{}}
\L{\LB{__\K{int}_\V{M},_\V{N},_\V{U};}}
\L{\LB{__}}
\L{\LB{__\V{complex}_\V{TEMP};}}
\L{\LB{__}}
\L{\LB{__\V{M}_=_\N{1}\<\<\V{Q};}}
\L{\LB{__\K{for}_(\V{N}_=_\N{1};_\V{N}_\<_(\V{M}\-\N{1});_\V{N}++)_\{}}
\L{\LB{____\V{U}_=_\V{br}(\V{N},_\V{Q});}}
\L{\LB{____\K{if}_(\V{U}_\>_\V{N})_\{}}
\L{\LB{______\V{TEMP}_=_\V{X}[\V{N}];}}
\L{\LB{______\V{X}[\V{N}]_=_\V{X}[\V{U}];}}
\L{\LB{______\V{X}[\V{U}]_=_\V{TEMP};}}
\L{\LB{____\}}}
\L{\LB{__\}}}
\L{\LB{\}}}
\L{\LB{}}
\L{\LB{\C{}\1\*_fftomega:_Compute_table_of_sines_and_cosines_for_DFT_\*\1\CE{}}}
\L{\LB{}}
\L{\LB{\Proc{fftomega}\K{void}_\V{fftomega}(\V{complex}_\*\V{W},_\K{int}_\V{M})}}
\L{\LB{\{}}
\L{\LB{__\K{int}_\V{K};}}
\L{\LB{__}}
\L{\LB{__\K{double}_\V{FACTOR};}}
\L{\LB{__}}
\L{\LB{__\V{FACTOR}_=_(\K{double})(\-\V{PI}\1\V{M});}}
\L{\LB{__\K{if}_(\V{M}_\<_\N{0})_\{}}
\L{\LB{____\V{M}_=_\-\V{M};}}
\L{\LB{__\}}}
\L{\LB{__\K{for}_(\V{K}_=_\N{0};_\V{K}_\<_\V{M};_\V{K}++)_\{}}
\L{\LB{____\V{W}[\V{K}].\V{RE}_=_\V{cos}(\V{K}\*\V{FACTOR});}}
\L{\LB{____\V{W}[\V{K}].\V{IM}_=_\V{sin}(\V{K}\*\V{FACTOR});}}
\L{\LB{__\}}}
\L{\LB{\}}}
\L{\LB{}}
\L{\LB{\C{}\1\*_fftproduct:_Product_of_sparse_matrices_for_DFT_\*\1\CE{}_}}
\L{\LB{}}
\L{\LB{\Proc{fftproduct}\K{void}_\V{fftproduct}(\V{complex}_\*\V{F},_\K{int}_\V{Q},_\V{complex}_\*\V{W})}}
\L{\LB{\{}}
\L{\LB{__\K{int}_\V{K},_\V{N},_\V{J},_\V{M},_\V{N1},_\V{B};}}
\L{\LB{__}}
\L{\LB{__\V{complex}_\V{TMP};}}
\L{\LB{__}}
\L{\LB{__\V{N}_=_\N{1}\<\<\V{Q};}}
\L{\LB{__\V{K}_=_\V{Q};}}
\L{\LB{__\K{while}_(\V{K}_\>_\N{0})_\{}}
\L{\LB{____\V{K}_\-=_\N{1};}}
\L{\LB{____\V{N1}_=_\V{N}\>\>\V{K};____\C{}\1\*_Block_size_\*\1\CE{}}}
\L{\LB{____\V{M}_=_\V{N1}\1\N{2};_____\C{}\1\*_Butterfly_size_\*\1\CE{}}}
\L{\LB{____\V{B}_=_\N{0};}}
\L{\LB{____\K{while}_(\V{B}_\<_\V{N})_\{}}
\L{\LB{______\V{TMP}.\V{RE}_=_\V{F}[\V{B}+\V{M}].\V{RE};}}
\L{\LB{______\V{TMP}.\V{IM}_=_\V{F}[\V{B}+\V{M}].\V{IM};}}
\L{\LB{______\V{F}[\V{B}+\V{M}].\V{RE}_=_\V{F}[\V{B}].\V{RE}_\-_\V{TMP}.\V{RE};}}
\L{\LB{______\V{F}[\V{B}+\V{M}].\V{IM}_=_\V{F}[\V{B}].\V{IM}_\-_\V{TMP}.\V{IM};}}
\L{\LB{______\V{F}[\V{B}].\V{RE}_+=_\V{TMP}.\V{RE};}}
\L{\LB{______\V{F}[\V{B}].\V{IM}_+=_\V{TMP}.\V{IM};}}
\L{\LB{______}}
\L{\LB{______\K{for}_(\V{J}_=_\N{1};_\V{J}_\<_\V{M};_\V{J}++)_\{}}
\L{\LB{}\Tab{8}{\V{TMP}.\V{RE}_=_\V{CCMULRE}(\V{F}[\V{B}+\V{M}+\V{J}],_\V{W}[\V{J}\*(\V{N}\1\V{N1})]);}}
\L{\LB{}\Tab{8}{\V{TMP}.\V{IM}_=_\V{CCMULIM}(\V{F}[\V{B}+\V{M}+\V{J}],_\V{W}[\V{J}\*(\V{N}\1\V{N1})]);}}
\L{\LB{}\Tab{8}{\V{F}[\V{B}+\V{M}+\V{J}].\V{RE}_=_\V{F}[\V{B}+\V{J}].\V{RE}_\-_\V{TMP}.\V{RE};}}
\L{\LB{}\Tab{8}{\V{F}[\V{B}+\V{M}+\V{J}].\V{IM}_=_\V{F}[\V{B}+\V{J}].\V{IM}_\-_\V{TMP}.\V{IM};}}
\L{\LB{}\Tab{8}{\V{F}[\V{B}+\V{J}].\V{RE}_+=_\V{TMP}.\V{RE};}}
\L{\LB{}\Tab{8}{\V{F}[\V{B}+\V{J}].\V{IM}_+=_\V{TMP}.\V{IM};}}
\L{\LB{______\}}}
\L{\LB{______\V{B}_+=_\V{N1};}}
\L{\LB{____\}}}
\L{\LB{__\}}}
\L{\LB{\}}}
\L{\LB{}}
\L{\LB{\C{}\1\*_dct4omega:_Compute_table_of_sines_and_cosines_for_}}
\L{\LB{___DCT-IV,_DST-IV_\*\1\CE{}}}
\L{\LB{}}
\L{\LB{\Proc{dct4omega}\K{void}_\V{dct4omega}(\K{double}_\*\V{COS},_\K{double}_\*\V{SIN},_\K{int}_\V{M})}}
\L{\LB{\{}}
\L{\LB{__\K{int}_\V{K};}}
\L{\LB{__}}
\L{\LB{__\K{double}_\V{FACTOR};}}
\L{\LB{__}}
\L{\LB{__\V{FACTOR}_=_\V{PI}\1(\N{2.0}\*\V{M});}}
\L{\LB{__\K{for}_(\V{K}_=_\N{0};_\V{K}_\<_\V{M};_\V{K}++)_\{}}
\L{\LB{____\V{COS}[\V{K}]_=_\V{cos}(\V{K}\*\V{FACTOR});}}
\L{\LB{____\V{SIN}[\V{K}]_=_\V{sin}(\V{K}\*\V{FACTOR});}}
\L{\LB{__\}}}
\L{\LB{\}}}
\L{\LB{}}
\L{\LB{\C{}\1\*_dctnormal:_Normalization_for_unitary_L-point_DCT_}}
\L{\LB{___and_DST_(T_is_log_L)_\*\1\CE{}}}
\L{\LB{}}
\L{\LB{\Proc{dctnormal}\K{void}_\V{dctnormal}(\K{double}_\*\V{Z},_\K{int}_\V{L},_\K{int}_\V{T})}}
\L{\LB{\{}}
\L{\LB{__\K{double}_\V{NORM};}}
\L{\LB{__}}
\L{\LB{__\K{int}_\V{K};}}
\L{\LB{__}}
\L{\LB{__\K{if}_((\V{T}\%\N{2})_!=_\N{0})}}
\L{\LB{____\V{NORM}_=_\N{0.5}\1(\N{1}\<\<((\V{T}+\N{1})\1\N{2}))_;}}
\L{\LB{__\K{else}}}
\L{\LB{____\V{NORM}_=_\N{0.5}\1\V{sqrt}(\N{1}\<\<(\V{T}+\N{1}));}}
\L{\LB{__\K{for}_(\V{K}_=_\N{0};_\V{K}_\<_\V{L};_\V{K}++)_\{}}
\L{\LB{____\V{Z}[\V{K}]_\*=_\V{NORM};}}
\L{\LB{__\}}}
\L{\LB{\}}}
\L{\LB{}}
\L{\LB{\C{}\1\*_dct4:_(in_place)_Unitary_DCT-IV._}}
\L{\LB{___OBS!_Q_is_log_of_vectorlength_\*\1\CE{}}}
\L{\LB{_}}
\L{\LB{\Proc{dct4}\K{void}_\V{dct4}(\K{double}_\*\V{X},_\K{int}_\V{Q})_}}
\L{\LB{\{}}
\L{\LB{__\K{int}_\V{N},_\V{K};}}
\L{\LB{__}}
\L{\LB{__\V{complex}_\*\V{F},_\*\V{W},_\V{U},_\V{TMP};}}
\L{\LB{__}}
\L{\LB{__\K{double}_\*\V{C},_\*\V{S},_\*\V{Y};}}
\L{\LB{__}}
\L{\LB{__\V{N}_=_\N{1}\<\<\V{Q};___}}
\L{\LB{__\V{F}_=_(\V{complex}_\*)\V{calloc}(\N{2}\*\V{N},_\K{sizeof}_(\V{complex}));}}
\L{\LB{__\V{W}_=_(\V{complex}_\*)\V{calloc}(\V{N},_\K{sizeof}_(\V{complex}));}}
\L{\LB{__\V{C}_=_(\K{double}_\*)\V{calloc}(\V{N},_\K{sizeof}_(\K{double}));}}
\L{\LB{__\V{S}_=_(\K{double}_\*)\V{calloc}(\V{N},_\K{sizeof}_(\K{double}));_}}
\L{\LB{__\C{}\1\*_Y_=_(double_\*)calloc(N,_sizeof_(double));_\*\1\CE{}}}
\L{\LB{__}}
\L{\LB{__\V{dct4omega}(\V{C},_\V{S},_\V{N});}}
\L{\LB{__\V{F}[\N{0}].\V{RE}_=_\V{X}[\N{0}];}}
\L{\LB{__\V{F}[\V{N}].\V{IM}_=_\V{X}[\V{N}\-\N{1}];}}
\L{\LB{__}}
\L{\LB{__\K{for}_(\V{K}_=_\N{1};_\V{K}_\<_\V{N};_\V{K}++)_\{}}
\L{\LB{____\V{F}[\V{K}].\V{RE}_=_\V{X}[\V{K}]_\*_\V{C}[\V{K}];}}
\L{\LB{____\V{F}[\V{K}].\V{IM}_=_\-\V{X}[\V{K}]_\*_\V{S}[\V{K}];}}
\L{\LB{____\V{F}[\N{2}\*\V{N}\-\V{K}].\V{RE}_=_\V{X}[\V{K}\-\N{1}]\*\V{C}[\V{K}];}}
\L{\LB{____\V{F}[\N{2}\*\V{N}\-\V{K}].\V{IM}_=_\V{X}[\V{K}\-\N{1}]\*\V{S}[\V{K}];_}}
\L{\LB{__\}}}
\L{\LB{__\V{bitrevi}(\V{F},_\V{Q}+\N{1});}}
\L{\LB{__\V{fftomega}(\V{W},_\V{N});}}
\L{\LB{__\V{fftproduct}(\V{F},_\V{Q}+\N{1},_\V{W});}}
\L{\LB{__\V{U}.\V{RE}_=_\V{cos}(\-\V{PI}\1(\N{4.0}\*\V{N}));}}
\L{\LB{__\V{U}.\V{IM}_=_\V{sin}(\-\V{PI}\1(\N{4.0}\*\V{N}));}}
\L{\LB{__\V{TMP}.\V{RE}_=_\V{F}[\N{0}].\V{RE}_+_\V{CRRMULRE}(\V{F}[\N{2}\*\V{N}\-\N{1}],_\V{C}[\N{1}],_\V{S}[\N{1}]);}}
\L{\LB{__\V{TMP}.\V{IM}_=_\V{F}[\N{0}].\V{IM}_+_\V{CRRMULIM}(\V{F}[\N{2}\*\V{N}\-\N{1}],_\V{C}[\N{1}],_\V{S}[\N{1}]);}}
\L{\LB{__\V{X}[\N{0}]_=_\V{CCMULRE}(\V{TMP},_\V{U});}}
\L{\LB{__\V{TMP}.\V{RE}_=_\V{CRRMULRE}(\V{F}[\V{N}\-\N{1}],_\V{C}[\V{N}\-\N{1}],_\-\V{S}[\V{N}\-\N{1}])_\-_\V{F}[\V{N}].\V{IM};}}
\L{\LB{__\V{TMP}.\V{IM}_=_\V{CRRMULIM}(\V{F}[\V{N}\-\N{1}],_\V{C}[\V{N}\-\N{1}],_\-\V{S}[\V{N}\-\N{1}])_+_\V{F}[\V{N}].\V{RE};}}
\L{\LB{__\V{X}[\V{N}\-\N{1}]_=_\V{CCMULRE}(\V{TMP},_\V{U});}}
\L{\LB{__}}
\L{\LB{__\K{for}_(\V{K}_=_\N{1};_\V{K}_\<_(\V{N}\-\N{1});_\V{K}++)_\{}}
\L{\LB{____\V{TMP}.\V{RE}_=_\V{CRRMULRE}(\V{F}[\V{K}],_\V{C}[\V{K}],_\-\V{S}[\V{K}])_+}}
\L{\LB{______\V{CRRMULRE}(\V{F}[\N{2}\*\V{N}\-\V{K}\-\N{1}],_\V{C}[\V{K}+\N{1}],_\V{S}[\V{K}+\N{1}]);}}
\L{\LB{____\V{TMP}.\V{IM}_=_\V{CRRMULIM}(\V{F}[\V{K}],_\V{C}[\V{K}],_\-\V{S}[\V{K}])_+}}
\L{\LB{______\V{CRRMULIM}(\V{F}[\N{2}\*\V{N}\-\V{K}\-\N{1}],_\V{C}[\V{K}+\N{1}],_\V{S}[\V{K}+\N{1}]);}}
\L{\LB{____\V{X}[\V{K}]_=_\V{CCMULRE}(\V{TMP},_\V{U});}}
\L{\LB{__\}}}
\L{\LB{__\V{free}(\V{W});}}
\L{\LB{__\V{free}(\V{C});}}
\L{\LB{__\V{free}(\V{S});}}
\L{\LB{__\V{free}(\V{F});}}
\L{\LB{__\V{dctnormal}(\V{X},_\V{N},_\V{Q});_}}
\L{\LB{__\C{}\1\*_return(Y);_\*\1\CE{}}}
\L{\LB{}}
\L{\LB{\}}}
\L{\LB{}}
\L{\LB{\C{}\1\*_rcfis:_Iterated_sine_rising_cutoff_function_\*\1\CE{}}}
\L{\LB{}}
\L{\LB{\Proc{rcfis}\K{double}_\V{rcfis}(\K{int}_\V{N},_\K{double}_\V{T})}}
\L{\LB{\{}}
\L{\LB{__\K{int}_\V{I};}}
\L{\LB{__}}
\L{\LB{__\K{if}_(\V{T}_\>_\-\N{1.0})_\{}}
\L{\LB{____\K{if}_(\V{T}_\<_\N{1.0})_\{}}
\L{\LB{______\K{for}_(\V{I}_=_\N{0};_\V{I}_\<_\V{N};_\V{I}++)_\{}}
\L{\LB{}\Tab{8}{\V{T}_=_\V{sin}(\N{0.5}\*\V{PI}\*\V{T});}}
\L{\LB{______\}}}
\L{\LB{______\V{T}_=_\V{sin}(\N{0.25}\*\V{PI}\*(\N{1.0}_+_\V{T}));}}
\L{\LB{____\}}}
\L{\LB{____\K{else}}}
\L{\LB{______\V{T}_=_\N{1.0};}}
\L{\LB{__\}}}
\L{\LB{__\K{else}}}
\L{\LB{____\V{T}_=_\N{0.0};}}
\L{\LB{__\K{return}_\V{T};}}
\L{\LB{\}}}
\L{\LB{}}
\L{\LB{\C{}\1\*_rcfmidp:_Rising_cutoff_function_sampled_between_gridpoints_\*\1\CE{}}}
\L{\LB{}}
\L{\LB{\Proc{rcfmidp}\K{void}_\V{rcfmidp}(\K{struct}_\V{interval}_\V{R})}}
\L{\LB{\{}}
\L{\LB{__\K{int}_\V{J};}}
\L{\LB{__}}
\L{\LB{__\K{double}_\V{X},_\V{DX};}}
\L{\LB{__}}
\L{\LB{__\V{X}_=_\N{0.5}\1(\V{R}.\V{FINAL}_+_\N{1.0});}}
\L{\LB{__\V{DX}_=_\N{1.0}\1(\V{R}.\V{FINAL}_+_\N{1.0});}}
\L{\LB{__\K{for}_(\V{J}_=_\N{0};_\V{J}_\<=_\V{R}.\V{FINAL};_\V{J}++)_\{}}
\L{\LB{____\V{R}.\V{ORIGIN}[\V{J}]_=_\V{rcf}(\V{X});}}
\L{\LB{____\V{R}.\V{ORIGIN}[\-\V{J}\-\N{1}]_=_\V{rcf}(\-\V{X});}}
\L{\LB{____\V{X}_+=_\V{DX};}}
\L{\LB{__\}}}
\L{\LB{\}}}
\L{\LB{}}
\L{\LB{\C{}\1\*_fdcn:_(midpoint)_Fold_disjoint_cosine_negative_\*\1\CE{}}}
\L{\LB{}}
\L{\LB{\Proc{fdcn}\K{void}_\V{fdcn}(\K{double}_\*\V{ONEG},_\K{int}_\V{STEP},_\K{double}_\*\V{INEG},_}}
\L{\LB{}\Tab{8}{__\K{double}_\*\V{IPOS},_\K{int}_\V{N},_\K{struct}_\V{interval}_\V{RISE})}}
\L{\LB{\{}}
\L{\LB{__\K{int}_\V{K};}}
\L{\LB{__}}
\L{\LB{__\K{for}_(\V{K}_=_\-\V{N};_\V{K}_\<=_(\V{RISE}.\V{LEAST}_\-_\N{1});_\V{K}++)_\{}}
\L{\LB{____\*(\V{ONEG}+\V{K}\*\V{STEP})_=_\*(\V{INEG}+\V{K});}}
\L{\LB{__\}}}
\L{\LB{__\K{for}_(\V{K}_=_\V{RISE}.\V{LEAST};_\V{K}_\<=_\-\N{1};_\V{K}++)_\{}}
\L{\LB{____\*(\V{ONEG}+\V{K}\*\V{STEP})_=_\V{RISE}.\V{ORIGIN}[\-\N{1}\-\V{K}]\*(\*(\V{INEG}+\V{K}))_\-}}
\L{\LB{______\V{RISE}.\V{ORIGIN}[\V{K}]\*(\*(\V{IPOS}\-\N{1}\-\V{K}));}}
\L{\LB{__\}}}
\L{\LB{\}}}
\L{\LB{}}
\L{\LB{}}
\L{\LB{\C{}\1\*_fdcp:_(midpoint)_Fold_disjoint_cosine_positive_\*\1\CE{}}}
\L{\LB{}}
\L{\LB{\Proc{fdcp}\K{void}_\V{fdcp}(\K{double}_\*\V{OPOS},_\K{int}_\V{STEP},_\K{double}_\*\V{INEG},_}}
\L{\LB{}\Tab{8}{__\K{double}_\*\V{IPOS},_\K{int}_\V{N},_\K{struct}_\V{interval}_\V{RISE})}}
\L{\LB{\{}}
\L{\LB{__\K{int}_\V{K};}}
\L{\LB{__}}
\L{\LB{__\K{for}_(\V{K}_=_\N{0};_\V{K}_\<=_\V{RISE}.\V{FINAL};_\V{K}++)_\{}}
\L{\LB{____\*(\V{OPOS}+\V{K}\*\V{STEP})_=_\V{RISE}.\V{ORIGIN}[\V{K}]\*(\*(\V{IPOS}+\V{K}))_+}}
\L{\LB{______\V{RISE}.\V{ORIGIN}[\-\N{1}\-\V{K}]\*(\*(\V{INEG}\-\N{1}\-\V{K}));}}
\L{\LB{__\}}}
\L{\LB{__\K{for}_(\V{K}_=_(\V{RISE}.\V{FINAL}_+_\N{1});_\V{K}_\<=_(\V{N}\-\N{1});_\V{K}++)_\{}}
\L{\LB{____\*(\V{OPOS}+\V{K}\*\V{STEP})_=_\*(\V{IPOS}+\V{K});}}
\L{\LB{__\}}}
\L{\LB{\}}}
\L{\LB{__}}
\L{\LB{\C{}\1\*_fdsn:_(midpoint)_Fold_disjoint_sine_negative_\*\1\CE{}}}
\L{\LB{}}
\L{\LB{\Proc{fdsn}\K{void}_\V{fdsn}(\K{double}_\*\V{ONEG},_\K{int}_\V{STEP},_\K{double}_\*\V{INEG},_}}
\L{\LB{}\Tab{8}{__\K{double}_\*\V{IPOS},_\K{int}_\V{N},_\K{struct}_\V{interval}_\V{RISE})}}
\L{\LB{\{}}
\L{\LB{__\K{int}_\V{K};}}
\L{\LB{__}}
\L{\LB{__\K{for}_(\V{K}_=_\-\V{N};_\V{K}_\<=_(\V{RISE}.\V{LEAST}_\-_\N{1});_\V{K}++)_\{}}
\L{\LB{____\*(\V{ONEG}+\V{K}\*\V{STEP})_=_\*(\V{INEG}+\V{K});}}
\L{\LB{__\}}}
\L{\LB{__\K{for}_(\V{K}_=_\V{RISE}.\V{LEAST};_\V{K}_\<=_\-\N{1};_\V{K}++)_\{}}
\L{\LB{____\*(\V{ONEG}+\V{K}\*\V{STEP})_=_\V{RISE}.\V{ORIGIN}[\-\N{1}\-\V{K}]\*(\*(\V{INEG}+\V{K}))_+}}
\L{\LB{______\V{RISE}.\V{ORIGIN}[\V{K}]\*(\*(\V{IPOS}\-\N{1}\-\V{K}));}}
\L{\LB{__\}}}
\L{\LB{\}}}
\L{\LB{}}
\L{\LB{\C{}\1\*_fdsp:_(midpoint)_Fold_disjoint_sine_positive_\*\1\CE{}}}
\L{\LB{}}
\L{\LB{\Proc{fdsp}\K{void}_\V{fdsp}(\K{double}_\*\V{OPOS},_\K{int}_\V{STEP},_\K{double}_\*\V{INEG},_}}
\L{\LB{}\Tab{8}{___\K{double}_\*\V{IPOS},_\K{int}_\V{N},_\K{struct}_\V{interval}_\V{RISE})}}
\L{\LB{\{}}
\L{\LB{__\K{int}_\V{K};}}
\L{\LB{__}}
\L{\LB{__\K{for}_(\V{K}_=_\N{0};_\V{K}_\<=_\V{RISE}.\V{FINAL};_\V{K}++)_\{}}
\L{\LB{____\*(\V{OPOS}+\V{K}\*\V{STEP})_=_\V{RISE}.\V{ORIGIN}[\V{K}]\*(\*(\V{IPOS}+\V{K}))_\-}}
\L{\LB{______\V{RISE}.\V{ORIGIN}[\-\N{1}\-\V{K}]\*(\*(\V{INEG}\-\N{1}\-\V{K}));}}
\L{\LB{__\}}}
\L{\LB{__\K{for}_(\V{K}_=_(\V{RISE}.\V{FINAL}_+_\N{1});_\V{K}_\<=_(\V{N}\-\N{1});_\V{K}++)_\{}}
\L{\LB{____\*(\V{OPOS}+\V{K}\*\V{STEP})_=_\*(\V{IPOS}+\V{K});}}
\L{\LB{__\}}}
\L{\LB{\}}}
\L{\LB{}}
\L{\LB{\C{}\1\*_localcosine:_(in_place)_(adapted_dyadic)_local_cosine_analysis_}}
\L{\LB{___(with)_fixed_folding_applying_a_fixed_cutoff_function_\3RISE\3_}}
\L{\LB{___sampled_between_gridpoints_resulting_in_L+1_different_}}
\L{\LB{___representations_of_the_signal.}}
\L{\LB{___}}
\L{\LB{___We_have:}}
\L{\LB{}}
\L{\LB{___N_=_The_length_of_the_signal_to_be_analysed.}}
\L{\LB{}}
\L{\LB{___L_=_The_depth_of_expansion_in_the_dictionary_of_bases.}}
\L{\LB{_}}
\L{\LB{___PARENT_=_Pointer_to_array_of_length_N\*(L+1)_containing_}}
\L{\LB{___the_signal_to_be_analysed_in_its_first_N_locations._\*\1\CE{}___}}
\L{\LB{_}}
\L{\LB{\Proc{localcosine}\K{void}_\V{localcosine}(\K{double}_\*\V{PARENT},_\K{int}_\V{N},_\K{int}_\V{L})}}
\L{\LB{\{}}
\L{\LB{__\K{int}_\V{NP},_\V{NC},_\V{LEVEL},_\V{PBLOCK},_\V{RADIUS};}}
\L{\LB{__}}
\L{\LB{__\K{double}_\*\V{MIDP},_\*\V{CHILD};}}
\L{\LB{__}}
\L{\LB{__\K{struct}_\V{interval}_\V{RISE};}}
\L{\LB{__}}
\L{\LB{__\V{RADIUS}_=_(\V{N}\>\>(\V{L}+\N{1}))\-\N{1};}}
\L{\LB{__\K{if}_(\V{RADIUS}_\<=_\N{0})}}
\L{\LB{____\V{printf}(\S{}\3Negative_folding_radius_in_lcaff\2n\3\SE{});}}
\L{\LB{_}}
\L{\LB{__\V{RISE}_=_\V{makeinterval}(\V{NULL},\-\V{RADIUS},_\V{RADIUS}\-\N{1});}}
\L{\LB{__\V{rcfmidp}(\V{RISE});}}
\L{\LB{__\V{NP}_=_\V{N};}}
\L{\LB{__}}
\L{\LB{__\K{for}_(\V{LEVEL}_=_\N{0};_\V{LEVEL}_\<_\V{L};_\V{LEVEL}++)_\{__}}
\L{\LB{____\V{NC}_=_\V{NP}\1\N{2};}}
\L{\LB{____}}
\L{\LB{____\K{for}_(\V{PBLOCK}_=_\N{1};_\V{PBLOCK}_\<=_(\N{1}\<\<\V{LEVEL});_\V{PBLOCK}++)_\{}}
\L{\LB{______\V{MIDP}_=_\V{PARENT}_+_\V{NC};}}
\L{\LB{______\V{CHILD}_=_\V{MIDP}_+_\V{N};}}
\L{\LB{______\V{fdcn}(\V{CHILD},_\N{1},_\V{MIDP},_\V{MIDP},_\V{NC},_\V{RISE});}}
\L{\LB{______\V{fdcp}(\V{CHILD},_\N{1},_\V{MIDP},_\V{MIDP},_\V{NC},_\V{RISE});}}
\L{\LB{______\V{dct4}(\V{PARENT},_\V{intlog2}(\V{NP}));}}
\L{\LB{______\V{PARENT}_+=_\V{NP};}}
\L{\LB{____\}}}
\L{\LB{____\V{NP}_=_\V{NC};}}
\L{\LB{__\}}}
\L{\LB{__\K{for}_(\V{PBLOCK}_=_\N{1};_\V{PBLOCK}_\<=_(\N{1}\<\<\V{L});_\V{PBLOCK}++)_\{_}}
\L{\LB{____\V{dct4}(\V{PARENT},_\V{intlog2}(\V{NP}));}}
\L{\LB{____\V{PARENT}_+=_\V{NP};}}
\L{\LB{__\}}}
\L{\LB{__\V{freeinterval}(\&\V{RISE});}}
\L{\LB{}}
\L{\LB{\}}}
\L{\LB{}}
\L{\LB{}}
\L{\LB{}}
\L{\LB{}}
\L{\LB{}}
\L{\LB{}}
\L{\LB{}}

\File{wickprograms.c}{1997}{Oct}{30}{23:13}{9487}

%% file: myprograms.tex

\File{myprograms.c}{1997}{Nov}{15}{18:54}{26126}
\L{\LB{\K{\#include}_\<\V{stdio}.\V{h}\>}}
\L{\LB{\K{\#include}_\<\V{math}.\V{h}\>}}
\L{\LB{\K{\#include}_\<\V{stdlib}.\V{h}\>}}
\L{\LB{\K{\#include}_\<\V{time}.\V{h}\>}}
\L{\LB{\K{\#include}_\<\V{assert}.\V{h}\>}}
\L{\LB{}}
\L{\LB{\C{}\1\*_basisnode:_Define_a_new_data_structure_of_type_basisnode}}
\L{\LB{___containing_five_members:}}
\L{\LB{___}}
\L{\LB{___LEVEL_=_The_level_coordinate_of_a_subspace_in_the_dictionary_tree.}}
\L{\LB{___BLOCK_=_The_position_of_a_subspace_counted_from_the_left.}}
\L{\LB{___POS_=_The_coordinate_of_a_basisfunction_in_the_subspace.}}
\L{\LB{___COST_=_The_cost\1discrimination_number.}}
\L{\LB{___TAG_=_YES_or_NO._\*\1\CE{}}}
\L{\LB{__}}
\L{\LB{\K{struct}_\V{basisnode}_\{}}
\L{\LB{__\K{int}_\V{LEVEL};}}
\L{\LB{__\K{int}_\V{BLOCK};}}
\L{\LB{__\K{int}_\V{POS};}}
\L{\LB{__\K{double}_\V{COST};}}
\L{\LB{__\K{int}_\V{TAG};}}
\L{\LB{\};}}
\L{\LB{}}
\L{\LB{\K{\#define}__\V{YES}_____________________________\N{1}}}
\L{\LB{\K{\#define}__\V{NO}______________________________\N{0}}}
\L{\LB{\K{\#define}__\V{PI}______________________________\N{3.1415926535}}}
\L{\LB{\K{\#define}__\V{SQ2}_____________________________\N{1.4142135623}}}
\L{\LB{\K{\#define}__\V{SQH}_____________________________\N{0.7071067811}}}
\L{\LB{\K{\#define}__\V{MAXINT}__________________________\N{2147483647}}}
\L{\LB{\K{\#define}__\V{DIVISOR}_________________________\N{127773}}}
\L{\LB{\K{\#define}__\V{HIGH}____________________________\N{2836}}}
\L{\LB{\K{\#define}__\V{LOW}_____________________________\N{16807}}}
\L{\LB{\K{\#define}__\V{SEEDFACTOR}______________________\N{76928675}}}
\L{\LB{\K{\#define}__\V{CLASS\_1}_________________________\N{10}}}
\L{\LB{\K{\#define}__\V{CLASS\_2}_________________________\N{20}}}
\L{\LB{\K{\#define}__\V{K\_1}_____________________________\V{pow}(\N{10.0},\N{2.0})}}
\L{\LB{\K{\#define}__\V{K\_2}_____________________________\V{pow}(\N{10.0},\N{2.0})}}
\L{\LB{\K{\#define}__\V{A\_1}_____________________________\N{0.100}}}
\L{\LB{\K{\#define}__\V{A\_2}_____________________________\N{0.050}}}
\L{\LB{\K{\#define}__\V{RDIST}___________________________\N{10000}}}
\L{\LB{\K{\#define}__\V{expand}__________________________\V{localcosine}}}
\L{\LB{\K{\#define}__\V{s\_measure}(\V{IN\_1},\V{IN\_2},\V{N})__________\V{lp\_norm\_p}(\V{IN\_1},\V{IN\_2},\V{N},\N{2.0})_}}
\L{\LB{\K{\#define}__\V{makesignal}(\V{IN},\V{N},\V{K},\V{A},\V{M})__________\V{make\_samples\_of\_cosine\_signal}(}}
\L{\LB{\V{IN},\V{N},\V{K},\V{A},\V{RDIST},\N{16},\V{M},\N{1.0}\1\N{8.0},\V{PI}\1\N{4})}}
\L{\LB{}}
\L{\LB{\C{}\1\*_Allocate_a_basisnode_data_structure_an_assign_its_content_\*\1\CE{}}}
\L{\LB{}}
\L{\LB{\Proc{makebasisnode}\K{struct}_\V{basisnode}_\*\V{makebasisnode}(\K{int}_\V{LEVEL},_\K{int}_\V{BLOCK},_}}
\L{\LB{}\Tab{24}{_______\K{int}_\V{POS},_\K{double}_\V{COST},_\K{int}_\V{TAG})}}
\L{\LB{}\Tab{24}{}}
\L{\LB{\{}}
\L{\LB{__\K{struct}_\V{basisnode}_\*\V{BASISNODE};}}
\L{\LB{__}}
\L{\LB{__\V{BASISNODE}_=_(\K{struct}_\V{basisnode}_\*)\V{malloc}(\K{sizeof}(\K{struct}_\V{basisnode}));}}
\L{\LB{_}}
\L{\LB{__\V{BASISNODE}\-\!\>\V{LEVEL}_=_\V{LEVEL};}}
\L{\LB{__\V{BASISNODE}\-\!\>\V{BLOCK}_=_\V{BLOCK};}}
\L{\LB{__\V{BASISNODE}\-\!\>\V{POS}_=_\V{POS};}}
\L{\LB{__\V{BASISNODE}\-\!\>\V{COST}_=_\V{COST};}}
\L{\LB{__\V{BASISNODE}\-\!\>\V{TAG}_=_\V{TAG};}}
\L{\LB{}}
\L{\LB{__\K{return}(\V{BASISNODE});}}
\L{\LB{\}}}
\L{\LB{}}
\L{\LB{\C{}\1\*_Deallocate_an_array_of_length_N_of_pointers_}}
\L{\LB{___to_basisnode_data_structures_\*\1\CE{}}}
\L{\LB{}}
\L{\LB{\Proc{freebasisnodes}\K{void}_\V{freebasisnodes}(\K{struct}_\V{basisnode}_\*\*\V{A},_\K{int}_\V{N})}}
\L{\LB{\{}}
\L{\LB{__\K{int}_\V{K};}}
\L{\LB{__}}
\L{\LB{__\K{for}_(\V{K}_=_\N{0};_\V{K}_\<_\V{N};_\V{K}++)_\{}}
\L{\LB{____\V{free}(\V{A}[\V{K}]);}}
\L{\LB{__\}}}
\L{\LB{__\V{free}(\V{A});}}
\L{\LB{\}}}
\L{\LB{}}
\L{\LB{\C{}\1\*_ladder:_(disjoint)_filter_a_real_array_of_length_N_into_}}
\L{\LB{___lowpass_and_highpass-coefficients_by_the_\3rotation\3_algorithm_}}
\L{\LB{___applying_an_orthonormal_filter_of_length_2\*L,_periodizing}}
\L{\LB{___at_endpoints.__}}
\L{\LB{}}
\L{\LB{___We_have:}}
\L{\LB{___}}
\L{\LB{___IN_=_Pointer_to_array_of_numbers_to_be_filtered.}}
\L{\LB{___}}
\L{\LB{___OUT_=_Pointer_to_array_that_results_from_filtering_IN.}}
\L{\LB{}}
\L{\LB{___ALPHA_=_Pointer_to_array_of_\3rotation\3_coefficients_including_}}
\L{\LB{___a_normalization_factor.}}
\L{\LB{}}
\L{\LB{___N_=_The_length_of_IN_=_The_length_of_OUT.}}
\L{\LB{}}
\L{\LB{___L_=_Half_the_length_of_the_filter._\*\1\CE{}}}
\L{\LB{}}
\L{\LB{\Proc{ladder}\K{void}_\V{ladder}(\K{double}_\*\V{OUT},_\K{double}_\*\V{IN},_}}
\L{\LB{}\Tab{8}{____\K{double}_\*\V{ALPHA},_\K{int}_\V{N},_\K{int}_\V{L})}}
\L{\LB{\{}}
\L{\LB{__\K{int}_\V{LEVEL},_\V{TRANSLATION},_\V{K};}}
\L{\LB{__}}
\L{\LB{__\K{double}_\*\V{WORK};}}
\L{\LB{___}}
\L{\LB{__\V{WORK}_=_(\K{double}_\*)\V{calloc}(\V{N},_\K{sizeof}(\K{double}));}}
\L{\LB{__\K{for}_(\V{K}_=_\N{0};_\V{K}_\<_\V{N};_\V{K}++)_}}
\L{\LB{____\V{WORK}[\V{K}]_=_\V{IN}[\V{K}];}}
\L{\LB{__}}
\L{\LB{__\K{for}_(\V{LEVEL}_=_\N{0};_\V{LEVEL}_\<_\V{L};_\V{LEVEL}++)_\{}}
\L{\LB{____\V{TRANSLATION}_=_(\V{LEVEL}\%\N{2});}}
\L{\LB{____}}
\L{\LB{____\K{for}_(\V{K}_=_\N{0};_\V{K}_\<_\V{N}\1\N{2};_\V{K}++)_\{}}
\L{\LB{______\*(\V{OUT}+\V{TRANSLATION}_+_\N{2}\*\V{K})_=}}
\L{\LB{}\Tab{8}{\V{WORK}[\V{TRANSLATION}_+_\N{2}\*\V{K}]_\-_\V{ALPHA}[\V{LEVEL}]\*}}
\L{\LB{}\Tab{8}{\V{WORK}[(\V{TRANSLATION}_+_\N{1}_+_\N{2}\*\V{K})\%\V{N}];}}
\L{\LB{______\*(\V{OUT}+(\V{TRANSLATION}_+_\N{1}_+_\N{2}\*\V{K})\%\V{N})_=}}
\L{\LB{}\Tab{8}{\V{ALPHA}[\V{LEVEL}]\*\V{WORK}[\V{TRANSLATION}_+_\N{2}\*\V{K}]_+}}
\L{\LB{}\Tab{8}{\V{WORK}[(\V{TRANSLATION}_+_\N{1}_+_\N{2}\*\V{K})\%\V{N}];}}
\L{\LB{______}}
\L{\LB{______\V{WORK}[\V{TRANSLATION}_+_\N{2}\*\V{K}]_=_}}
\L{\LB{}\Tab{8}{\*(\V{OUT}+\V{TRANSLATION}_+_\N{2}\*\V{K});}}
\L{\LB{______\V{WORK}[(\V{TRANSLATION}_+_\N{1}_+_\N{2}\*\V{K})\%\V{N}]_=_}}
\L{\LB{}\Tab{8}{\*(\V{OUT}+(\V{TRANSLATION}_+_\N{1}_+_\N{2}\*\V{K})\%\V{N});}}
\L{\LB{____\}}}
\L{\LB{__\}}}
\L{\LB{__\K{for}_(\V{K}_=_\N{0};_\V{K}_\<_\V{N}\1\N{2};_\V{K}++)_\{}}
\L{\LB{____}}
\L{\LB{____\C{}\1\*_Store_the_lowpass_and_highpass_coefficients_in_}}
\L{\LB{_______separate_blocks._\*\1\CE{}}}
\L{\LB{____}}
\L{\LB{____\K{if}_(\V{L}\%\N{2}_==_\N{0})_\{_}}
\L{\LB{______\*(\V{OUT}+\V{K})_=_\V{WORK}[\N{2}\*\V{K}_+\N{1}]\*\V{ALPHA}[\V{L}];_______\C{}\1\*_c\4s_\*\1\CE{}}}
\L{\LB{______\*(\V{OUT}+\V{K}+\V{N}\1\N{2})_=_\V{WORK}[\N{2}\*\V{K}]\*\V{ALPHA}[\V{L}];______\C{}\1\*_d\4s_\*\1\CE{}}}
\L{\LB{____\}}}
\L{\LB{____}}
\L{\LB{____\K{else}_\{}}
\L{\LB{______\*(\V{OUT}+\V{K})_=_\V{WORK}[\N{2}\*\V{K}]\*\V{ALPHA}[\V{L}];__________\C{}\1\*_c\4s_\*\1\CE{}}}
\L{\LB{______\*(\V{OUT}+\V{K}+\V{N}\1\N{2})_=_\V{WORK}[\N{2}\*\V{K}+\N{1}]\*\V{ALPHA}[\V{L}];____\C{}\1\*_d\4s_\*\1\CE{}}}
\L{\LB{____\}}}
\L{\LB{__\}}}
\L{\LB{__\V{free}(\V{WORK});}}
\L{\LB{__}}
\L{\LB{\}}}
\L{\LB{}}
\L{\LB{\C{}\1\*_waveletpacket:_(in_place)_waveletpacket_analysis_resulting_in_L+1_}}
\L{\LB{___different_representations_of_the_signal.}}
\L{\LB{___}}
\L{\LB{___We_have:}}
\L{\LB{}}
\L{\LB{___N_=_The_length_of_the_signal_to_be_analysed.}}
\L{\LB{}}
\L{\LB{___L_=_The_depth_of_expansion_in_the_dictionary_of_bases.}}
\L{\LB{}}
\L{\LB{___PARENT_=_Pointer_to_array_of_length_N\*(L+1)_containing_}}
\L{\LB{___the_sampled_lowpass_coefficients_of_the_signal_to_be_analysed_}}
\L{\LB{___in_its_first_N_locations.__}}
\L{\LB{}}
\L{\LB{___ALPHA_=_Pointer_to_array_containing_in_the_following_order:}}
\L{\LB{___}}
\L{\LB{___1._Half_the_length_of_the_orthonormal_filter.}}
\L{\LB{___2._The_rotation_coefficients_corresponding_to_the_filter.}}
\L{\LB{___3._A_scaling_factor._\*\1\CE{}}}
\L{\LB{}}
\L{\LB{\Proc{waveletpacket}\K{void}_\V{waveletpacket}(\K{double}_\*\V{PARENT},_\K{int}_\V{N},_\K{int}_\V{L})}}
\L{\LB{\{_}}
\L{\LB{}}
\L{\LB{__\V{FILE}_\*\V{FILTER};}}
\L{\LB{__}}
\L{\LB{__\K{int}_\V{LEVEL},_\V{BLOCK},_\V{FILEN},_\V{NP},_\V{M},_\V{K};}}
\L{\LB{__}}
\L{\LB{__\K{double}_\*\V{CHILD},_\*\V{ALPHA};_}}
\L{\LB{__}}
\L{\LB{__\K{if}((_\V{FILTER}_=_\V{fopen}(\S{}\3\1users\1stud\1eirikf\1programs\1hovedfag\1}}
\L{\LB{__filters\1laddercoif18\3\SE{},\S{}\3r\3\SE{}))}}
\L{\LB{_____==_\V{NULL})_\{}}
\L{\LB{____}}
\L{\LB{____\V{printf}(\S{}\3ERROR_in_opening_FILTERfile\2n\3\SE{});}}
\L{\LB{____\V{exit}(\N{1});}}
\L{\LB{__\}_}}
\L{\LB{}}
\L{\LB{__\V{fscanf}(\V{FILTER},_\S{}\3\%d\3\SE{},_\&\V{FILEN});}}
\L{\LB{__\V{ALPHA}_=_(\K{double}_\*)\V{calloc}(\V{FILEN}+\N{1},_\K{sizeof}(\K{double}));_}}
\L{\LB{__\K{for}_(\V{K}_=_\N{0};_\V{K}_\<=_\V{FILEN};_\V{K}++)_\{}}
\L{\LB{____\V{fscanf}(\V{FILTER},_\S{}\3\%lf\3\SE{},_\&\V{ALPHA}[\V{K}]);}}
\L{\LB{__\}}}
\L{\LB{__\V{fclose}(\V{FILTER});}}
\L{\LB{__}}
\L{\LB{__\K{for}_(\V{LEVEL}_=_\N{0};_\V{LEVEL}_\<_\V{L};_\V{LEVEL}++)_\{}}
\L{\LB{____\V{NP}_=_\V{N}\>\>\V{LEVEL};}}
\L{\LB{____}}
\L{\LB{____\K{for}_(\V{BLOCK}_=_\N{0};_\V{BLOCK}_\<_(\N{1}\<\<\V{LEVEL});_\V{BLOCK}++)_\{}}
\L{\LB{______\V{CHILD}_=_\V{PARENT}+\V{N};}}
\L{\LB{______\V{ladder}(\V{CHILD},_\V{PARENT},_\V{ALPHA},_\V{NP},_\V{FILEN});}}
\L{\LB{______\V{PARENT}_+=_\V{NP};____________________________________}}
\L{\LB{____\}}}
\L{\LB{__\}}}
\L{\LB{__\V{free}(\V{ALPHA});}}
\L{\LB{__}}
\L{\LB{\}}}
\L{\LB{}}
\L{\LB{\C{}\1\*_my\_qsort:_sort_v[left].\,.\,.v[right]_into_decreasing_order.}}
\L{\LB{___(This_function_is_copied_from_Kernighan\1Ritchie:_}}
\L{\LB{___\3The_Ansi_C_programming_language\3_second_edition_p.120.)_\*\1\CE{}}}
\L{\LB{}}
\L{\LB{\Proc{my\_qsort}\K{void}_\V{my\_qsort}(\K{void}_\*\V{v}[\,],_\K{int}_\V{left},_\K{int}_\V{right},_}}
\L{\LB{\Proc{int}}\Tab{8}{___\K{int}_(\*\V{comp})(\K{void}_\*,_\K{void}_\*))}}
\L{\LB{\{}}
\L{\LB{__\K{int}_\V{i},_\V{last};}}
\L{\LB{__}}
\L{\LB{__\K{void}_\V{my\_swap}(\K{void}_\*\V{v}[\,],_\K{int},_\K{int});}}
\L{\LB{_}}
\L{\LB{__\K{if}_(\V{left}_\>=_\V{right})_}}
\L{\LB{____\K{return};}}
\L{\LB{__}}
\L{\LB{__\V{my\_swap}(\V{v},_\V{left},_(\V{left}_+_\V{right})\1\N{2});}}
\L{\LB{__\V{last}_=_\V{left};}}
\L{\LB{__\K{for}_(\V{i}_=_\V{left}+\N{1};_\V{i}_\<=_\V{right};_\V{i}++)_\{}}
\L{\LB{____\K{if}_((\*\V{comp})(\V{v}[\V{i}],_\V{v}[\V{left}])_\<_\N{0})}}
\L{\LB{______\V{my\_swap}(\V{v},_++\V{last},_\V{i});}}
\L{\LB{__\}}}
\L{\LB{__\V{my\_swap}(\V{v},_\V{left},_\V{last});}}
\L{\LB{__\V{my\_qsort}(\V{v},_\V{left},_\V{last}\-\N{1},_\V{comp});}}
\L{\LB{__\V{my\_qsort}(\V{v},_\V{last}+\N{1},_\V{right},_\V{comp});}}
\L{\LB{\}}}
\ProcCont{my\_qsort}\L{\LB{}}
\L{\LB{\C{}\1\*_my\_swap:_Exchange_pointers_within_an_array_\*\1\CE{}}}
\L{\LB{}}
\L{\LB{\Proc{my\_swap}\K{void}_\V{my\_swap}(\K{void}_\*\V{v}[\,],_\K{int}_\V{i},_\K{int}_\V{j})}}
\L{\LB{\{}}
\L{\LB{__\K{void}_\*\V{temp};}}
\L{\LB{__}}
\L{\LB{__\V{temp}_=_\V{v}[\V{i}];}}
\L{\LB{__\V{v}[\V{i}]_=_\V{v}[\V{j}];}}
\L{\LB{__\V{v}[\V{j}]_=_\V{temp};}}
\L{\LB{\}}}
\ProcCont{my\_qsort}\L{\LB{}}
\L{\LB{\C{}\1\*_comp\_basisnode\_by\_cost:_Compare_the_discrimination_measures_}}
\L{\LB{___of_two_basisnode_data_structures_\*\1\CE{}}}
\L{\LB{}}
\L{\LB{\Proc{comp\_basisnode\_by\_cost}\K{int}_\V{comp\_basisnode\_by\_cost}(\K{struct}_\V{basisnode}_\*\V{X},_}}
\L{\LB{}\Tab{24}{___\K{struct}_\V{basisnode}_\*\V{Y})}}
\L{\LB{\{}}
\L{\LB{__\K{if}_((\V{X}\-\!\>\V{COST})_\>_(\V{Y}\-\!\>\V{COST}))}}
\L{\LB{____\K{return}(\-\N{1});}}
\L{\LB{__\K{else}_\K{if}_((\V{X}\-\!\>\V{COST})_\<_(\V{Y}\-\!\>\V{COST}))}}
\L{\LB{____\K{return}(\N{1});}}
\L{\LB{__\K{else}_}}
\L{\LB{____\K{return}(\N{0});}}
\L{\LB{\}}}
\ProcCont{my\_qsort}\L{\LB{}}
\L{\LB{\C{}\1\*_my\_rand:_Generate_a_pseudo_random_floating_point_}}
\L{\LB{___number_between_0_and_1._\*\1\CE{}}}
\L{\LB{}}
\L{\LB{\Proc{my\_rand}\K{double}_\V{my\_rand}(\K{void})}}
\L{\LB{\{}}
\L{\LB{__\K{int}_\V{high},_\V{low},_\V{temp};_}}
\L{\LB{__}}
\L{\LB{__\K{static}_\K{int}_\V{SEED};}}
\L{\LB{__}}
\L{\LB{__\K{if}_(\V{SEED}_\<_\N{1}_\|\,\|_\V{SEED}_\>_\V{MAXINT})_}}
\L{\LB{____\V{SEED}_=_(_(\K{int})_\V{time}(\V{NULL})\*\V{SEEDFACTOR}_)_\%_\V{MAXINT};}}
\L{\LB{_}}
\L{\LB{__\V{high}_=_\V{SEED}\1\V{DIVISOR};}}
\L{\LB{__\V{low}_=_\V{SEED}_\%_\V{DIVISOR};}}
\L{\LB{__\V{temp}_=_\V{LOW}\*\V{low}_\-_\V{HIGH}\*\V{high};_}}
\L{\LB{__\V{SEED}_=_(\V{temp}_\>_\N{0}_?_\V{temp}_:_\V{temp}_+_\V{MAXINT});}}
\L{\LB{__\K{return}((\K{double})\V{SEED}\1\V{MAXINT});}}
\L{\LB{\}}}
\ProcCont{my\_qsort}\L{\LB{}}
\L{\LB{\C{}\1\*_lp\_norm\_p:_Compute_the_(l\5P_difference)\5P_of_two_arrays_}}
\L{\LB{___of_length_N_\*\1\CE{}}}
\L{\LB{}}
\L{\LB{\Proc{lp\_norm\_p}\K{double}_\V{lp\_norm\_p}(\K{double}_\*\V{IN\_1},_\K{double}_\*\V{IN\_2},_\K{int}_\V{N},_\K{double}_\V{P})}}
\L{\LB{\{}}
\L{\LB{__\K{double}_\V{T},_\V{X};}}
\L{\LB{__}}
\L{\LB{__\K{int}_\V{K};}}
\L{\LB{__}}
\L{\LB{__\V{T}_=_\N{0};}}
\L{\LB{__\K{if}_(\V{IN\_2}_==_\V{NULL})_\{}}
\L{\LB{____\K{for}_(\V{K}_=_\N{0};_\V{K}_\<_\V{N};_\V{K}++)_\{}}
\L{\LB{______\V{X}_=_\V{fabs}(\*(\V{IN\_1}+\V{K}));}}
\L{\LB{______\K{if}_(\V{X}_\>_\N{0})_}}
\L{\LB{}\Tab{8}{\V{T}_+=\V{exp}(\V{P}\*\V{log}(\V{X}));}}
\L{\LB{____\}}}
\L{\LB{__\}}}
\L{\LB{}}
\L{\LB{__\K{else}_\{}}
\L{\LB{____\K{for}_(\V{K}_=_\N{0};_\V{K}_\<_\V{N};_\V{K}++)_\{}}
\L{\LB{______\V{X}_=_\V{fabs}(\*(\V{IN\_1}+\V{K})\-(\*(\V{IN\_2}+\V{K})));}}
\L{\LB{______\K{if}_(_\V{X}_\>_\N{0})}}
\L{\LB{}\Tab{8}{\V{T}_+=_\V{exp}(\V{P}\*\V{log}(\V{X}));}}
\L{\LB{____\}}}
\L{\LB{__\}}}
\L{\LB{__\K{return}_\V{T};}}
\L{\LB{\}}}
\ProcCont{my\_qsort}\L{\LB{}}
\L{\LB{\C{}\1\*_modf\_2PI:_Compute_remainder_of_input_modulo_2\*PI._}}
\L{\LB{___(This_is_because_of_the_bad_accuracy_of_the_}}
\L{\LB{___standard_math_trigonometric_functions_for_large_aruments)_\*\1\CE{}}}
\L{\LB{}}
\L{\LB{\Proc{modf\_2PI}\K{double}_\V{modf\_2PI}(\K{double}_\V{X})}}
\L{\LB{\{}}
\L{\LB{__\K{int}_\V{K};}}
\L{\LB{__}}
\L{\LB{__\K{double}_\V{T},_\V{S};}}
\L{\LB{__}}
\L{\LB{__\V{T}_=_\V{X};}}
\L{\LB{}}
\L{\LB{__\V{S}_=_\N{2.0}\*\V{PI};}}
\L{\LB{__\K{if}_(\V{T}_\>_\N{0})_\{}}
\L{\LB{____\K{while}_(\V{T}_\>_\N{0})}}
\L{\LB{______\V{T}_\-=_\V{S};_}}
\L{\LB{____}}
\L{\LB{____\V{T}_+=_\V{S};}}
\L{\LB{__\}}}
\L{\LB{__\K{else}_\{}}
\L{\LB{____\K{while}_(\V{T}_\<_\N{0})}}
\L{\LB{______\V{T}_+=_\V{S};_}}
\L{\LB{__\}}}
\L{\LB{__\K{return}(\V{T});}}
\L{\LB{}}
\L{\LB{\}}}
\ProcCont{my\_qsort}\L{\LB{}}
\L{\LB{\C{}\1\*_set\_tag\_to\_zero:_Set_the_tags_of_all_children_nodes_of_a_}}
\L{\LB{___basisnode_data_type,_to_NO._\*\1\CE{}}}
\L{\LB{}}
\L{\LB{\Proc{set\_tag\_to\_zero}\K{void}_\V{set\_tag\_to\_zero}(\K{struct}_\V{basisnode}_\*\*\V{BASIS},_\K{int}_\V{L},_}}
\L{\LB{}\Tab{16}{_____\K{int}_\V{LEVEL},_\K{int}_\V{BLOCK})}}
\L{\LB{\{}}
\L{\LB{__\K{int}_\V{PARENT},_\V{LEFTCHILD},_\V{RIGHTCHILD};}}
\L{\LB{__}}
\L{\LB{__\V{PARENT}_=_(\N{1}\<\<\V{LEVEL})_\-_\N{1}_+_\V{BLOCK};}}
\L{\LB{__\V{LEFTCHILD}_=_\N{2}\*\V{PARENT}_+_\N{1};}}
\L{\LB{__\V{RIGHTCHILD}_=_\V{LEFTCHILD}_+_\N{1};}}
\L{\LB{__\V{BASIS}[\V{LEFTCHILD}]\-\!\>\V{TAG}_=_\V{NO};}}
\L{\LB{__\V{BASIS}[\V{RIGHTCHILD}]\-\!\>\V{TAG}_=_\V{NO};}}
\L{\LB{__\K{if}_(\V{LEVEL}_\<_(\V{L}_\-_\N{1}))_\{}}
\L{\LB{____\V{set\_tag\_to\_zero}(\V{BASIS},_\V{L},_\V{LEVEL}+\N{1},_\N{2}\*\V{BLOCK});}}
\L{\LB{____\V{set\_tag\_to\_zero}(\V{BASIS},_\V{L},_\V{LEVEL}+\N{1},_\N{2}\*\V{BLOCK}_+_\N{1});}}
\L{\LB{__\}}}
\L{\LB{}}
\L{\LB{\}}}
\ProcCont{my\_qsort}\L{\LB{____}}
\L{\LB{\C{}\1\*_make\_samples\_of\_cosines:_Returns_an_array_of_samples_of_}}
\L{\LB{___a_specific_(cosine)function._}}
\L{\LB{}}
\L{\LB{___We_have:}}
\L{\LB{}}
\L{\LB{___SIGNAL_=_Array_of_length_N\*L}}
\L{\LB{}}
\L{\LB{___N_=_The_length_of_the_sampling.}}
\L{\LB{___}}
\L{\LB{___K_=_The_wavenumber_of_the_transmitted_wave(s).}}
\L{\LB{}}
\L{\LB{___A_=_Amplitude_(factor)_of_signal.}}
\L{\LB{___}}
\L{\LB{___R_=_The_distance_from_origo_to_the_interval_arc_of_sampling.}}
\L{\LB{___}}
\L{\LB{___P_=_The_number_of_samples_pr._wavelength._}}
\L{\LB{_}}
\L{\LB{___M_=_The_number_of_signal_transmitters}}
\L{\LB{}}
\L{\LB{___T_=_Starting_point_of_sampling._}}
\L{\LB{}}
\L{\LB{___W_=_Width_of_sector_containing_the_signal_emitter}}
\L{\LB{___as_a_fraction_of_2.0PI._\*\1\CE{}}}
\L{\LB{}}
\L{\LB{\Proc{make\_samples\_of\_cosine\_signal}\K{void}_\V{make\_samples\_of\_cosine\_signal}(\K{double}_\*\V{SIGNAL},_\K{int}_\V{N},_\K{double}_\V{K},_}}
\L{\LB{}\Tab{32}{___\K{double}_\V{A},_\K{double}_\V{R},_\K{int}_\V{P},_\K{int}_\V{M},_}}
\L{\LB{}\Tab{32}{___\K{double}_\V{W},_\K{double}_\V{T})}}
\L{\LB{\{__}}
\L{\LB{__\K{int}_\V{I},_\V{J},_\V{L};}}
\L{\LB{__}}
\L{\LB{__\K{double}_\V{RADIUS},_\V{ANGLE},_\V{SAMPL},_\V{AMP},_\V{ARG},_\V{X},_\V{Y},_\*\V{WORK};}}
\L{\LB{}}
\L{\LB{__\V{SAMPL}_=_\N{2.0}\*\V{PI}\1(\V{K}\*\V{P});}}
\L{\LB{__}}
\L{\LB{__\V{WORK}_=_(\K{double}_\*)\V{calloc}(\V{N},_\K{sizeof}(\K{double}));}}
\L{\LB{}}
\L{\LB{__\V{RADIUS}_=_\N{0};}}
\L{\LB{__\V{L}_=_\N{1};}}
\L{\LB{__}}
\L{\LB{__\K{for}_(\V{J}_=_\N{1};_\V{J}_\<=_\V{M};_\V{J}++)_\{}}
\L{\LB{____\K{while}_(\V{RADIUS}_\<_\N{1.0})__}}
\L{\LB{______\V{RADIUS}_=_\N{10.0}\*\V{my\_rand}();_}}
\L{\LB{_____}}
\L{\LB{____\V{ANGLE}_=_\N{2.0}\*\V{PI}\*(\V{L}\*\N{1.0}\1\V{M}_+_\V{my\_rand}()\*\V{W});_}}
\L{\LB{____\V{AMP}____=_\V{A};}}
\L{\LB{____\V{X}_=_\V{K}\*\V{RADIUS}\*\V{RADIUS}\1(\N{2.0}\*\V{R});}}
\L{\LB{____\V{Y}_=_\V{K}\*\V{RADIUS};}}
\L{\LB{____}}
\L{\LB{____\K{for}_(\V{I}_=_\N{0};_\V{I}_\<_\V{N};_\V{I}++)_\{}}
\L{\LB{______\V{ARG}_=_\V{T}_+_\V{I}\*\V{SAMPL};}}
\L{\LB{______\C{}\1\*__WORK[I]_+=_AMP\*(1.0\1R)\*cos(modf\_2PI(X_-_}}
\L{\LB{}\Tab{8}{__Y\*cos(ARG_-_ANGLE)));_\*\1\CE{}__}}
\L{\LB{______\V{WORK}[\V{I}]_+=_\V{AMP}\*\V{cos}(\V{modf\_2PI}(\V{X}_\-_\V{Y}\*\V{cos}(\V{ARG}_\-_\V{ANGLE})));__}}
\L{\LB{______\V{SIGNAL}[\V{I}]_=_\V{WORK}[\V{I}];}}
\L{\LB{____\}}}
\L{\LB{____\V{L}_+=_\N{1};}}
\L{\LB{____\V{RADIUS}_=_\N{0};}}
\L{\LB{__\}}}
\L{\LB{__\V{free}(\V{WORK});}}
\L{\LB{__}}
\L{\LB{\}}}
\ProcCont{my\_qsort}\L{\LB{}}
\L{\LB{\C{}\1\*_time\_frequency\_energy\_map:_Constructs_the_training_}}
\L{\LB{___signals_and_their_expansions_into_some_dictionary_of_orthonormal}}
\L{\LB{___wavelet_bases_or_local_trigonometric_bases._The_resulting_}}
\L{\LB{___\3time-frequency_energy_map\3_for_each_class,_and_the_basis_}}
\L{\LB{___coordinates_of_each_training_signal_are_stored_in_files._}}
\L{\LB{___}}
\L{\LB{___We_have:}}
\L{\LB{______}}
\L{\LB{___N_=_The_length_of_the_training_signals.(Power_of_2)}}
\L{\LB{___}}
\L{\LB{___L_=_The_depth_of_expansion_in_the_dictionary_tree_of_bases.}}
\L{\LB{___}}
\L{\LB{___NS_=_The_number_of_training_signals_in_each_class.}}
\L{\LB{___}}
\L{\LB{___makesignal_=_Function_that_constructs_a_signal_}}
\L{\LB{___by_some_sampling_procedure._}}
\L{\LB{}}
\L{\LB{___expand_=_Function_that_expands_the_signal_into_}}
\L{\LB{___a_dictionary_of_the_desired_orthonormal_basis_functions._\*\1\CE{}}}
\L{\LB{}}
\L{\LB{\Proc{time\_frequency\_energy\_map}\K{void}_\V{time\_frequency\_energy\_map}(\K{int}_\V{N},_\K{int}_\V{L},_\K{int}_\V{NS},_\V{FILE}_\*\V{TF1},_}}
\L{\LB{}\Tab{24}{_______\V{FILE}_\*\V{TF2},_\V{FILE}_\*\V{SIG1},_\V{FILE}_\*\V{SIG2})_}}
\L{\LB{\{__}}
\L{\LB{__\K{double}_\V{LOCENERGY\_1},_\V{LOCENERGY\_2},_\V{CLASSNORMP\_1},_}}
\L{\LB{____\V{CLASSNORMP\_2},\*\V{WORK\_1},_\*\V{WORK\_2},_}}
\L{\LB{____\*\V{GAMMA\_1},_\*\V{GAMMA\_2};}}
\L{\LB{__}}
\L{\LB{__\K{int}_\V{LEVEL},_\V{BLOCK},_\V{POS},_\V{K},_\V{J};}}
\L{\LB{____}}
\L{\LB{__\V{GAMMA\_1}_=_(\K{double}_\*)\V{calloc}(\V{N}\*(\V{L}+\N{1}),_\K{sizeof}(\K{double}));}}
\L{\LB{__\V{GAMMA\_2}_=_(\K{double}_\*)\V{calloc}(\V{N}\*(\V{L}+\N{1}),_\K{sizeof}(\K{double}));}}
\L{\LB{__\V{WORK\_1}_=_(\K{double}_\*)\V{calloc}(\V{N}\*(\V{L}+\N{1}),_\K{sizeof}(\K{double}));}}
\L{\LB{__\V{WORK\_2}_=_(\K{double}_\*)\V{calloc}(\V{N}\*(\V{L}+\N{1}),_\K{sizeof}(\K{double}));}}
\L{\LB{__\V{LOCENERGY\_1}__=_\N{0.0};}}
\L{\LB{__\V{LOCENERGY\_2}__=_\N{0.0};}}
\L{\LB{__\V{CLASSNORMP\_1}_=_\N{0.0};}}
\L{\LB{__\V{CLASSNORMP\_2}_=_\N{0.0};}}
\L{\LB{__\V{fprintf}(\V{SIG1},_\S{}\3\%d\2n\3\SE{},_\V{N});}}
\L{\LB{__\V{fprintf}(\V{SIG1},_\S{}\3\%d\2n\3\SE{},_\V{L});}}
\L{\LB{__\V{fprintf}(\V{SIG1},_\S{}\3\%d\2n\3\SE{},_\V{NS});}}
\L{\LB{__\V{fprintf}(\V{SIG2},_\S{}\3\%d\2n\3\SE{},_\V{N});}}
\L{\LB{__\V{fprintf}(\V{SIG2},_\S{}\3\%d\2n\3\SE{},_\V{L});}}
\L{\LB{__\V{fprintf}(\V{SIG2},_\S{}\3\%d\2n\3\SE{},_\V{NS});}}
\L{\LB{__}}
\L{\LB{__\K{for}_(\V{J}_=_\N{1};_\V{J}_\<=_\V{NS};_\V{J}++)_\{}}
\L{\LB{____\V{makesignal}(\V{WORK\_1},_\V{N},_\V{K\_1},_\V{A\_1},_\V{CLASS\_1});_}}
\L{\LB{____\V{makesignal}(\V{WORK\_2},_\V{N},_\V{K\_2},_\V{A\_2},_\V{CLASS\_2});_}}
\L{\LB{}}
\L{\LB{____\C{}\1\*_make\_samples\_of\_1(WORK\_1,_N\*(L+1));_}}
\L{\LB{____}}
\L{\LB{____make\_samples\_of\_oscillating\_fcn(WORK\_2,_N\*(L+1),_32);_\*\1\CE{}}}
\L{\LB{____}}
\L{\LB{____\V{CLASSNORMP\_1}_+=_\V{lp\_norm\_p}(\V{WORK\_1},_\V{NULL},_\V{N},_\N{2.0});_}}
\L{\LB{____\V{CLASSNORMP\_2}_+=_\V{lp\_norm\_p}(\V{WORK\_2},_\V{NULL},_\V{N},_\N{2.0});}}
\L{\LB{____\V{expand}(\V{WORK\_1},_\V{N},_\V{L});__}}
\L{\LB{____\V{expand}(\V{WORK\_2},_\V{N},_\V{L});_}}
\L{\LB{____}}
\L{\LB{____\K{for}_(\V{K}_=_\N{0};_\V{K}_\<_\V{N}\*(\V{L}+\N{1});_\V{K}++)_\{_}}
\L{\LB{______\V{fprintf}(\V{SIG1},_\S{}\3\%le\2n\3\SE{},_\V{WORK\_1}[\V{K}]);}}
\L{\LB{______\V{fprintf}(\V{SIG2},_\S{}\3\%le\2n\3\SE{},_\V{WORK\_2}[\V{K}]);}}
\L{\LB{____\}}}
\L{\LB{}}
\L{\LB{___\K{for}_(\V{LEVEL}_=_\N{0};_\V{LEVEL}_\<=_\V{L};_\V{LEVEL}++)_\{}}
\L{\LB{_____\K{for}_(\V{BLOCK}_=_\N{0};_\V{BLOCK}_\<_(\N{1}\<\<\V{LEVEL});_\V{BLOCK}++)_\{}}
\L{\LB{_______\K{for}_(\V{POS}_=_\N{0};_\V{POS}_\<_\V{abtlength}(\V{N},_\V{LEVEL});_\V{POS}++)_\{}}
\L{\LB{___}}
\L{\LB{}\Tab{8}{_\V{LOCENERGY\_1}_=_}}
\L{\LB{}\Tab{8}{___(\*(\V{WORK\_1}_+_\V{abtblock}(\V{N},\V{LEVEL},\V{BLOCK})_+_\V{POS}))\*}}
\L{\LB{}\Tab{8}{___(\*(\V{WORK\_1}_+_\V{abtblock}(\V{N},\V{LEVEL},\V{BLOCK})_+_\V{POS}));}}
\L{\LB{}\Tab{8}{_}}
\L{\LB{}\Tab{8}{_\V{LOCENERGY\_2}_=_}}
\L{\LB{}\Tab{8}{___(\*(\V{WORK\_2}_+_\V{abtblock}(\V{N},\V{LEVEL},\V{BLOCK})_+_\V{POS}))\*}}
\L{\LB{}\Tab{8}{___(\*(\V{WORK\_2}_+_\V{abtblock}(\V{N},\V{LEVEL},\V{BLOCK})_+_\V{POS}));}}
\L{\LB{}\Tab{8}{_}}
\L{\LB{}\Tab{8}{_\*(\V{GAMMA\_1}_+_\V{abtblock}(\V{N},\V{LEVEL},\V{BLOCK})_+_\V{POS})_}}
\L{\LB{}\Tab{8}{___+=_\V{LOCENERGY\_1};}}
\L{\LB{}\Tab{8}{_\*(\V{GAMMA\_2}_+_\V{abtblock}(\V{N},\V{LEVEL},\V{BLOCK})_+_\V{POS})_}}
\L{\LB{}\Tab{8}{___+=_\V{LOCENERGY\_2};}}
\L{\LB{}\Tab{8}{_}}
\L{\LB{}\Tab{8}{_\K{if}_(\V{J}_==_\V{NS})_\{__}}
\L{\LB{}\Tab{8}{___\*(\V{GAMMA\_1}_+_\V{abtblock}(\V{N},\V{LEVEL},\V{BLOCK})_+_\V{POS})_}}
\L{\LB{}\Tab{8}{_____\1=_\V{CLASSNORMP\_1};_}}
\L{\LB{}\Tab{8}{___\*(\V{GAMMA\_2}_+_\V{abtblock}(\V{N},\V{LEVEL},\V{BLOCK})_+_\V{POS})_}}
\L{\LB{}\Tab{8}{_____\1=_\V{CLASSNORMP\_2};}}
\L{\LB{}}
\L{\LB{}\Tab{8}{_\}}}
\L{\LB{_______\}}}
\L{\LB{_____\}}}
\L{\LB{___\}}}
\L{\LB{__\}}}
\L{\LB{__\V{fprintf}(\V{TF1},_\S{}\3\%d\2n\3\SE{},_\V{N});}}
\L{\LB{__\V{fprintf}(\V{TF1},_\S{}\3\%d\2n\3\SE{},_\V{L});}}
\L{\LB{__\V{fprintf}(\V{TF2},_\S{}\3\%d\2n\3\SE{},_\V{N});}}
\L{\LB{__\V{fprintf}(\V{TF2},_\S{}\3\%d\2n\3\SE{},_\V{L});}}
\L{\LB{__}}
\L{\LB{__\K{for}_(\V{K}_=_\N{0};_\V{K}_\<_\V{N}\*(\V{L}+\N{1});_\V{K}++)_\{__}}
\L{\LB{____\V{fprintf}(\V{TF1},_\S{}\3\%le\2n\3\SE{},_\V{GAMMA\_1}[\V{K}]);}}
\L{\LB{____\V{fprintf}(\V{TF2},_\S{}\3\%le\2n\3\SE{},_\V{GAMMA\_2}[\V{K}]);}}
\L{\LB{____}}
\L{\LB{__\}}}
\L{\LB{__\V{free}(\V{WORK\_1});}}
\L{\LB{__\V{free}(\V{WORK\_2});}}
\L{\LB{__\V{free}(\V{GAMMA\_1});}}
\L{\LB{__\V{free}(\V{GAMMA\_2});}}
\L{\LB{__\V{fclose}(\V{TF1});}}
\L{\LB{__\V{fclose}(\V{TF2});}}
\L{\LB{__\V{fclose}(\V{SIG1});}}
\L{\LB{__\V{fclose}(\V{SIG2});}}
\L{\LB{\}}}
\ProcCont{my\_qsort}\L{\LB{}}
\L{\LB{\C{}\1\*_ldb\_search\_and\_sort:_Do_a_LDB_search_and_sorting_procedure_}}
\L{\LB{___on_two_given_binary_tree_arrays_X_and_Y_of_time-frequency_}}
\L{\LB{___energy_maps_resulting_from_a_collection_of_wavelet_packet_}}
\L{\LB{___dictionarys_or_local_cosine\1sine_dictionarys_of_training_}}
\L{\LB{___signals_belonging_to_two_distinct_classes,_maximizing_some_}}
\L{\LB{___given_discriminant_measure_on_the_time-frequency_energy_maps.}}
\L{\LB{___}}
\L{\LB{___We_have:}}
\L{\LB{}}
\L{\LB{___TF\_1_=_Pointer_to_the_file_of_time-frequency_energy_}}
\L{\LB{___maps_resulting_from_the_set_of_expanded_training_signals_}}
\L{\LB{___belonging_to_a_certain_class,_arranged_level_by_level,_}}
\L{\LB{___block_by_block.}}
\L{\LB{___}}
\L{\LB{___TF\_2_=_Pointer_to_the_file_of__time-frequency_energy_}}
\L{\LB{___maps_resulting_from_the_set_of_expanded_training_signals_}}
\L{\LB{___belonging_to_some_other_class,_arranged_level_by_level,_}}
\L{\LB{___block_by_block.}}
\L{\LB{___}}
\L{\LB{___OUT_=_Pointer_to_file_of_output_data.}}
\L{\LB{}}
\L{\LB{___N_=_The_length_of_each_training_signal.}}
\L{\LB{___}}
\L{\LB{___L_=_The_depth_of_expansion_of_each_signal_in_}}
\L{\LB{___the_dictionarys._\*\1\CE{}}}
\L{\LB{}}
\L{\LB{\Proc{ldb\_search\_and\_sort}\K{void}_\V{ldb\_search\_and\_sort}(\V{FILE}_\*\V{TF\_1},_\V{FILE}_\*\V{TF\_2},_\V{FILE}_\*\V{OUT})}}
\L{\LB{\{}}
\L{\LB{}}
\L{\LB{__\K{int}_\V{LEVEL},_\V{BLOCK},_\V{BLOCKLEN},_\V{POS},_\V{PARENT},_}}
\L{\LB{____\V{NODE},_\V{CHILDNODE},_\V{N},_\V{L},_\V{K};}}
\L{\LB{__}}
\L{\LB{__\K{double}_\V{DELTA},_\V{DELTACHI},_\*\V{WORK\_1},_\*\V{WORK\_2};}}
\L{\LB{__}}
\L{\LB{__\K{struct}_\V{basisnode}_\*\*\V{BASIS},_\*\*\V{TEMP};}}
\L{\LB{}}
\L{\LB{__\V{fscanf}(\V{TF\_1},_\S{}\3\%d\3\SE{},_\&\V{N});}}
\L{\LB{__\V{fscanf}(\V{TF\_1},_\S{}\3\%d\3\SE{},_\&\V{L});}}
\L{\LB{__\V{fscanf}(\V{TF\_2},_\S{}\3\%d\3\SE{},_\&\V{N});__\C{}\1\*_Advancement_of_FILE_pointer_\*\1\CE{}}}
\L{\LB{__\V{fscanf}(\V{TF\_2},_\S{}\3\%d\3\SE{},_\&\V{L});__\C{}\1\*_Advancement_of_FILE_pointer_\*\1\CE{}}}
\L{\LB{}}
\L{\LB{__\V{WORK\_1}_=_(\K{double}_\*)\V{calloc}(\V{N}\*(\V{L}+\N{1}),_\K{sizeof}(\K{double}));__}}
\L{\LB{__\V{WORK\_2}_=_(\K{double}_\*)\V{calloc}(\V{N}\*(\V{L}+\N{1}),_\K{sizeof}(\K{double}));}}
\L{\LB{__\V{TEMP}_=_(\K{struct}_\V{basisnode}_\*\*)\V{calloc}(\N{1}\<\<(\V{L}+\N{1}),_}}
\L{\LB{}\Tab{32}{_____\K{sizeof}(\K{struct}_\V{basisnode}_\*));}}
\L{\LB{__\V{BASIS}_=_(\K{struct}_\V{basisnode}_\*\*)\V{calloc}(\V{N},_}}
\L{\LB{}\Tab{32}{______\K{sizeof}(\K{struct}_\V{basisnode}_\*));}}
\L{\LB{__}}
\L{\LB{__\K{for}_(\V{K}_=_\N{0};_\V{K}_\<_\V{N}\*(\V{L}+\N{1});_\V{K}++)_\{}}
\L{\LB{____\V{fscanf}(\V{TF\_1},_\S{}\3\%lf\3\SE{},_\&\V{WORK\_1}[\V{K}]);}}
\L{\LB{____\V{fscanf}(\V{TF\_2},_\S{}\3\%lf\3\SE{},_\&\V{WORK\_2}[\V{K}]);}}
\L{\LB{__\}__________}}
\L{\LB{__\V{BLOCKLEN}_=_\V{abtlength}(\V{N},\V{L});}}
\L{\LB{__}}
\L{\LB{__\K{for}_(\V{BLOCK}_=_\N{0};_\V{BLOCK}_\<_(\N{1}\<\<\V{L});_\V{BLOCK}++)_\{}}
\L{\LB{____\V{NODE}_=_(\N{1}\<\<\V{L})_\-_\N{1}_+_\V{BLOCK};}}
\L{\LB{____\V{DELTA}_=__\V{s\_measure}(\V{WORK\_1}_+_\V{L}\*\V{N}_+_\V{BLOCK}\*\V{BLOCKLEN},_}}
\L{\LB{}\Tab{16}{_______\V{WORK\_2}_+_\V{L}\*\V{N}_+_\V{BLOCK}\*\V{BLOCKLEN},_}}
\L{\LB{}\Tab{16}{_______\V{BLOCKLEN});}}
\L{\LB{____\V{TEMP}[\V{NODE}]_=_\V{makebasisnode}(\V{L},_\V{BLOCK},_\N{0},_\V{DELTA},_\V{YES});}}
\L{\LB{__\}}}
\L{\LB{__}}
\L{\LB{__\K{for}_(\V{LEVEL}_=_\V{L}\-\N{1};_\V{LEVEL}_\>=_\N{0};_\V{LEVEL}\-\-)_\{________}}
\L{\LB{____\V{BLOCKLEN}_=_\V{abtlength}(\V{N},\V{LEVEL});}}
\L{\LB{____}}
\L{\LB{____\K{for}_(\V{BLOCK}_=_\N{0};_\V{BLOCK}_\<_(\N{1}\<\<\V{LEVEL});_\V{BLOCK}++)_\{}}
\L{\LB{______\V{PARENT}_=_\V{abtblock}(\V{N},_\V{LEVEL},_\V{BLOCK});}}
\L{\LB{______\V{NODE}_=_(\N{1}\<\<\V{LEVEL})_\-_\N{1}_+_\V{BLOCK};}}
\L{\LB{______\V{CHILDNODE}_=_\N{2}\*\V{NODE}_+_\N{1};}}
\L{\LB{______\V{DELTA}_=_\V{s\_measure}(\V{WORK\_1}+\V{PARENT},_}}
\L{\LB{}\Tab{24}{\V{WORK\_2}+\V{PARENT},_\V{BLOCKLEN});}}
\L{\LB{______\V{DELTACHI}_=_\V{TEMP}[\V{CHILDNODE}]\-\!\>\V{COST}_+_}}
\L{\LB{}\Tab{8}{\V{TEMP}[\V{CHILDNODE}_+_\N{1}]\-\!\>\V{COST};}}
\L{\LB{______\K{if}_(\V{DELTA}_\>=_\V{DELTACHI})_\{}}
\L{\LB{}\Tab{8}{\V{set\_tag\_to\_zero}(\V{TEMP},_\V{L},_\V{LEVEL},_\V{BLOCK});}}
\L{\LB{}\Tab{8}{\V{TEMP}[\V{NODE}]_=_\V{makebasisnode}(\V{LEVEL},_\V{BLOCK},_\N{0},_}}
\L{\LB{}\Tab{32}{___\V{DELTA},_\V{YES});}}
\L{\LB{______\}}}
\L{\LB{______\K{else}}}
\L{\LB{}\Tab{8}{\V{TEMP}[\V{NODE}]_=_\V{makebasisnode}(\V{LEVEL},_\V{BLOCK},_\N{0},_}}
\L{\LB{}\Tab{32}{___\V{DELTACHI},_\V{NO});}}
\L{\LB{____\}}}
\L{\LB{__\}}}
\L{\LB{__\V{K}_=_\N{0};}}
\L{\LB{__}}
\L{\LB{__\K{for}_(\V{NODE}_=_\N{0};_\V{NODE}_\<_((\N{1}\<\<(\V{L}+\N{1}))\-\N{1});_\V{NODE}++)_\{}}
\L{\LB{____\K{if}_(\V{K}_\<_\V{N})_\{}}
\L{\LB{______\K{if}_(\V{TEMP}[\V{NODE}]\-\!\>\V{TAG}_==_\V{YES})_\{}}
\L{\LB{}\Tab{8}{\V{LEVEL}_=_\V{TEMP}[\V{NODE}]\-\!\>\V{LEVEL};}}
\L{\LB{}\Tab{8}{\V{BLOCK}_=_\V{TEMP}[\V{NODE}]\-\!\>\V{BLOCK};}}
\L{\LB{}\Tab{8}{\K{for}_(\V{POS}_=_\N{0};_\V{POS}_\<_(\V{N}\>\>\V{LEVEL});_\V{POS}++)_\{}}
\L{\LB{}\Tab{8}{__\V{DELTA}_=_\V{s\_measure}(\V{WORK\_1}_+_}}
\L{\LB{}\Tab{24}{____\V{abtblock}(\V{N},_\V{LEVEL},_\V{BLOCK})_+_\V{POS},_}}
\L{\LB{}\Tab{24}{____\V{WORK\_2}_+_}}
\L{\LB{}\Tab{24}{____\V{abtblock}(\V{N},_\V{LEVEL},_\V{BLOCK})_+_\V{POS},_\N{1});}}
\L{\LB{}\Tab{8}{__\V{BASIS}[\V{K}]_=_\V{makebasisnode}(\V{LEVEL},_\V{BLOCK},_\V{POS},_}}
\L{\LB{}\Tab{32}{___\V{DELTA},_\V{YES});}}
\L{\LB{}\Tab{8}{__\V{K}_+=_\N{1};}}
\L{\LB{}\Tab{8}{\}}}
\L{\LB{______\}}}
\L{\LB{____\}}}
\L{\LB{__\}}}
\L{\LB{__\V{my\_qsort}((\K{void}_\*\*)\V{BASIS},_\N{0},_\V{N}\-\N{1},_}}
\L{\LB{}\Tab{8}{___(\K{int}_(\*)(\K{void}_\*,_\K{void}_\*))\V{comp\_basisnode\_by\_cost});}}
\L{\LB{__\V{fprintf}(\V{OUT},_\S{}\3\%d\2n\%d\2n\3\SE{},_\V{N},_\V{L});}}
\L{\LB{__}}
\L{\LB{__\K{for}_(\V{K}_=_\N{0};_\V{K}_\<_\V{N};_\V{K}++)_}}
\L{\LB{____\V{fprintf}(\V{OUT},_\S{}\3\%d_\%d_\%d_\%le\2n\2n\3\SE{},_}}
\L{\LB{}\Tab{8}{____\V{BASIS}[\V{K}]\-\!\>\V{LEVEL},_\V{BASIS}[\V{K}]\-\!\>\V{BLOCK},}}
\L{\LB{}\Tab{8}{____\V{BASIS}[\V{K}]\-\!\>\V{POS},_\V{BASIS}[\V{K}]\-\!\>\V{COST});}}
\L{\LB{__}}
\L{\LB{__\V{free}(\V{WORK\_1});}}
\L{\LB{__\V{free}(\V{WORK\_2});}}
\L{\LB{__\V{freebasisnodes}(\V{TEMP},_\N{1}\<\<(\V{L}+\N{1}));_}}
\L{\LB{__\V{freebasisnodes}(\V{BASIS},_\V{N});}}
\L{\LB{__\V{fclose}(\V{TF\_1});}}
\L{\LB{__\V{fclose}(\V{TF\_2});}}
\L{\LB{__\V{fclose}(\V{OUT});}}
\L{\LB{__}}
\L{\LB{\}}}
\ProcCont{my\_qsort}\L{\LB{}}
\L{\LB{\C{}\1\*_make2bestcoordinates:_Make_a_set_of_signals_from_each_class}}
\L{\LB{___and_write_out_the_coordinates_of_the_two_most_discriminating}}
\L{\LB{___LDB_basis_elements.}}
\L{\LB{}}
\L{\LB{___We_have:}}
\L{\LB{}}
\L{\LB{___NS_=_The_number_of_signals_to_be_generated.}}
\L{\LB{___}}
\L{\LB{___LDB_=_Pointer_to_file_containing_the_LDB_data.}}
\L{\LB{}}
\L{\LB{___CL1C1_=_Pointer_to_file_containing_the_coordinates_of_the_}}
\L{\LB{___best_LDB_basis_element_of_CLASS_1.}}
\L{\LB{__}}
\L{\LB{___CL1C2_=_Pointer_to_file_containing_the_coordinates_of_the_}}
\L{\LB{___second_best_LDB_basis_element_of_CLASS_1.}}
\L{\LB{___}}
\L{\LB{___CL2C1_=_Pointer_to_file_containing_the_coordinates_of_the_}}
\L{\LB{___best_LDB_basis_element_of_CLASS_2.}}
\L{\LB{}}
\L{\LB{___CL2C2_=_Pointer_to_file_containing_the_coordinates_of_the_}}
\L{\LB{___second_best_LDB_basis_element_of_CLASS_2.___\*\1\CE{}}}
\L{\LB{}}
\L{\LB{\Proc{make2bestcoordinates}\K{void}_\V{make2bestcoordinates}(\K{int}_\V{NS},_\V{FILE}_\*\V{LDB},_\V{FILE}_\*\V{CL1C1},_}}
\L{\LB{}\Tab{24}{___\V{FILE}_\*\V{CL1C2},_\V{FILE}_\*\V{CL2C1},_\V{FILE}_\*\V{CL2C2})}}
\L{\LB{\{}}
\L{\LB{__\K{double}_\*\V{SIGNAL\_1},_\*\V{SIGNAL\_2},_\V{XCOORDCL1},_\V{YCOORDCL1},_}}
\L{\LB{____\V{XCOORDCL2},_\V{YCOORDCL2},_\V{NORMP1},_\V{NORMP2},_\V{DISC};_}}
\L{\LB{}}
\L{\LB{__\K{struct}_\V{basisnode}_\*\*\V{BASIS};}}
\L{\LB{}}
\L{\LB{__\K{int}_\V{N},_\V{L},_\V{LEVEL},_\V{BLOCK},_\V{POS},_\V{K},_\V{J};}}
\L{\LB{_____}}
\L{\LB{__\V{fscanf}(\V{LDB},_\S{}\3\%d\3\SE{},_\&\V{N});_}}
\L{\LB{__\V{fscanf}(\V{LDB},_\S{}\3\%d\3\SE{},_\&\V{L});__}}
\L{\LB{__\V{BASIS}_=_(\K{struct}_\V{basisnode}_\*\*)\V{calloc}(\N{2},_\K{sizeof}(\K{struct}_\V{basisnode}_\*));}}
\L{\LB{__\V{SIGNAL\_1}_=_(\K{double}_\*)\V{calloc}(\V{N}\*(\V{L}+\N{1}),_\K{sizeof}(\K{double}));}}
\L{\LB{__\V{SIGNAL\_2}_=_(\K{double}_\*)\V{calloc}(\V{N}\*(\V{L}+\N{1}),_\K{sizeof}(\K{double}));}}
\L{\LB{__}}
\L{\LB{__\K{for}_(\V{K}_=_\N{0};_\V{K}_\<_\N{2};_\V{K}++)_\{}}
\L{\LB{____\K{if}_(\V{K}_\<_\V{N})_\{}}
\L{\LB{______\V{fscanf}(\V{LDB},_\S{}\3\%d\3\SE{},_\&\V{LEVEL});}}
\L{\LB{______\V{fscanf}(\V{LDB},_\S{}\3\%d\3\SE{},_\&\V{BLOCK});}}
\L{\LB{______\V{fscanf}(\V{LDB},_\S{}\3\%d\3\SE{},_\&\V{POS});}}
\L{\LB{______\V{fscanf}(\V{LDB},_\S{}\3\%lf\3\SE{},_\&\V{DISC});}}
\L{\LB{______\V{BASIS}[\V{K}]_=_\V{makebasisnode}(\V{LEVEL},_\V{BLOCK},_\V{POS},_\V{DISC},_\N{1});}}
\L{\LB{____\}}}
\L{\LB{__\}}}
\L{\LB{__\V{fprintf}(\V{CL1C1},_\S{}\3\%d\2n\3\SE{},_\V{NS});}}
\L{\LB{__\V{fprintf}(\V{CL1C2},_\S{}\3\%d\2n\3\SE{},_\V{NS});}}
\L{\LB{__\V{fprintf}(\V{CL2C1},_\S{}\3\%d\2n\3\SE{},_\V{NS});}}
\L{\LB{__\V{fprintf}(\V{CL2C2},_\S{}\3\%d\2n\3\SE{},_\V{NS});}}
\L{\LB{__}}
\L{\LB{__\K{for}_(\V{J}_=_\N{0};_\V{J}_\<_\V{NS};_\V{J}++)_\{}}
\L{\LB{____\V{makesignal}(\V{SIGNAL\_1},_\V{N},_\V{K\_1},_\V{A\_1},_\V{CLASS\_1});_}}
\L{\LB{____\V{makesignal}(\V{SIGNAL\_2},_\V{N},_\V{K\_2},_\V{A\_2},_\V{CLASS\_2});_}}
\L{\LB{____\V{NORMP1}_=_\V{lp\_norm\_p}(\V{SIGNAL\_1},_\V{NULL},_\V{N},_\N{2.0});}}
\L{\LB{____\V{NORMP2}_=_\V{lp\_norm\_p}(\V{SIGNAL\_2},_\V{NULL},_\V{N},_\N{2.0});}}
\L{\LB{____\V{expand}(\V{SIGNAL\_1},_\V{N},_\V{L});__}}
\L{\LB{____\V{expand}(\V{SIGNAL\_2},_\V{N},_\V{L});__}}
\L{\LB{____}}
\L{\LB{____\V{XCOORDCL1}_=_\*(\V{SIGNAL\_1}_+_}}
\L{\LB{}\Tab{16}{__\V{abtblock}(\V{N},_\V{BASIS}[\N{0}]\-\!\>\V{LEVEL},_\V{BASIS}[\N{0}]\-\!\>\V{BLOCK})_+_}}
\L{\LB{}\Tab{16}{__\V{BASIS}[\N{0}]\-\!\>\V{POS});}}
\L{\LB{____\V{XCOORDCL2}_=_\*(\V{SIGNAL\_2}_+_}}
\L{\LB{}\Tab{16}{__\V{abtblock}(\V{N},_\V{BASIS}[\N{0}]\-\!\>\V{LEVEL},_\V{BASIS}[\N{0}]\-\!\>\V{BLOCK})_+_}}
\L{\LB{}\Tab{16}{__\V{BASIS}[\N{0}]\-\!\>\V{POS});}}
\L{\LB{____\V{YCOORDCL1}_=_\*(\V{SIGNAL\_1}_+_}}
\L{\LB{}\Tab{16}{__\V{abtblock}(\V{N},_\V{BASIS}[\N{1}]\-\!\>\V{LEVEL},_\V{BASIS}[\N{1}]\-\!\>\V{BLOCK})_+_}}
\L{\LB{}\Tab{16}{__\V{BASIS}[\N{1}]\-\!\>\V{POS});}}
\L{\LB{____\V{YCOORDCL2}_=_\*(\V{SIGNAL\_2}_+_}}
\L{\LB{}\Tab{16}{__\V{abtblock}(\V{N},_\V{BASIS}[\N{1}]\-\!\>\V{LEVEL},_\V{BASIS}[\N{1}]\-\!\>\V{BLOCK})_+_}}
\L{\LB{}\Tab{16}{__\V{BASIS}[\N{1}]\-\!\>\V{POS});_}}
\L{\LB{____\V{fprintf}(\V{CL1C1},_\S{}\3\%lf\2n\3\SE{},_\V{XCOORDCL1});}}
\L{\LB{____\V{fprintf}(\V{CL1C2},_\S{}\3\%lf\2n\3\SE{},_\V{YCOORDCL1});}}
\L{\LB{____\V{fprintf}(\V{CL2C1},_\S{}\3\%lf\2n\3\SE{},_\V{XCOORDCL2});}}
\L{\LB{____\V{fprintf}(\V{CL2C2},_\S{}\3\%lf\2n\3\SE{},_\V{YCOORDCL2});_____}}
\L{\LB{__\}}}
\L{\LB{__\V{freebasisnodes}(\V{BASIS},_\N{2});}}
\L{\LB{__\V{free}(\V{SIGNAL\_1});}}
\L{\LB{__\V{free}(\V{SIGNAL\_2});}}
\L{\LB{__\V{fclose}(\V{LDB});}}
\L{\LB{__\V{fclose}(\V{CL1C1});}}
\L{\LB{__\V{fclose}(\V{CL1C2});}}
\L{\LB{__\V{fclose}(\V{CL2C1});}}
\L{\LB{__\V{fclose}(\V{CL2C2});}}
\L{\LB{\}}}
\ProcCont{my\_qsort}\L{\LB{}}
\L{\LB{\C{}\1\*_Write_a_signal_of_each_class_to_a_file_for_plotting._\*\1\CE{}}}
\L{\LB{}}
\L{\LB{\Proc{write\_signals\_to\_file}\K{void}_\V{write\_signals\_to\_file}(\K{int}_\V{N})}}
\L{\LB{}}
\L{\LB{\{}}
\L{\LB{__\K{int}_\V{K};}}
\L{\LB{}}
\L{\LB{__\K{double}_\*\V{WORK1},_\*\V{WORK2};}}
\L{\LB{}}
\L{\LB{__\V{FILE}_\*\V{P1},_\*\V{P2};}}
\L{\LB{}}
\L{\LB{__\K{if}((_\V{P1}_=_\V{fopen}(\S{}\3\1users\1stud\1eirikf\1programs\1hovedfag\1idlplot\1}}
\L{\LB{__signalplot\1signal1\3\SE{},\S{}\3w\3\SE{}))}}
\L{\LB{_____==_\V{NULL})_\{_}}
\L{\LB{____\V{printf}(\S{}\3ERROR_in_opening_signal1\2n\3\SE{});}}
\L{\LB{____\V{exit}(\N{1});}}
\L{\LB{__\}}}
\L{\LB{__\K{if}((_\V{P2}_=_\V{fopen}(\S{}\3\1users\1stud\1eirikf\1programs\1hovedfag\1idlplot\1}}
\L{\LB{__signalplot\1signal2\3\SE{},\S{}\3w\3\SE{}))}}
\L{\LB{_____==_\V{NULL})_\{_}}
\L{\LB{____\V{printf}(\S{}\3ERROR_in_opening_F2\2n\3\SE{});}}
\L{\LB{____\V{exit}(\N{1});}}
\L{\LB{__\}}}
\L{\LB{__\V{WORK1}_=_(\K{double}_\*)\V{calloc}(\V{N},_\K{sizeof}(\K{double}));}}
\L{\LB{__\V{WORK2}_=_(\K{double}_\*)\V{calloc}(\V{N},_\K{sizeof}(\K{double}));}}
\L{\LB{___\V{makesignal}(\V{WORK1},_\V{N},_\V{K\_1},_\V{A\_1},_\V{CLASS\_1});_}}
\L{\LB{___\V{makesignal}(\V{WORK2},_\V{N},_\V{K\_2},_\V{A\_2},_\V{CLASS\_2});_}}
\L{\LB{___\V{fprintf}(\V{P1},_\S{}\3\%d\2n\3\SE{},_\V{N});}}
\L{\LB{___\V{fprintf}(\V{P2},_\S{}\3\%d\2n\3\SE{},_\V{N});}}
\L{\LB{__}}
\L{\LB{___\K{for}_(\V{K}_=_\N{0};_\V{K}_\<_\V{N};_\V{K}++)_\{_}}
\L{\LB{_____\V{fprintf}(\V{P1},_\S{}\3\%le\2n\3\SE{},_\V{WORK1}[\V{K}]);}}
\L{\LB{_____\V{fprintf}(\V{P2},_\S{}\3\%le\2n\3\SE{},_\V{WORK2}[\V{K}]);}}
\L{\LB{___\}___}}
\L{\LB{___\V{free}(\V{WORK1});}}
\L{\LB{___\V{free}(\V{WORK2});}}
\L{\LB{___\V{fclose}(\V{P1});}}
\L{\LB{___\V{fclose}(\V{P2});}}
\L{\LB{\}}}
\ProcCont{my\_qsort}\L{\LB{}}
\L{\LB{\C{}\1\*_Compute_the_best_decision_surface_of_our_two_classes_of_signals}}
\L{\LB{___that_can_be_obtained_from_the_shape_of_a_\3hyper-sphere\3.}}
\L{\LB{}}
\L{\LB{____M_is_the_maximum_number_of_coordinates_to_be_used_in_constructing_}}
\L{\LB{____the_\3hypersphere\3.}}
\L{\LB{}}
\L{\LB{____1\11\<\<POWER_is_the_step_size_in_the_search_grid.}}
\L{\LB{}}
\L{\LB{____LIMIT_is_the_upper_search_limit_for_the_best_radius.}}
\L{\LB{}}
\L{\LB{____LDB_is_pointer_to_file_containing_the_coordinates_}}
\L{\LB{____of_the_most_discriminating_basis_elements.}}
\L{\LB{}}
\L{\LB{____BESTSPHERE_is_pointer_to_file_of_output_data.}}
\L{\LB{____}}
\L{\LB{____OPTION_is_a_\3switch\3_to_reverse_the_definition_of_}}
\L{\LB{____the_classifier_if_needed._\*\1\CE{}}}
\L{\LB{____}}
\L{\LB{\Proc{find\_decision\_surface}\K{void}_\V{find\_decision\_surface}(\K{int}_\V{M},_\K{int}_\V{POWER},_\K{double}_\V{LIMIT},_}}
\L{\LB{}\Tab{24}{___\V{FILE}_\*\V{LDB},_\V{FILE}_\*\V{BESTSPHERE},_\K{int}_\V{OPTION})}}
\L{\LB{\{}}
\L{\LB{__\K{int}_\V{N},_\V{L},_\V{NS},_\V{LEVEL},_\V{BLOCK},_\V{POS},_\V{MISPLACED},_}}
\L{\LB{____\V{LEASTMISPLACED},_\V{BESTDIMENSION},_\V{COUNT},_\V{LENGTH},_\V{J},_\V{K},_\V{I};}}
\L{\LB{}}
\L{\LB{__\K{double}_\*\V{WORK1},_\*\V{WORK2},_\V{RADIUS},_\V{BESTRADIUS},_\V{DELTA},_}}
\L{\LB{____\V{STEP},_\*\V{TRAINPOINT1},_\*\V{TRAINPOINT2},_\V{DISTANCE1},_\V{DISTANCE2},_}}
\L{\LB{____\V{MISCLASSERROR};}}
\L{\LB{__}}
\L{\LB{__\K{struct}_\V{basisnode}_\*\*\V{LDBDATA};}}
\L{\LB{__}}
\L{\LB{__\V{FILE}_\*\V{SIG1},_\*\V{SIG2};}}
\L{\LB{__}}
\L{\LB{__\V{SIG1}_=_\V{fopen}(\S{}\3\1users\1stud\1eirikf\1work\1timefrequency\1sig1\3\SE{},\S{}\3r\3\SE{});}}
\L{\LB{__\V{SIG2}_=_\V{fopen}(\S{}\3\1users\1stud\1eirikf\1work\1timefrequency\1sig2\3\SE{},\S{}\3r\3\SE{});}}
\L{\LB{__\V{fscanf}(\V{LDB},_\S{}\3\%d\3\SE{},_\&\V{N});}}
\L{\LB{__\V{fscanf}(\V{LDB},_\S{}\3\%d\3\SE{},_\&\V{L});_}\Tab{32}{}}
\L{\LB{__\V{fscanf}(\V{SIG1},_\S{}\3\%d\3\SE{},_\&\V{N});}}
\L{\LB{__\V{fscanf}(\V{SIG1},_\S{}\3\%d\3\SE{},_\&\V{L});}}
\L{\LB{__\V{fscanf}(\V{SIG1},_\S{}\3\%d\3\SE{},_\&\V{NS});}}
\L{\LB{__\V{fscanf}(\V{SIG2},_\S{}\3\%d\3\SE{},_\&\V{N});}}
\L{\LB{__\V{fscanf}(\V{SIG2},_\S{}\3\%d\3\SE{},_\&\V{L});}}
\L{\LB{__\V{fscanf}(\V{SIG2},_\S{}\3\%d\3\SE{},_\&\V{NS});}\Tab{32}{}}
\L{\LB{__\V{LENGTH}_=_\V{NS}\*\V{N}\*(\V{L}+\N{1});}}
\L{\LB{__\V{LDBDATA}_=_(\K{struct}_\V{basisnode}_\*\*)\V{calloc}(\V{M},_\K{sizeof}(\K{struct}_\V{basisnode}_\*));}}
\L{\LB{__\V{WORK1}_=_(\K{double}_\*)\V{calloc}(\V{LENGTH},_\K{sizeof}(\K{double}));}}
\L{\LB{__\V{WORK2}_=_(\K{double}_\*)\V{calloc}(\V{LENGTH},_\K{sizeof}(\K{double}));}}
\L{\LB{__}}
\L{\LB{__\K{for}_(\V{I}_=_\N{0};_\V{I}_\<_\V{LENGTH};_\V{I}++)_\{}}
\L{\LB{____\V{fscanf}(\V{SIG1},_\S{}\3\%lf\3\SE{},_\&\V{WORK1}[\V{I}]);}}
\L{\LB{____\V{fscanf}(\V{SIG2},_\S{}\3\%lf\3\SE{},_\&\V{WORK2}[\V{I}]);}}
\L{\LB{__\}}}
\L{\LB{__\V{TRAINPOINT1}_=_(\K{double}_\*)\V{calloc}(\V{M},_\K{sizeof}(\K{double}));}}
\L{\LB{__\V{TRAINPOINT2}_=_(\K{double}_\*)\V{calloc}(\V{M},_\K{sizeof}(\K{double}));}}
\L{\LB{__}}
\L{\LB{__\K{for}_(\V{K}_=_\N{0};_\V{K}_\<_\V{M};_\V{K}++)_\{}}
\L{\LB{____\V{fscanf}(\V{LDB},_\S{}\3\%d\3\SE{},_\&\V{LEVEL});}}
\L{\LB{____\V{fscanf}(\V{LDB},_\S{}\3\%d\3\SE{},_\&\V{BLOCK});}}
\L{\LB{____\V{fscanf}(\V{LDB},_\S{}\3\%d\3\SE{},_\&\V{POS});}}
\L{\LB{____\V{fscanf}(\V{LDB},_\S{}\3\%lf\3\SE{},_\&\V{DELTA});}}
\L{\LB{____\V{LDBDATA}[\V{K}]_=_\V{makebasisnode}(\V{LEVEL},_\V{BLOCK},_\V{POS},_\V{DELTA},_\N{0});_}}
\L{\LB{__\}}}
\L{\LB{_}}
\L{\LB{__\V{STEP}_=_\N{1.0}\1(\N{1}\<\<\V{POWER});_}}
\L{\LB{__\V{LEASTMISPLACED}_=_\N{2}\*\V{NS};}}
\L{\LB{__}}
\L{\LB{__\K{for}_(\V{J}_=_\N{2};_\V{J}_\<=_\V{M};_\V{J}++)_\{}}
\L{\LB{____\V{COUNT}_=_\N{1};}}
\L{\LB{____\K{while}_(\V{STEP}\*\V{COUNT}_\<=_\V{LIMIT})_\{}}
\L{\LB{______\V{MISPLACED}_=_\N{0};}}
\L{\LB{______\V{RADIUS}_=_\V{STEP}\*\V{COUNT};_}}
\L{\LB{__}}
\L{\LB{______\K{for}_(\V{K}_=_\N{0};_\V{K}_\<_\V{NS};_\V{K}++)_\{}}
\L{\LB{}\Tab{8}{\V{DISTANCE1}_=_\N{0};}}
\L{\LB{}\Tab{8}{\V{DISTANCE2}_=_\N{0};}}
\L{\LB{}\Tab{8}{}}
\L{\LB{}\Tab{8}{\K{for}_(\V{I}_=_\N{0};_\V{I}_\<_\V{J};_\V{I}++)_\{}}
\L{\LB{}\Tab{8}{__\V{LEVEL}_=_\V{LDBDATA}[\V{I}]\-\!\>\V{LEVEL};}}
\L{\LB{}\Tab{8}{__\V{BLOCK}_=_\V{LDBDATA}[\V{I}]\-\!\>\V{BLOCK};}}
\L{\LB{}\Tab{8}{__\V{POS}___=_\V{LDBDATA}[\V{I}]\-\!\>\V{POS};}}
\L{\LB{}\Tab{8}{__\V{TRAINPOINT1}[\V{I}]_=_}}
\L{\LB{}\Tab{8}{____\*(\V{WORK1}_+_\V{K}\*\V{N}\*(\V{L}+\N{1})_+_}}
\L{\LB{}\Tab{8}{______\V{abtblock}(\V{N},\V{LEVEL},\V{BLOCK})_+_\V{POS});}}
\L{\LB{}\Tab{8}{__\V{TRAINPOINT2}[\V{I}]_=_}}
\L{\LB{}\Tab{8}{____\*(\V{WORK2}_+_\V{K}\*\V{N}\*(\V{L}+\N{1})_+_}}
\L{\LB{}\Tab{8}{______\V{abtblock}(\V{N},\V{LEVEL},\V{BLOCK})_+_\V{POS});}}
\L{\LB{}\Tab{8}{__\V{TRAINPOINT1}[\V{I}]_\*=_\V{TRAINPOINT1}[\V{I}];}}
\L{\LB{}\Tab{8}{__\V{TRAINPOINT2}[\V{I}]_\*=_\V{TRAINPOINT2}[\V{I}];}}
\L{\LB{}\Tab{8}{__\V{DISTANCE1}_+=_\V{TRAINPOINT1}[\V{I}];}}
\L{\LB{}\Tab{8}{__\V{DISTANCE2}_+=_\V{TRAINPOINT2}[\V{I}];}}
\L{\LB{}\Tab{8}{\}}}
\L{\LB{}\Tab{8}{\V{DISTANCE1}_=_\V{sqrt}(\V{DISTANCE1});}}
\L{\LB{}\Tab{8}{\V{DISTANCE2}_=_\V{sqrt}(\V{DISTANCE2});}}
\L{\LB{}\Tab{8}{\K{if}_(\V{OPTION}_==_\N{0})_\{}}
\L{\LB{}\Tab{8}{__\K{if}_(\V{DISTANCE1}_\<_\V{RADIUS})}}
\L{\LB{}\Tab{8}{____\V{MISPLACED}_+=_\N{1};}}
\L{\LB{}\Tab{8}{__\K{if}_(\V{DISTANCE2}_\>_\V{RADIUS})}}
\L{\LB{}\Tab{8}{____\V{MISPLACED}_+=_\N{1};}}
\L{\LB{}\Tab{8}{\}}}
\L{\LB{}\Tab{8}{\K{else}_\{_}}
\L{\LB{}\Tab{8}{__\K{if}_(\V{DISTANCE1}_\>_\V{RADIUS})}}
\L{\LB{}\Tab{8}{____\V{MISPLACED}_+=_\N{1};}}
\L{\LB{}\Tab{8}{__\K{if}_(\V{DISTANCE2}_\<_\V{RADIUS})}}
\L{\LB{}\Tab{8}{____\V{MISPLACED}_+=_\N{1};}}
\L{\LB{}\Tab{8}{\}}}
\L{\LB{______\}}}
\L{\LB{______\K{if}_(\V{MISPLACED}_\<_\V{LEASTMISPLACED})_\{}}
\L{\LB{}\Tab{8}{\V{LEASTMISPLACED}_=_\V{MISPLACED};}}
\L{\LB{}\Tab{8}{\V{BESTDIMENSION}_=_\V{J};}}
\L{\LB{}\Tab{8}{\V{BESTRADIUS}_=_\V{RADIUS};}}
\L{\LB{______\}}}
\L{\LB{______\V{COUNT}_+=_\N{1};}}
\L{\LB{____\}}}
\L{\LB{__\}}}
\L{\LB{__\V{MISCLASSERROR}_=_(\N{1.0}\*\V{LEASTMISPLACED})\1(\N{2.0}\*\V{NS});__}}
\L{\LB{__\V{fprintf}(\V{BESTSPHERE},_\S{}\3\%d\2n\3\SE{},_\V{BESTDIMENSION});}}
\L{\LB{__\V{fprintf}(\V{BESTSPHERE},_\S{}\3\%lf\2n\3\SE{},_\V{MISCLASSERROR});}}
\L{\LB{__\V{fprintf}(\V{BESTSPHERE},_\S{}\3\%lf\2n\3\SE{},_\V{BESTRADIUS});}}
\L{\LB{__\V{free}(\V{WORK1});}}
\L{\LB{__\V{free}(\V{WORK2});}}
\L{\LB{__\V{free}(\V{TRAINPOINT1});}}
\L{\LB{__\V{free}(\V{TRAINPOINT2});}}
\L{\LB{__\V{freebasisnodes}(\V{LDBDATA},_\V{M});}}
\L{\LB{__\V{fclose}(\V{LDB});}}
\L{\LB{__\V{fclose}(\V{BESTSPHERE});}}
\L{\LB{__\V{fclose}(\V{SIG1});}}
\L{\LB{__\V{fclose}(\V{SIG2});}}
\L{\LB{\}______}}
\ProcCont{my\_qsort}\L{\LB{}}
\L{\LB{\C{}\1\*_Given_a_decision_surface_to_separate_our_two_classes_of_}}
\L{\LB{___signals,_generate_a_collection_of_test_signals_and_measure_}}
\L{\LB{___the_misclassification_rate_on_this_collection_using_}}
\L{\LB{___the_given_surface._Also_compute_the_misclassification_rate_}}
\L{\LB{___on_the_collection_of_training_signals.}}
\L{\LB{}}
\L{\LB{___NS_is_the_number_of_test_signals_to_be_generated_from_each_class.}}
\L{\LB{___}}
\L{\LB{___BESTSPHERE_is_a_pointer_to_file_containing_the_data_}}
\L{\LB{___determining_the_best_classifier_using_a_hyperspherical_surface.}}
\L{\LB{}}
\L{\LB{___LDB_is_a_pointer_to_file_containing_the_coordinates_of_}}
\L{\LB{___the_most_discriminating_basis_functions.}}
\L{\LB{}}
\L{\LB{___DATA_is_a_pointer_to_the_output_file.}}
\L{\LB{}}
\L{\LB{___OPTION_is_a_\3switch\3_to_reverse_the_definition_of_the_}}
\L{\LB{___classifier._\*\1\CE{}}}
\L{\LB{}}
\L{\LB{\Proc{test\_decision\_surface}\K{void}_\V{test\_decision\_surface}(\K{int}_\V{NS},_\V{FILE}_\*\V{BESTSPHERE},_}}
\L{\LB{}\Tab{16}{______\V{FILE}_\*\V{LDB},_\V{FILE}_\*\V{DATA},_\K{int}_\V{OPTION})}}
\L{\LB{\{}}
\L{\LB{__\K{int}_\V{LEVEL},_\V{BLOCK},_\V{POS},_\V{BESTDIMENSION},_\V{MISPLACED},_}}
\L{\LB{____\V{N},_\V{L},_\V{J},_\V{K},_\V{I};}}
\L{\LB{}}
\L{\LB{__\K{double}_\*\V{WORK1},_\*\V{WORK2},_\V{BESTRADIUS},_\*\V{TESTPOINT1},_}}
\L{\LB{____\*\V{TESTPOINT2},_\V{MCRTEST},_\V{MCRTRAIN},_\V{DISTANCE1},_}}
\L{\LB{____\V{DISTANCE2},_\V{DELTA};}}
\L{\LB{}}
\L{\LB{__\K{struct}_\V{basisnode}_\*\*\V{LDBDATA};}}
\L{\LB{}}
\L{\LB{__\V{fscanf}(\V{BESTSPHERE},_\S{}\3\%d\3\SE{},_\&\V{BESTDIMENSION});}}
\L{\LB{__\V{fscanf}(\V{BESTSPHERE},_\S{}\3\%lf\3\SE{},_\&\V{MCRTRAIN});}}
\L{\LB{__\V{fscanf}(\V{BESTSPHERE},_\S{}\3\%lf\3\SE{},_\&\V{BESTRADIUS});}}
\L{\LB{__\V{LDBDATA}_=_(\K{struct}_\V{basisnode}_\*\*)\V{calloc}(\V{BESTDIMENSION},_}}
\L{\LB{}\Tab{40}{\K{sizeof}(\K{struct}_\V{basisnode}));}}
\L{\LB{__\V{fscanf}(\V{LDB},_\S{}\3\%d\3\SE{},_\&\V{N});}}
\L{\LB{__\V{fscanf}(\V{LDB},_\S{}\3\%d\3\SE{},_\&\V{L});}}
\L{\LB{__\V{WORK1}_=_(\K{double}_\*)\V{calloc}(\V{N}\*(\V{L}+\N{1}),_\K{sizeof}(\K{double}));}}
\L{\LB{__\V{WORK2}_=_(\K{double}_\*)\V{calloc}(\V{N}\*(\V{L}+\N{1}),_\K{sizeof}(\K{double}));_}}
\L{\LB{__\V{TESTPOINT1}_=_(\K{double}_\*)\V{calloc}(\V{BESTDIMENSION},_\K{sizeof}(\K{double}));}}
\L{\LB{__\V{TESTPOINT2}_=_(\K{double}_\*)\V{calloc}(\V{BESTDIMENSION},_\K{sizeof}(\K{double}));_}}
\L{\LB{__}}
\L{\LB{__\K{for}_(\V{K}_=_\N{0};_\V{K}_\<_\V{BESTDIMENSION};_\V{K}++)_\{}}
\L{\LB{____\V{fscanf}(\V{LDB},_\S{}\3\%d\3\SE{},_\&\V{LEVEL});}}
\L{\LB{____\V{fscanf}(\V{LDB},_\S{}\3\%d\3\SE{},_\&\V{BLOCK});}}
\L{\LB{____\V{fscanf}(\V{LDB},_\S{}\3\%d\3\SE{},_\&\V{POS});}}
\L{\LB{____\V{fscanf}(\V{LDB},_\S{}\3\%lf\3\SE{},_\&\V{DELTA});}}
\L{\LB{____\V{LDBDATA}[\V{K}]_=_\V{makebasisnode}(\V{LEVEL},_\V{BLOCK},_\V{POS},_\V{DELTA},_\N{0});_}}
\L{\LB{__\}}}
\L{\LB{__\V{MISPLACED}_=_\N{0};}}
\L{\LB{__}}
\L{\LB{__\K{for}_(\V{J}_=_\N{1};_\V{J}_\<=_\V{NS};_\V{J}++)_\{}}
\L{\LB{____\V{makesignal}(\V{WORK1},_\V{N},_\V{K\_1},_\V{A\_1},_\V{CLASS\_1});_}}
\L{\LB{____\V{makesignal}(\V{WORK2},_\V{N},_\V{K\_2},_\V{A\_2},_\V{CLASS\_2});_}}
\L{\LB{____\V{expand}(\V{WORK1},_\V{N},_\V{L});__}}
\L{\LB{____\V{expand}(\V{WORK2},_\V{N},_\V{L});_}}
\L{\LB{____\V{DISTANCE1}_=_\N{0};}}
\L{\LB{____\V{DISTANCE2}_=_\N{0};}}
\L{\LB{____}}
\L{\LB{____\K{for}_(\V{I}_=_\N{0};_\V{I}_\<_\V{BESTDIMENSION};_\V{I}++)_\{}}
\L{\LB{______\V{LEVEL}_=_\V{LDBDATA}[\V{I}]\-\!\>\V{LEVEL};}}
\L{\LB{______\V{BLOCK}_=_\V{LDBDATA}[\V{I}]\-\!\>\V{BLOCK};}}
\L{\LB{______\V{POS}___=_\V{LDBDATA}[\V{I}]\-\!\>\V{POS};}}
\L{\LB{______\V{TESTPOINT1}[\V{I}]_=_}}
\L{\LB{}\Tab{8}{\*(\V{WORK1}_+_\V{abtblock}(\V{N},\V{LEVEL},\V{BLOCK})_+_\V{POS});}}
\L{\LB{______\V{TESTPOINT2}[\V{I}]_=_}}
\L{\LB{}\Tab{8}{\*(\V{WORK2}_+_\V{abtblock}(\V{N},\V{LEVEL},\V{BLOCK})_+_\V{POS});}}
\L{\LB{______\V{TESTPOINT1}[\V{I}]_\*=_\V{TESTPOINT1}[\V{I}];}}
\L{\LB{______\V{TESTPOINT2}[\V{I}]_\*=_\V{TESTPOINT2}[\V{I}];}}
\L{\LB{______\V{DISTANCE1}_+=_\V{TESTPOINT1}[\V{I}];}}
\L{\LB{______\V{DISTANCE2}_+=_\V{TESTPOINT2}[\V{I}];}}
\L{\LB{____\}}}
\L{\LB{____\V{DISTANCE1}_=_\V{sqrt}(\V{DISTANCE1});}}
\L{\LB{____\V{DISTANCE2}_=_\V{sqrt}(\V{DISTANCE2});}}
\L{\LB{____\K{if}_(\V{OPTION}_==_\N{0})_\{}}
\L{\LB{______\K{if}_(\V{DISTANCE1}_\<_\V{BESTRADIUS})}}
\L{\LB{}\Tab{8}{\V{MISPLACED}_+=_\N{1};}}
\L{\LB{______\K{if}_(\V{DISTANCE2}_\>_\V{BESTRADIUS})}}
\L{\LB{}\Tab{8}{\V{MISPLACED}_+=_\N{1};}}
\L{\LB{____\}}}
\L{\LB{____\K{else}_\{_}}
\L{\LB{______\K{if}_(\V{DISTANCE1}_\>_\V{BESTRADIUS})}}
\L{\LB{}\Tab{8}{\V{MISPLACED}_+=_\N{1};}}
\L{\LB{______\K{if}_(\V{DISTANCE2}_\<_\V{BESTRADIUS})}}
\L{\LB{}\Tab{8}{\V{MISPLACED}_+=_\N{1};}}
\L{\LB{____\}}}
\L{\LB{__\}}}
\L{\LB{__\V{MCRTEST}_=_(\N{1.0}\*\V{MISPLACED})\1(\N{2.0}\*\V{NS});}}
\L{\LB{__\V{fprintf}(\V{DATA},_\S{}\3\%d\2n\3\SE{},_\V{NS});}}
\L{\LB{__\V{fprintf}(\V{DATA},_\S{}\3\%d\2n\3\SE{},_\V{BESTDIMENSION});}}
\L{\LB{__\V{fprintf}(\V{DATA},_\S{}\3\%lf\2n\3\SE{},_\V{MCRTRAIN});}}
\L{\LB{__\V{fprintf}(\V{DATA},_\S{}\3\%lf\2n\3\SE{},_\V{MCRTEST});}}
\L{\LB{__\V{fprintf}(\V{DATA},_\S{}\3\%lf\2n\3\SE{},_\V{BESTRADIUS});}}
\L{\LB{__\V{free}(\V{WORK1});}}
\L{\LB{__\V{free}(\V{WORK2});}}
\L{\LB{__\V{free}(\V{TESTPOINT1});}}
\L{\LB{__\V{free}(\V{TESTPOINT2});}}
\L{\LB{__\V{freebasisnodes}(\V{LDBDATA},_\V{BESTDIMENSION});}}
\L{\LB{__\V{fclose}(\V{LDB});}}
\L{\LB{__\V{fclose}(\V{BESTSPHERE});}}
\L{\LB{__\V{fclose}(\V{DATA});}}
\L{\LB{\}}}
\ProcCont{my\_qsort}

\File{myprograms.c}{1997}{Nov}{15}{18:54}{26126}

%% file: mapleprograms.tex

\File{mapleprograms.c}{1997}{Nov}{14}{13:19}{5324}
\L{\LB{\K{\#}_\V{Compute}_\V{the}_\V{SO2}(\V{R})_\V{elements}_\V{from}_\V{the}_\V{filter}_\V{coefficients}._\K{\#}}}
\L{\LB{}}
\L{\LB{\V{lagevinkler}_:=_}}
\L{\LB{___\V{proc}_(\V{k})_;_}}
\L{\LB{_______\K{for}_\V{i}_\V{from}_\N{1}_\V{to}_\V{N}__\K{do}_}}
\L{\LB{\Proc{x}___________\K{if}_\V{i}_\<_\V{N}_\V{then}__\V{x}(\V{i})_:=_\V{evalf}_(_\V{h}(\V{i}\-\N{1},\V{i}\-\N{1})_\1_\V{h}(\V{i}\-\N{1},\V{i})_)_\V{fi}_;_}}
\L{\LB{___________\K{for}_\V{j}_\V{from}_\V{i}_\V{to}_\N{2}_\*_\V{N}_\-_\V{i}_\K{do}_}}
\L{\LB{\Proc{type}_______________\K{if}_\V{type}_(_\V{i}_+_\V{j},_\V{odd}_)_\V{then}_}}
\L{\LB{\Proc{h}_____________\V{h}(\V{i},\V{j})_:=_\V{evalf}_(_\-_\V{x}(\V{i})_\*_\V{h}(\V{i}\-\N{1},\V{j}+\N{1})_+_\V{h}(\V{i}\-\N{1},\V{j})_)_}}
\L{\LB{________\K{else}_}}
\L{\LB{\Proc{h}_____________\V{h}(\V{i},\V{j})_:=_\V{evalf}_(_\V{x}(\V{i})_\*_\V{h}(\V{i}\-\N{1},\V{j}\-\N{1})_+_\V{h}(\V{i}\-\N{1},\V{j})_)_}}
\L{\LB{_______________\V{fi}_;_}}
\L{\LB{___________\V{od}_;}}
\L{\LB{\Proc{x}_______\K{if}_\V{i}_=_\V{N}_\V{then}_\V{x}(\V{N})_:=_\V{evalf}_(_\-_\V{h}(\V{N}\-\N{1},\V{N})_\1_\V{h}(\V{N}\-\N{1},\V{N}\-\N{1})_)_}}
\L{\LB{_______\V{fi}_;}}
\L{\LB{_______\V{od}_;}}
\L{\LB{\Proc{print}_______\K{for}_\V{i}__\V{from}_\N{1}_\V{to}_\V{N}_\K{do}_\V{print}_(_\V{i},_\V{x}(\V{i})_)_}}
\L{\LB{_______\V{od}_;_}}
\L{\LB{___\V{end}_;}}
\L{\LB{}}
\L{\LB{\K{\#}_\V{Compute}_\V{the}_\V{polynomials}_\V{P}\{\V{n}\}(\V{m})_\V{of}_\V{degree}_\V{n}_\V{that}_\V{result}_\V{from}_\V{lowpass}}}
\ProcCont{x}\L{\LB{\V{filtering}_\V{of}_\V{the}_\V{monomials}_\V{m}\5\{\V{n}\}._\K{\#}}}
\ProcCont{h}\L{\LB{}}
\L{\LB{}}
\L{\LB{\V{f}_:=_\V{proc}(_\V{j},\V{i},\V{m}_)_;}}
\L{\LB{}}
\L{\LB{\Proc{RETURN}\K{if}_\V{m}_=_\N{0}_\V{then}_\V{RETURN}(\N{1}\1\N{10})_\K{else}}}
\L{\LB{\Proc{RETURN}\V{RETURN}(_(_(\V{j}+\N{2}\*\V{i})\1\N{10})\5\V{m}_)_\V{fi}_;}}
\L{\LB{\V{end}_;}}
\L{\LB{}}
\L{\LB{}}
\L{\LB{}}
\L{\LB{\Proc{z}\V{z}(\N{1})_:=___.\N{4122865951}_;}}
\L{\LB{\Proc{z}\V{z}(\N{2})_:=_____\N{1.831178514}_;}}
\L{\LB{\Proc{z}\V{z}(\N{3})_:=___\-.\N{1058894200}_;}}
\L{\LB{\Proc{z}\V{z}(\N{4})_:=___\N{7.508378888}_;}}
\L{\LB{\Proc{z}\V{z}(\N{5})_:=____\-.\N{02083494630}_;}}
\L{\LB{\Proc{z}\V{z}(\N{6})_:=___\-.\N{04396989341}__;}}
\L{\LB{\Proc{z}\V{z}(\N{7})_:=___\-.\N{004543409641}__;}}
\L{\LB{\Proc{z}\V{z}(\N{8})_:=___\N{15.03636438}__;}}
\L{\LB{\Proc{z}\V{z}(\N{9})_:=___\-.\N{009120773147}__;}}
\L{\LB{\Proc{z}\V{z}(\N{10})_:=_\N{4.100416655}_;}}
\L{\LB{\Proc{z}\V{z}(\N{11})_:=_\N{15.61099604}_;}}
\L{\LB{\Proc{z}\V{z}(\N{12})_:=_\N{11.59905847}_;}}
\L{\LB{\Proc{z}\V{z}(\N{13})_:=_\N{37.56973541}_;}}
\L{\LB{\Proc{z}\V{z}(\N{14})_:=_\N{197.1316159}_;}}
\L{\LB{\Proc{z}\V{z}(\N{15})_:=_\-.\N{0004543371650}_;}}
\L{\LB{}}
\L{\LB{\V{N}_:=_\N{3}_;}}
\L{\LB{}}
\L{\LB{\Proc{proc}\V{javel}_:=_\V{proc}(_\V{l}_)_}}
\L{\LB{}}
\L{\LB{}}
\L{\LB{\K{for}_\V{m}_\V{from}_\N{0}_\V{to}_\N{6}_\K{do}}}
\L{\LB{\K{for}_\V{i}_\V{from}_\-\N{4}_\V{to}_\N{4}_\K{do}}}
\L{\LB{\Proc{a}\K{for}_\V{j}_\V{from}_\N{0}_\V{to}_\N{2}\*\V{N}\-\N{1}_\K{do}_\V{a}(\V{m},\N{0},\V{j},\V{i})_:=_\V{evalf}(_\V{f}(\V{j},\V{i},\V{m})_)_\V{od}_;}}
\L{\LB{\K{for}_\V{k}_\V{from}_\N{1}_\V{to}_\V{N}_\K{do}_}}
\L{\LB{\K{for}_\V{j}_\V{from}_\V{k}\-\N{1}_\V{to}_\N{2}\*\V{N}\-\V{k}_\K{do}_}}
\L{\LB{\Proc{type}\K{if}_\V{type}_(_\V{k}+\V{j},\V{odd}_)_\V{then}_}}
\L{\LB{\Proc{a}\V{a}(\V{m},\V{k},\V{j},\V{i})_:=_\V{evalf}(_\-\V{z}(\V{k})\*\V{a}(\V{m},\V{k}\-\N{1},\V{j}+\N{1},\V{i})+\V{a}(\V{m},\V{k}\-\N{1},\V{j},\V{i})_)_\K{else}}}
\L{\LB{\Proc{a}\V{a}(\V{m},\V{k},\V{j},\V{i})_:=_\V{evalf}(__\V{z}(\V{k})\*\V{a}(\V{m},\V{k}\-\N{1},\V{j}\-\N{1},\V{i})+\V{a}(\V{m},\V{k}\-\N{1},\V{j},\V{i})_)_}}
\L{\LB{\V{fi}_;}}
\L{\LB{\V{od}_;}}
\L{\LB{\V{od}_;}}
\L{\LB{\V{od}_;}}
\L{\LB{\V{od}_;}}
\L{\LB{}}
\L{\LB{}}
\L{\LB{\V{p0}_:=_\V{a}(\N{0},\V{N},\V{N}\-\N{1},\N{0})_;}}
\L{\LB{}}
\L{\LB{\V{p1}_:=_\V{n}_\-\!\>_\V{B1}\*\V{n}_+_\V{C1}_;}}
\L{\LB{}}
\L{\LB{\V{p2}_:=_\V{n}_\-\!\>_\V{B2}\*\V{n}\5\N{2}_+_\V{C2}\*\V{n}_+_\V{D2}_;}}
\L{\LB{}}
\L{\LB{\V{p3}_:=_\V{n}_\-\!\>_\V{B3}\*\V{n}\5\N{3}_+_\V{C3}\*\V{n}\5\N{2}_+\V{D3}\*\V{n}_+_\V{E3}_;}}
\L{\LB{}}
\L{\LB{\V{p4}_:=_\V{n}_\-\!\>_\V{B4}\*\V{n}\5\N{4}_+_\V{C4}\*\V{n}\5\N{3}_+_\V{D4}\*\V{n}\5\N{2}_+_\V{E4}\*\V{n}_+_\V{F4}_;}}
\L{\LB{}}
\L{\LB{\Proc{solve}\V{s}_:=_\V{solve}(_\{_\V{p1}(\V{N}\-\N{1})=\V{a}(\N{1},\V{N},\V{N}\-\N{1},\N{0}),}}
\L{\LB{\V{p1}(\V{N}+\N{1})=\V{a}(\N{1},\V{N},\V{N}\-\N{1},\N{1})_\},\{_\V{B1},\V{C1}_\}_)_;}}
\ProcCont{z}\L{\LB{\V{assign}(_\V{s}_)_;}}
\L{\LB{}}
\L{\LB{\Proc{solve}\V{s}_:=_\V{solve}(_\{_\V{p2}(\V{N}\-\N{1})=\V{a}(\N{2},\V{N},\V{N}\-\N{1},\N{0}),}}
\L{\LB{\V{p2}(\V{N}+\N{1})=\V{a}(\N{2},\V{N},\V{N}\-\N{1},\N{1}),}}
\L{\LB{\V{p2}(\V{N}+\N{3})=\V{a}(\N{2},\V{N},\V{N}\-\N{1},\N{2})_\},\{_\V{B2},\V{C2},\V{D2}_\}_)_;}}
\ProcCont{z}\L{\LB{\V{assign}(_\V{s}_)_;}}
\L{\LB{}}
\L{\LB{\Proc{solve}\V{s}_:=_\V{solve}(_\{_\V{p3}(\V{N}\-\N{3})=\V{a}(\N{3},\V{N},\V{N}\-\N{1},\-\N{1}),}}
\L{\LB{\V{p3}(\V{N}\-\N{1})=\V{a}(\N{3},\V{N},\V{N}\-\N{1},\N{0}),\V{p3}(\V{N}+\N{1})=\V{a}(\N{3},\V{N},\V{N}\-\N{1},\N{1}),}}
\L{\LB{\V{p3}(\V{N}+\N{3})=\V{a}(\N{3},\V{N},\V{N}\-\N{1},\N{2})_\},\{_\V{B3},\V{C3},\V{D3},\V{E3}_\}_)_;}}
\ProcCont{z}\L{\LB{\V{assign}(_\V{s}_)_;}}
\L{\LB{}}
\L{\LB{\Proc{solve}\V{s}_:=_\V{solve}(_\{_\V{p4}(\V{N}\-\N{5})=\V{a}(\N{4},\V{N},\V{N}\-\N{1},\-\N{2}),}}
\L{\LB{\V{p4}(\V{N}\-\N{3})=\V{a}(\N{4},\V{N},\V{N}\-\N{1},\-\N{1}),\V{p4}(\V{N}\-\N{1})=\V{a}(\N{4},\V{N},\V{N}\-\N{1},\N{0}),}}
\L{\LB{\V{p4}(\V{N}+\N{1})=\V{a}(\N{4},\V{N},\V{N}\-\N{1},\N{1}),\V{p4}(\V{N}+\N{3})=\V{a}(\N{4},\V{N},\V{N}\-\N{1},\N{2})_\},\{_\V{B4},\V{C4},\V{D4},\V{E4},\V{F4}_\}_)_;}}
\ProcCont{z}\L{\LB{\V{assign}(_\V{s}_)_;}}
\L{\LB{}}
\L{\LB{\V{print}(_\V{p0}_)_;}}
\L{\LB{}}
\L{\LB{\V{end}_;}}
\L{\LB{}}
\L{\LB{}}
\L{\LB{\K{\#}_\V{Compute}_\V{the}_\V{row}_\V{vectors}_\V{of}_\V{the}_\V{edge}_\V{matrices}_\V{by}_\V{claiming}_}}
\L{\LB{\V{some}_\V{vanishing}_\V{moments}_\V{on}_\V{the}_\V{highpass}_\V{edge}_\V{coefficients}_\V{and}}}
\L{\LB{\V{certain}_\V{polynomial}_\V{conditions}_\V{on}_\V{the}_\V{lowpass}_\V{edge}_\V{coefficients}_\V{ensuring}_}}
\L{\LB{\V{polynomials}_\V{of}_\V{degree}_\V{up}_\V{to}_\V{n}_\V{mapping}_\V{continuously}_\V{to}_\V{polynomials}_\V{of}_}}
\L{\LB{\V{the}_\V{same}_\V{degree}._\K{\#}}}
\L{\LB{}}
\L{\LB{\V{with}(_\V{linalg}_)_;}}
\L{\LB{}}
\L{\LB{\Proc{proc}\V{jafs}_:=_\V{proc}(_\V{o}_)_}}
\L{\LB{}}
\L{\LB{\V{v}(\N{0})_:=_\V{array}(_\N{1.\,.5},[\V{a}(\N{0},\N{0},\N{0},\N{0}),\V{a}(\N{0},\N{1},\N{1},\N{0}),\V{a}(\N{0},\N{2},\N{2},\N{0}),\V{a}(\N{0},\N{3},\N{3},\N{0}),\N{0}]_)_;}}
\L{\LB{\V{v}(\N{1})_:=_\V{array}(_\N{1.\,.5},[\V{a}(\N{1},\N{0},\N{0},\N{0}),\V{a}(\N{1},\N{1},\N{1},\N{0}),\V{a}(\N{1},\N{2},\N{2},\N{0}),\V{a}(\N{1},\N{3},\N{3},\N{0}),\N{0}]_)_;}}
\L{\LB{\V{v}(\N{2})_:=_\V{array}(_\N{1.\,.5},[\V{a}(\N{2},\N{0},\N{0},\N{0}),\V{a}(\N{2},\N{1},\N{1},\N{0}),\V{a}(\N{2},\N{2},\N{2},\N{0}),\V{a}(\N{2},\N{3},\N{3},\N{0}),\N{0}]_)_;}}
\L{\LB{\V{v}(\N{3})_:=_\V{array}(_\N{1.\,.5},[\V{a}(\N{3},\N{0},\N{0},\N{0}),\V{a}(\N{3},\N{1},\N{1},\N{0}),\V{a}(\N{3},\N{2},\N{2},\N{0}),\V{a}(\N{3},\N{3},\N{3},\N{0}),\N{0}]_)_;}}
\L{\LB{\Proc{v}\V{v}(\N{4})_:=_\V{array}(_\N{1.\,.5},[\V{a}(\N{4},\N{1},\N{0},\N{0}),\V{a}(\N{4},\N{2},\N{1},\N{0}),\V{a}(\N{4},\N{3},\N{2},\N{0}),\V{a}(\N{4},\N{4},\N{3},\N{0}),}}
\L{\LB{\V{a}(\N{4},\N{5},\N{4},\N{0})]_)_;}}
\L{\LB{}}
\L{\LB{}}
\L{\LB{\V{c0}_:=_\V{array}(_\N{1.\,.5},[\,]_)_;_}}
\L{\LB{\V{c0}[\N{5}]_:=_\N{0}_;}}
\L{\LB{}}
\L{\LB{\V{d0}_:=_\V{array}(_\N{1.\,.5},[\,]_)_;}}
\L{\LB{\V{d0}[\N{4}]_:=\N{0}_;}}
\L{\LB{\V{d0}[\N{5}]_:=_\N{0}_;}}
\L{\LB{}}
\L{\LB{\V{c1}_:=_\V{array}(_\N{1.\,.5},[\,]_)_;}}
\L{\LB{}}
\L{\LB{\V{d1}_:=_\V{array}(_\N{1.\,.5},[\,]_)_;}}
\L{\LB{\V{d1}[\N{5}]_:=_\N{0}_;}}
\L{\LB{}}
\L{\LB{\V{d2}_:=_\V{array}(_\N{1.\,.5},[\,]_)_;}}
\L{\LB{}}
\L{\LB{\Proc{solve}\V{s}_:=_\V{solve}(_\{_\V{dotprod}(_\V{c0},\V{v}(\N{0})_)_=_\V{p0},_\V{dotprod}(_\V{c0},\V{v}(\N{1})_)_=_\V{p1}(\N{1}),}}
\L{\LB{\V{dotprod}(_\V{c0},\V{v}(\N{2})_)_=_\V{p2}(\N{1}),_\V{dotprod}(_\V{c0},\V{v}(\N{3})_)_=_\V{p3}(\N{1})_\},_}}
\ProcCont{v}\L{\LB{\{_\V{c0}[\N{1}],\V{c0}[\N{2}],\V{c0}[\N{3}],\V{c0}[\N{4}]_\}_)_;}}
\ProcCont{proc}\L{\LB{\V{assign}(_\V{s}_)_;}}
\L{\LB{}}
\L{\LB{\Proc{solve}\V{s}_:=_\V{solve}(_\{_\V{dotprod}(_\V{d0},\V{v}(\N{1})_)_=_\N{0},_\V{dotprod}(_\V{d0},\V{v}(\N{0})_)_=_\N{0}_\},_}}
\ProcCont{proc}\L{\LB{\{_\V{d0}[\N{1}],\V{d0}[\N{2}]_\}_)_;}}
\ProcCont{z}\L{\LB{\V{assign}(_\V{s}_)_;}}
\L{\LB{}}
\L{\LB{\Proc{solve}\V{s}_:=_\V{solve}(_\{_\V{dotprod}(_\V{c1},\V{v}(\N{0})_)_=_\V{p0},_\V{dotprod}(_\V{c1},\V{v}(\N{1})_)_=_\V{p1}(\N{3}),}}
\L{\LB{\V{dotprod}(_\V{c1},\V{v}(\N{2})_)_=_\V{p2}(\N{3}),_\V{dotprod}(_\V{c1},\V{v}(\N{3})_)_=_\V{p3}(\N{3})_\},_}}
\ProcCont{z}\L{\LB{\{_\V{c1}[\N{1}],\V{c1}[\N{2}],\V{c1}[\N{3}],\V{c1}[\N{4}]_\}_)_;}}
\ProcCont{z}\L{\LB{\V{assign}(_\V{s}_)_;}}
\L{\LB{}}
\L{\LB{\Proc{solve}\V{s}_:=_\V{solve}(_\{_\V{dotprod}(_\V{d1},\V{v}(\N{1})_)_=_\N{0},_\V{dotprod}(_\V{d1},\V{v}(\N{2})_)_=_\N{0},}}
\L{\LB{\V{dotprod}(_\V{d1},\V{v}(\N{0})_)_=_\N{0}_\},_}}
\ProcCont{z}\L{\LB{\{_\V{d1}[\N{1}],\V{d1}[\N{2}],\V{d1}[\N{3}]_\}_)_;}}
\ProcCont{z}\L{\LB{\V{assign}(_\V{s}_)_;}}
\L{\LB{}}
\L{\LB{}}
\L{\LB{\V{i0}_:=_\-\N{10.5}_;}}
\L{\LB{\V{j0}_:=_\-\N{10.5}_;}}
\L{\LB{\V{l0}_:=_\-\N{10.5}_;}}
\L{\LB{}}
\L{\LB{\V{d0}[\N{3}]_:=_\V{i0}_;}}
\L{\LB{\V{c1}[\N{5}]_:=_\V{j0}_;}}
\L{\LB{\V{d1}[\N{4}]_:=_\V{l0}_;}}
\L{\LB{}}
\L{\LB{\Proc{evalm}\V{A}_:=_\V{evalm}(_\V{array}(_\N{1.\,.5},\N{1.\,.5},_[(\N{1},\N{1})=_\V{d0}[\N{1}],(\N{1},\N{2})=\V{d0}[\N{2}],(\N{1},\N{3})=\V{d0}[\N{3}],}}
\L{\LB{(\N{1},\N{4})=\V{d0}[\N{4}],(\N{1},\N{5})=\V{d0}[\N{5}],(\N{2},\N{1})=\V{c0}[\N{1}],(\N{2},\N{2})=\V{c0}[\N{2}],(\N{2},\N{3})=\V{c0}[\N{3}],(\N{2},\N{4})=\V{c0}[\N{4}],}}
\L{\LB{(\N{2},\N{5})=\V{c0}[\N{5}],(\N{3},\N{1})=\V{d1}[\N{1}],(\N{3},\N{2})=\V{d1}[\N{2}],(\N{3},\N{3})=\V{d1}[\N{3}],(\N{3},\N{4})=\V{d1}[\N{4}],(\N{3},\N{5})=\V{d1}[\N{5}],}}
\L{\LB{(\N{4},\N{1})=\V{c1}[\N{1}],(\N{4},\N{2})=\V{c1}[\N{2}],(\N{4},\N{3})=\V{c1}[\N{3}],(\N{4},\N{4})=\V{c1}[\N{4}],(\N{4},\N{5})=\V{c1}[\N{5}],(\N{5},\N{1})=\N{0},}}
\L{\LB{(\N{5},\N{2})=\N{0},(\N{5},\N{3})=\N{0},(\N{5},\N{4})=\N{0},(\N{5},\N{5})=\N{1}_]_)_)_;}}
\L{\LB{}}
\L{\LB{\V{B}_:=_\V{evalm}(_\V{inverse}(_\V{A}_)_)_;}}
\L{\LB{}}
\L{\LB{\V{z}_:=_\V{evalf}(_\V{norm}(_\V{B},\N{1}_)_+_\V{norm}(_\V{A},\N{1}_)_)_;}}
\L{\LB{}}
\L{\LB{\K{for}_\V{i}_\V{from}_\V{i0}_\V{by}_\N{1}_\V{to}_\N{10.5}_\K{do}}}
\L{\LB{\K{for}_\V{j}_\V{from}_\V{j0}_\V{by}_\N{1}_\V{to}_\N{10.5}_\K{do}}}
\L{\LB{\K{for}_\V{l}_\V{from}_\V{j0}_\V{by}_\N{1}_\V{to}_\N{10.5}_\K{do}}}
\L{\LB{}}
\L{\LB{\V{d0}[\N{3}]_:=_\V{i}_;}}
\L{\LB{\V{c1}[\N{5}]_:=_\V{j}_;}}
\L{\LB{\V{d1}[\N{4}]_:=_\V{l}_;}}
\L{\LB{}}
\L{\LB{\Proc{evalm}\V{A}_:=_\V{evalm}(_\V{array}(_\N{1.\,.5},\N{1.\,.5},_[(\N{1},\N{1})=_\V{d0}[\N{1}],(\N{1},\N{2})=\V{d0}[\N{2}],(\N{1},\N{3})=\V{d0}[\N{3}],}}
\L{\LB{(\N{1},\N{4})=\V{d0}[\N{4}],(\N{1},\N{5})=\V{d0}[\N{5}],(\N{2},\N{1})=\V{c0}[\N{1}],(\N{2},\N{2})=\V{c0}[\N{2}],(\N{2},\N{3})=\V{c0}[\N{3}],(\N{2},\N{4})=\V{c0}[\N{4}],}}
\L{\LB{(\N{2},\N{5})=\V{c0}[\N{5}],(\N{3},\N{1})=\V{d1}[\N{1}],(\N{3},\N{2})=\V{d1}[\N{2}],(\N{3},\N{3})=\V{d1}[\N{3}],(\N{3},\N{4})=\V{d1}[\N{4}],(\N{3},\N{5})=\V{d1}[\N{5}],}}
\L{\LB{(\N{4},\N{1})=\V{c1}[\N{1}],(\N{4},\N{2})=\V{c1}[\N{2}],(\N{4},\N{3})=\V{c1}[\N{3}],(\N{4},\N{4})=\V{c1}[\N{4}],(\N{4},\N{5})=\V{c1}[\N{5}],(\N{5},\N{1})=\N{0},}}
\L{\LB{(\N{5},\N{2})=\N{0},(\N{5},\N{3})=\N{0},(\N{5},\N{4})=\N{0},(\N{5},\N{5})=\N{1}_]_)_)_;}}
\L{\LB{}}
\L{\LB{\V{B}_:=_\V{evalm}(_\V{inverse}(_\V{A}_)_)_;}}
\L{\LB{}}
\L{\LB{\V{w}_:=_\V{evalf}(_\V{norm}(_\V{B},\N{1}_)_+_\V{norm}(_\V{A},\N{1}_)_)_;}}
\L{\LB{}}
\L{\LB{\K{if}_\V{w}_\<_\V{z}_\V{then}_\V{z}_:=_\V{w}_;}}
\L{\LB{\V{i0}_:=_\V{i}_;}}
\L{\LB{\V{j0}_:=_\V{j}_;}}
\L{\LB{\V{l0}_:=_\V{l}_;}}
\L{\LB{\V{fi}_;}}
\L{\LB{\V{od}_;}}
\L{\LB{\V{od}_;}}
\L{\LB{\V{od}_;}}
\L{\LB{}}
\L{\LB{\V{d0}[\N{3}]_:=_\V{i0}_;}}
\L{\LB{\V{c1}[\N{5}]_:=_\V{j0}_;}}
\L{\LB{\V{d1}[\N{4}]_:=_\V{l0}_;}}
\L{\LB{}}
\L{\LB{\Proc{evalm}\V{A}_:=_\V{evalm}(_\V{array}(_\N{1.\,.5},\N{1.\,.5},_[(\N{1},\N{1})=_\V{d0}[\N{1}],(\N{1},\N{2})=\V{d0}[\N{2}],(\N{1},\N{3})=\V{d0}[\N{3}],}}
\L{\LB{(\N{1},\N{4})=\V{d0}[\N{4}],(\N{1},\N{5})=\V{d0}[\N{5}],(\N{2},\N{1})=\V{c0}[\N{1}],(\N{2},\N{2})=\V{c0}[\N{2}],(\N{2},\N{3})=\V{c0}[\N{3}],(\N{2},\N{4})=\V{c0}[\N{4}],}}
\L{\LB{(\N{2},\N{5})=\V{c0}[\N{5}],(\N{3},\N{1})=\V{d1}[\N{1}],(\N{3},\N{2})=\V{d1}[\N{2}],(\N{3},\N{3})=\V{d1}[\N{3}],(\N{3},\N{4})=\V{d1}[\N{4}],(\N{3},\N{5})=\V{d1}[\N{5}],}}
\L{\LB{(\N{4},\N{1})=\V{c1}[\N{1}],(\N{4},\N{2})=\V{c1}[\N{2}],(\N{4},\N{3})=\V{c1}[\N{3}],(\N{4},\N{4})=\V{c1}[\N{4}],(\N{4},\N{5})=\V{c1}[\N{5}],(\N{5},\N{1})=\N{0},}}
\L{\LB{(\N{5},\N{2})=\N{0},(\N{5},\N{3})=\N{0},(\N{5},\N{4})=\N{0},(\N{5},\N{5})=\N{1}_]_)_)_;}}
\L{\LB{}}
\L{\LB{\V{B}_:=_\V{evalm}(_\V{inverse}(_\V{A}_)_)_;}}
\L{\LB{}}
\L{\LB{\V{print}(_\V{z}_)_;}}
\L{\LB{\V{print}(_\N{1},\V{norm}(_\V{A},\N{1}_)_)_;}}
\L{\LB{\V{print}(_\N{2},\V{norm}(_\V{B},\N{1}_)_)_;_}}
\L{\LB{\V{print}(_\N{1},\V{A}_)_;}}
\L{\LB{\V{print}(_\N{2},\V{B}_)_;}}
\L{\LB{\V{print}(_\V{evalm}(_\V{B}_\&\*_\V{A}_)_)_;}}
\L{\LB{}}
\L{\LB{}}
\L{\LB{}}
\L{\LB{}}
\L{\LB{\V{end}_;}}